\documentclass[a4paper,12pt,twoside,openright]{book}

\usepackage[T1]{fontenc}
\usepackage[utf8]{inputenc}
\usepackage[english,spanish]{babel}
\usepackage{emptypage}
\usepackage{graphicx,color}
\usepackage{amssymb,bm}
\usepackage{dcolumn}
\usepackage{comment}
\usepackage{slashed}
\usepackage{amsmath}
\usepackage{hyperref}
\usepackage[toc,page]{appendix}

\usepackage{booktabs}
\usepackage{fancyhdr}
\usepackage{etoolbox}
\usepackage{siunitx}  

\newif\ifinappendix

\newcommand{\bea}{\begin{eqnarray}}
\newcommand{\eea}{\end{eqnarray}}
\newcommand{\be}{\begin{equation}}
\newcommand{\ee}{\end{equation}}
\newcommand{\np}{{\bf p}}
\newcommand{\nP}{{\bf P}}
\newcommand{\nr}{{\bf r}}

\newcommand{\nh}{{\bf h}}
\newcommand{\nk}{{\bf k}}
\newcommand{\nl}{{\bf l}}
\newcommand{\na}{{\bf a}}
\newcommand{\nA}{{\bf A}}
\newcommand{\nB}{{\bf B}}
\newcommand{\nb}{{\bf b}}
\newcommand{\nq}{{\bf q}}

\newcommand{\nj}{{\bf j}}
\newcommand{\nJ}{{\bf J}}
\newcommand{\nL}{{\bf L}}
\newcommand{\nS}{{\bf S}}
\newcommand{\nR}{{\bf R}}
\newcommand{\nK}{{\bf K}}
\newcommand{\Qbar}{\not{\!Q}}

\newcommand{\kbar}{\not{\!k}}
\newcommand{\Pbar}{\not{\!P}}
\newcommand{\pbar}{\not{\!p}}

\newcommand{\ntau}{\mbox{\boldmath $\tau$}}
\newcommand{\nsigma}{\mbox{\boldmath $\sigma$}}
\newcommand{\neta}{\mbox{\boldmath $\eta$}}

\begin{document}

\begin{titlepage}
    \centering
    \vspace*{2cm}

    {\scshape\LARGE University of Granada \par}
    \vspace{1.0cm}

    {\scshape\Large PhD Program in Physics and Mathematics \par}
    \vspace{1.0cm}

     \includegraphics[width=0.3\textwidth]{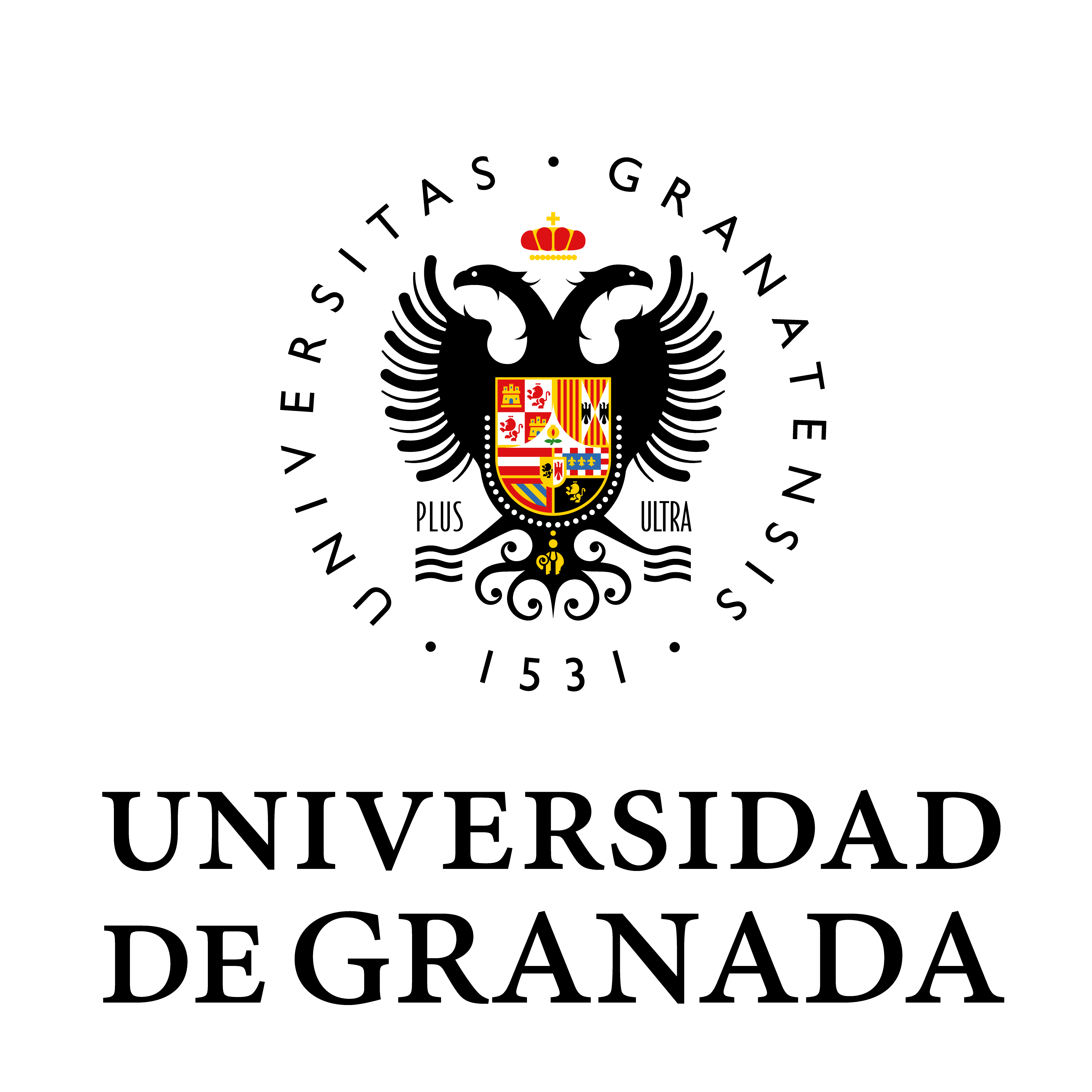}\par

    {\Huge\bfseries Meson-exchange currents and nuclear correlations
      in neutrino and electron scattering with nuclei\par}
    \vspace{1cm}

    {\Large Submitted by :\\
     Paloma Rodríguez Casalé\par}
     \vspace{0.5cm}
     {\Large In partial fulfillment of the requirements for the degree of
     Doctor of Philosophy in Physics\par}
     \vspace{0.5cm}
     {\Large Supervised by:\\
      Dr. José Enrique Amaro Soriano\par}
    \vfill

    \vspace{0.8cm}

    {\large Granada, 2025}

    \vspace{1cm}
\end{titlepage}

\begin{titlepage}
    \centering
    \vspace*{2cm}

    {\scshape\LARGE Universidad de Granada \par}
    \vspace{1.0cm}

    {\scshape\Large Programa de doctorado en Física y Matemáticas \par}
    \vspace{1.0cm}

     \includegraphics[width=0.3\textwidth]{ugr.png}\par

    {\Huge\bfseries Corrientes de intercambio de mesones y
      correlaciones nucleares en la dispersión de neutrinos y
      electrones con núcleos \par}
    \vspace{2cm}
  
    {\Large Memoria presentada por \\
      Paloma Rodríguez Casalé\par}
     \vspace{0.5cm}
    {\Large Para optar al título de Doctora en Física \par}
     \vspace{0.5cm}
    {\Large Bajo la dirección del Dr. \\
      José Enrique Amaro Soriano\par} \vfill

    \vspace{0.5cm}

    {\large Granada, 2025}

    \vspace{1cm}
\end{titlepage}

\thispagestyle{empty}
\mbox{}
\newpage

\thispagestyle{plain}
\selectlanguage{spanish}

\section*{\Huge Agradecimientos}
\vspace{2cm}
Estos cuatro años de tesis han sido una auténtica aventura, llena de
desafíos, aprendizaje y momentos inolvidables. Desde que empecé Física
siempre supe que quería doctorarme y ahora que lo he conseguido no
puedo sino estar orgullosa de mí misma.

Este trabajo refleja también el apoyo y la generosidad de quienes me
han acompañado durante estos años. Quiero agradecer especialmente a mi
director de tesis, Quique, por guiarme con paciencia y sabiduría, por
nuestros intensos debates llenos de ``esa será tu opinión'', que sin
duda me han hecho ser mejor física. Pero, sobre todo, por su fe en
mí. Porque sin ella nada de esto habría sido posible.

También le estoy agradecida a Nacho, por contratarme por primera vez,
dándome la oportunidad de empezar a crecer profesionalmente. A María,
Enrique, Victor, Pablo, Maria G, Maria B., Marco, Valerio y Arturo, por su
colaboración y apoyo durante el desarrollo de esta tesis.

A Alberto, por aguantarme cuando el cortisol se me disparaba por culpa
de la tesis y por escucharme sin descanso mientras practicaba mis
charlas una y otra vez. Gracias por estar a mi lado incluso en los
días más caóticos. 

Y finalmente, a mi familia, mi refugio. Gracias por cada palabra de
aliento que me inspiró a seguir mis metas, por enseñarme a no
rendirme y por recordarme siempre que los sueños no se cumplen solos y
que detrás de cada logro hay esfuerzo, constancia y dedicación.

\vspace{1cm}
\begin{center}
\end{center}

\vspace{1cm}

\newpage
\thispagestyle{empty}
\mbox{}
\newpage

\thispagestyle{plain}
\selectlanguage{spanish}

\section*{\Huge Resumen}
\vspace{2cm}
Esta tesis está dedicada al estudio de la dispersión de electrones y
neutrinos en núcleos, con especial atención a las corrientes de
intercambio de mesones (MEC) y a las correlaciones de corto alcance
(SRC) entre pares de nucleones, en procesos de emisión de una
partícula.

En el capítulo 2 se mejora el modelo de superscaling con masa efectiva
(SuSAM*) mediante la redefinición del tensor hadrónico de un nucleón
promediado sobre una distribución de Fermi, en lugar de la
extrapolación del gas de Fermi relativista utilizada hasta ahora. Esta
formulación elimina la inconsistencia de la negatividad en cinemáticas
alejadas del pico cuasielástico. La nueva definición del nucleón
efectivo permite extender el formalismo de superscaling para incluir
las MEC, que es el foco del capítulo 3.

En el capítulo 3 se hace un análisis de scaling de los datos del
$^{12}$C incorporando de manera explícita las MEC en el
single-nucleon, obteniendo una nueva función de scaling
fenomenológica. Con este modelo se estudia el efecto de las MEC en la
respuestas electromagnéticas, comparando con el gas de Fermi
relativista (RFG)y el modelo de campo medio relativista (RMF).

En el capítulo 4 se analiza en profundidad la interferencia entre las
corrientes a un cuerpo (OB) y a dos cuerpos en la respuesta transversal,
demostrando con distintos modelos, incluido el de la función
espectral, que dicha interferencia resulta siempre negativa en la
aproximación  de partícula independiente.

El capítulo 5 extiende el estudio de las MEC al caso de la dispersión
cuasielástica de neutrinos con corrientes cargadas (CCQE) en los
modelos RFG, RMF y SuSAM*. Se encuentra que la interferencia OB-MEC
reduce las respuestas transversales y la sección eficaz de neutrinos.

En el capítulo 6 se aborda el estudio de las correlaciones de corto
alcance en la función de onda de un par de nucleones en materia
nuclear. Para ello, se resuelve la ecuación de Bethe-Goldstone
utilizando el potencial nucleón--nucleón desarrollado por el grupo de
Granada, lo que permite calcular las componentes de alto momento.

Finalmente, en el capítulo 7 se estudia el efecto conjunto de las
MEC y SRC en aproximación de pares independientes,
que extiende el gas de Fermi incluyendo las componentes de alto
momento de los pares de nucleones sobre los que actúan las MEC. Se
encuentra que las SRC generan un aumento de la respuesta transversal,
en contraste con lo observado en modelos no correlacionados.

\newpage
\thispagestyle{empty}
\mbox{}
\newpage

\thispagestyle{plain}
\selectlanguage{english}

\section*{\Huge Abstract}
\vspace{2cm}

This thesis is dedicated to the study of electron and neutrino
scattering on nuclei, with special emphasis on meson-exchange currents
(MEC) and short-range correlations (SRC) between nucleon pairs in
one-particle emission processes.

In chapter 2, the superscaling model with effective mass (SuSAM*) is
improved by redefining the single-nucleon hadronic tensor, averaging
it over a Fermi distribution instead of using the previous
extrapolation from the relativistic Fermi gas. This formulation
removes the inconsistency associated with negative contributions in
kinematics far from the quasielastic peak. The new definition of the
effective nucleon allows extending the superscaling formalism to
include MEC, which is the focus of chapter 3.

In chapter 3, a scaling analysis of $^{12}$C data is performed
incorporating MEC explicitly at the single-nucleon level, leading to a
new phenomenological scaling function. Using this model, the effect of
MEC on electromagnetic responses is studied and compared with the
relativistic Fermi gas (RFG) and the relativistic mean field (RMF)
models.

Chapter 4 presents a detailed analysis of the interference between
one-body (OB) and two-body currents in the transverse response. It is
shown, using several approaches including the spectral function model,
that this interference is always negative within the
independent-particle approximation.

Chapter 5 extends the study of MEC to the quasielastic charged-current
neutrino scattering (CCQE) within the RFG, RMF, and SuSAM*
frameworks. It is found that OB-MEC interference reduces both the
transverse responses and the neutrino cross section.

Chapter 6 addresses short-range correlations in the wave function of a
nucleon pair in nuclear matter. The Bethe-Goldstone equation is solved
using the nucleon--nucleon potential developed by the Granada group,
allowing for the calculation of high-momentum components.

Finally, chapter 7 studies the combined effect of MEC and SRC in the
independent-pair approximation, which extends the Fermi gas by
including the high-momentum components of nucleon pairs affected by
MEC. It is found that SRC enhance the transverse response, in contrast
with what is observed in uncorrelated models.

\tableofcontents

\chapter{Introduction}

This Thesis is organized in two main parts. The first part is devoted
to the study of meson-exchange currents (MEC) in single-nucleon
emission reactions induced by electrons and neutrinos in uncorrelated
nuclear models.  The second part extends the analysis to systems that
include short–range correlations (SRC), exploring how their interplay
with MEC affects the transverse electromagnetic response, which serves
as a necessary step toward a more complete description of neutrino
interactions with nuclei.

This analysis is particularly relevant for accelerator-based neutrino
experiments, whose increasing precision demands equally accurate
theoretical descriptions of neutrino-nucleus interactions
\cite{Alv14,Mos16,Kat18,Ank17,Ben17,Alv18,Ama20,Ank22,Alv25}. Reliable
modeling of these processes is crucial to extract neutrino mixing
parameters and understand neutrino oscillations. Thus, providing
cross-section predictions for realistic nuclear targets over wide
energy ranges is essential
\cite{Benhar2015,Nie11,Martini2010}. Electron scattering serves as a
key benchmark: its abundant data \cite{archive,archive2} and the link
between electromagnetic and weak isovector currents allow nuclear
models validated for electrons to be adapted to neutrinos by including
an axial current. In the first part of this work, we first examine the case of electron
scattering in detail and then extend the analysis to neutrino
interactions.

In this context of lepton--nucleus scattering, the general objective
of this Thesis is to address two related but distinct problems. The
first concerns the role of MEC in the 1p1h nuclear response to
electron and neutrino probes, an effect that has not yet been
systematically evaluated in the case of neutrino scattering. The
second focuses on the influence of SRC on the transverse
electromagnetic response in single--particle emission, which are
essential to understand the transverse enhancement observed in
electron reactions. In this introduction, we develop the context in
which this research has been carried out, including the main
antecedents, previous models, and calculations relevant to the present
work.

Inclusive lepton--nucleus cross sections vary with the energy
transferred to the nucleus, which determines the dominant reaction
mechanisms across different excitation regions. From elastic and
collective excitations at low energies to quasielastic (QE)
scattering, resonance production, and deep-inelastic scattering at
higher energies, each regime involves distinct nuclear processes. This
thesis focuses on the quasielastic region, specifically on
1p-1h excitations, excluding two-nucleon
emission, pion production, and other inelastic channels. Accurate
treatment of this domain requires reliable theoretical models and
many-body approaches capable of describing the underlying nuclear
dynamics \cite{Ank17,Ben17,Pan24,Sajj23,Morf12,coloma,Ama20}.

One of the simplest and most widely used approaches to describe
quasielastic lepton-nucleus scattering is the relativistic Fermi gas
(RFG) model \cite{Moni69,Moni71,Moni75}. Although it provides a useful
framework for interpreting data, the RFG neglects finite-size effects,
nuclear interactions, and correlations. The Relativistic Mean Field
(RMF) model \cite{Wal74,Ser86,Ros80} improves upon these limitations
by including scalar and vector potentials that yield an effective
nucleon mass $m_N^*$ and a more realistic relativistic dynamics. Since
the RFG assumes a sharp Fermi surface, it also fails to reproduce the
the nuclear momentum distribution \cite{Ama05,Meg16b,Ama18}.

To achieve a more realistic description of lepton--nucleus scattering,
models beyond the RFG have been developed. These include
independent--particle approaches such as the plane--wave impulse
approximation (PWIA) and spectral function (SF) methods \cite{Ben94},
Dirac Equation Based (DEB) models \cite{Hor91,Udi95}, the shell model
with Woods--Saxon potentials \cite{Cap91,Ama94a,Ama94}, and Random
Phase Approximation (RPA) frameworks, as well as microscopic
approaches like Green's Function Monte Carlo (GFMC) calculations for
light nuclei \cite{Lov16a} and coupled--cluster methods
\cite{Fab97}. Local Fermi Gas (LFG) models, such as those developed by
the Valencia group \cite{Nie04} and by the Lyon group \cite{Mar09},
represent sophisticated many--body frameworks, 
including RPA correlations, effective interactions, and
final--state interactions (FSI), and are widely used for describing
neutrino-nucleus scattering. 
Some of these models have been revisited in this thesis,
specifically the PWIA, SF, DEB, and shell model with Woods–Saxon
potential in chapter 4.

In addition to microscopic and independent-particle approaches,
phenomenological scaling models provide an alternative description of
the nuclear response. These frameworks, such as the super-scaling
approach SuSA, SuSA-v2, and SuSAM* \cite{Ama05,Meg16b,Ama18}
factorize the response into a single-nucleon contribution multiplied
by an universal scaling function fitted to electron scattering
data. SuSAM* further incorporates medium effects via the relativistic
effective nucleon mass inspired by the RMF.  In chapter~2, this Thesis
employs an improved version of the SuSAM* model, replacing the
single-nucleon prefactor previously used in the SuSA approach, which
was based on a simple extrapolation of the Fermi gas average. We
demonstrate that this extrapolation leads to problems for extreme
kinematics outside the Fermi gas region, producing nonphysical negative
results in certain cases. This issue is resolved by introducing an
alternative averaging procedure, replacing the sharp RFG momentum
distribution with a smoother function of Fermi kind.
 This new definition of the
single-nucleon contribution also provides a natural framework to
consistently include MEC, as developed in chapter~3.

One of the main objectives of this Thesis is to study in detail the
effect of MEC in the 1p-1h channel for both electron and neutrino
scattering. MEC naturally arise from the nucleon--nucleon interaction
\cite{Ris89}, with pion-exchange processes providing the dominant
contribution. While MEC are known to play a significant role in the
two-particle-two-hole (2p--2h) channel in both electron and neutrino
scattering \cite{Mar09,Ama11,Nie11}, they are typically neglected in
the 1p--1h quasielastic response, particularly in the case of neutrino
scattering. Previous studies have analyzed MEC effects in electron
scattering on medium nuclei using various models, such as the Fermi
gas, RFG, and shell model \cite{Koh81,Alb90,Ama94,Ama03}. Chapter~3
focuses on the consistent inclusion of MEC within the superscaling
formalism, a novel development made possible by the approach
introduced in chapter~2, which averages the single-nucleon response
over a diffused Fermi surface.

In addition to the improved SuSAM*+MEC model developed in chapter~3,
independent-particle models, including the RFG, mean-field, and
spectral function approaches, typically predict that MEC reduce the
transverse response due to partial cancellations among the seagull,
$\Delta$, and pion-in-flight contributions. This indicates that MEC
alone cannot account for the enhancement observed in experimental data
relative to theoretical predictions based solely on one-body currents
\cite{Ama94,Jou96,Bod22}. However, some studies report results that
disagree with these models \cite{Van95,Fra23,Lov23}. These
discrepancies motivate the systematic analysis presented in chapter~4,
where the interference between one-body and two-body currents is
studied in detail using various nuclear models. In particular, in the
low-momentum limit, we rigorously demonstrate that within the
non-relativistic Fermi gas model, the interference of the $\Delta$ and
pion-in-flight currents with the one-body current is negative, a
result that is consistent with all models examined in the chapter.

As final step of part 1 of this Thesis, in chapter~5 we address the
case of neutrino scattering. Building on the formalism developed in
the previous chapters, chapter~5 presents the first comprehensive
analysis of charged-current quasielastic (CCQE) neutrino scattering
that includes MEC in the one-nucleon knockout channel. As previously
discussed, the existing neutrino studies have incorporated MEC only in
the 2p-2h channel \cite{Nie11,Mar09,Ama11}, which typically accounts
for about 15--20\% of the total cross section, depending on the
kinematics and on the specific model used
\cite{Meg16b,Ben15,Gra13,Sob20,Meg17,Gon23}. This is particularly
relevant since precision in neutrino cross-section modeling is
essential for the accurate determination of oscillation parameters. 
Yet it is also necessary to quantify the effect of MEC in the
1p-1h channel. In this chapter, we perform this analysis for the
first time using simple models (RFG, RMF, and SuSAM*). All five response
functions relevant to neutrino scattering are computed, including
contributions from both one- and two-body axial currents. Special
attention is given to the axial current with $\Delta$ excitation,
which dominates weak MEC, while the seagull axial contribution is
found to be negligible—an important difference compared to electron
scattering. Finally, the total cross section is also calculated and
compared with neutrino experimental data, allowing us to evaluate the
relative importance of MEC in the 1p-1h channel compared to the 2p-2h
contribution.

In the second part of this Thesis, we go beyond uncorrelated nuclear
models and address the case of correlated nuclei. The main motivation,
as mentioned above, is to investigate the long-standing problem of the
transverse electromagnetic response enhancement, which is the central
focus of this part.  Meson--exchange currents are widely regarded as
the primary mechanism behind the transverse enhancement, a view
supported by \textit{ab initio} calculations in light
nuclei~\cite{Car02,Lov16}.  However, as demonstrated in the first part
of this thesis, single--particle approaches fail to reproduce this
enhancement.  Instead, they generally predict a negative interference
between one-- and two--body currents, mainly due to the dominant
destructive contribution of the $\Delta$ current at higher momentum
transfers.  An important element absent in these independent-particle
frameworks and potentially responsible for the transverse enhancement,
as suggested by Fabrocini \cite{Fab97} and Leidemann and Orlandini
\cite{Lei13}, are the short-range correlations in the nuclear wave
function.  Since MEC are two–body operators, it is natural to expect
them to be sensitive to the short–distance structure of correlated
nucleon pairs, and therefore to investigate whether the enhancement of
the 1p1h transverse response may be connected to such
correlations. Studying these correlations and quantifying their effect
on the 1p1h transverse response is the main goal of chapters 6 and 7,
which we address within the simplest possible framework: the
independent–pair approximation in nuclear matter.

Short-range correlations, studied in chapter 6, arise when two
nucleons interact strongly at short distances within the nuclear
medium. These correlations generate high-momentum components in the
nuclear wave function that play an essential role in understanding the
transverse enhancement discussed above. The interest in SRC has grown
substantially in recent years, largely driven by the development of
high-energy electron scattering experiments. Facilities such as the
Continuous Electron Beam Accelerator Facility (CEBAF) at Jefferson Lab
\cite{Wal95} have played a central role in this progress, providing
high-intensity, high-resolution electron beams that enable
unprecedented exploration of nuclear structure. These experimental
advances have opened the way to systematic studies of SRC and have
spurred renewed theoretical efforts to understand their role in the
nuclear wave function and in various nuclear processes
\cite{sube,Hen14}.

From a theoretical perspective, SRC play a crucial role in nuclear
physics, spanning from fundamental studies to applied contexts. They
significantly affect the properties of nuclear
matter~\cite{Bethe:1971xm, Jeukenne:1976uy, Ramos:1989hqs,
  Dewulf:2003nj, Vonderfecht:1991zz}, contribute to the high-momentum
components of nuclear wave functions~\cite{Fantoni:1984zz,
  Muther:1995zz, Wiringa:2013ala, Benhar:1986jha, VanOrden:1979mt,
  Sargsian:2012sm}, and have important implications in nuclear
astrophysics, including the equation of state and evolution of neutron
stars~\cite{Riffert:1996jf, Frankfurt:2008zv, Mukherjee:2008un,
  Shen:2011kr, Ropke:2014fia}. They also impact calculations of
symmetry energy and pairing gaps in nuclear and neutron
matter~\cite{Hen:2014yfa, Ding:2016oxp, Rios:2017muz}, modeling of
relativistic heavy-ion collisions~\cite{Broniowski:2010jd}, and
nuclear matrix elements relevant for neutrinoless double-beta
decay~\cite{Simkovic:2009pp, Kortelainen:2007rn}. Moreover, SRCs are
essential in describing electron scattering reactions, including
$(e,e^\prime)$, $(e,e^\prime N)$, and $(e,e^\prime NN)$
processes~\cite{Frankfurt:1993sp, Weinstein:2010rt, Giusti:1999sv,
  Ryckebusch:1995usx, Colle:2015ena}. More recently, the universality
of N-N SRCs has been explored in connection with factorization
properties of nuclear wave functions and momentum distributions, as
well as with nuclear contacts~\cite{Tan:2008a, Tan:2008b,Tan:2008c,
  Alvioli:2011aa, Alvioli:2013qyz, Weiss:2015mba, Alvioli:2016wwp,
  Weiss:2016obx, Weiss:2017huz}.

The main approaches traditionally used to address SRC fall into two
categories. The first employs Jastrow correlation functions with
appropriately tailored short- and long-range behaviors, applied to
Slater determinants of single-particle wave functions within
variational frameworks~\cite{Ryckebusch:1995usx,Jastrow:1955zz,
  Fantoni:1974jv, Fantoni:1975a, Guardiola:1980ma, Guardiola:1981ujn,
  Benhar:1991iw, Stoitsov:1993zz, Benhar:1994hw, Guardiola:1996dq,
  Bishop:1998za, Vanhalst:2012ur}. The second is the Brueckner theory
of nuclear matter~\cite{Brueckner:1958zz, Brueckner:1954zz,
  Brueckner:1955zze}, based on solving the Bethe--Goldstone (BG)
equation~\cite{Bethe:1956zz, Goldstone:1957zz, Dahll:1969hmo} or the
effective interaction represented by the G-matrix
formalism~\cite{Kohler:1961a, Haftel:1970zz, Jeukenne:1974zz,
  Nakayama:1984xgi, Hosaka:1985xwy, Nakayama:1987czg,
  Boersma:1993yy}. Several other powerful methods have also been
developed. Similarity renormalization group (SRG) techniques provide
phase-equivalent potentials that soften the short-range interaction,
thereby circumventing the complications associated with the hard
core~\cite{Bogner:2006pc, Timoteo:2011tt, Neff:2015xda}. In addition,
\textit{ab initio} variational Monte Carlo approaches can solve the
non-relativistic many-body problem exactly for light nuclei, given a
specific nucleon-nucleon interaction
\cite{Carlson:1993zz,Forest:1995zz, Quaglioni:2009mn, Hagen:2010gd,
  Leidemann:2012hr,Barrett:2013nh}.

The goal of chapter 6 is to construct the wave function of a
correlated nucleon pair and analyze its properties, providing the
foundation for the study of combined MEC and SRC effects in chapter
7. Within the independent-pair approximation~\cite{Viollier:1976ab},
SRC are incorporated by solving the Bethe--Goldstone (BG) equation in
nuclear matter. This chapter extends the previous work of the Granada
group~\cite{RuizSimo:2016vsh, RuizSimo:2017tcb} to the case in which
the total center-of-mass (CM) momentum of the nucleon pair is
non-zero.  For this purpose, we employ the realistic coarse-grained
Granada 2013 nucleon-nucleon potential, obtained from a partial-wave
analysis of nucleon-nucleon scattering below the pion production
threshold~\cite{Perez:2013mwa}. The angular average of the
Pauli-blocking operator appearing in the BG equation is used, an
approximation widely applied in the past~\cite{Brueckner:1958zz,
  Haftel:1970zz,Jeukenne:1974zz,Bhargava:1967a, Kallio:1969mis}.
Alternative approaches that solve the BG equation without this
angular-average approximation have also been
explored~\cite{werner1959solution, Cheon:1988hn, Schiller:1998ff,
  Suzuki:1999jb, Sammarruca:2000dd, Stephenson:2004xs, White:2014oca}.
With the Granada 2013 potential and the angular-average approximation,
the BG equation simplifies to an algebraic form via a multipole
expansion. The resulting correlated radial wave functions are analyzed
in both coordinate and momentum space, and their dependence on the CM
momentum is studied as a preparatory step for chapter 7.

Finally, in chapter 7, the formalism developed in chapter 6 is applied
to incorporate the high-momentum components of the correlated
nucleon-pair wave functions into the MEC matrix elements for 1p-1h
excitations. These modified matrix elements are then embedded within
the response function formalism of the Fermi gas to evaluate the
impact of short-range correlations on the transverse nuclear
response. Proceeding in this way places us in an optimal position to
explore, in a novel and systematic manner, whether SRCs lead to an
enhancement of the transverse response. In particular, we aim to
investigate whether the high-momentum components of the correlated
nucleon pairs can reverse the negative interference effects identified
in chapters 3 and 4, potentially changing their sign and producing an
enhancement of the transverse channel, as previously observed in the
correlated basis function (CBF) calculations of
Fabrocini~\cite{Fab97}.

From the experimental point of view, the enhancement of the transverse
response has been clearly established in inclusive electron scattering
data. Analyses of world data by Jourdan~\cite{Jou96} demonstrated
that the transverse response systematically exceeds the predictions of
independent-particle models, particularly in the quasielastic
region. Similar conclusions were reached by Bodek and
Christy~\cite{Bod22}, who introduced phenomenological scaling
functions to reproduce this enhancement across a wide range of
nuclei. On the theoretical side, \textit{ab initio} Green's Function
Monte Carlo (GFMC) calculations 
\cite{Car02,Sch07,Lov16} have confirmed the same
trend in light nuclei, showing that two-body currents and correlations
lead to an increase of the transverse strength consistent with the
experimental findings. These results provide a solid empirical and
theoretical motivation for the study of chapter 7, which aims to identify
the microscopic origin of the transverse enhancement in terms of
short-range correlations and meson-exchange currents within the
framework of the Fermi gas.

\section*{Objectives and methodology }
The objectives of this Thesis, as were presented in the original
Thesis project are:
\begin{enumerate}
\item Improvement of the SuSAM* model by extending the definition of the
averaged single-nucleon response to all values of the scaling
variable, thus avoiding unphysical extrapolations and ensuring a more
accurate and reliable foundation for the superscaling model.

\item Inclusion of MEC within the superscaling formalism, introducing
  their interference with one-body currents directly in the
  single-nucleon prefactor and comparing the results with the
  Relativistic Fermi Gas and Relativistic Mean Field 
  models for electron and neutrino scattering.

\item Study of short-range correlations, obtained by solving the
  Bethe–Goldstone equation in nuclear matter using the coarse-grained
  Granada NN potential to extract high-momentum components of the
  correlated wave function.

\item Evaluation of one-body–two-body interference in the presence of
  SRC, to assess their possible role in the enhancement of the
  transverse response.
\end{enumerate}
The methodology combines well-established nuclear models with several
new developments introduced in this work. The SuSAM* model is refined
by implementing a realistic Fermi-type momentum distribution and by
explicitly including MEC contributions, which are validated through
scaling analyses with electron scattering data. Additional theoretical
frameworks employed in this thesis include the shell model, the
semirelativistic shell model, and the relativistic mean-field (RMF)
model with Dirac-equation--based potentials, using existing
computational codes. We also analyze the RFG, the RMF in nuclear
matter and the spectral function (SF) models.

To compute the flux-integrated neutrino cross sections in chapter 5,
including MEC effects, numerical integration were
performed on the \textbf{PROTEUS} supercomputer at the Instituto
Carlos I de F\'isica Te\'orica y Computacional. The Bethe--Goldstone
equation is solved in a partial-wave expansion using the Granada 2013
nucleon--nucleon potential, leading to a system of linear equations
that are solved via the Gauss method. The resulting
correlated wave functions are obtained in momentum space and
then incorporated into the Fermi gas model. Finally, to
evaluate the interference between one- and two-body currents in the
presence of SRC, a seven-dimensional
integral is computed numerically, which can be simplified under the
frozen approximation for the spectator nucleon.

\chapter{Improved Superscaling model with relativistic effective mass}

This chapter focuses on the nuclear quasielastic response in electron
scattering, and more specifically, on the superscaling model
\cite{Don99,Don99a}, whose basic theoretical foundations we aim to
examine. Scaling assumes that the response factorizes and is expressed
as proportional to an average single-nucleon response. The behavior of
the single-nucleon responses will be investigated when averaged over a
Fermi gas and extrapolated outside of the kinematic range allowed by
Pauli blocking. We show that the extrapolation leads to unphysical
results in extreme kinematics where the nucleon can not be on shell. We
propose a new averaging method that removes the need for
extrapolation. This leads to a new definition of the single-nucleon
response, averaged over momentum space, with a momentum distribution
where the Fermi surface is smeared out instead of using the sharp
Fermi gas distribution. This average therefore has a theoretical
justification, in contrast to the extrapolation approach
\cite{Ama04,Meg16b,Ama18}, and produces results that are similar to
those of the traditional superscaling models. This approach provides a
solid argument that justifies the choice of the single-nucleon
response and does not suffer from the previous issues. It will be
shown that the use of the new averaged single-nucleon or the
extrapolated one is indifferent in the scaling region, and then this
work improves the superscaling formalism from the theoretical point of
view by providing a physical justification for its use, which
strengthens the applicability of such phenomenological models.
See Ref. \cite{Cas23a} for more details.

\section{Superscaling formalism}

The theory of the RFG response function and
its connection with the theory of superscaling will be briefly
reviewed. The scaling variable $\psi$ was first introduced in
ref. \cite{Alb88}. The scaling formalism was refined in subsequent
works \cite{Cen97,Don99,Don99a} until reaching the most up-to-date
version of the SuSA-v2 model \cite{Ama20}.

The formalism in this chapter is an extension of the SuSA 
to the SuSAM* approach based on the
equations of nuclear matter interacting with a relativistic mean field
(RMF) \cite{Ros80,Ser86,Dre89,Weh93}, which allows for the incorporation of dynamic effects. The simplest approximation in this
framework is to introduce scalar and vector potentials
with which the nucleons interact.
The scalar potential is attractive,
while the vector potential is repulsive.
The on shell energy is defined as
\begin{equation}
E = \sqrt{m_N^ {*2}  + p^2},
\end{equation}
where \(m_N^*\) is the relativistic effective mass of the nucleon,
\begin{equation} \label{mefec}
m_N^*=m_N-g_s\phi_0 = M^*m_N.
\end{equation}
Here  $\phi_0$ is the scalar potential energy of the RMF and $g_s$ the corresponding coupling constant. To account for the interaction with the
vector potential, a positive energy term needs to be added to the
on-shell energy. Therefore, the total energy of the nucleon can be
expressed as:
\begin{equation}  \label{energyrmf}
E_{RMF}=E+E_v,
\end{equation}
being $E_v$ the called vector energy.

\subsection{Electromagnetic response functions}

We consider the inclusive electron scattering process where an
incident electron with energy $\epsilon$ scatters off a nucleus with
scattering angle $\theta$. The final electron energy is
$\epsilon'$. The momentum transfer is $q$, the energy transfer is
$\omega$, and $Q^2=\omega^2-q^2<0$. 
The cross section in plane-wave Born approximation with one
photon-exchange is written
\begin{equation}  \label{crosseq}
\frac{d\sigma}{d\Omega d\epsilon'}
= \sigma_{\rm Mott}
(v_L R_L(q,\omega) +  v_T  R_T(q,\omega)),
\end{equation}
were $\Omega$ is the final electron solid angle, 
$\sigma_{\rm Mott}$ is the Mott cross section, 
\begin{equation}
\sigma_{\rm Mott}=
\left(\frac{\alpha\cos\theta/2}{2\epsilon\sin^2\theta/2}\right)^2,
\end{equation} 
$v_L$ and $v_T$ are the kinematic factors 
\begin{equation}
v_L = 
\frac{Q^4}{q^4}, 
\kern 1cm
v_T =  
\tan^2\frac{\theta}{2}-\frac{Q^2}{2q^2},
\end{equation}
and finally, $R_K(q,\omega)$, $K=L,T$, are the longitudinal and
transverse response functions defined below.  We focus on the
description of the nuclear response functions resulting from the
interaction of the electron with the one-body (OB) electromagnetic
current, giving rise to one-particle one-hole excitations of the Fermi
gas. They are defined in a similar way to the usual RFG formalism
\cite{Ama20}, with the difference that in our case the nucleons have
an effective mass $m^*_N<m_N$.

The hole momentum is $\nh$ with $h<k_F$ and 
on-shell energy $E_h=\sqrt{h^2+(m_N^*)^2}$.
By momentum conservation, the final particle momentum is $\np=\nh+\nq$ 
with on-shell energy $E_p=\sqrt{p{}^2+(m_N^*)^2}$. Pauli blocking implies
$p>k_F$.
 The nuclear response functions are then  given by
 \begin{equation}
R_K^{QE}(q,\omega)
= 
 \frac{V}{(2 \pi)^3}
\int
d^3h
\frac{(m^*_N)^2}{E_hE_p}  2w_K(\np,\nh)  \,
\theta(p-k_F)
\theta(k_F-h) 
\delta(E_p-E_h-\omega),
\label{rmf}
\end{equation}
where $w_K$ are the single-nucleon responses for the 1p1h excitation
\begin{eqnarray} \label{snresponses}
w_L =  w^{00},
\kern 1cm
w_T  =  w^{11}+w^{22},
\end{eqnarray}
corresponding to the single-nucleon hadronic tensor
\begin{equation} \label{traza}
w^{\mu\nu}=\frac12 \sum_{s_ps_h} j^{\mu}(\np,\nh)^{*}j^{\nu}(\np,\nh)
\end{equation}
and $j^\mu$ is the electromagnetic current matrix element
\begin{equation} \label{corriente}
j^{\mu}(\np,\nh)= \overline{u}_{s_p}(\np)
\left[F_1\gamma^\mu+i\frac{F_2}{2m_N}\sigma^{\mu\nu}Q_{\nu}
\right]u_{s_h}(\nh),
\end{equation}
where $F_1$ and $F_2$, are the Dirac and Pauli form factors of the
nucleon.  Note that we use the current operator in the vacuum, but the
spinors correspond to nucleons with effective mass $m_N^*$.

To compute the integral (\ref{rmf}), we change to the variables $E_h, E_p,
\phi$, with Jacobian $h^2 dh d\cos\theta= (E_hE_p/q)dE_hdE_p$.  Then
the integral over $E_p$ is made using the Dirac delta.  This fixes the
angle between $\nq$ and $\nh$ to the value
\begin{equation} \label{angulo}
\cos\theta_h= \frac{2E_h\omega+Q^2}{2hq}, 
\end{equation}
 and the integration over the angle $\phi$ gives
$2\pi$ by symmetry of the responses when $\nq$ is on the $z$-axis
\cite{Ama20}. We are left with an integral over the initial nucleon
energy
 \begin{equation}
R_K^{QE}(q,\omega)
= 
 \frac{V}{(2 \pi)^3}
 \frac{2\pi m_N^{*3}}{q}
\int_{\epsilon_0}^{\infty}d\epsilon\, n(\epsilon)\, 2w_K(\epsilon,q,\omega),
\label{respuesta}
\end{equation}
where $\epsilon=E_h/m^*_N$ is the initial nucleon energy in units of
$m_N^*$, and $\epsilon_F=E_F/m_N^*$ is the (relativistic) Fermi energy
in the same units.  Moreover we have introduced the energy
distribution of the Fermi gas $n(\epsilon)=
\theta(\epsilon_F-\epsilon)$.  The lower limit, $\epsilon_0$ of the
integral in Eq. (\ref{respuesta}) corresponds to the minimum energy
for a initial nucleon that absorbs energy $\omega$ and momentum $q$.
It can be written as (see Appendix C of ref. \cite{Ama20})
\begin{equation}
\epsilon_0={\rm Max}
\left\{ 
       \kappa\sqrt{1+\frac{1}{\tau}}-\lambda, \epsilon_F-2\lambda
\right\},
\end{equation}
where we have
introduced the
dimensionless variables 
\begin{eqnarray}
\lambda  = \omega/2m_N^* & 
\kappa   =  q/2m_N^* &
\tau  =  \kappa^2-\lambda^2. 
\end{eqnarray}

\begin{figure}
\centering
\includegraphics[width=15cm,bb=20 600 540 770]{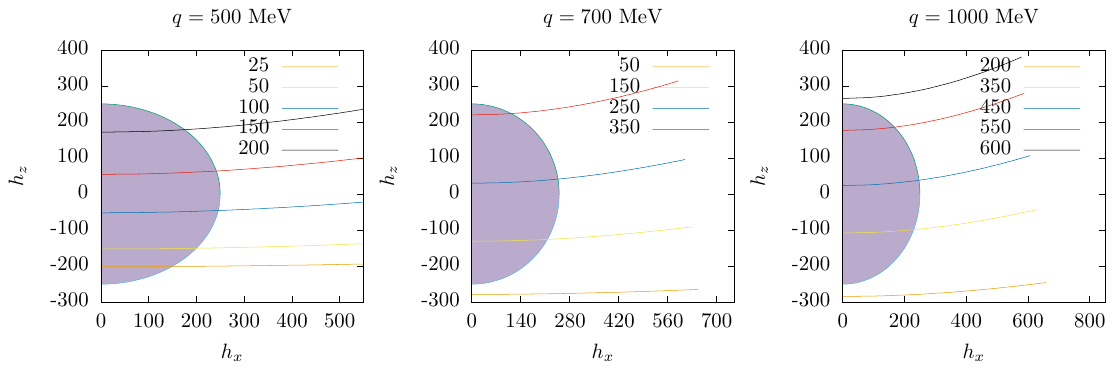}
\caption{Integration path in momentum space of the initial nucleon for
  different values of the energy transfer $\omega$ (indicated in MeV
  in the key for each panel) and for three values of the momentum
  transfer.}
  \label{esfera}
\end{figure}

For a fixed value of $\phi, q, \omega$, the integral over energy
$\epsilon$ in Eq. (\ref{respuesta}) corresponds to integrating the
single nucleon response over a path in the momentum space of the hole
$\nh$, weighted with the momentum distribution.  This curve is easily
obtained from Eq. (\ref{angulo}), giving the angle $\theta_h$ as a
function of the hole energy.  Some examples are shown in
Fig. \ref{esfera} for three values of $q$.  For each $q$ we plot the
integration trajectories in the $(h_x,h_z)$-plane for several values
of $\omega$. The semicircles indicate the moment distribution for
$k_F=250$ MeV.  The nuclear response function, $R_K(q,\omega)$, therefore
correspond to the integral of the single-nucleon responses
along one path.  The minimum momentum $h_0$, and therefore the minimum
energy $\epsilon_0$, correspond to the intersection of each curve with
the $h_z$ axis. The curves for different values of $\omega$ do not
intersect. The case $h_0=0$ only occurs for a certain
value of $\omega$, which is precisely the position of the
quasielastic peak; this corresponds also to $\epsilon_0=1$ (or
$\psi^*=0$ for the scaling variable). 
For very large or very small $\omega$-values, the curves lie
in the region where the momentum distribution is zero, and therefore
the corresponding response function is also zero.

\subsection{Scaling}

 Scaling is based on the approximated factorization of an averaged
 single-nucleon response from the nuclear cross section. This
 factorization is exact in the RMF model with the OB current, where
 analytical expressions are obtained  by
 explicit integration of the one-body responses \cite{Ama20}.


 For our purpose, we define a mean value of the single-nucleon
 responses by averaging with the energy distribution $n(\epsilon)$,
\begin{eqnarray}\label{def_sn}
\overline{w}_K(q,\omega)
 = \frac{\int^{\infty}_{\epsilon_0} d \epsilon \,  n (\epsilon)
w_K(\epsilon,q,\omega)}{\int_{\epsilon_0}^{\infty} d\epsilon \, n (\epsilon)}.
\end{eqnarray}
This corresponds to the average of the single-nucleon response
$w_K(\epsilon,q,\omega)$ over one of the paths in Fig. \ref{esfera}.
Using these averaged single-nucleon responses
we can rewrite Eq. (\ref{respuesta}) in the form
 \begin{eqnarray}
R_K^{QE}(q,\omega)
&=& 
 \frac{V}{(2 \pi)^3}
 \frac{2\pi m_N^{*3}}{q}  2\overline{w}_K(q,\omega)
\int_{\epsilon_0}^{\infty}d\epsilon\, n(\epsilon).
\label{respuesta2}
\end{eqnarray}
This last integral depends only on the variable $\epsilon_0$, which in turn
depends on $(q, \omega)$. In the superscaling approach the
$\psi^*$-scaling variable is used instead of the minimum energy of the
nucleon, $\epsilon_0$. This energy is transformed by a change of
variable into the scaling variable, $\psi^*$, defined as
\begin{equation}
\psi^* = \sqrt{\frac{\epsilon_0-1}{\epsilon_F-1}} {\rm sgn} (\lambda-\tau),
\end{equation}
where $\psi^*$ is negative (positive) for $\lambda<\tau$ ($\lambda>\tau$).

The superscaling function is defined as
\begin{equation} \label{scaling}
 f^*(\psi^*)=\frac{3}{4}\frac{1}{\epsilon_F-1}
\int_{\epsilon_0}^{\infty} n (\epsilon) d\epsilon,
\end{equation}
where $\epsilon_F-1 \ll 1$ is the kinetic Fermi energy in units of
$m_N^*$.  The definition (\ref{scaling}) is, except for a factor,
similar to that of the $y$-scaling function $f(y)$ \cite{Wes75,Sar93},
where the scaling variable $y$ was the minimum momentum of the initial
nucleon.

In RFG and nuclear matter with RMF, Eq. (\ref{scaling}) is easily evaluated
(remember that the RFG is recovered as the particular case $M^*=1$) as
\begin{equation} 
\int_{\epsilon_0}^{\infty} \theta(\epsilon_F-\epsilon) d\epsilon
= \theta(\epsilon_F-\epsilon_0) (\epsilon_F-\epsilon_0)
= (\epsilon_F-1)(1-\psi^*{}^2)\theta(1-\psi^*{}^2).
\end{equation}
Therefore the scaling function of nuclear matter is
\begin{equation} \label{scalingRFG}
 f^*(\psi^*)=\frac{3}{4}(1-\psi^*{}^2)\theta(1-\psi^*{}^2).
\end{equation}
Note that  the scaling function of nuclear matter is zero for
$\epsilon_0 > \epsilon_F$, and this is equivalent to $|\psi^*| > 1$.
This is a consequence of the maximum momentum $k_F$ for the nucleons
in nuclear matter, which implies that $\epsilon_0< \epsilon_F$.

Using $V/(2\pi)^3= N/(\frac83 \pi k_F^3)$ for nuclear matter
we can write the response
functions (\ref{respuesta2}) as
\begin{equation}
R^{QE}_K(q,\omega) = 
\frac{\epsilon_F-1}{m_N^* \eta_F^3 \kappa}
 (Z \overline{w}^p_K(q,\omega)+N \overline{w}^n_K(q,\omega))
f^*(\psi^*),
\label{susam} 
\end{equation}
where we have added the contribution of $Z$ protons and $N$ neutrons
to the response functions, and $\eta_F=k_F/m_N^*$.

\subsection{${\rm SuSAM^*}$}

\begin{figure}[h]
  \centering
\includegraphics[width=8.5cm,bb=110 270 460 770]{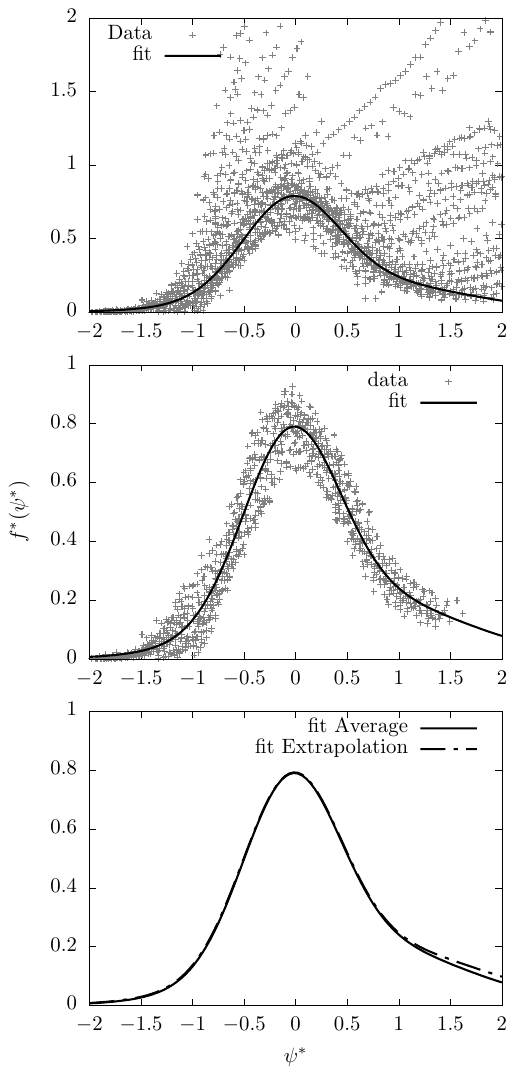}
\caption{
Super scaling analysis with relativistic effective mass
  (SuSAM*) of $^{12}$C data. Top panel: experimental scaling data
$f^*_{exp}$ plotted against $\psi^*$.  Middle panel: data surviving
after cleanup of non-quasielastic sparse points.  The black curve is
Gaussian fit made in this work, $f^*_{QE}(\psi^*)$.  In the bottom panel
we compare the two scaling functions obtained with two different
definitions of the averaged single-nucleon responses: 
using the extrapolated Fermi gas responses and performing the average
with a Fermi distribution defined in Section \ref{aver}.
}\label{fig0}
\end{figure}

The SuSAM* approach extends the formula (\ref{susam}) by replacing
$f^*(\psi^*)$ by a phenomenological scaling function obtained from
experimental data of $(e,e')$. 
In a real, finite nucleus the momentum is not limited by $k_F$
(in particular correlated nucleons can greatly exceed the Fermi
momentum, see chapter 6). This has the effect that the {\em phenomenological}
superscaling function is not zero for $|\psi^*|>1$,
and therefore takes into account that the nucleons are
not limited by a maximum Fermi momentum.  

Several approaches have been used in the past to obtain a
phenomenological scaling function. In the original SuSA model, based
on the RFG without effective mass, the scaling function was obtained
from the longitudinal response data. In the SuSAv2 model, a scaling
function for the transverse response was also introduced by means of
a RMF theoretical model in finite nuclei. In this chapter we will focus
on the SuSAM* model with effective mass where the
phenomenological scaling function is obtained directly from the
quasielastic data of the inclusive cross section.  Different scaling
models with effective mass and without effective mass provide
different scaling functions, but all may reproduce the quasielastic cross
section reasonably well, since they have been fitted to experimental
data.

In the procedure followed in ref \cite{Mar17,Ama18,Rui18} the
inclusive cross section data are divided by the contribution of the
single nucleon. But the SuSAM* model was extended in
refs. \cite{Mar21,Mar21b}, by subtracting the theoretical contribution
of MEC in the 2p2h channel from the
experimental data before dividing by the single nucleon, that is
\begin{equation} \label{pheno}
f_{exp}^* =
\frac{\displaystyle \left(\frac{d\sigma}{d\Omega d\omega}\right)_{exp}
  -\left(\frac{d\sigma}{d\Omega d\omega}\right)_{2p2h}}{\sigma_M ( v_L r_L + v_T
  r_T) }.
\end{equation}
where 
\begin{equation} \label{rsn}
r_K = \frac{\epsilon_F-1}{m_N^* \eta_F^3 \kappa} 
(Z \overline{w}^p_K(q,\omega)+N \overline{w}^n_K(q,\omega)).
\end{equation}
In Fig. \ref{fig0} these experimental data, $f^*_{exp}$, are plotted
against $\psi^*$ in the interval $-2<\psi^*<2$,
 which
largely encompasses the quasielastic domain, extending well beyond the Fermi region $-1<\psi^*<1$.
It is observed that about
half of them roughly collapse forming a thin band around the
quasielastic peak.  This band constitutes the set of selected data
that can be considered QE and we reject the rest, which mainly
contribute to inelastic processes.  The selected quasielastic data are
well parameterized with a sum of two Gaussians, thus obtaining the
phenomenological quasielastic function $f^*_{QE}$, shown also in
Fig. \ref{fig0}.

\section{Averaged single-nucleon response functions}
\label{aver}

One of the most confusing aspects in the
superscaling formalism is the definition and meaning of the averaged
single-nucleon response functions for $|\psi^*| > 1$ or, equivalently,
$\epsilon_0> \epsilon_F$, i.e., outside the allowed $\omega$-range of
the Fermi gas. Traditionally an extrapolation of the Fermi gas formula has
often been used. In this section we expose the intrinsic theoretical
problems of the Fermi gas extrapolation, and  propose
an alternative definition that is more satisfactory from the
theoretical point of view.

\subsection{RFG extrapolation}

In the traditional superscaling 
approach, first the averaged single-nucleon responses 
$\overline{w}_K$ are calculated 
for $\epsilon_0<\epsilon_F$ (or $|\psi^*|<1$) using the Fermi gas
momentum distribution, 
\begin{eqnarray}
\overline{w}_K(q,\omega)
&=&
\frac{\int_{\epsilon_0}^{\infty}w_K(\epsilon,q,\omega)
      \theta(\epsilon_F-\epsilon)d\epsilon }
{ \int_{\epsilon_0}^{\infty}
\theta(\epsilon_F-\epsilon)d\epsilon} 
\nonumber\\
&=&
\frac{ \theta(\epsilon_F -\epsilon_0) 
\int_{\epsilon_0}^{\epsilon_F}
w_K(\epsilon,q,\omega)}
{\theta(\epsilon_F-\epsilon_0) \int_{\epsilon_0}^{\epsilon_F}d\epsilon}. 
\end{eqnarray}
Note that this expression is only defined for $\epsilon_0 < \epsilon_F$,
in which case the step functions cancel and  we obtain
\begin{equation} \label{wkextrapolated}
\overline{w}_K(q,\omega)=
\frac{1}{\epsilon_F-\epsilon_0}
\int_{\epsilon_0}^{\epsilon_F}
w_K(\epsilon,q,\omega)
d\epsilon
,\kern 1cm
(\epsilon_0 < \epsilon_F).
\end{equation}
The function $w_K(\epsilon,q,\omega)$ inside the integral is well
defined and positive only if $\epsilon>\epsilon_0$,
because it corresponds to the response of a single nucleon with energy
$\epsilon$, that absorbs momentum $q$ and energy $\omega$. 
In the traditional SuSA and SuSAM*
approaches the function (\ref{wkextrapolated}) is extended
analytically for $\epsilon_0> \epsilon_F$ in the obvious way.  This is
called in this work the {\em extrapolated} single nucleon response
function, and it can be written equivalently in the way
\begin{equation}
\overline{w}_K(q,\omega)=
\frac{1}{\epsilon_0-\epsilon_F}
\int_{\epsilon_F}^{\epsilon_0}
w_K(\epsilon,q,\omega)
d\epsilon.
\end{equation}
From this expression it is clear that, for $\epsilon_0 > \epsilon_F$,
the function $w_K(\epsilon,q,\omega)$ inside the integral must be
evaluated for $\epsilon<\epsilon_0$.  But this is not possible for a
nucleon on-shell that absorbs $(q,\omega)$, because its minimum energy
is $\epsilon_0$.  Therefore it is not guaranteed that the function
$w_K(\epsilon,q,\omega)$ inside the integral is positive if is
evaluated for $\epsilon<\epsilon_0$.
This is a fundamental problem of the single
nucleon extrapolation. Next we will study some particular cases where
the extrapolated responses are 
explicitly negative for $\epsilon_0>\epsilon_F$, that
is, for $|\psi^*|>1$.

\subsection*{Longitudinal single-nucleon response}
We use the analytical formulas of the single nucleon responses from
Appendix \ref{appA},
\begin{equation} \label{wl}
w_L= 
\frac{(G_M^*)^2}{1+\tau}
[\tau(\epsilon+\lambda)^2-(1+\tau)\kappa^2]
+\frac{(G_E^*)^2}{1+\tau}
(\epsilon+\lambda)^2.
\end{equation}
To better understand the kinematic dependence of this response function 
it is convenient to express it in terms of the minimal nucleon 
energy $\epsilon_0$ using
\begin{equation}\label{e0lambda}
\epsilon_0+\lambda = \kappa\sqrt{\frac{1+\tau}{\tau}}
\Longrightarrow
\kappa^2(1+\tau)= \tau(\epsilon_0+\lambda)^2.
\end{equation}
in the regime without Pauli blocking. Then Eq. (\ref{wl}) becomes
\begin{equation}
w_L= 
\frac{(G_M^*)^2\tau}{1+\tau}
[(\epsilon+\lambda)^2-(\epsilon_0+\lambda)^2]
+\frac{(G_E^*)^2}{1+\tau}
(\epsilon+\lambda)^2.
\end{equation}
In this equation it is evident that the electric term is always
positive.  However the magnetic term is positive only for $\epsilon >
\epsilon_0$.  For this reason, if $w_L$ is calculated using the Fermi
gas momentum distribution and then extrapolated to values
$\epsilon_0>\epsilon_F$ (or $\psi^*>1$), the magnetic term becomes
negative. This does not make physical sense because the longitudinal
response must be positive, by definition, regardless of the value of
the form factors. In fact if we artificially turn off the electric
contribution, a negative averaged response $\overline{w_L}$ is
obtained for $\epsilon_0> \epsilon_F$.  Let suppose for
simplicity that $G_E^*= 0$.  Then the extrapolated
single-nucleon longitudinal response would be
\begin{equation}\label{wlM}
\overline{w}_L=
\frac{(G_M^*)^2}{\epsilon_0-\epsilon_F}
\frac{\tau}{1+\tau}
\int_{\epsilon_F}^{\epsilon_0}
[(\epsilon+\lambda)^2-(\epsilon_0+\lambda)^2]d\epsilon,
\end{equation}
that 
is negative for $\epsilon_0 > \epsilon_F$.

\subsection*{Transverse single-nucleon response}

We find a similar situation in the case 
of the transverse response  from Eq. (\ref{wtfinal}) in Appendix
\ref{appA},
\begin{equation}
w_T=
2\tau (G_M^*)^2
+\frac{(G_E^*)^2+\tau (G_M^*)^2}{1+\tau}
\frac{\tau}{\kappa^2}
\left[(\epsilon+\lambda)^2-\kappa^2\frac{1+\tau}{\tau}\right].
\end{equation}
Again we can rewrite this response as a function of the minimum 
nucleon energy, $\epsilon_0$, 
using $\kappa^2(1+\tau)/\tau= (\epsilon_0+\lambda)^2$
\begin{equation}
w_T= 
2\tau (G_M^*)^2
+\left[(G_E^*)^2+\tau (G_M^*)^2\right]
\left[
\left(\frac{\epsilon+\lambda}{\epsilon_0+\lambda}\right)^2-1
\right].
\end{equation}
Rearranging terms containing $G_E^*$ and $G_M^*$ the single-nucleon transverse 
response becomes finally
\begin{equation} \label{wtexplicit}
w_T= 
(G_E^*)^2
\left[
\left(\frac{\epsilon+\lambda}{\epsilon_0+\lambda}\right)^2-1
\right]
+\tau (G_M^*)^2
\left[
\left(\frac{\epsilon+\lambda}{\epsilon_0+\lambda}\right)^2+1
\right].
\end{equation}
Written in this way, it is evident that the magnetic contribution of
$w_T$ is always positive. While the electrical term is positive only for
$\epsilon > \epsilon_0$.
The situation is similar to what we found with the longitudinal
response, but in the transverse response it is the electrical term
that becomes negative in the extrapolation to $\epsilon_0>\epsilon_F$. 
We now turn off the
magnetic contribution and suppose that $G_M^*=0$.  Then
the averaged T response in RFG would be, with analogy to
Eq. (\ref{wlM})
\begin{equation} \label{wtE}
\overline{w}_T=
\frac{(G_E^*)^2}{\epsilon_0-\epsilon_F}
\int_{\epsilon_F}^{\epsilon_0}
\left[
\left(\frac{\epsilon+\lambda}{\epsilon_0+\lambda}\right)^2-1
\right]
d\epsilon.
\end{equation}
From this expression it is clear that the extrapolated
$\overline{w}_T$ is negative for $\epsilon_0 > \epsilon_F$ because the
function inside the integral is negative, which is not physically acceptable:
the transverse response should be
positive by definition regardless of the form
factors values. In other words, the electrical contribution to the transverse
response, although small, cannot be negative.

\subsection{Alternative to the extrapolated single-nucleon responses}

In this work we propose an alternative definition of the averaged
single-nucleon responses that solves the extrapolation problem in the
superscaling model.  As we have seen, the problem is a consequence of
the fact that in the Fermi gas there is a maximum momentum for the
nucleons. If this momentum is exceeded by extrapolation, i.e.
$\epsilon_0 > \epsilon_F$, mathematically this is equivalent to
assuming nucleons with energy less than $\epsilon_0$, which is
impossible in the Fermi gas because nucleons are on-shell.  Hence
results without physical sense, such as negative responses, are
obtained if the extrapolated formula is applied.

The proposed solution involves using equation (\ref{def_sn})
 for the averaged
single-nucleon responses, but introducing a momentum distribution
without a maximum momentum, and that at the same time does not differ
much from the Fermi gas distribution, for $h<k_F$. An appropriate
function is a distribution of Fermi type
\begin{equation}\label{distribucion}
n(h)= \frac{a}{1+e^{(h-k_F)/b}}.
\end{equation}
Where $b$ is a smearing parameter for the Fermi surface, which is no
longer restricted to a sphere as in figure 1. Then the integrals by
averaging in Eq. (\ref{def_sn}) extend to infinity and therefore there is no
longer an upper limit for $\epsilon_0$, which can take any value up to
infinity. The single-nucleon responses 
of the integrand always are evaluated for $\epsilon
> \epsilon_0$ and they are therefore positive definite

In addition, for $\epsilon < \epsilon_F$, the momentum distribution is
similar to the Fermi Gas distribution, $\theta(k_F-h)$, and then it is
expected that the averaged single-nucleon be similar to that of the
RFG. Now, for $\epsilon > \epsilon_F$ the integration in the eq.
(\ref{def_sn}) extends in momentum space along one of the paths
outside the Fermi sphere of Figure (\ref{esfera}). Then the average
has the physical sense of coming from regions above the Fermi sphere,
that is to say, from the high momentum zone that the Fermi gas cannot
describe. This is in accordance with the meaning attached to the
experimental scaling function for $|\psi^*|>1$, which comes mainly
from high momentum nucleons.

\section{Results}

\begin{figure}
  \centering
\includegraphics[width=11cm,bb=110 270 460 770]{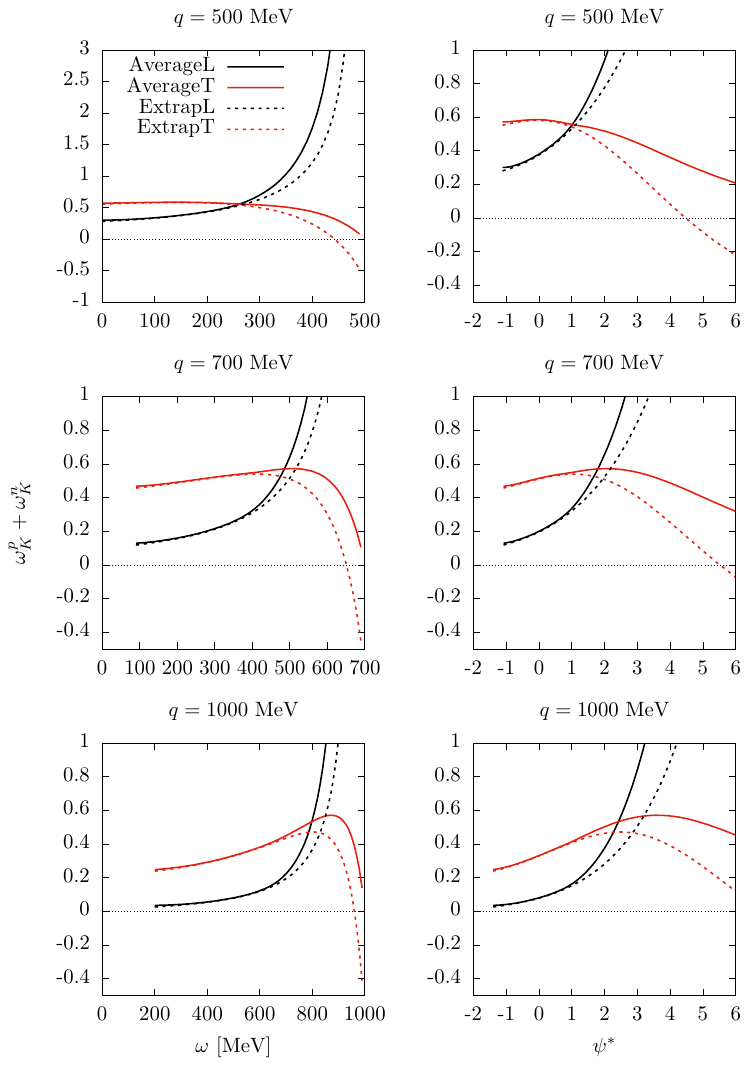}
  \caption{Averaged and extrapolated longitudinal and transverse
    response functions for proton plus neutron, as a function of
    $\omega$ and of the scaling variable $\psi^*$, for three values of
    the momentum transfer.}
  \label{qfij2}
\end{figure}

We present results for the averaged nucleon responses
and for the total nuclear responses in the SuSAM* model. The
calculations are made for electron scattering off the nucleus $^{12}$C
with Fermi momentum $k_F=225$ MeV/c and effective mass
$m_N^*=0.8m_N$. These values were fitted to the quasielastic data of
$f^*_{exp}$ to obtain the best possible scaling \cite{Mar17,Ama18}.
We evaluate the validity of the scaling model when using the Fermi gas
extrapolation for the nucleon response function. Specifically,
the results obtained by averaging the single-nucleon response
function over a smeared Fermi momentum distribution,
Eq. (\ref{distribucion}) are compared with the extrapolated response function
obtained from the Fermi gas model.

In Fig. \ref{qfij2} we compare the averaged nucleon responses with the
extrapolated ones. The sum of proton plus neutron is shown. The
averaged responses have been calculated with a Fermi distribution
using a smearing parameter $b=50$ MeV/c.  The responses do not depend
much on the precise value of this parameter for small variations. We
see that the averaged responses are practically the same as the
extrapolated responses of the Fermi gas 
in the quasielastic scaling region, $-2<\psi^*<2$,.
 But both results start to diverge for large $\omega$ or
$\psi^*>2$. The extrapolated transverse response
becomes negative for $\psi^* > 4$, 5 and 7, for $q=500$, 700, and
1000 MeV/c, respectively, very close to the photon line. 
This is easily explained because in Eq. \ref{wtexplicit},
the magnetic term is multiplied by $\tau$.  Therefore the $w_T$
response is dominated by the electric term for $\tau\rightarrow 0$,
that is, for large $\omega$, and in Eq. \ref{wtE} we have seen that this
term is negative when extrapolated to $\epsilon_0 > \epsilon_F$.

\begin{figure}
\centering
\includegraphics[width=7.5cm,bb=110 270 460 650]{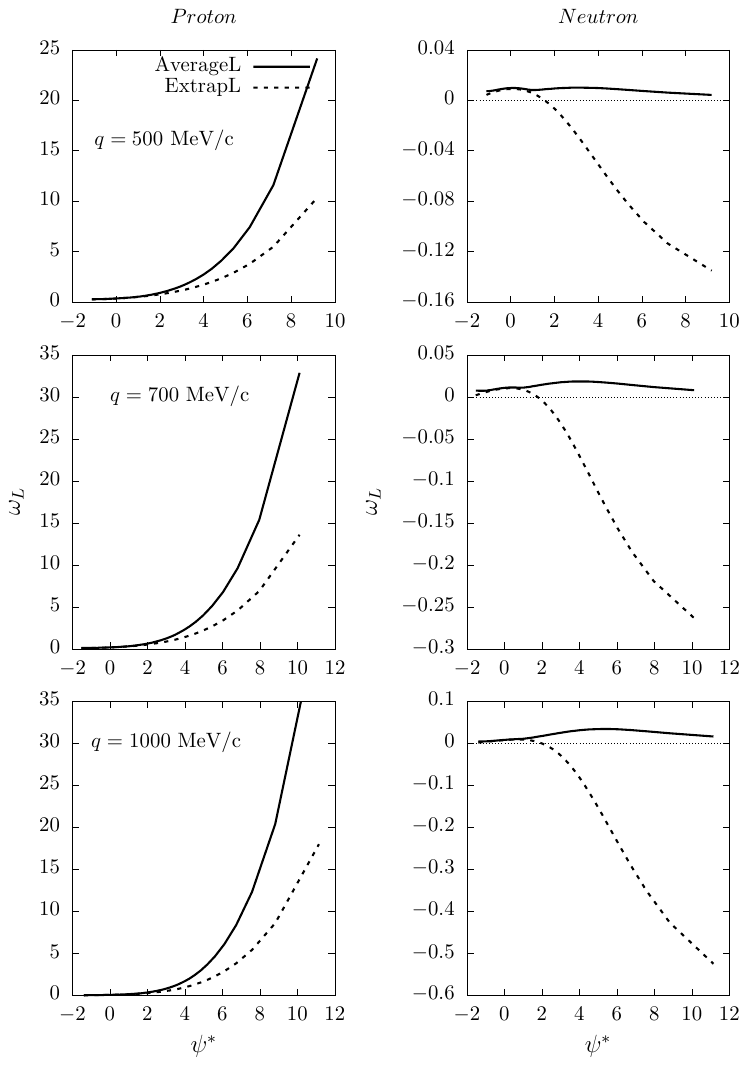}
\kern 3,5mm
\includegraphics[width=7.5cm,bb=110 270 460 790]{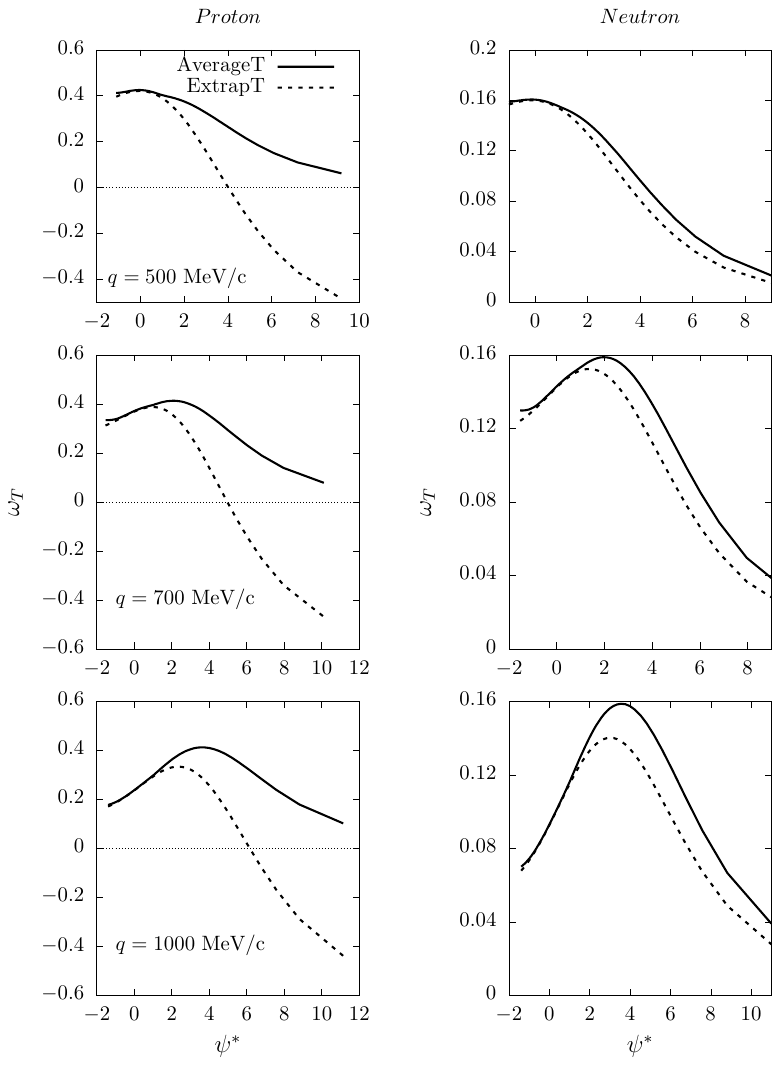}
  \caption{Averaged and extrapolated longitudinal and transverse 
response functions for
    protons and neutrons, as a function of the scaling
    variable and for three values of the momentum transfer.}
\label{qfij4}
\end{figure}

More details can be seen in Fig. \ref{qfij4} where we show the
averaged and extrapolated response functions separated for protons and
neutrons, as a function of the scaling variable.  The extrapolated and
averaged responses start to differ in the region $\psi^*>2$ and the
discrepancy increases with $\psi^*$.  The extrapolated longitudinal
response of neutrons is negative for $\psi^* >2$. This agrees with
what was seen analytically in the previous section, because the
extrapolation of the longitudinal magnetic response is negative and
the electric form factor of the neutron is negligible.  This does not
affect the results of the SuSAM* model in the scaling region
 because the longitudinal
response of the neutron is much smaller than that of the proton.

In Fig.\ref{qfij4} we also can see see that the averaged proton
transverse response is very similar to the extrapolation in the
scaling region and differ for $\psi^*>2$.  
They also start to differ in the
$\psi^*$-negative region for $\psi^* < -2$.  The extrapolated
transverse response of protons is negative from $\psi^*\sim 4$--6
depending on the value of $q$.  Again this is because the electrical
term of the proton dominates this response for large $\omega$ since
the magnetic term carries a factor $\tau$, which tends to zero for
$\omega \rightarrow q$. In contrast the averaged proton transverse
responses are always positive.

The averaged transverse neutron response shown in Fig. \ref{qfij4} is
similar in shape to the Fermi gas extrapolation in the scaling region. 
But again they differ for
$|\psi^*|>2$, where the averaged one is the largest, and the difference
between the two increases with the momentum transfer.

\begin{figure}
\centering
\includegraphics[width=7.5cm,bb=110 270 460 650]{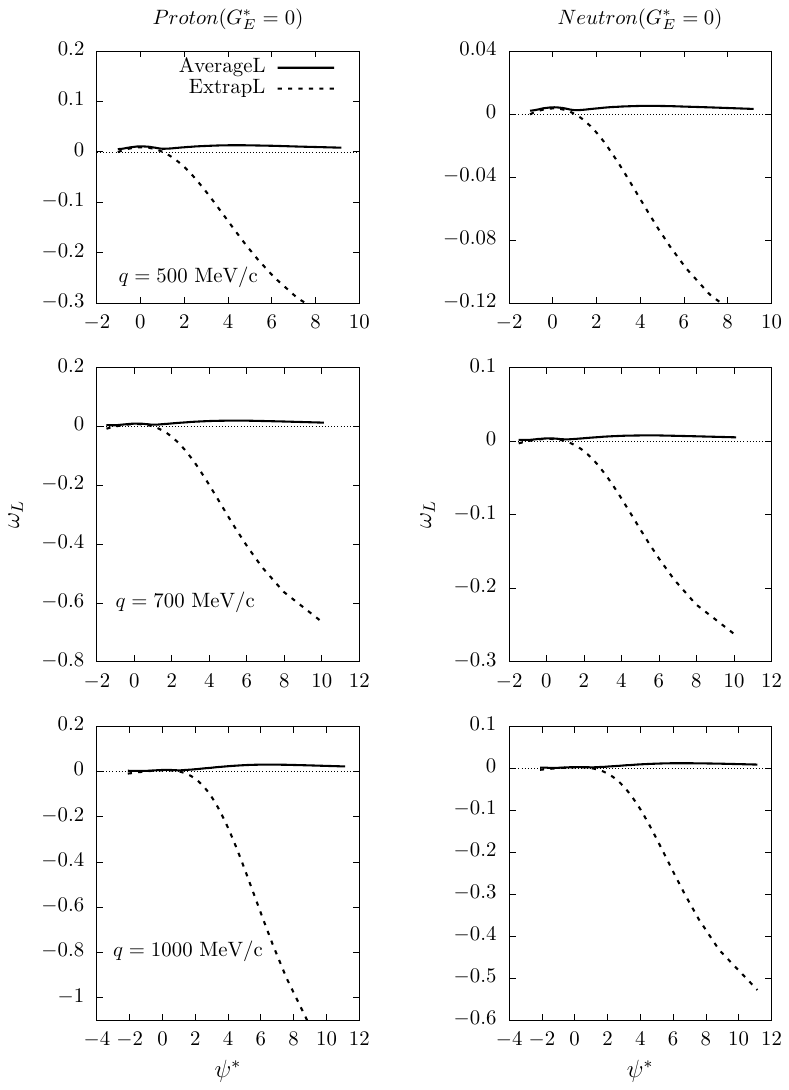}
\kern 3.5mm
\includegraphics[width=7.5cm,bb=110 270 460 830]{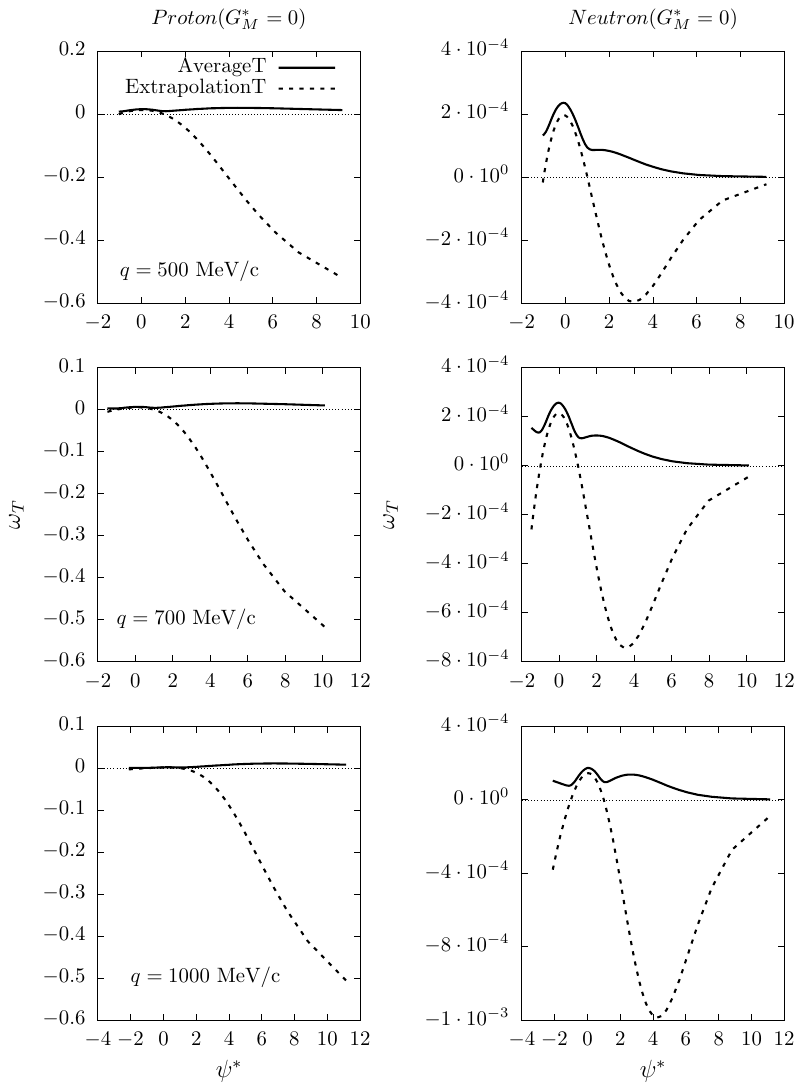}
  \caption{Averaged and extrapolated transverse response functions for
    protons and neutrons, for $G_M^*=0$, as a function of the scaling
    variable and for three values of the momentum transfer.
     Averaged and extrapolated longitudinal response functions for
    protons and neutrons, for $G_E^*=0$, as a function of the scaling
    variable and for three values of the momentum transfer.}
  \label{fig5}
\end{figure}

We have seen in the extrapolation formulas, Eqs.(\ref{wlM}) and
(\ref{wtE}), that the magnetic contribution to the longitudinal
response and the electrical contribution to the transverse response
become both negative for $\epsilon_0> \epsilon_F$, This can be
explicitly seen in the results in Fig. \ref{fig5}, where we plot the
longitudinal responses computed for $G^*_E=0$ and the transverse
responses computed for $G^*_M=0$, for protons and neutrons.  In fact,
in all cases of Fig. \ref{fig5} the extrapolated responses are
negative for $|\psi^*|>1$. On the contrary, the averaged responses are
always positive.

\begin{figure}
\centering
\includegraphics[width=9cm,bb=110 435 460 785]{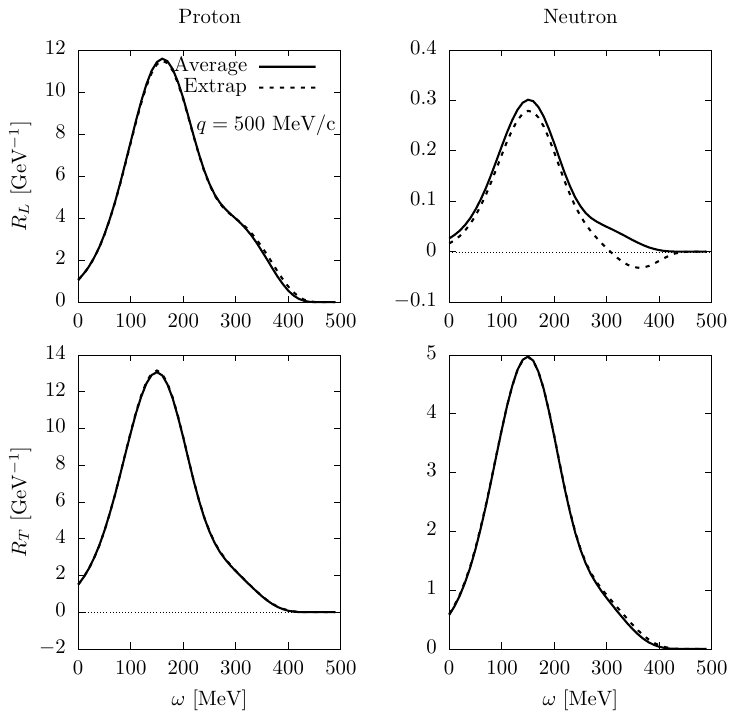}
  \caption{Longitudinal and transverse response functions separated
    for protons and neutrons in the SuSAM* model using the averaged
    and extrapolated single nucleon responses for $q=500$ Mev/c}
  \label{qfij6}
\end{figure}

\begin{figure}
\centering
\includegraphics[width=9cm,bb=110 435 460 785]{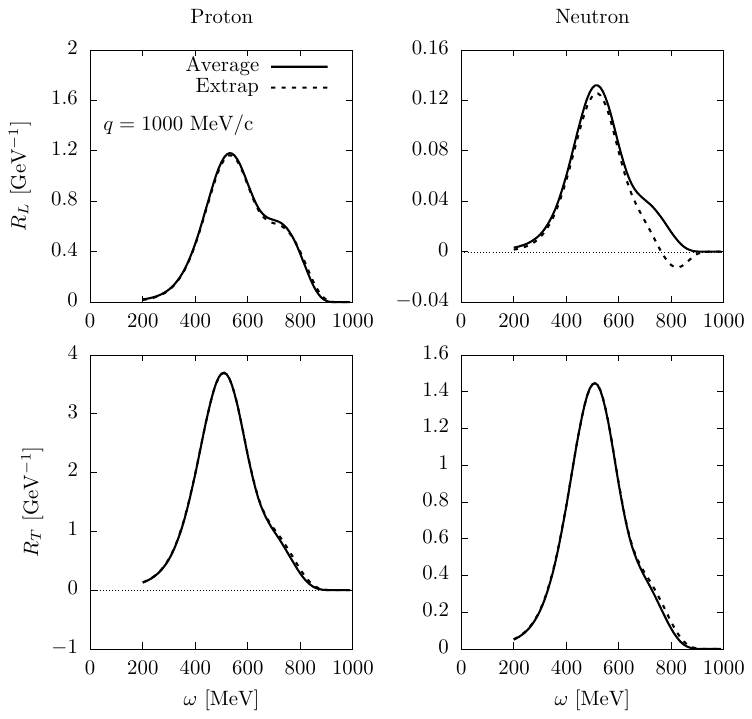}
\caption{The same as in Fig \ref{qfij6} for $q=1000$ MeV/c}
  \label{qfij10}
\end{figure}

In Figs. \ref{qfij6} and \ref{qfij10} we use the superscaling model to
investigate the nuclear responses under various inputs for the
single-nucleon. The nuclear response is computed from the product of
the averaged nucleon-responses and a phenomenological scaling function
obtained from the data, using Eq. (\ref{susam}).

The results in Figs. \ref{qfij6} and \ref{qfij10} demonstrate that
there are no significant differences in the separate responses of
protons and neutrons when computed with the averaged single-nucleon
compared to the extrapolation. The only difference is seen in the
longitudinal neutron response for high $\omega$, which becomes
negative in the extrapolated model.  However this is not relevant for
the total nuclear response, as the neutron contribution is negligible
in the longitudinal response as compared to the proton one.

\begin{figure}
  \centering
\includegraphics[width=9.5cm,bb=110 270 460 770]{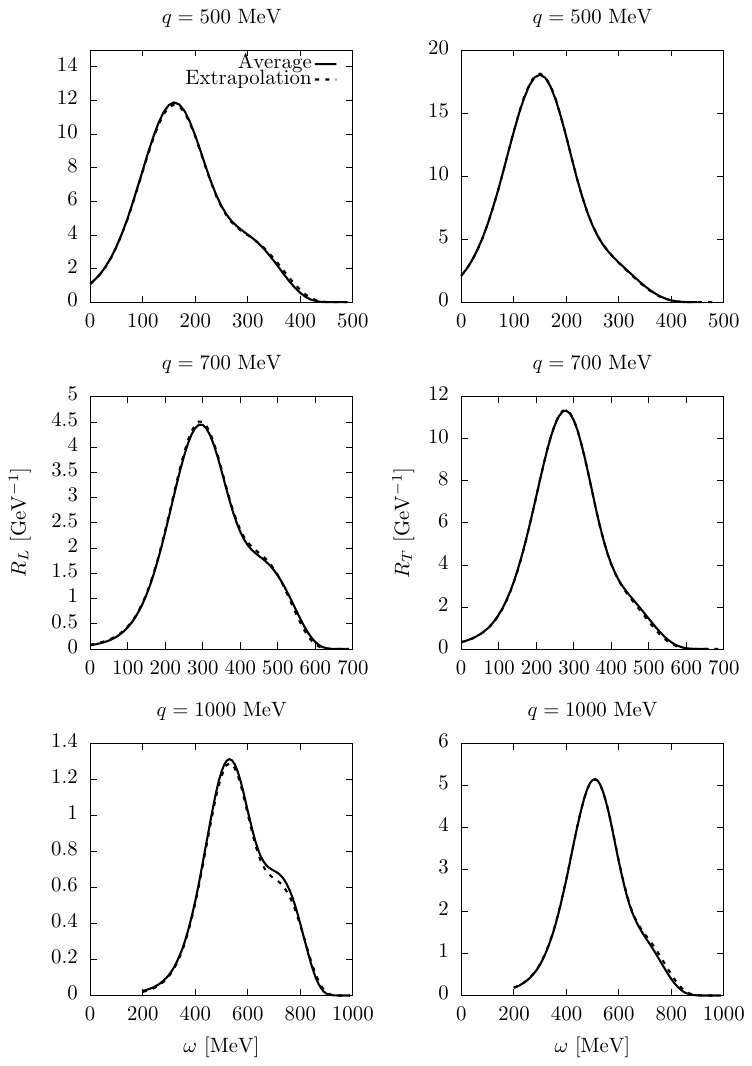}
  \caption{Longitudinal and transverse response functions in the SuSAM* 
model using the averaged and extrapolated single nucleon responses}
  \label{qfij7}
\end{figure}

This is verified in the results of Fig. \ref{qfij7} for the total
responses. Both the averaged and the extrapolated single-nucleon
responses give essentially the same result.  The results obtained 
have two important implications. Firstly, they provide
support for the validity of using the single-nucleon response
extrapolated from the Fermi gas, as this approach yields the same
results as using a response averaged with a nuclear momentum
distribution that does not have a maximum momentum. Secondly, they
justify the use of the averaged response as a means of avoiding the
potential issues that we have identified with the extrapolation
method.

\begin{figure}
\includegraphics[width=12cm,bb=40 275 530 775]{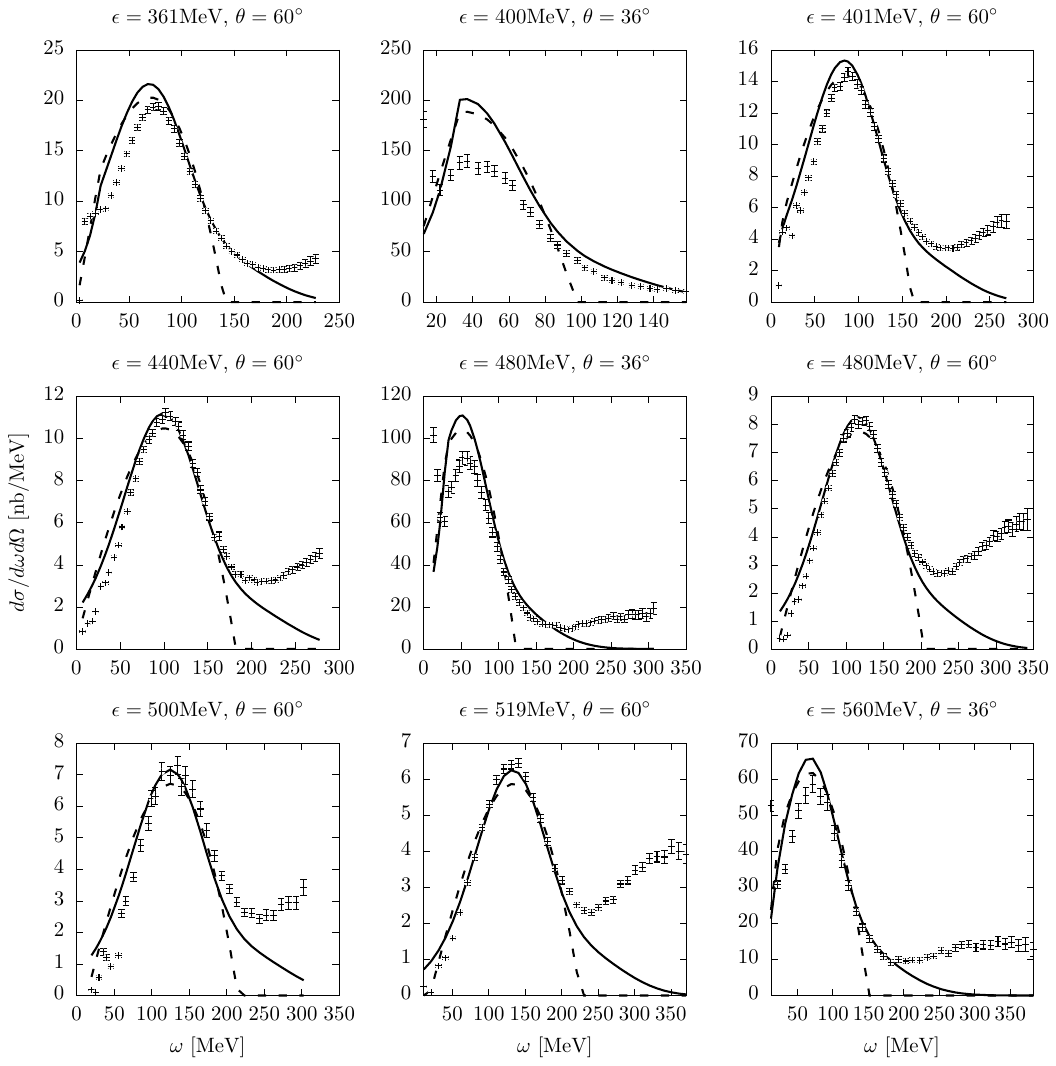}
  \caption{Quasielastic $(e,e')$ cross section of $^{12}C$ as a
    function of $\omega$ for several values of the electron energy,
    $\epsilon$, and scattering angles $\theta$, computed with the
    present SuSAM* model (black lines) compared to the RFG with
    effective mass (dashed lines). Experimental data are from
    refs. \cite{archive,archive2}}
  \label{qfij12}
\end{figure}

\begin{figure}
\includegraphics[width=12cm,bb=40 275 530 775]{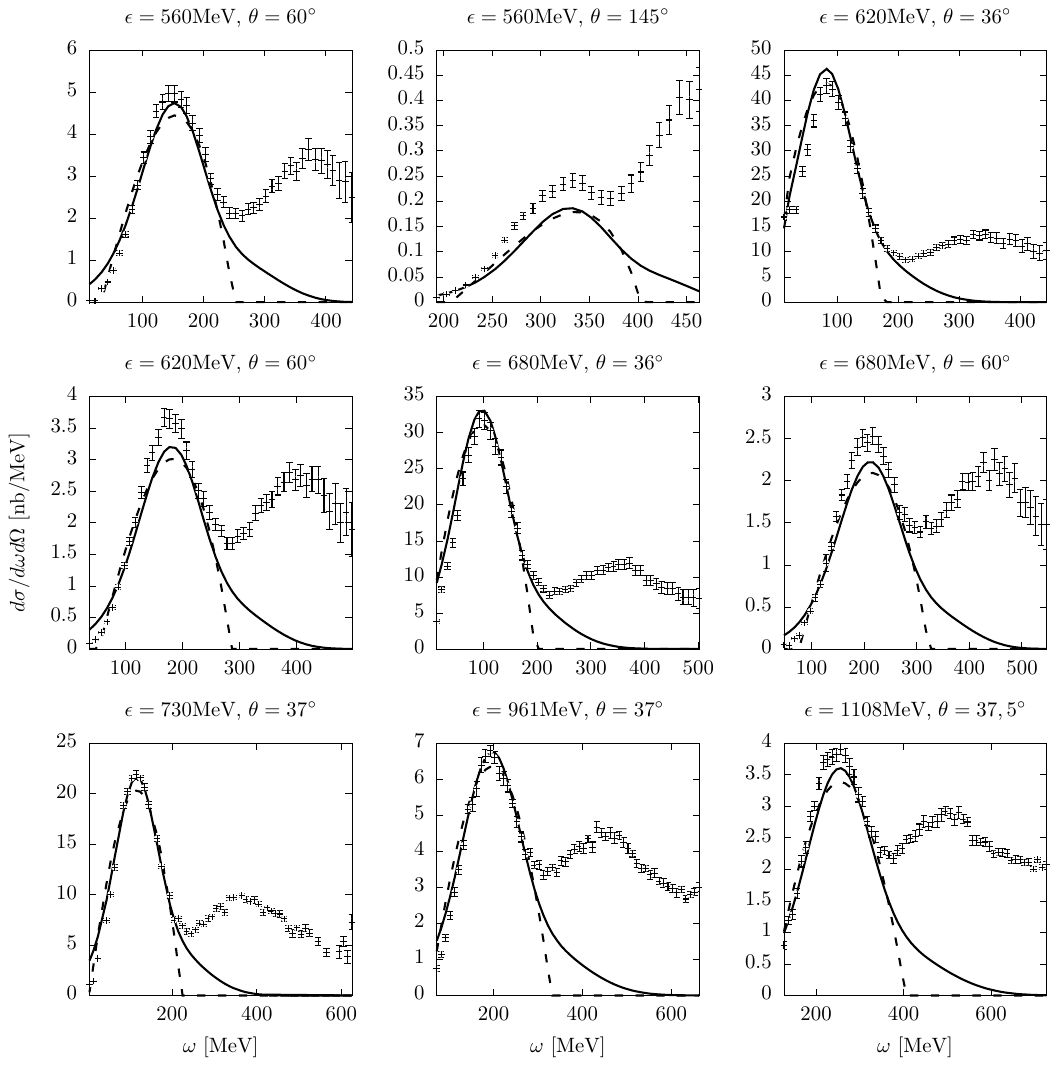}
  \caption{Quasielastic $(e,e')$ cross section of $^{12}C$ as a
    function of $\omega$ for several values of the electron energy,
    $\epsilon$, and scattering angles $\theta$, computed with the
    present SuSAM* model (black lines) compared to the RFG with effective
    mass (dashed lines). Experimental data are from refs. \cite{archive,archive2}}
  \label{qfij13}
\end{figure}

Finally we have conducted a new scaling analysis of the $^{12}C$ data
using the single-nucleon response averaged with the Fermi
distribution. The results, as shown in Figure \ref{fig0},
demonstrate that the scaling function obtained using this approach is
virtually indistinguishable from the one obtained through
extrapolation. These findings highlight the robustness of the scaling
approach and suggest that using the averaged response may be a viable
alternative to extrapolation in certain cases.  Furthermore, in
Figures \ref{qfij12} and \ref{qfij13}, we compare the cross section
of $^{12}$C using the SuSAM* model and the RMF model of nuclear matter
for a selected set of kinematics. The SuSAM* model still proves to be
an excellent method to parameterize the quasielastic cross-section
through a single scaling function.

\section{Final remarks}

The findings found in this chapter demonstrate the robustness and
versatility of the superscaling models with respect to the choice of
the averaged single-responses, and its potential applications in a
variety of situations in electron and neutrino scattering.  The
updated single-nucleon responses provide a well-defined
theoretical basis for the scaling function that is compatible with the
traditional extrapolation in the scaling region. This reinforces the
universality of the scaling function because it is independent of the
way in which the average response of the nucleon is defined. This
means that the scaling function can be used to describe the
electromagnetic response of nucleons in different types of nuclei,
regardless of their size or composition.

The averaged single-nucleon model has promising applications in other
situations outside the scaling region for high-energy transfer. For
instance in two-particle emission reactions, two-particle two-hole
(2p2h) excitation can be produced by the one-body current due to
nuclear short-range correlations.  The electromagnetic interaction
with a nucleon belonging to a correlated pair can result in the
emission of both nucleons because the correlated nucleons acquire
high-momentum components that allow the overlap of the wave function
with states above the Fermi momentum.  A simple model of emission of
two correlated nucleons has been proposed in ref. \cite{Mar22} to
explain phenomenologically the tail of the scaling function at high
energies.

Another direct application of this method concerns the calculation of
the contribution of MEC to the quasielastic
1p1h response in the superscaling model. This calculation was
performed in the RFG for instance in Refs. \cite{Ama01,Ama03} and
involves computing an effective one-body current as the sum of
one-body plus MEC, 1p1h matrix elements.  The traditional scaling
model with extrapolation is not trivial to apply in this case, as the
single-nucleon responses of the MEC must be computed
numerically. However, the averaged single-nucleon responses of the
OB+MEC operator can be directly computed as we have done in this work.
The analysis of this application constitutes the main content of the
next chapter.

\chapter{Improved Superscaling model with Meson Exchange Currents}
\label{art2}

In this chapter, the effect of meson exchange currents will be
incorporated consistently into the framework of the relativistic
effective mass superscaling model that was discussed in chapter 2. See
Ref. \cite{Cas23} for more details.

Until now, an unified model that incorporates 1p1h MEC in the
superscaling function had not been proposed. This was primarily due to
the violation of scaling properties by MEC, even at the Fermi gas
level \cite{Ama02}. Additionally, the 1p1h matrix element of MEC is
not easily extrapolated to the $|\psi|>1$ region outside the range
where the Fermi gas response is zero, as nucleons are constrained by
the Fermi momentum. In this chapter, we address both of these points
by using the new approach where the single nucleon response is
averaged with a smeared momentum distribution around the Fermi
surface, defined in Eq. (\ref{distribucion}). As a result, the
averaged single nucleon responses are well defined for all the values
of $\psi$.

Let us begin by recalling the formalism of QE electron scattering
within the RFG model of nuclear matter, now incorporating the MEC
effect. In the independent particle models, the main contribution to
the hadronic tensor in the quasielastic peak comes from the
one-particle one-hole (1p1h) final states. As the transferred energy
increases, there are contributions from two-particle two-hole (2p2h)
emission, the inelastic contribution of pion emission above the pion
mass threshold, and the deep inelastic scattering at higher
energies. Therefore, the hadronic tensor can be generally decomposed
as the sum of the 1p1h contribution and other contributions:
\begin{equation}
W^{\mu\nu}=W^{\mu\nu}_{1p1h}+ W^{\mu\nu}_{2p2h}+ \ldots
\end{equation}

\begin{figure}[t]
\centering
\includegraphics[width=10cm,bb=120 460 495 700]{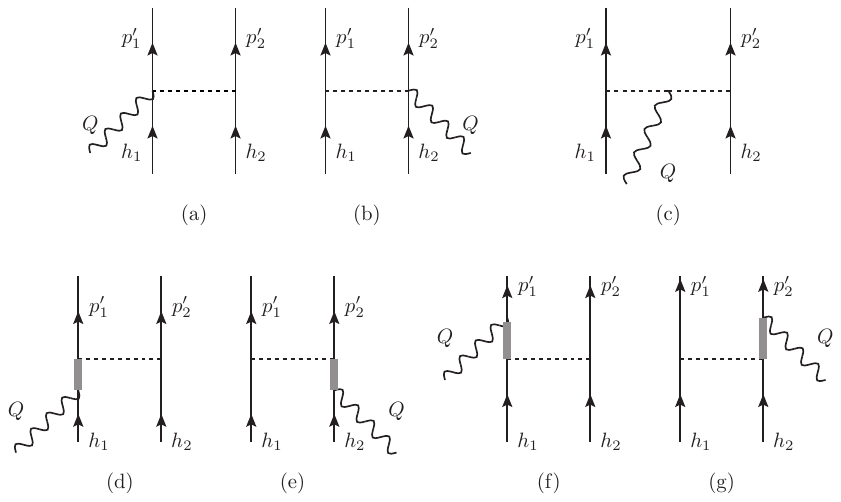}
\caption{Feynman diagrams for the 2p2h MEC model used in 
this work.}\label{feynman}
\end{figure}

We focus on the 1p1h response which, in the RFG model, reads
\begin{eqnarray}
W^{\mu\nu}_{1p1h}&=& \sum_{ph}
\left\langle
ph^{-1} \right|\hat{J}^{\mu} |\left. F \right\rangle^{*}
\left\langle
ph^{-1} \right|\hat{J}^{\nu} |\left. F \right\rangle 
 \delta(E_{p}-E_{h}-\omega)
\theta(p-k_F)\theta(k_F-h)
\label{hadronicc2}
\end{eqnarray}
where $|p\rangle \equiv |\np s_p t_p\rangle$ and $|h\rangle \equiv
|\nh s_h t_h\rangle$ are plane wave states for particles and holes,
respectively, and $|F\rangle$ is the RFG ground state with all momenta
occupied below the Fermi momentum $k_F$.  The novelty compared to
previous chapter is that we start from a current operator
that is a sum of one-body and two-body operators. This approach allows
us to consider the contributions of both the usual electromagnetic
current of the nucleon and the meson-exchange currents to
the 1p1h response:
\begin{equation}
\hat{J}^\mu = 
\hat{J}^\mu_{1b} 
+\hat{J}^\mu_{2b},
\end{equation}
where $\hat{J}_{1b}$ represents the OB electromagnetic current
of the nucleon and $\hat J_{2b}$ is the two-body MEC.  Both currents
can generate non-zero matrix elements for 1p1h excitation. MEC are
two-body operators and they can induce 1p1h excitation due to the
interaction of the hit nucleon with a second nucleon acting as a
spectator.  The many-body matrix elements of these operators are given
by
\begin{equation}
\left\langle ph^{-1} \right|\hat{J}_{1b}^{\mu} |\left. F \right\rangle
=
\left\langle p \right|\hat{J}_{1b}^{\mu} |\left. h \right\rangle
\end{equation}
for the OB current and
\begin{equation}
\left\langle ph^{-1} \right|\hat{J}_{2b}^{\mu} |\left. F \right\rangle 
=
\sum_{k<k_F}\left[
\left\langle pk \right|\hat{J}_{2b}^{\mu} |\left. hk \right\rangle 
- \left\langle pk \right|\hat{J}_{2b}^{\mu} |\left. kh \right\rangle
\right]
\label{melementc2}
\end{equation}
for the two-body current, where the sum over 
spectator states $(k)$ is performed over the occupied states 
in the Fermi gas, considering both the direct and exchange matrix elements.
Due to momentum conservation, the  matrix element of the OB current
between plane waves can be written as
 \begin{equation}
 \langle p |\hat{J}_{1b}^{\mu} | h\rangle =
  \frac{(2\pi)^{3}}{V}\delta^{3}(\nq+\nh-\np)
\frac{m_{N}}{\sqrt{E_{p}E_{h}}}
j_{1b}^{\mu}(\np,\nh), 
\label{OBmatrixc2}
\end{equation}
where $V$ is the volume of the system, $m_{N}$ is the nucleon mass,
$E_p=\sqrt{p^2+m_N^2}$ and
$E_h=\sqrt{h^2+m_N^2}$ are the on-shell energies 
of the nucleons involved
in the process, and
 $j_{1b}^{\mu}(\np,\nh)$ is the OB current (spin-isospin) matrix
\begin{equation} \label{1bcur}
  j^{\mu}_{1b}(\np,\nh)
=\bar{u}(\np)
\left(F_{1}\gamma^{\mu}+i\frac{F_{2}}{2m_{N}}\sigma^{\mu\nu}Q_{\nu}
\right)u(\nh),
\end{equation}
being
$F_{1}$ and $F_{2}$ 
the Dirac and Pauli form factors of the
nucleon.  In the case of the two-body current, the elementary 
matrix element can
be written in a similar form:
\begin{eqnarray}
\kern-1cm
\langle p'_{1}p'_{2}|\hat{J}_{2b}^{\mu}|p_{1}p_{2}\rangle &=&
\frac{(2\pi)^{3}}{V^{2}}\delta^{3}(\np_1+\np_2+\nq-\np'_1-\np'_2)
\frac{m_{N}^{2}}{\sqrt{E'_1E'_2E_1E_2}}
j_{2b}^{\mu}(\np'_1,\np'_2,\np_1,\np_2).
\label{two-body-matrix}
\end{eqnarray}
Here $j_{2b}^{\mu}(\np'_1,\np'_2,\np_1,\np_2)$ is a spin-isospin matrix
and it depends on the momenta of the two nucleons in the initial and
final state.  The two-body current contains the sum of the diagrams
shown in Figure \ref{feynman}, including the seagull, pionic, and
$\Delta$ isobar currents.  The specific form of the two-body current
function will be given later when we discuss the MEC model.  By
inserting (\ref{two-body-matrix}) into Eq. (\ref{melementc2}) we obtain
an expression similar to (\ref{OBmatrixc2}) that resembles the matrix
element of an effective OB current for the MEC:
 \begin{equation}
\left\langle ph^{-1} \right|\hat{J}_{2b}^{\mu} |\left. F \right\rangle 
=
  \frac{(2\pi)^{3}}{V}\delta^{3}(\nq+\nh-\np)
\frac{m_{N}}{\sqrt{E_{p}E_{h}}}
j_{2b}^{\mu}(\np,\nh).
\end{equation}
The effective OB current generated by the MEC involves a sum over
the spectator nucleons and is defined  by
\begin{eqnarray}
j_{2b}^{\mu}(\np,\nh) 
&\equiv &
\sum_{k<k_F}
\frac{m_{N}}{VE_k}
\left[ j_{2b}^{\mu}(\np,\nk,\nh,\nk)-j_{2b}^{\mu}(\np,\nk,\nk,\nh)\right] .
\label{effectiveOBc1}
\end{eqnarray}
Note that in the  thermodynamic limit 
$V \rightarrow \infty$ the above sum will be transformed into 
 an integral over the momenta occupied in the Fermi gas:
\begin{equation}
\frac{1}{V}\sum_{k<k_F}
\rightarrow 
\sum_{s_kt_k}\int \frac{d^3k}{(2\pi)^3} \theta(k_F-k) .
\end{equation}
Finally, we can write the transition matrix element of the total
current between the ground state and the 1p1h state as
 \begin{equation}
\left\langle ph^{-1} \right|\hat{J}^{\mu} |\left. F \right\rangle 
=
  \frac{(2\pi)^{3}}{V}\delta^{3}(\nq+\nh-\np)
\frac{m_{N}}{\sqrt{E_{p}E_{h}}}
j^{\mu}(\np,\nh), 
\label{totalc2}
\end{equation}
where the effective total current for the 1p1h excitation 
includes contributions from both the OB current and MEC:
\begin{equation}
j^{\mu}(\np,\nh)= j_{1b}^{\mu}(\np,\nh)+ j_{2b}^{\mu}(\np,\nh). 
\end{equation}
By inserting (\ref{totalc2}) into Eq. (\ref{hadronicc2}) and taking the
thermodynamic limit, we obtain the following expression for the
hadronic tensor:
\begin{eqnarray}
W^{\mu\nu}&=&
\frac{V}{(2\pi)^{3}}\int{d^3h\delta(E_{p}-E_{h}-\omega)
\frac{m_{N}^{2}}{E_{p}E_{h}}}2w^{\mu\nu}(\np,\nh) 
\theta(p-k_{F})\theta(k_{F}-h),\label{integralw}
\end{eqnarray}  
where $\np=\nh+\nq$ by momentum conservation after integration over $\np$.
The  function $w^{\mu\nu}$ is the effective single-nucleon hadronic tensor 
in the transition 
\begin{eqnarray}
w^{\mu\nu}(\np,\nh)=\frac{1}{2}\sum_{s_ps_h}
j^\mu(\np,\nh)^*j^\nu(\np,\nh) .
\end{eqnarray}
 In this equation, we did not include the sum over isospin $t_p=t_h$.
 Therefore, $w^{\mu\nu}$ refers to the tensor of either proton or
 neutron emission, and the total tensor would be the sum of the two
 contributions.  Note that the effective single-nucleon tensor
 $w^{\mu\nu}$ includes the contribution of MEC, thus encompassing an
 interference between the one-body and two-body currents.  Indeed, the
 relevant diagonal components of the effective single-nucleon hadronic
 tensor for the longitudinal and transverse responses can be expanded
 as
\begin{eqnarray}
w^{\mu\mu}(\np,\nh)
&=& \frac{1}{2}\sum_{s_ps_h} |j^\mu_{1b}+j^\mu_{2b}|^2
\nonumber\\
&=& 
\frac{1}{2}\sum |j^\mu_{1b}|^2
+\mbox{Re}\sum (j^\mu_{1b})^*j^\mu_{2b}
+\frac{1}{2}\sum |j^\mu_{2b}|^2
\nonumber\\
&\equiv&
w^{\mu\mu}_{1b}+ w^{\mu\mu}_{1b2b}+w^{\mu\mu}_{2b},
\label{snucleonc2}
\end{eqnarray}
where \(w^{\mu\mu}_{1b}\) is the tensor corresponding to the one-body
current, \(w^{\mu\mu}_{1b2b}\) represents the interference between the
one-body and two-body currents, and \(w^{\mu\mu}_{2b}\) corresponds to
the contribution of the two-body current alone.  The one-body part is
the leading contribution in the quasielastic peak, while the dominant
contribution of the MEC corresponds to the interference with the
one-body current \cite{Ama94,Ama03}, being the pure contribution of
the two-body current generally smaller.

\section{Meson-exchange currents}

In this section, we present the relativistic meson exchange currents
model developed in Ref. \cite{Rui17}. The Feynman diagrams shown
in Fig. \ref{feynman} illustrate the different components of the MEC
model. Diagrams (a) and (b) correspond to the seagull current, diagram
(c) represents the pion-in-flight current, and diagrams (d,e) and
(f,g) depict the forward- and backward- $\Delta(1232)$ currents,
respectively. The specific treatment of the $\Delta$ current is
model-dependent, and various versions exist with possible corrections
to the off-shell relativistic interaction of the
$\Delta$. While these different models may exhibit slight variations and
corrections to the $\Delta$ off-shell interaction, they generally yield
similar results for the dominant transverse response at the
quasielastic peak. In particular, in the results section, we will
compare our findings with the model presented in Refs. \cite{Pas95,Ama03},
which was previously employed to assess the impact of MEC on the 1p1h
response.

In our model the MEC functions defined in Eq. (\ref{two-body-matrix})
correspond to the sum of diagrams of Fig. \ref{feynman}
\begin{equation}
j_{2b}^\mu(\np'_1,\np'_2,\np_1,\np_2) 
=  j^{\mu}_{sea}+  j^{\mu}_{\pi}+  j^{\mu}_{\Delta},
\end{equation}
where the $\Delta$ current is the sum of forward and backward terms
\begin{equation}
 j^{\mu}_{\Delta}= j^{\mu}_{\Delta F}+ j^{\mu}_{\Delta B}.
\end{equation}
These functions are defined by
\begin{eqnarray} \label{2p2h}
  j^{\mu}_{sea}
  &=&
  \kern -0.2cm
i[\ntau^{(1)} \times \ntau^{(2)}]_z
\frac{f^{2}}{m_{\pi}^{2}}
V_{s'_{1}s_{1}}(p'_{1},p_{1})
F_{\pi NN}(k_{1}^{2})
\bar{u}_{s'_2}(p_{2}^{'})F_{1}^{V}\gamma^{5}\gamma^{\mu}u_{s_2}(p_{2}) 
+ (1 \leftrightarrow 2) \label{mec1}
\\
  j^{\mu}_{\pi}
&=& i[\ntau^{(1)} \times \ntau^{(2)}]_z
\frac{f^{2}}{m_{\pi}^{2}}F_{1}^{V}
 V_{s'_{1}s_{1}}(p'_{1},p_{1})
 V_{s'_{2}s_{2}}(p'_{2},p_{2})(k_{1}^{\mu}-k_{2}^{\mu}) 
\\  
  j^{\mu}_{\Delta F}
&=&
U_{F}(1,2)_z
\frac{ff^{*}}{m_{\pi}^{2}}
V_{s'_{2}s_{2}}(p'_{2},p_{2})F_{\pi N \Delta}(k_{2}^{2})
\bar{u}_{s'_1}(p'_{1})k_{2}^{\alpha}G_{\alpha\beta}(p_{1}+Q)
\Gamma^{\beta\mu}(Q)u_{s_1}(p_{1})  \nonumber \\
&&
+ (1 \leftrightarrow 2)
\label{delta1} 
\\
  j^{\mu}_{\Delta B}
&=&
U_{B}(1,2)_z
\frac{ff^{*}}{m_{\pi}^{2}}
V_{s'_{2}s_{2}}(p'_{2},p_{2})
F_{\pi N \Delta}(k_{2}^{2})
\bar{u}_{s'_1}(p'_{1})k_{2}^{\beta}
\hat{\Gamma}^{\mu\alpha}(Q)G_{\alpha\beta}(p'_{1}-Q)u_{s_1}(p_{1}) \nonumber \\
&&
+ (1 \leftrightarrow 2)  
\label{delta2} 
\end{eqnarray}
We will evaluate these matrix elements in the framework of the RMF
model, where the spinors \(u(p)\) are the solutions of the Dirac equation
with relativistic effective mass \(m_N^*\).  The four-vectors
$k_{i}^{\mu}=p'_{i}{}^{\mu}-p_{i}^{\mu}$ with $i=1,2$ are the momenta
transferred to the nucleons 1,2.  We have defined the following
function that includes the $\pi NN$ vertex, a form factor, and the
pion propagator
\begin{equation} \label{Vfun}
  V_{s'_{1}s_{1}}(p'_{1},p_{1})
= F_{\pi NN}(k_{1}^{2})
\frac{ \bar{u}_{s'_{1}}(p'_{1})\gamma^{5}\slashed{k}_{1}u_{s_{1}}(p_{1})}{k_1^2-m_{\pi}^2}.
\end{equation}
We apply strong form factors at the pion absorption/emission 
vertices given by  \cite{Alb84,Som78} 
\begin{equation}
  F_{\pi NN}(k^2)=   F_{\pi N\Delta}(k^2)
=\frac{\Lambda^{2}-m_{\pi}^{2}}{\Lambda^{2}-k^{2}}.
\label{pinnff}
\end{equation}
The charge structure of the $\Delta$ current involves the isospin matrix element
of the operators
\begin{eqnarray}
 U_{F}(1,2)_z 
&=&
 \sqrt{\frac{3}{2}}\sum_{i=1}^{3}{(T_{i}^{(1)}T_{z}^{(1)\dagger}) \otimes \tau_{i}^{(2)}},
 \label{ufc1}
\\ 
  U_{B}(1,2)_z 
  &=& \sqrt{\frac{3}{2}}\sum_{i}^{3}{(T_{z}^{(1)}T_{i}^{(1)\dagger}) \otimes \tau_{i}^{(2)}},
  \label{ubc1}
\end{eqnarray}
where we denote by $T^{\dagger}_i$ the Cartesian coordinates of the
$\frac{1}{2} \rightarrow \frac{3}{2}$ transition isospin
operator. This transition can be represented using the Clebsch-Gordan
coefficients,
\begin{equation}
\textstyle
\langle \frac32 t_\Delta | T^\dagger_\mu | \frac12 t \rangle
= \langle \frac12 t 1 \mu | \frac32 t_\Delta \rangle
\end{equation}  
with $T^\dagger_\mu$ being the spherical components of the vector
$\vec{T}^\dagger$.  With the aid of the expression
$T_{i}T^{\dagger}_{j}=(2/3)\delta_{ij}-\frac{i}{3}\epsilon_{ijk}\tau_{k}$
and making the summation, we can rewrite the isospin operators in the
forward and backward $\Delta$ current as
\begin{align}
  U_{F}(1,2)_z&=\sqrt{\frac{3}{2}}\left(\frac{2}{3}\tau_{z}^{(2)}-\frac{i}{3}\left[\ntau^{(1)}\times\ntau^{(2)}\right]_{z}\right)\\
 U_{B}(1,2)_z&=\sqrt{\frac{3}{2}}\left(\frac{2}{3}\tau_{z}^{(2)}+\frac{i}{3}\left[\ntau^{(1)}\times\ntau^{(2)}\right]_{z}\right).    
\end{align}
The $\gamma N\rightarrow \Delta$ transition vertex in the
forward $\Delta$ current is defined as \cite{Lle72,Her07}
\begin{equation} \label{gammabetamu}
  \Gamma^{\beta\mu}(Q)=
\frac{C_3^V}{m_N}
(g^{\beta\mu}\slashed{Q}-Q^{\beta}\gamma^{\mu})\gamma_5
\end{equation}
while for the backward $\Delta$ current
\begin{equation}
 \hat{ \Gamma}^{\mu\alpha}(Q)=\gamma^{0}[\Gamma^{\alpha\mu}(-Q)]^{\dagger}\gamma^{0}.
\end{equation}
In this vertex we neglect the contributions of order $O(1/m_N^2)$.
Note that the $\Gamma^{\beta\mu}$ operator is a spin matrix 
and depends on the vector form factor $C_3^V$. 
In our work, we use the vector form factor in $\Delta$ current from
Refs. \cite{Her07}:
\begin{equation}
  C_{3}^{V}(Q^{2})
=\frac{2.13}{(1-\frac{Q^{2}}{M_{V}^{2}})^{2}}
\frac{1}{1-\frac{Q^{2}}{4M_{V}^{2}}} .
\end{equation}
Various alternative
approximations to the propagator have been proposed \cite{Qui17}.
 However, in the
case of the quasielastic peak, the typical kinematics are  of
the order
of 1 GeV, and these issues are not expected to be relevant. They are
overshadowed by other more significant nuclear effects that dominate
in this energy regime.
Here we use the $\Delta$ propagator commonly used for the spin-3/2 field
\begin{equation}
  G_{\alpha\beta}(P)=
\frac{{\cal P}_{\alpha\beta}(P)}{
P^{2}-M_{\Delta}^{2}+iM_{\Delta}\Gamma(P^{2})+\frac{\Gamma(P^{2})^{2}}{4}}
\end{equation} 
where $M_{\Delta}$ and $\Gamma$ are the $\Delta$ mass and width
respectively. The projector ${\cal P}_{\alpha\beta}(P)$ over spin-3/2 on-shell
particles is given by
\begin{eqnarray}
{\cal  P}_{\alpha\beta}(P)
&=&
-(\slashed{P}+M_{\Delta})
\left[
g_{\alpha\beta}-\frac{\gamma_{\alpha}\gamma_{\beta}}{3}-\frac{2P_{\alpha}P_{\beta}}{3M_{\Delta}^{2}}+\frac{P_{\alpha}\gamma_{\beta}-P_{\beta}\gamma_{\alpha}}{3M_{\Delta}}
\right] .
\end{eqnarray}  
Finally, the 
$\Delta$ width $\Gamma(P^{2})$ is given by
\begin{equation}
  \Gamma(P^{2})=\Gamma_{0}\frac{m_{\Delta}}{\sqrt{P^{2}}}
\left(\frac{p_{\pi}}{p^{res}_{\pi}}\right)^{3}.
\label{widthc1}
\end{equation}
In the above equation, $p_{\pi}$ is the momentum of the final pion
resulting from the $\Delta$ decay an $p^{res}_{\pi}$ is its value at
resonance ($P^2=m_\Delta^2$), and $\Gamma_{0}=120$ MeV is the width at
rest.
The width (\ref{widthc1}) corresponds to the $\Delta$ in a vacuum,
and it is expected to be slightly different in the medium depending on
the kinematics.
 
\begin{figure}
  \centering
\includegraphics[width=10cm,bb=110 460 500 690]{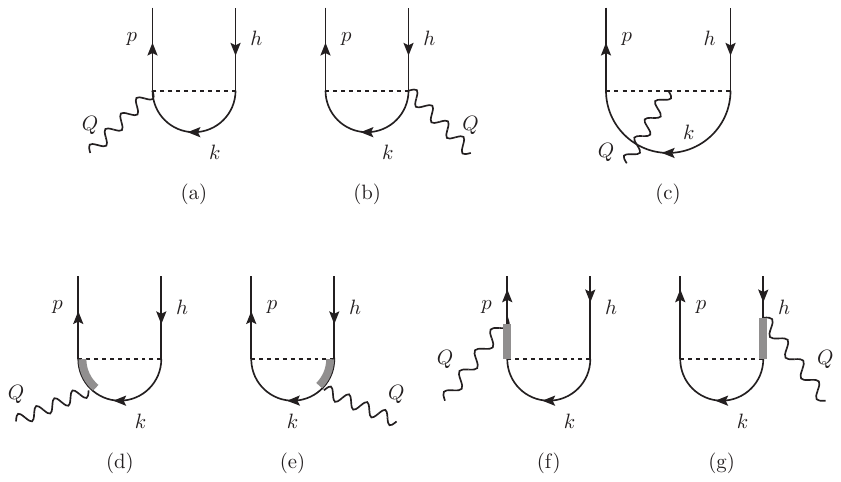}
\caption{Diagrams for the 1p1h MEC matrix elements}
\label{feynman2}
\end{figure}

In the relativistic mean field description used in this work, we
consider that $\Delta$ also interacts with scalar and vector
fields, acquiring an effective mass and vector energy. To treat this
case, we make the following substitutions in the $\Delta$ propagator 
for the $\Delta$ mass and momentum
\cite{Weh93,Kim96}: 
\begin{equation} 
M_\Delta \rightarrow M_\Delta^*,
\kern 1cm
P^{*\mu} = P^\mu - \delta_{\mu0}E_v^\Delta.
\end{equation}
   We use the value $M^*_\Delta=1042$ MeV, taken from
\cite{Mar21b}, and the universal vector coupling $E_v^\Delta=E_v=141$ MeV.

With the MEC model defined in Eqs. (\ref{2p2h})-(\ref{delta2}), the
effective one-body current \(j_{2b}^\mu(\np,\nh)\) is generated by summing
over the spin, isospin and momentum of the spectator nucleon, as in
Eq. (\ref{effectiveOBc1}). Due to the sum over isospin \(t_k\) and
spin, the direct part is zero and two of the four exchange diagrams
contributing to the $\Delta$ current in the 1p1h matrix element are
zero. Specifically, in the forward $\Delta$ current, the diagram
containing the isospin operator $U_F(2,1)$ yields zero (diagram (g) of
Fig. \ref{feynman2}. Similarly, for the backward $\Delta$ current,
the diagram involving the isospin operator $U_B(1,2)$ also vanishes
(diagram (f) of Fig. \ref{feynman2}. Therefore only diagrams (d),
forward, and (e), backward, contribute in the case of the $\Delta$
current. These results are demonstrated in Appendix \ref{appB}.

\section{Results}

In this section, we present results for the effects of MEC on the 1p1h
response functions using several models: RFG, the relativistic Fermi
gas with effective mass (RMF), and the generalized SuSAM* model of
chapter 2. By employing these different models, we take into account
relativity and we can analyze the impact of including the relativistic
effective mass of the nucleon and the $\Delta$ resonance appearing in
the MEC. The scaling analysis described in chapter 2 will allow us to
study the influence of MEC on the generalized scaling function also in
the region $|\psi^*|>1$ where the RFG and RMF responses are
zero. Moreover, we can investigate how the inclusion of MEC affects
the 1p1h response functions and compare it with the predictions of the
RFG and RMF models.

Unless stated otherwise, we present the results for $^{12}$C with a
Fermi momentum of $k_F = 225$ MeV/c. We use an effective mass of $M^*
= 0.8$, following the same choice of parameters as in reference
\cite{Mar21,Mar21b}.  The calculation of 1p1h responses involves
evaluating the 1p1h matrix element of the MEC, as given by
Eq. (\ref{effectiveOBc1}). This requires performing a numerical
three-dimensional integration to account for the momentum
dependence. Subsequently, a one-dimensional integration is carried out
to calculate the averaged single-nucleon responses, as described in
Eq.  (\ref{def_sn}).

First, since this work is an extension of the MEC model from
Ref. \cite{Ama03} to the superscaling formalism, we will compare with
the OB-MEC interference responses presented in \cite{Ama03} within the
framework of the RFG. It should be noted that in \cite{Ama03} a
different version of the $\Delta$ current was used.  The $\Delta$
current was obtained from the \(\gamma N \Delta\) Lagrangian proposed
by Pascalutsa \cite{Pas95}
\begin{equation} \label{pascalutsa}
{\cal L}_{\gamma N \Delta} = ie\frac{G_1}{2m_N}
\overline{\psi}^\alpha \Theta_{\alpha\mu}\gamma_\nu\gamma_5T_3^\dagger 
N F^{\nu\mu} + \mbox{h.c.},
\end{equation}

\begin{figure}[t]
  \centering
\includegraphics[width=7cm,bb=147 475 380 798]{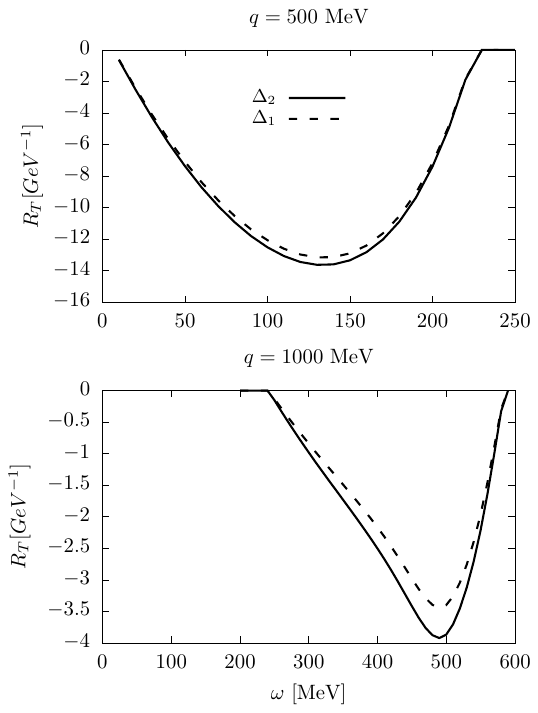}
\caption{Interference OB-MEC in the transverse response of $^{40}$Ca
  for two values of the momentum transfer, with $k_F=237$ MeV/c.  In
  the graph, the curve labeled $\Delta_1$ corresponds to using the $\Delta$
  current of the present work in RFG. The curve $\Delta_2$ corresponds to the
  calculation from reference \cite{Ama03}.}
\label{figpas}
\end{figure}

plus $O(1/m_N^2)$ terms that give negligible contribution 
 in  the quasielastic energy region.  The tensor
$\Theta_{\alpha\mu}$ may contain an off-shell parameter and another
arbitrary parameter related to the contact invariance of the
Lagrangian.
In this work we use the simplest form
\begin{equation}
\Theta_{\alpha\mu}=g_{\alpha\mu}-\frac14\gamma_\alpha\gamma_\mu.
\end{equation}
The coupling constant $G_1$ was determined in \cite{Pas95} by fitting
Compton scattering on the nucleon. However, there is a detail that
needs to be clarified: the isospin operator used by Pascalutsa is
normalized differently from the standard convention. That is,
\(T_i^{\text{Pascalutsa}} = \sqrt{\frac{3}{2}} T_i\), where \(T_i\) is
the operator used in our calculation. This means that if we use the
standard \(T_i\) in the Lagrangian (\ref{pascalutsa}), it should be
multiplied by \(\sqrt{\frac{3}{2}}\). This is equivalent to
multiplying Pascalutsa's coupling constant $G_1=4.2$ by the factor
\(\sqrt{\frac{3}{2}}\). In reference \cite{Ama03}, this detail went
unnoticed, and the $\sqrt{3/2}$ factor was not included in the
calculations.

Using the Lagrangian given by Eq. (\ref{pascalutsa}), the following
$\Delta$ current was obtained in \cite{Ama03}
\begin{eqnarray}
  j^{\mu}_{\Delta F}
&=&
[(T_iT_3^\dagger)\otimes\tau_i]_{t'_1t'_2,t_1t_2}
\frac{ff^{*}}{m_{\pi}^{2}} F_\Delta(Q^2)
V^{s'_{2}s_{2}}_{\pi NN}(p'_{2},p_{2})F_{\pi N \Delta}(k_{2}^{2})
\nonumber\\
&&
\bar{u}_{s'_1}(p'_{1})k_{2}^{\alpha}
\left[
\Theta^{\alpha\beta}G_{\beta\rho}(p_{1}+Q)
\frac{G_1}{2m_N}
[ \Theta^{\rho\mu}\gamma^\nu-\Theta^{\rho\nu}\gamma^\mu]\gamma_5Q_\nu
\right]u_{s_1}(p_{1}) 
+ (1 \leftrightarrow 2), \nonumber \\
\end{eqnarray}
and a similar expression for the $\Delta$ backward current. 
This current was used in Ref. \cite{Ama03} to compute the  OB-MEC interference
with the following form factor
\begin{equation}
  F_\Delta(Q^{2})=
G^{p}_{E}(Q^{2})\left(1-\frac{Q^{2}}{3.5(GeV/c)^{2}}\right)^{-1/2}  
\end{equation}  
where $G_{E}^{p}$ is the electric form factor of the proton.

\begin{figure}
  \centering
  \includegraphics[width=11cm,bb=100 350 450 800]{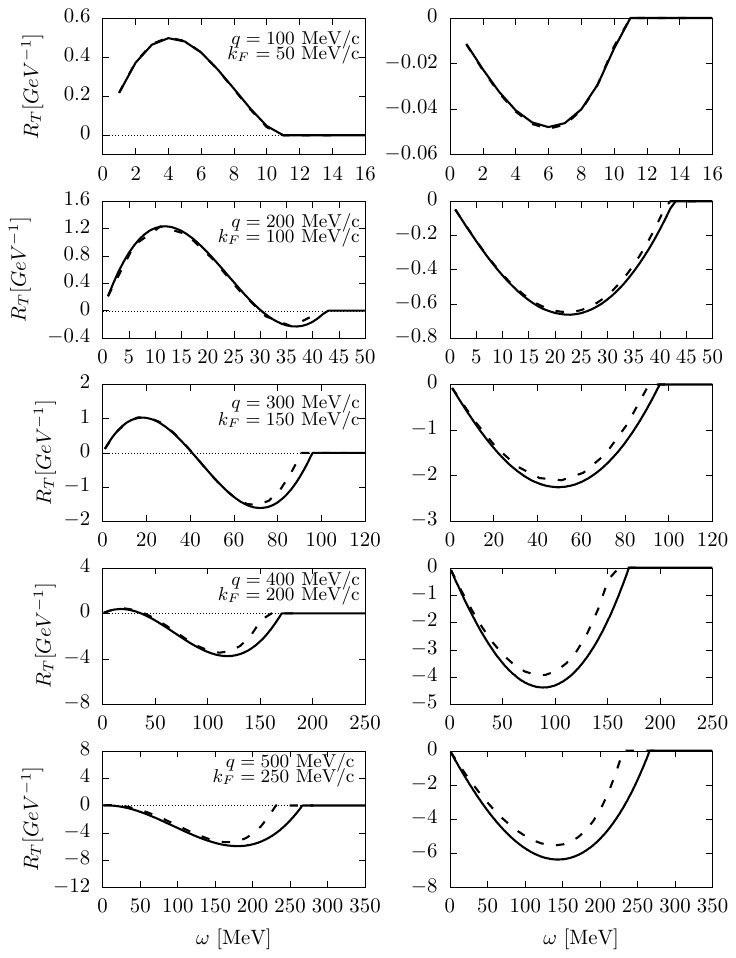}
\caption{Comparison between relativistic and non relativistic MEC
  transverse responses in $^{12}$C. Dashed lines: RFG. Solid lines:
  non-relativistic Fermi gas. Left panels show the interference
  OB-$\pi$ (seagull and pionic), and left panels the interference OB-$\Delta$.
  In these calculations the strong form factors in the pion vertices are set to
  one.}
\label{fig_scaling}
\end{figure}
In Figure \ref{figpas}, we present the interference between the OB and
$\Delta$ currents in the transverse response of $^{40}$Ca. We compare
our results with the model of reference \cite{Ama03} in RFG, where the
Lagrangian of Pascalutsa was used. The results of \cite{Ama03} have
been corrected with the factor of \(\sqrt{\frac{3}{2}}\) mentioned
earlier.  For \(q = 500\) MeV/c, there is little difference between
the two models. However, for \(q = 1\) GeV/c, the difference becomes
more noticeable.

The results of Fig. \ref{figpas} show that the $\Delta$ current
model used in this work does not differ significantly from the model
in reference \cite{Ama03}, providing similar results. The small
differences observed can be attributed to the different form factor
and coupling constants,
and can be understood as a model dependence in these results.
From here on, all the results refer to the $\Delta$ 
current model described in equations (\ref{delta1},\ref{delta2}).

It is expected that any relativistic model should reproduce the
results of the well-established non-relativistic model for small
values of energy and momentum in the non-relativistic limit
\cite{Hok73}.  As a check in this regard, in Fig. \ref{fig_scaling} we
compare the present model with the non-relativistic Fermi gas model
from ref. \cite{Ama94}. The non relativistic $\Delta$ current used is
taken from \cite{Ama03}. This current is similar to the non
relativistic $\Delta$-current of the vector MEC performed in Appendix
\ref{appC}.  To perform this comparison the same form factors and
coupling constants are used in the relativistic and non relativistic
models. To take this limit in Fig. \ref{fig_scaling}, we follow the
procedure as follows: q is small and $k_F=q/2$. We show the comparison
between the two models for various values of $q$ ranging from 100 to
500 MeV/c.  In the left panels, we present the contribution of the
transverse response stemming from the interference OB-$\pi$ (including
seagull and pionic) between the pure pionic MEC (diagrams a–c in
Fig. \ref{feynman2}), and in the right panels, we show the OB-$\Delta$
interference (diagrams d–g in Fig. \ref{feynman2}) for the same values of
$q$.  As expected, we observe that for $q=100$ MeV/c, the relativistic
and non-relativistic models practically coincide, demonstrating the
consistency between the two models in the non-relativistic limit.

\begin{figure}
\centering
\includegraphics[width=8cm,bb=110 270 460 770]{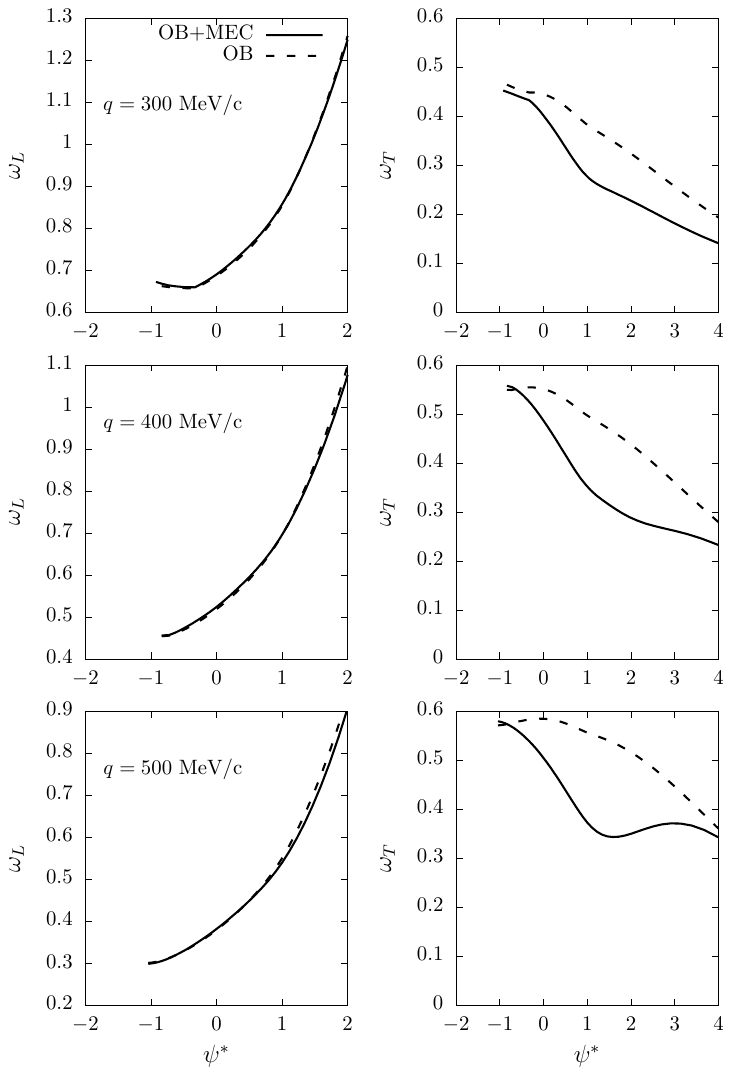}
\kern 0.6mm
\includegraphics[width=8cm,bb=110 270 460 770]{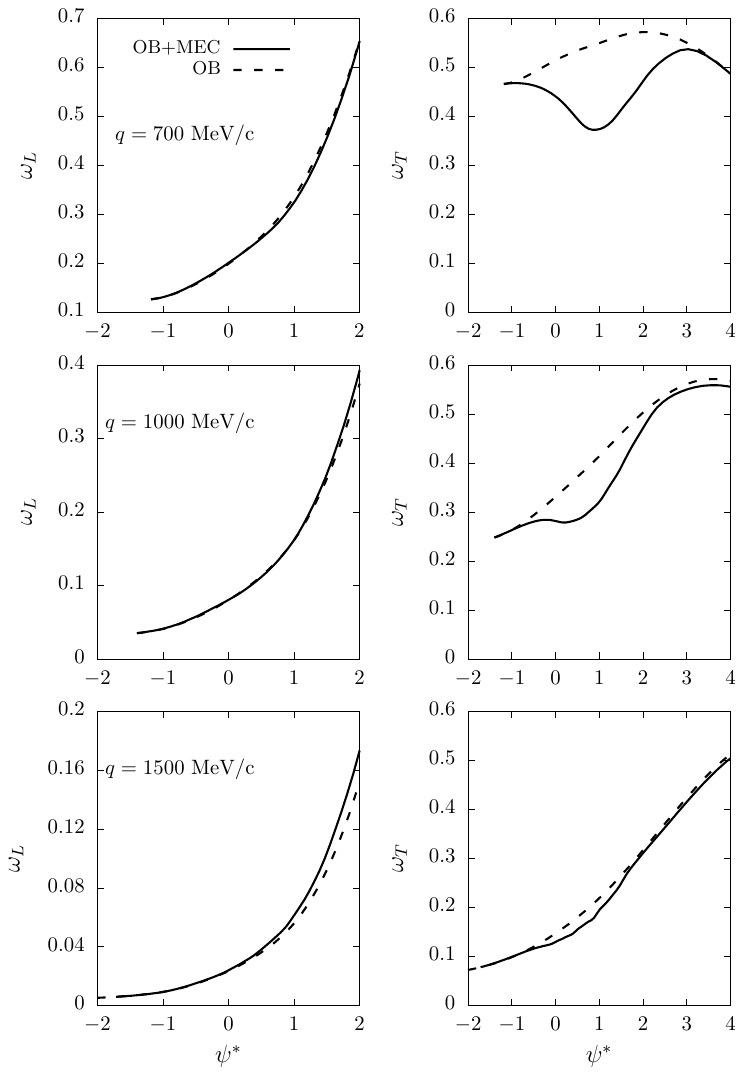}
  \caption{Averaged single nucleon responses computed with and without
    MEC, for several values of the momentum transfer as a function of
    the scaling variable $\psi^*$.  }
\label{fsingle}
\end{figure}

In Fig. \ref{fig_scaling} one can also observe that for low values of \(q\) the
dominant contributions to the MEC are from the seagull and
pion-in-flight diagrams, with the seagull diagram playing a
particularly important role. These diagrams contribute positively to
the MEC, enhancing the overall response. On the other hand, the
contribution from the $\Delta$ resonance is 
 negative. As \(q\) increases, the
influence of the $\Delta$ resonance becomes more significant, and it
starts to dominate the MEC contribution for \(q\) values around 400
MeV/c.

Before performing the scaling analysis, we examine the averaged
single-nucleon responses that will be used to scale the data. In
Figure \ref{fsingle}, we display the longitudinal and transverse effective
single-nucleon responses for various values of \(q\) as a function of
the scaling variable. The calculated responses are shown separately
for the OB current and the total responses including the MEC and
taking into account the sum of protons and neutrons.  The total
response, which we have defined in equation (\ref{susam}), comes from
the product of the single nucleon with the phenomenological scaling
function obtained from the $(e,e')$ data as shown below (note that the
single-nucleon tensor in equation (\ref{susam}) now includes the
contribution from meson exchange currents). We have used
the Fermi distribution, Eq. (\ref{distribucion}), with a smearing parameter
$b=50$ MeV/c, although the single nucleon responses do not depend much
on this specific value.  It is observed that the effect of the MEC is
negligible in the longitudinal response, as the curves for the OB
current and total response overlap. However, in the transverse
response, the effect of the MEC becomes appreciable, resulting in a
reduction of the \(w_T\) response compared to the OB current. This
reduction can be attributed to the interference between the one-body
and two-body currents, which leads to a modified transverse
response. Note that the center of the quasielastic peak corresponds to
\(\psi^*=0\), where the energy and momentum can be transferred to a
nucleon at rest.  We see that MEC have a larger impact in the region
$\psi^*> 0$, that is, the right-hand side of the peak, corresponding
to higher energy transfers.

\begin{figure}
  \centering
\includegraphics[width=15cm,bb=10 630 560 790]{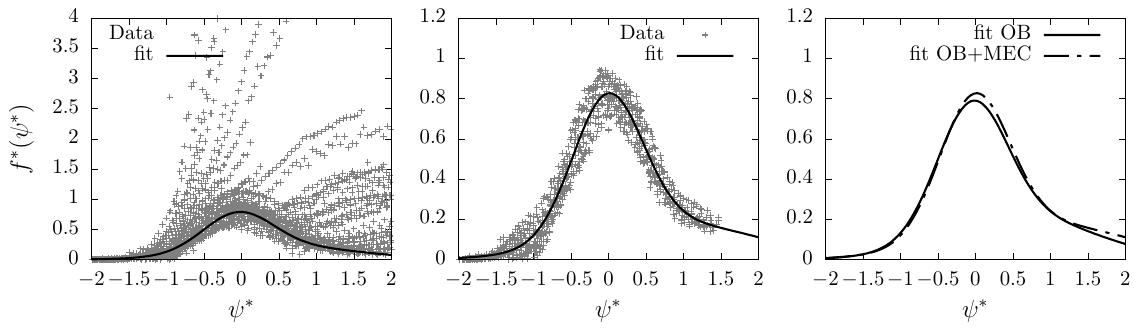}
\caption{Scaling analysis of ${}^{12}$C data including MEC and
  relativistic effective mass $M^{*}=0.8$. The Fermi momentum is
  $k_F=225$ MeV/c.  In the left panel, we show the data points after
  scaling, representing the overall distribution. In the middle panel,
  we display the selected data points, which have been chosen after
  eliminating those that do not exhibit clear scaling behavior.  In
  the right panel, we present the phenomenological scaling function,
  which has been fitted to the selected data points, compared to the
  scaling function obtained in a similar analysis without MEC.
  Experimental data are taken from Refs. \cite{archive,archive2}.  }
\label{fig-scaling}
\end{figure}

In Figure \ref{fig-scaling}, we present the scaling analysis of the
$^{12}$C data using the formalism of scaling explained in the previous
chapter but including MEC in the single-nucleon prefactors. In the
left panel, the experimental data, $f^*_{exp}$, are plotted against
$\psi^*$ in the interval $-2<\psi^*<2$. Experimental data are from
Refs. \cite{archive,archive2} and cover a wide electron energy range,
from 160 MeV up to 5.8 GeV. We observe, as in chapter 2, a significant
dispersion of many data points, indicating a wide range of inelastic
scattering events. However, we also notice that a portion of the data
points cluster together and collapse into a thick band. These data
points can be considered as associated to quasielastic (1p1h)
events. To select these quasielastic data, we apply a density
criterion. For each point, we count the number of points, \(n\),
within a neighborhood of radius \(r=0.1\), and eliminate the point if
\(n\) is less than 25.  Points that have been disregarded are likely
to correspond to inelastic excitations and low energy processes that
violate scaling and cannot be considered within quasielastic
processes. We observe that the remaining selected points, about half
of the total, shown in the middle panel of Fig. \ref{fig-scaling},
form a distinct thick band. These points represent the ones that best
describe the quasielastic region and approximately exhibit scaling
behavior. The black curve represents the phenomenological quasielastic
function $f^*(\psi^*)$, that provides the best fit to the selected
data using a sum of two Gaussian functions:
\begin{equation}
  f^{*}(\psi^{*})=
a_{3}e^{-(\psi^{*}-a_{1})^{2}/(2a_{2}^{2})}
+b_{3}e^{-(\psi^{*}-b_{1})^{2}/(2b_{2}^{2})}.
\end{equation}
The parameters found are shown in table \ref{table:1}.

In the right panel of Fig. \ref{fig-scaling} we compare the scaling
function obtained in our analysis with the scaling function obtained
without including the MEC contributions in chapter 2. When including the MEC, the
scaling function appears slightly higher since the single-nucleon
response with MEC is slightly smaller than without them. However, both
analyses provide a similarly acceptable description of the data. This
suggests that while the MEC do have an impact on the scaling behavior,
their effect is relatively small and does not significantly alter the
overall scaling pattern observed in the data.

\begin{table}[b]
  \centering
  \begin{tabular}{|c|c|c|c|c|c|}
    \hline
$a_{1}$ & $a_{2}$ & $a_{3}$ & $b_{1}$ & $b_{2}$ & $b_{3}$ \\
\hline
$-0.01015$ & 0.46499 & 0.69118  & 0.86952 & 1.16236  & 0.17921 \\ 
\hline
\end{tabular}
\caption{Table of fitted parameters of the scaling function.}
\label{table:1}
\end{table}

\begin{figure}
\centering
\includegraphics[width=16cm,bb=10 550 520 800]{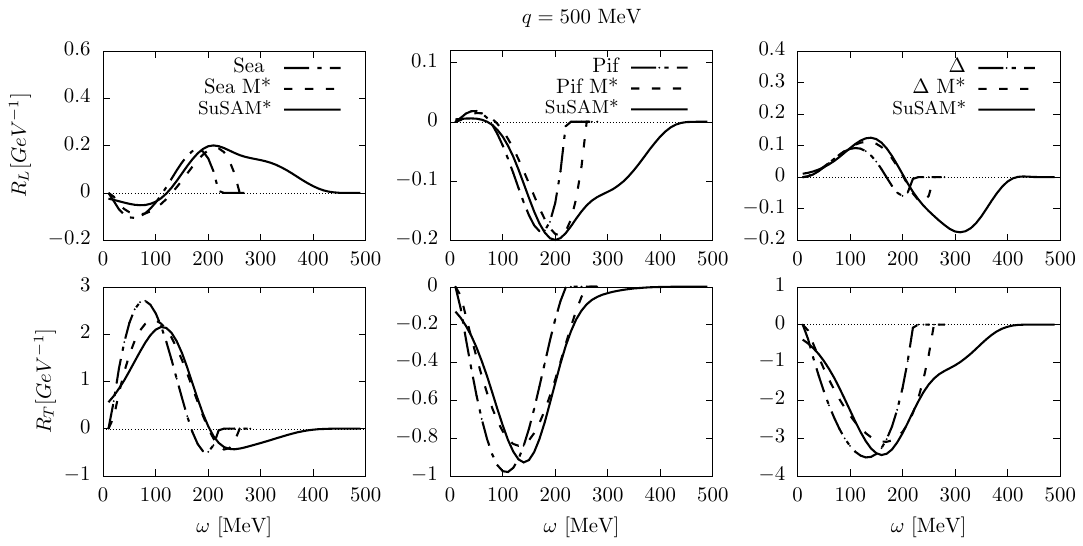}
\caption{Interference OB-MEC responses separated in seagull,
  pion-in-flight, and $\Delta$ contributions for $^{12}$C and $q=500$ MeV/c. In
  each panel we compare the results of RFG (with $M^*=1$,  dot-dashed), with the
  RMF (with $M^*=0.8$, dashed) and the SuSAM* model (solid). }
\label{fig1}
\end{figure}

\begin{figure}
\centering
\includegraphics[width=16cm,bb=10 550 520 800]{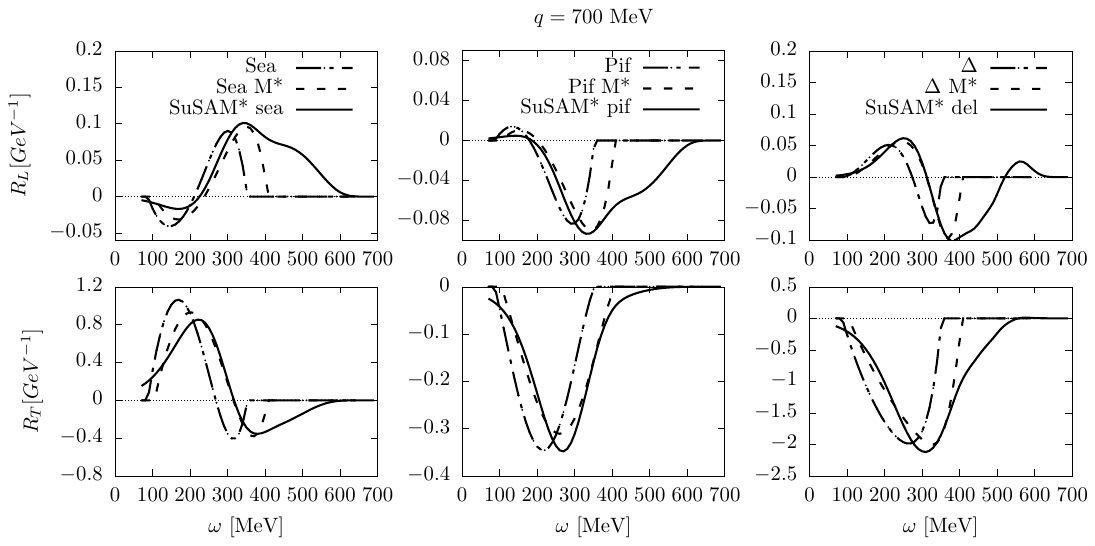}
\caption{
The same as Fig. 7 for $q=700$ MeV/c.
}
\label{fig2}
\end{figure}

\begin{figure}
\centering
\includegraphics[width=16cm,bb=10 550 520 800]{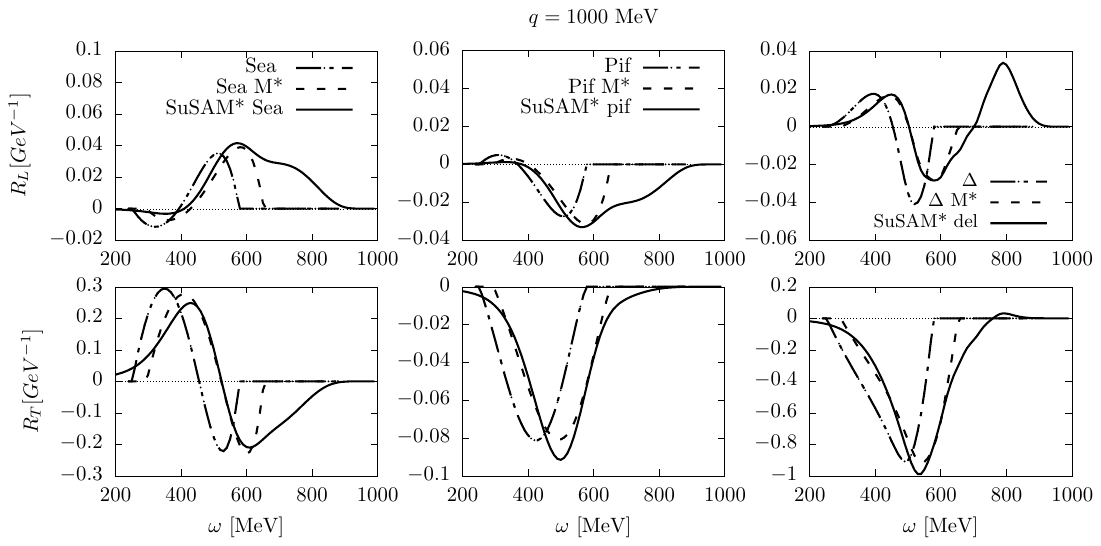}
\caption{The same as Fig. 7 for $q=1000$ MeV/c.
}
\label{fig3}
\end{figure}

\begin{figure}
\centering
\includegraphics[width=16cm,bb=10 550 520 800]{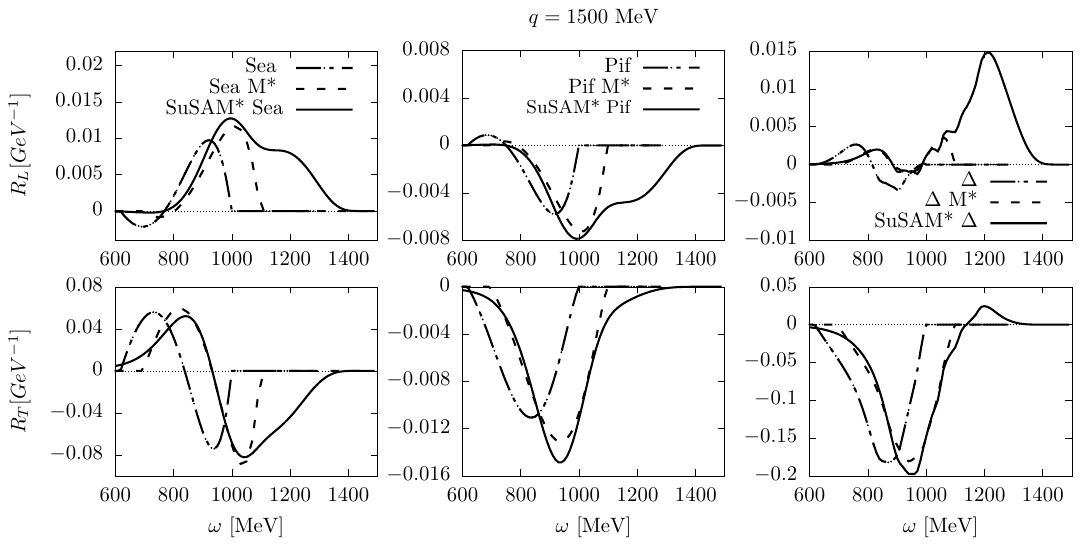}
\caption{The same as Fig. 7 for $q=1500$ MeV/c.
}
\label{figw}
\end{figure}

Now that we have obtained the phenomenological scaling function
through the scaling analysis, we can utilize this function to
calculate the response functions of the model beyond the RMF. By
 multiplying
the scaling function by
the averaged single nucleon
responses, as stated in Eq. (\ref{susam}), we can extend our calculations to
different kinematic regimes and explore the behavior of the responses
beyond the relativistic mean field description. This allows us to
investigate the influence of various factors, such as the MEC and
relativistic effects, on the response functions and cross sections.

In Figures \ref{fig1}-\ref{figw}, we present the interference's of
the OB-MEC in the response functions for different values of \(q\)
(500, 700, 1000, and 1500 MeV/c). We separate the interferences into
OB-seagull, OB-pionic, and OB-$\Delta$ contributions for both the
longitudinal and transverse responses as functions of \(\omega\). Each
panel displays three curves corresponding to the free RFG (with
effective mass \(M^*=1\)), the RMF (with effective mass \(M^*=0.8\)),
and the present SuSAM* model.  These figures allow us to analyze the
relative contributions of the different OB-MEC interferences in the
response functions at various kinematic regimes. By comparing the
results obtained from the RFG, RMF, and SuSAM* models, we can observe
the effects of including the relativistic interaction through the
effective mass and the scaling function on the interferences.

First we observe that the introduction of the effective mass $M^{*}=0.8$
shifts the responses to the right, towards higher energy values.
The effective mass takes into account the binding of
the nucleon in the nucleus, which causes the quasielastic peak to
approximately coincide with the maximum of the experimental cross
section. In the RFG, this is traditionally taken into account by
subtracting a binding energy of approximately 20 MeV from \(\omega\)
to account for the average separation energy of the nucleons. In the
RMF, this is automatically included by considering the effective mass
of the nucleon, \(M^*=0.8\), which was adjusted for \(^{12}\)C
precisely to achieve this effect.

In the transition from the RMF to the SuSAM* model, we replace the
scaling function of the RFG with the phenomenological scaling function
that we have adjusted. This new scaling function extends beyond the
region of $-1 < \psi^* < 1$, where the RFG scaling function is
zero. As a result, we observe in Figures \ref{fig1}-\ref{figw} that
the interferences acquire a tail towards high energies, similar to the
behavior of the scaling function. The tail effect is more pronounced
in the longitudinal responses because the single-nucleon longitudinal
response, as shown in Figure \ref{fsingle}, increases with
$\omega$. This amplifies the tail when multiplied by the scaling
function. However, it is important to note that the contribution of
the MEC to the longitudinal response is relatively small compared to
the dominant transverse response. Therefore, while the tail effect is
observed in the longitudinal responses, its impact on the cross
section is not as significant as in the transverse channel, if not
negligible.

In the dominant transverse response, the seagull contribution from the
MEC is positive, leading to an enhancement of the response, while the
pionic and $\Delta$ contributions are negative, causing a reduction in
the overall response when including the MEC. This is in line with
pioneering calculations by Kohno and Otsuka \cite{Koh81} and by
Alberico {\it et al.} \cite{Alb90} in the non-relativistic Fermi
gas. Moreover, in shell model calculations, similar results have been
obtained \cite{Ama94}. It is worth noting that the relative
importance of the MEC contributions depend on the momentum transfer
$q$ and the energy transfer $\omega$. For the values considered in
Figures \ref{fig1}-\ref{figw}, the $\Delta$ current is found to be the
dominant contribution, leading to a net negative effect from the MEC.

The observation in Fig. \ref{figw} of a sign change and a small bump in the
OB-$\Delta$ transverse response for high values of $\omega$ is indeed
interesting. The change of sign is already observed for $q$=1 GeV/c in
Fig. \ref{fig3}.  This connects with the findings in reference \cite{Ama10},
where a pronounced bump and sign change were reported in a
semi-relativistic shell model calculation based on the Dirac equation
with a relativistic energy-dependent potential.  
In the present calculation the bump is observed but it is very small
compared to the results of Ref, \cite{Ama10}. Note
that, in the present work, the fully relativistic SuSAM* approach is
employed, which takes into account the dynamical properties of both
nucleons and the $\Delta$, as well as the scaling function. This
differs from the approach in reference \cite{Ama10}, where a static
propagator for the $\Delta$ was used.  To definitively clarify the
difference with the present results, a fully relativistic calculation
in finite nuclei, considering the dynamical properties of the $\Delta$
would be necessary.

\begin{figure}
  \centering
  \includegraphics[width=7cm,bb=170 370 380 800]{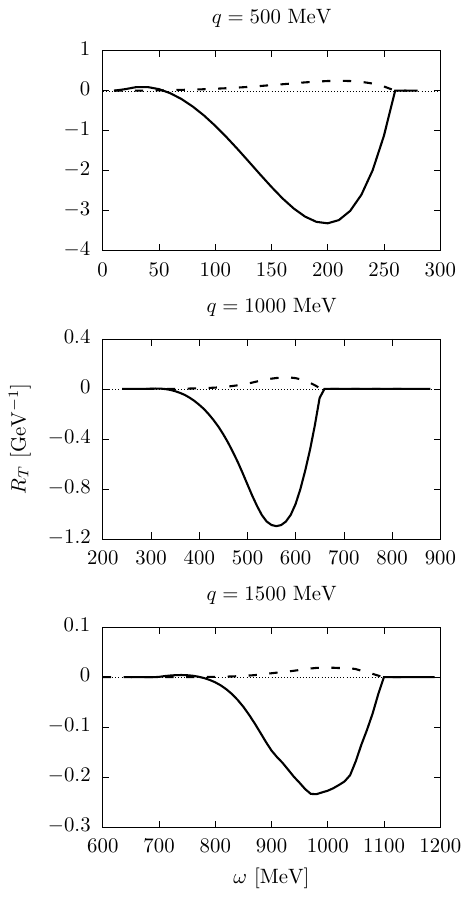}
  \includegraphics[width=7cm,bb=170 370 380 800]{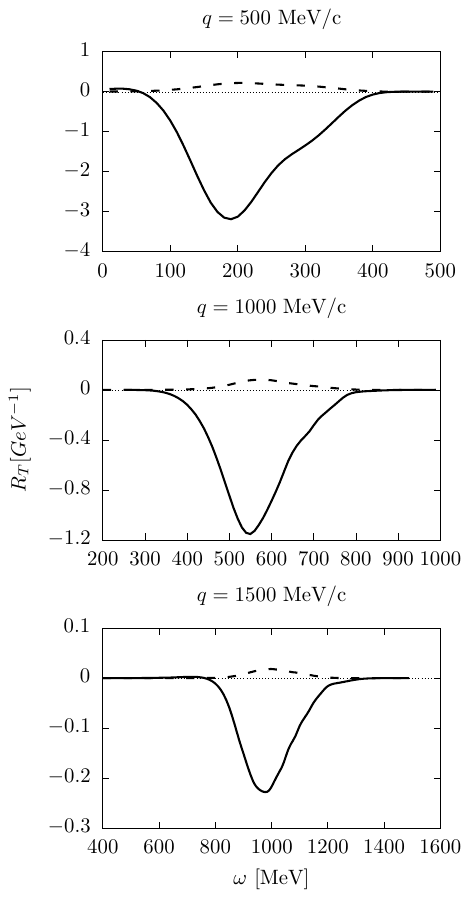}
  \caption{Left panel: Comparison of OB-MEC interference in the
    transverse response (black lines) with the pure MEC transverse
    response (dashed lines) for several values of $q$ in the RMF
    model. Right panel: The same but in the SuSAM* model. }
\label{fig4}
\end{figure}

The comparison of the OB-MEC interference with the MEC contribution
alone (represented by $w_{1b2b}^{\mu\nu}$ and $w_{2b}^{\mu\nu}$,
respectively in Eq. (\ref{snucleonc2}) in the transverse response is shown in
Fig. \ref{fig4}. We observe that the MEC contribution alone represents
a small and almost negligible contribution to the transverse
response. This justifies the previous calculations that focused only
on the OB-MEC interference (e.g., the semi-analytical calculations in
references \cite{Ama94a,Ama94} for the non-relativistic Fermi gas), as
it provides an excellent approximation.  This observation holds true
for both the RMF model and the SuSAM* model in Fig. \ref{fig4}. It
highlights the fact that the dominant contribution to the transverse
response arises from the interference between the OB and MEC, while
the pure MEC contribution is relatively small. 

\begin{figure}
\centering \includegraphics[width=10cm,bb=110 270 460 770]{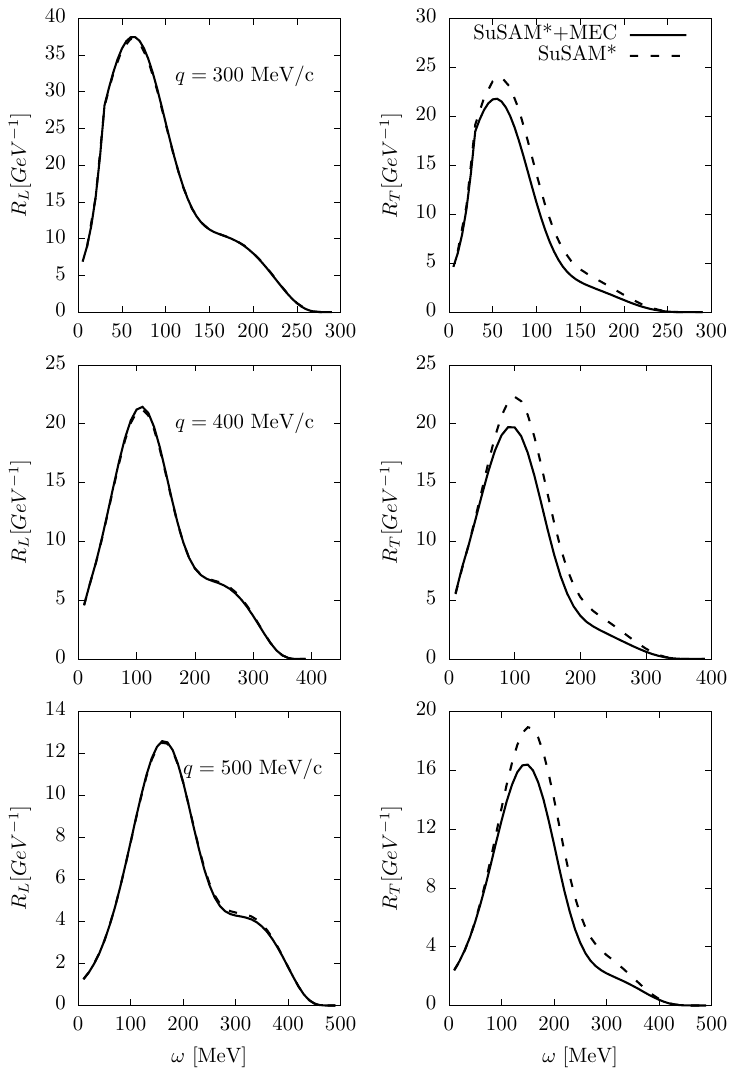}
\caption{ Response functions calculated in the generalized SuSAM*
  model (black curves) for $q=300,400,500$ MeV/c. The dashed curves do
  not include the MEC. }
 \label{fig-responses1}
\end{figure}

\begin{figure}[t]
\centering
\includegraphics[width=10cm,bb=110 270 460 770]{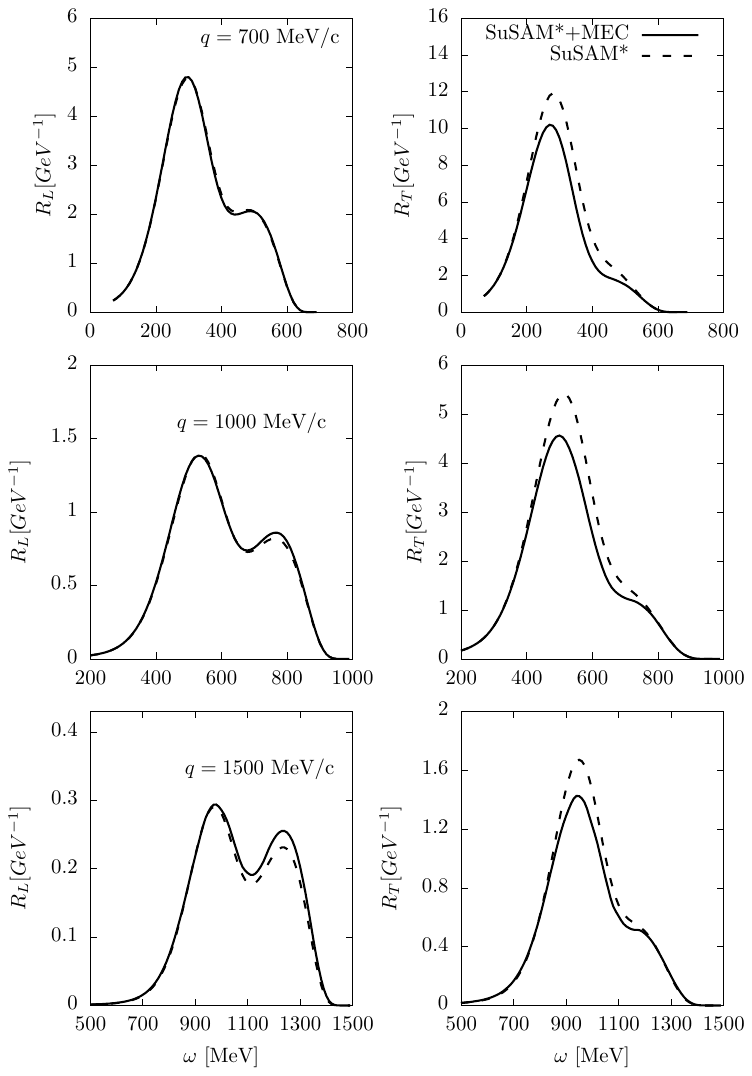}
\caption{Response functions calculated in the generalized SuSAM* model
  (black curves) for $q=700,1000,1500$ MeV/c. The dashed curves do not
  include the MEC.}
 \label{fig-responses2}
\end{figure}

In Fig. \ref{fig-responses1} and \ref{fig-responses2}, we present the
total responses of $^{12}$C computed using the generalized SuSAM*
model. These responses are obtained by multiplying the
phenomenological scaling function by the averaged single-nucleon
response and summing over protons and neutrons, as given by
Eq. (\ref{susam}). The responses are shown for different values of $q$
as a function of $\omega$. In the same figure, we also show the
results without including the MEC contributions, which corresponds to
setting the terms $w_{1b2b}+w_{2b}$ associated with the two-body current
(Eq. (\ref{snucleonc2})) to zero.

Comparing the results with and without MEC, we observe that the impact
of MEC is more significant in the transverse response compared to the
longitudinal response. This is expected since the corrections due to
MEC in the longitudinal response are higher-order effects in a
non-relativistic expansion in powers of $v/c$, as
known from previous studies \cite{Ris89}. Therefore, the MEC
contributions to the longitudinal response are minimal and only start
to become noticeable for $q \geq 1.5$ GeV in the high-energy region. However,
this high-energy region is dominated and overshadowed by pion emission
and inelastic processes, making it difficult to isolate
the 1p1h longitudinal response. 

The inclusion of MEC in the single-nucleon leads to a reduction of the
transverse response by around 10\% or even more for all studied values
of $q$. This is consistent with previous calculations in RFG and the
shell model \cite{Ama94,Fab97,Ama03,Ama10,Ama03b}. These calculations
have consistently shown that MEC in the 1p1h channel tend to decrease
the transverse response compared to the contribution from the one-body
current. Note that this reduction in the
transverse response is a direct consequence of the destructive
interference between the one-body current and MEC. The contribution of
MEC to the transverse response is negative because the direct two-body
matrix element is zero (in symmetric nuclear matter, $N=Z$) or almost
zero (in asymmetric nuclear matter, $N\ne Z$, or in finite nuclei)
after summing over isospin.

\begin{figure}
\centering
\includegraphics[width=6cm,bb=125 340 400 770]{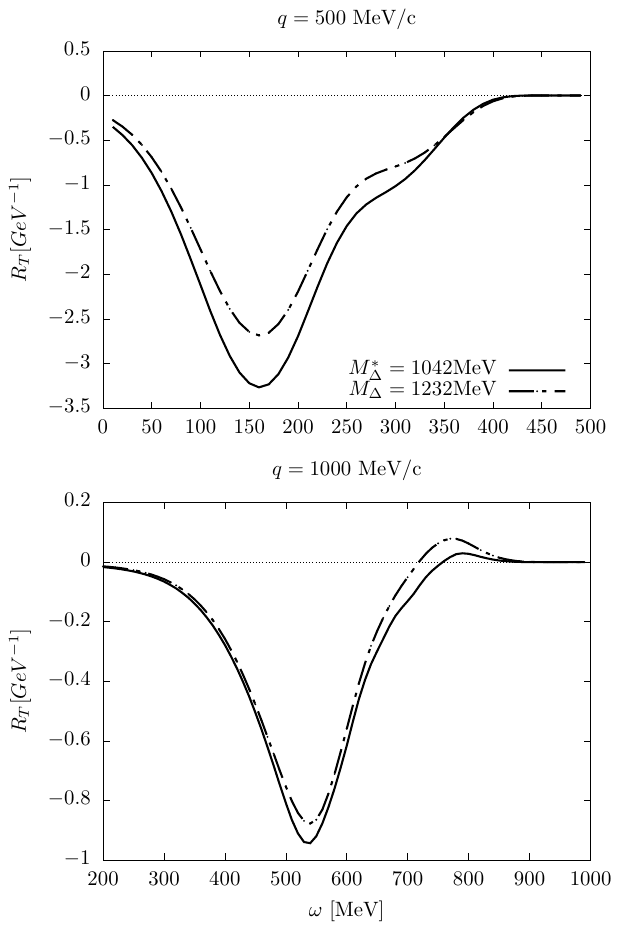}
  \caption{Comparison of the transverse interference OB-$\Delta$
    computed in the generalized SuSAM* model with and without
    relativistic effective mass and vector energy for the $\Delta$.}
  \label{fig10}
\end{figure}

The treatment of the $\Delta$ resonance in the medium is subject to
various ambiguities and uncertainties. In our generalized SuSAM*
model, we have assumed that the $\Delta$ resonance acquires an
effective mass $M_\Delta^*$ and vector energy $E_v^\Delta$ due to its
interaction with the RMF. This requires modifying the propagator
according to the formalism proposed in references \cite{Weh93,Kim96}.
To estimate the effect of this treatment, in Fig. \ref{fig10} we
compare the transverse response for the OB-$\Delta$ interference
calculated assuming that the $\Delta$ remains unchanged in the medium,
i.e., setting $M_\Delta^* = M_\Delta$ and $E_v^\Delta = 0$. The
response with the free $\Delta$ without medium modifications is
slightly smaller in absolute value, around 10\% depending on the
momentum transfer. This can be seen as an estimation of the
uncertainty associated with the $\Delta$ interaction in the medium.

Another related issue is the modification of the $\Delta$ width in the
medium, which we have not considered here assuming the free width
\eqref{widthc1}. This effect can also influence the results, but it is
expected to be of the same order as the observed effect in
Fig. \ref{fig10}. The treatment of the $\Delta$ resonance in the
medium is a complex topic, and further investigations and refinements
are needed to fully understand its effects and uncertainties.

\begin{figure}[t]
  \centering
\includegraphics[width=11cm,bb=90 450 450 800]{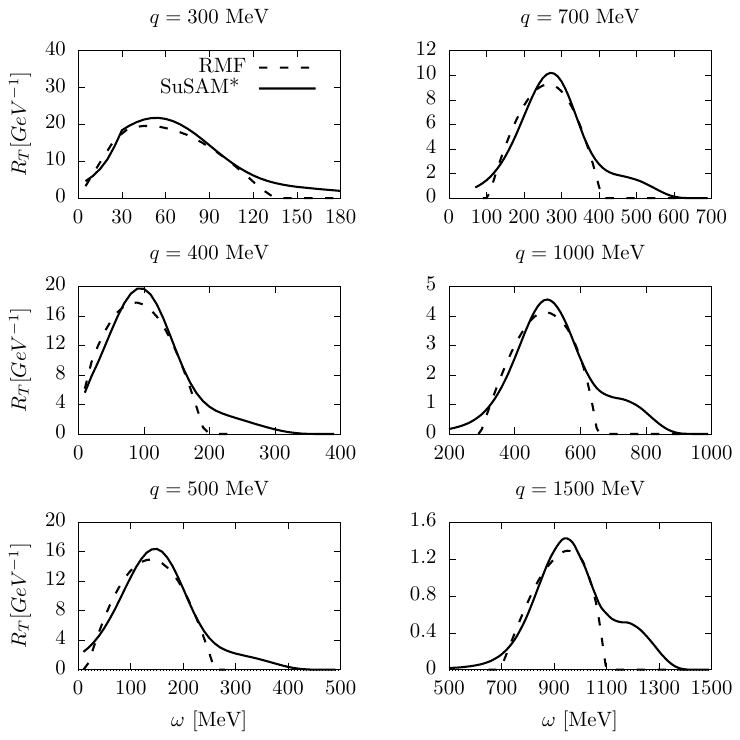}
\caption{Total transverse responses for $^{12}$C including MEC in the
  RMF model with $M^*=0.8$ compared to the generalized SuSAM* model.}
\label{fig-rt}
\end{figure}

In Fig. \ref{fig-rt}, we compare the total transverse response
calculated in the RMF model with an effective mass of $M^*=0.8$ to the
results obtained in the generalized SuSAM* approach for various
momentum transfers, ranging from $q=300$ MeV/c to $q=1500$ MeV/c. Both
calculations include the effects of MEC.  One notable difference
between the two approaches is the presence of a pronounced tail at
high energy transfer rates in the SuSAM* results. This tail extends
well beyond the upper limit of the RFG responses, reflecting the
effect of the phenomenological scaling function used in the SuSAM*
approach. Similar effects are found in the longitudinal response. 
Additionally, it is worth noting that the peak height
of the transverse response in the SuSAM* approach is generally higher
compared to the RMF model.

\begin{figure} 
\centering
\includegraphics[width=13cm,bb=40 275 525 770]{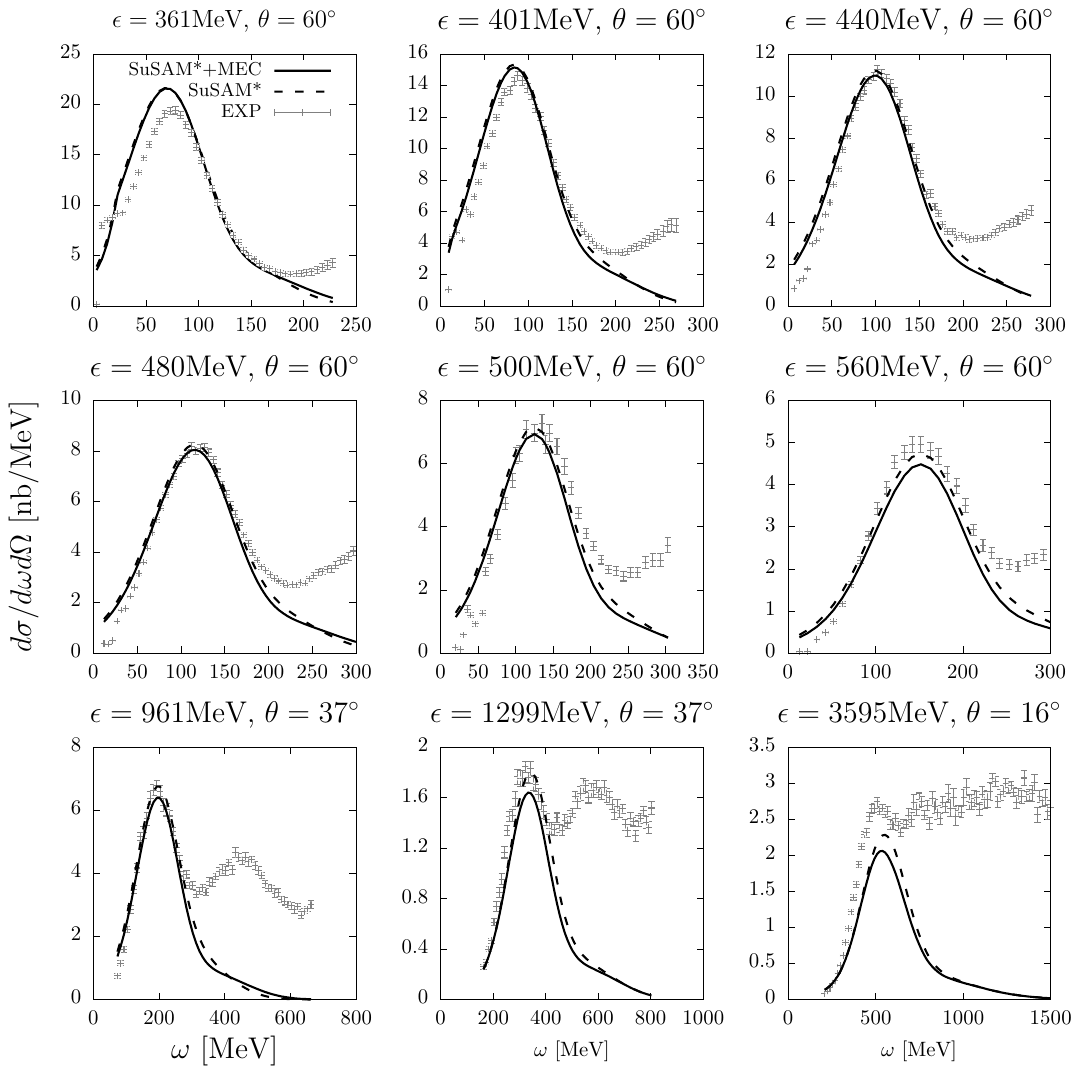}
  \caption{Cross section of $^{12}$C for several kinematics computed
    with the generalized SuSAM* model, including MEC, compared with
    the same calculation without MEC. Experimental data are from
    Refs. \cite{archive,archive2}.}
\label{cross}
\end{figure}

Finally, in Fig. \ref{cross}, we present the results for the (e,e')
double differential cross- section of $^{12}$C calculated with the
generalized SuSAM* model including MEC, compared to experimental data
for selected kinematics. We also compare these with the same model but
assuming that only the single-nucleon contribution is present, i.e.,
setting the MEC to zero. We observe that the inclusion of MEC in this
model leads to a small reduction in the cross-section compared to the
case without MEC. This reduction is a consequence of the decrease in
the transverse response due to the presence of MEC.

Figure \ref{cross} also provides an illustrative example of the
averaged description of the global (e, e') cross-section
data. Considering the scaling violation uncertainty (the scatter of
the points in the scaling band is larger than 20 \% ) significant
deviations from the QE data are expected for certain kinematic
conditions. In some cases, the predicted cross-section is above the
data, while in others, it is below. Note that the phenomenological
scaling function is obtained by excluding data points that
significantly deviate from scaling behavior. Specifically, kinematics
involving substantial contributions from pion emission are excluded as
non-quasielastic. Consequently, in these cases, the predicted
cross-section falls below the experimental data.

This work does not include comparisons with separate response function
experimental data or the Coulomb sum rule. While the scaling band
results from the clustering of data points with varying momentum and
energy transfers, the faithful reproduction of separate response
functions by a model is not necessarily guaranteed. To ensure the
accuracy of a scaling-based approach in describing response
functions, it would be imperative to somehow incorporate experimental
information about these response functions into the scaling
analysis. This is an endeavor that requires further investigation
beyond the scope of this chapter, which primarily serves to illustrate
how to modify the scaling approach to incorporate MEC within the
single-nucleon framework. The study of the response functions will be
addressed in chapter 7, in connection with MEC and short-range
correlations.

\section{Final remarks}

This chapter focused on the inclusion of meson exchange currents
within the framework of the superscaling analysis with effective mass,
presented in chapter 2. From the results, we observed that, in all the
models considered, the transverse response decreases when including
MEC in the 1p1h channel. This result is consistent with previous
independent calculations performed in the relativistic and
non-relativistic Fermi gas models as well as in the non-relativistic
and semi-relativistic shell models. The result is a consequence of the
fact that the main contribution arises from the interference of the OB
and $\Delta$ currents, in particular through the exchange diagram,
carrying a minus sign. The contribution from the direct part of the
MEC matrix element is zero in the Fermi gas, and this is one of the
reasons for the negative contribution.

We anticipate that the landscape may undergo transformation with the
development of a more realistic averaged single-nucleon framework,
possibly incorporating a Fermi gas description enhanced by short-range
correlations. In such a scenario, the meson-exchange currents are
expected to exert a more substantial influence, potentially leading to
a significant increase in the transverse response.  This issue will be
addressed in chapters 6 and 7, where the high-momentum components arising
from SRC are explicitly calculated \cite{Cas23b}, and subsequently
incorporated into our quasielastic scattering model.

\chapter{The low momentum OB-MEC interference}

Until now, we have studied the QE nuclear responses, first focusing on
the one-body current and later incorporating meson exchange
currents. Both cases have been examined within the framework of an
improved scaling model based on a mean value of the single-nucleon
tensor, which is averaged with the energy distribution $n(\epsilon)$.

In this chapter, we study in more detail the interference
responses with MEC at low momentum, with a particular focus on the
low-momentum proposition for single-particle emission. This proposition states
that, in the Fermi gas, the transverse interference response between
the OB current and the $\Delta$ and pionic currents is negative,
resulting in a partial cancellation with the seagull current. To
further investigate this behavior, we have compared our results with
several models, both relativistic and non-relativistic, and our
findings indicate that they all satisfy the proposition.

The motivation behind this study arises from the conflicting results
reported in previous works concerning the sign of the $\Delta$
interference response. In those cases, the calculations that use
independent-particle models \cite{Fra23,Lov23}, have reported a large
and positive interference of one-body and two-body currents in the
transverse response. These results contradict previous calculations,
including those presented in chapter 3, and their origin is not
entirely understood, necessitating further clarification. While a close
examination of \cite{Fra23,Lov23} suggest a $\Delta$ current with the
opposite sign to the commonly employed $\Delta$ current adopted in our
work \cite{Ris89}, a detailed study, such as the one presented in this
chapter, is required to systematically analyze the theoretical
foundations of the interference response and to assess whether the
observed discrepancies arise from fundamental differences in the
modeling of nuclear dynamics or from specific approximations used in
these calculations.
More details can be found in Ref. \cite{Cas25}.

\section{Non-relativistic OB-MEC transverse response}
In this section we calculate in detail the OB-MEC interference in the
T-response in the Fermi model, where we perform analytically the
spin-isospin sums. We provide detailed expressions from which the
low-momentum theorem follows trivially. First, we present the
non-relativistic (NR) reduction of the one-body, Eq.(\ref{1bcur}) and two-body relativistic
currents, Eqs.(\ref{mec1})-(\ref{delta2}). Since one of the main
objectives in this chapter is to demonstrate the low-momentum
proposition, we will first verify that the NR limit of
our current operators are consistent with
the well-known NR currents of Riska \cite{Ris89}, which
are the standard operators typically used in calculations including
seagull, pion-in-flight, and $\Delta$ currents. The one-body operator,
which corresponds to a transverse current, will be expressed as the
sum of a magnetization term $\nj_M$ and a convection term $\nj_C$
\begin{eqnarray}
  \nj_{1b}(\np,\nh)
&=& 
  \nj_{M}(\np,\nh)+  \nj_{C}(\np,\nh),
  \label{obvectorv}
\\
  \nj_{M}(\np,\nh)
&=&
-\delta_{t_pt_h}\frac{G_M^h}{2m_N}i\nq\times\nsigma_{s_ps_h},
\label{magnetizationv}
\\
  \nj_C(\np,\nh)
&=&
\delta_{t_pt_h}\delta_{s_ps_h}
\frac{G_E^h}{m_N}(\nh+\frac{\nq}{2}).
\label{convectionv}
\end{eqnarray}
with $\nq=\np-\nh$ by momentum conservation.  Here $G_M^h$ ($G_E^h$)
is the magnetic (electric) form factor of the nucleon with isospin
$t_h$.  In the quasielastic peak the convection current contribution
is much smaller that the magnetization and can be neglected.

For the NR reduction of the meson exchange currents, the
treatment is deliberately didactic, to ensure that our results are
easily reproducible. We aim to leave no ambiguous or unclear steps in
the derivation process. We will take the static limit in which the
momenta of the initial and final nucleons are very small.

In the NR limit, the lower component of the Dirac
spinors is neglected, and the $4\times 4$ Dirac matrices reduce to
$2\times 2$ Pauli matrices acting on the upper
components. Additionally, we will only consider the NR
reduction of the transverse current, i.e., \(J^i\) for \(i=1,2\), as
this is the dominant contribution in this limit. The contribution of
MEC to the longitudinal response is neglected as we have seen in
chapter 3. The NR approximation is further justified by
numerical calculations.

To achieve this reduction, we apply the following rules:
\begin{eqnarray}
\gamma^0 \longrightarrow 1, &
\gamma^i \longrightarrow 0, &
\gamma_5\gamma^0 \longrightarrow 0, 
\label{gammas1}\\
\gamma_5\gamma^i \longrightarrow -\sigma_i, &
\gamma^i\gamma^j \longrightarrow -\sigma_i\sigma_j, &
\gamma^0\gamma^j \longrightarrow 0. 
\label{gammas2v}
\end{eqnarray}
For a nucleon momentum: 
\begin{eqnarray}
 p^\mu \longrightarrow (m_N,p^i), &&
\pbar \longrightarrow p_0. 
\end{eqnarray}
For the momentum transfer to nucleon $i$:
\begin{eqnarray}
k^\mu \longrightarrow (0,k^i), &&
\gamma_5\kbar \longrightarrow \nk\cdot\nsigma.
\end{eqnarray}
To simplify the writing at this stage, we do not explicitly include
the strong form factors. These can be applied later on to the
NR currents. Most of the results will be obtained with
these form factors set to one, and we will see that the effect of
including them is small for the low momentum transfer values
considered here. Additionally, their inclusion does not alter the
general result of the proposition, which is our main objective. 

The $V$-function defined in Eq. (\ref{Vfun}), 
\begin{equation} \label{Vfun2}
  V_{s'_{1}s_{1}}(p'_{1},p_{1})
= F_{\pi NN}(k_{1}^{2})
\frac{ \bar{u}_{s'_{1}}(p'_{1})\gamma^{5}\slashed{k}_{1}u_{s_{1}}(p_{1})}{k_1^2-m_{\pi}^2}.
\end{equation}
is directly obtained in this limit
\begin{equation}
  V_{s'_{1}s_{1}}(p'_{1},p_{1}) \longrightarrow 
-\frac{\nk_1\cdot\nsigma^{(1)}}{\nk_1^2+m_{\pi}^2}
\label{Vnorelv}
\end{equation}
where a matrix element between initial and final spin states is
understood, i.e. $\langle s'_1| \cdot | s_1\rangle$, but is not
explicitly written for simplicity. The spin states $|s\rangle$ are NR,
two-component spinors, corresponding to the upper component of the
Dirac spinors.

\subsubsection{Seagull current}

In the seagull current, we start by using a notation to separate the
isospin part, which does not change in the NR limit,
\begin{eqnarray} 
  j^{\mu}_{s}&=&
i[\ntau^{(1)} \times \ntau^{(2)}]_z
j^\mu_{s3}
\end{eqnarray}
Here the function 
$j^\mu_{s3}$ is independent on isospin
and is defined by 
\begin{eqnarray}
  j^\mu_{s3}(1',2',1,2) &=& \frac{f^{2}}{m_{\pi}^{2}}V_{s'_{1}s_{1}}(p'_{1},p_{1})
  \bar{u}_{s_{2}}(p'_{2})F_{1}^{V}
  \gamma^{5}\gamma^{\mu}u_{s_{2}}(p_{2}) + (1\leftrightarrow 2)
\end{eqnarray}
where in the $(1\leftrightarrow 2)$ term we have used the property
\begin{equation}
[\ntau^{(2)} \times \ntau^{(1)}] =-[\ntau^{(1)} \times \ntau^{(2)}].
\end{equation}
The NR reduction is directly obtained
using Eqs. (\ref{gammas2v}) and (\ref{Vnorelv}). Therefore the seagull current
$j_{s3}^i$ in the NR limit becomes
\begin{equation}
\nj_{s3} \rightarrow 
\frac{f^{2}}{m_{\pi}^{2}}F_1^V
\left(
\frac{\nk_1\cdot\nsigma^{(1)}}{\nk_1^2+m_{\pi}^2}
\nsigma^{(2)}
-\frac{\nk_2\cdot\nsigma^{(2)}}{\nk_2^2+m_{\pi}^2}
\nsigma^{(1)}
\right)
\label{seagull}
\end{equation}

\subsubsection{Pionic current}

The pion-in-flight  current can be   written similarly to the seagull current as
\begin{equation} 
  j^{\mu}_{\pi}=
i[\ntau^{(1)} \times \ntau^{(2)}]_z
j^\mu_{\pi 3} 
\end{equation}
where  
$j^\mu_{\pi 3}$ is independent on isospin
and is defined by 
\begin{equation}
j^\mu_{\pi 3}(1',2',1,2)
=
\frac{f^{2}}{m_{\pi}^{2}}
F_{1}^{V}V_{s'_{1}s_{1}}(p'_{1},p_{1})V_{s'_{2}s_{2}}(p'_{2},p_{2})
(k_1^\mu-k_2^\mu)
\end{equation}
The NR reduction of the pionic current is directly obtained
from Eq. (\ref{Vnorelv})
\begin{equation}
\nj_{\pi 3} \rightarrow 
\frac{f^{2}}{m_{\pi}^{2}}F_1^V
\frac{\nk_1\cdot\nsigma^{(1)}}{\nk_1^2+m_{\pi}^2}
\frac{\nk_2\cdot\nsigma^{(2)}}{\nk_2^2+m_{\pi}^2}
(\nk_1-\nk_2)
\label{pionic}
\end{equation}
It is
straightforward to verify that
the NR reduction of the seagull plus pionic currents 
coincide with the usual expressions
found in the literature \cite{Ris89}.

\subsubsection{$\Delta$ current}

The NR reduction of the $\Delta$ current is somewhat more involved due
to its more complex spin and isospin structure. In order to simplify
the NR reduction, it is convenient to write the $\Delta$
current in an abbreviated form
\begin{eqnarray}
  j^{\mu}_{\Delta F}
&=&
U_F K_F^\mu 
+ (1 \leftrightarrow 2)
\\
  j^{\mu}_{\Delta B}
&=&
U_B K_B^\mu 
+ (1 \leftrightarrow 2)
\end{eqnarray}
where a matrix element is assumed to be taken between the initial and final
isospin states $\langle t'_1 t'_2 |\cdot |t_1t_2\rangle$, 
but we do not explicitly write this to simplify the
notation.

The functions $K^\mu_F$ and $K^\mu_B$ are independent on isospin
and can be written as
\begin{eqnarray}
  K^{\mu}_{ F}
&=&
\frac{ff^{*}}{m_{\pi}^{2}}
V_{s'_{2}s_{2}}(p'_{2},p_{2})
A^\mu,
\\
  K^{\mu}_{B}
&=&
\frac{ff^{*}}{m_{\pi}^{2}}
V_{s'_{2}s_{2}}(p'_{2},p_{2})
B^\mu.
\end{eqnarray}
where $A^\mu$ and $B^\mu$ are defined
\begin{eqnarray}
A^\mu
&=&
\bar{u}_{s_{1}}(p'_{1})k_{2}^{\alpha}G_{\alpha\beta}(p_{1}+Q)
\Gamma^{\beta\mu}(Q)u_{s_{1}}(p_{1})
\label{amu}\\
B^\mu
&=&
\bar{u}_{s_{1}}(p'_{1})k_{2}^{\beta}
\Gamma^{\alpha\mu}(-Q)G_{\alpha\beta}(p'_{1}-Q)u_{s_{1}}(p_{1}).
\label{bmu}
\end{eqnarray}
The NR reduction of the spatial components of the $\Delta$
current requires the NR reduction of the components $A^i$
and $B^i$. This detailed reduction process is carried out in Appendix
\ref{appC}. The result is
\begin{eqnarray}
\nA &\rightarrow & 
g_\Delta
\nq\times\left[\frac23 i \nk_2 - \frac13 \nk_2\times\nsigma^{(1)}\right]
\label{Anonrel}\\
\nB &\rightarrow & 
g_\Delta
\nq\times\left[\frac23 i \nk_2 + \frac13 \nk_2\times\nsigma^{(1)}\right]
\label{Bnonrel}
\end{eqnarray}
where we have defined the coefficient
\begin{equation}
g_\Delta \equiv \frac{C_3^V}{m_N}\frac{1}{m_\Delta-m_N}.
\end{equation}
Therefore
\begin{eqnarray}
\nK_F &\rightarrow & 
-\frac{ff^*}{m_\pi^2}
\frac{\nk_2\cdot\nsigma^{(2)}}{\nk_2^2+m_{\pi}^2}\nA
\\
\nK_B &\rightarrow & 
-\frac{ff^*}{m_\pi^2}
\frac{\nk_2\cdot\nsigma^{(2)}}{\nk_2^2+m_{\pi}^2}\nB
\end{eqnarray}
as before a matrix element is assumed to be taken between the initial and final
spin states, $\langle s'_1 s'_2 |\cdot |s_1s_2\rangle$, 
but we do not explicitly write this to simplify the
notation.

Using the result
\begin{equation} \label{titj}
T_{i}T^{\dagger}_{j}=
\frac23\delta_{ij}
-\frac{i}{3}\epsilon_{ijk}\tau_{k}
=\delta_{ij}-\frac{1}{3}\tau_{i}\tau_{j}
\end{equation}
the forward and backward isospin operators can be written as
\begin{eqnarray}
  U_F(1,2)
&=&
\frac{1}{\sqrt{6}}
\left(2\tau_{z}^{(2)}
      -i[\ntau^{(1)}\times\ntau^{(2)}]_{z}
\right)
\\
 U_B(1,2)
&=&
\frac{1}{\sqrt{6}}
\left(2\tau_{z}^{(2)}
      +i[\ntau^{(1)}\times\ntau^{(2)}]_{z}
\right).
\end{eqnarray}
Hence the $\Delta$ current can be written as
\begin{eqnarray}
\nj_\Delta 
&=&
\nj_{\Delta F} +\nj_{\Delta B}=
\frac{2}{\sqrt{6}}
\tau_{z}^{(2)}
[\nK_F+\nK_B]
+\frac{1}{\sqrt{6}}
i[\ntau^{(1)}\times\ntau^{(2)}]_{z}
[\nK_B-\nK_F]
+ (1\leftrightarrow 2).\nonumber \\
\label{jdeltanr}
\end{eqnarray} 
We see that the isospin operators in the $\Delta$ current, $\tau^{(2)}_z$ 
and $[\ntau^{(1)}\times \ntau^{(2)}]_z$, 
are multiplied by the sum and the difference between the backward and forward 
functions, \( \nK_B \) and \( \nK_F \), given in the NR limit by
\begin{eqnarray}
\nK_F+\nK_B 
&=&
-\frac{ff^*}{m_\pi^2}
\frac{\nk_2\cdot\nsigma^{(2)}}{\nk_2^2+m_{\pi}^2}(\nA+\nB)
\label{FpB}\\
\nK_B-\nK_F 
&=&
-\frac{ff^*}{m_\pi^2}
\frac{\nk_2\cdot\nsigma^{(2)}}{\nk_2^2+m_{\pi}^2}(\nB-\nA).
\label{BmF}
\end{eqnarray}
From Eqs. (\ref{Anonrel}) and (\ref{Bnonrel}),
\begin{eqnarray}
\nA+\nB &=& \frac43 g_\Delta i \nq\times\nk_2
\\
\nB-\nA &=& \frac 23 g_\Delta \nq\times(\nk_2\times\nsigma^{(1)})
\end{eqnarray}
Inserting this result into Eqs. (\ref{FpB}) and (\ref{BmF}) we have
\begin{eqnarray}
\nK_F+\nK_B 
&=&
-
\frac43 g_\Delta 
\frac{ff^*}{m_\pi^2}
\frac{\nk_2\cdot\nsigma^{(2)}}{\nk_2^2+m_{\pi}^2}
i \nq\times\nk_2,
\\
\nK_B-\nK_F 
&=&
-\frac 23 g_\Delta 
\frac{ff^*}{m_\pi^2}
\frac{\nk_2\cdot\nsigma^{(2)}}{\nk_2^2+m_{\pi}^2}
\nq\times(\nk_2\times\nsigma^{(1)}).
\end{eqnarray}
Using these results in Eq. (\ref{jdeltanr}), it is straightforward to obtain
the following expression for the NR $\Delta$ current
\begin{eqnarray}
\nj_\Delta 
&=&
i \sqrt{ \frac32 } \frac29 
\frac{ff^*}{m_\pi^2}
\frac{C_3^V}{m_N}\frac{1}{m_\Delta-m_N}
\left\{
\frac{\nk_2\cdot\nsigma^{(2)}}{\nk_2^2+m_{\pi}^2}
\left[
4\tau_{z}^{(2)}\nk_2+
[\ntau^{(1)}\times\ntau^{(2)}]_{z}
\nk_2\times\nsigma^{(1)}
\right]
\right.
\nonumber\\
&&
\kern 3cm 
\left. \mbox{}+
\frac{\nk_1\cdot\nsigma^{(1)}}{\nk_1^2+m_{\pi}^2}
\left[
4\tau_{z}^{(1)}\nk_1-
[\ntau^{(1)}\times\ntau^{(2)}]_{z}
\nk_1\times\nsigma^{(2)}
\right]
\right\}
\times\nq 
\label{deltafinalv}
\end{eqnarray} 

One can, in fact, verify that this expression coincides with the
$\Delta$ current appearing in the literature, particularly the
expression given in Refs. \cite{Ris89,Sch89}, that we use as
reference, except for the precise values of the coupling constants and
form factors. This assures us that the relativistic and NR
calculations in the low-momentum and low-energy limit should coincide
if identical values of coupling and form factors are used.

\subsection{MEC effective one-body transition currents}

With the NR MEC current obtained in the last section, in the NR limit,
the effective one-body transition current \(\nj_{2b}(\np,\nh)\) in the
Fermi gas is obtained by summing over spin, isospin, and integrating
over the momentum \(k\) of the spectator nucleon.  At leading order,
only the spatial part of the MEC survives, affecting solely the
transverse response, which is perpendicular to the transferred
momentum \(\nq\).  From Eq. (\ref{effectiveOBc1}) in the limit
$V\rightarrow \infty$ this current is
\begin{equation}
\nj_{2b}(\np,\nh) 
=
\int\frac{d^3k}{(2\pi)^3}
\sum_{t_ks_k}
\left[ \nj_{2b}(p,k,h,k)-\nj_{2b}(p,k,k,h)\right] .
\label{effective}
\end{equation}

\subsubsection{Sum over isospin}

 We begin by showing that the direct term in Eq. (\ref{effective})
 is zero.  Previously, we note that the MEC can be expanded in terms
 of the isospin operators \(\tau^{(1)}_z\) , \(\tau^{(2)}_z\) and
 \([\ntau^{(1)} \times \ntau^{(2)}]_z\)
\begin{equation}
\nj_{2b}= \tau^{(1)}_z \nj_1+ \tau^{(2)}_z \nj_2 + 
i[ \ntau^{(1)} \times \ntau^{(2)}]_z \ \nj_3 \label{descom}
\end{equation}
where $\nj_1$, $\nj_2$, $\nj_3$ are independent on isospin.

We first perform  the sum over isospin index $t_k$. 
The direct term is

\begin{eqnarray}
\sum_{t_k}\nj_{2b}(p,k,h,k)
&=&
\sum_{t_k} 
\langle t_pt_k|
\tau^{(1)}_z \nj_1+ \tau^{(2)}_z \nj_2 + +i[\ntau^{(1)} \times \ntau^{(2)}]_z\ \nj_3
| t_ht_k\rangle \nonumber \\
&=&
\delta_{t_pt_h}4t_h \nj_1(p,k,h,k), 
\end{eqnarray}

where we have used the elementary isospin sums performed in Appendix
\ref{appB}, Eqs. (\ref{iso1}--\ref{iso3}). Therefore the direct
term in the matrix element (\ref{effective}) is proportional to
\(\nj_1(p,k,h,k)\), which turns out to be zero. Indeed,
\(\nj_1\) can be obtained from equation (\ref{deltafinalv}) as
\begin{equation}
\nj_1 
=
iC_\Delta
\frac{\nk_1\cdot\nsigma^{(1)}}{\nk_1^2+m_{\pi}^2}
4\nk_1\times\nq,
\end{equation} 
with 
\begin{equation}  \label{cdelta}
C_\Delta\equiv
 \sqrt{ \frac32 } \frac29 
\frac{ff^*}{m_\pi^2}
\frac{C_3^V}{m_N}\frac{1}{m_\Delta-m_N}
\end{equation}
and $\nk_1=\np-\nh=\nq$. Therefore 
\begin{equation}
\sum_{t_k}\nj_{2b}(p,k,h,k)
=0.
\end{equation}
This results follows because he $\Delta$
current is transverse, i.e., perpendicular to $\nq$.  In the
relativistic case, a similar situation occurs, and the direct term is
zero although the demonstration is more involved. It requires summing
over the spin of \(\nk\) and handling a large number of terms that
involve many \(\gamma\) matrices.

The sum over isospin in the exchange part is obtained using the
isospin sums performed in Appendix \ref{appB}, Eqs
(\ref{iso4}--\ref{iso6})
\begin{eqnarray}
\sum_{t_k}\nj_{2b}(p,k,k,h)
&=&
\sum_{t_k=\pm 1/2} 
\langle t_pt_k|
\tau^{(1)}_z \nj_1+ \tau^{(2)}_z \nj_2 + 
i[\ntau^{(1)} \times \ntau^{(2)}]_z\ \nj_3
| t_kt_h\rangle
\nonumber\\
&=& \delta_{t_pt_h}2t_h [\nj_1(p,k,k,h)+\nj_2(p,k,k,h)-2\nj_3(p,k,k,h)]
\end{eqnarray}

Thus, in symmetric nuclear matter the direct matrix
element of the MEC vanishes, and only the exchange term survives in
the 1p1h matrix element.  Therefore, the many-body diagrams that
contribute to the MEC in the 1p1h channel are those shown in chapter 3
in Fig. \ref{feynman2}. Next, we proceed to perform the spin sums for
the different terms of the current.

\subsubsection{Sum over spin}

The resulting 1p1h matrix elements of the 2b current are computed as
\begin{equation}
\nj_{2b}(p,h)=
-\int \frac{d^3k}{(2\pi)^3}
\sum_{t_ks_k}\nj_{2b}(p,k,k,h)
=
\nj_{s}(p,h)+\nj_{\pi}(p,h)+\nj_{\Delta}(p,h),
\end{equation}
where only the exchange part contributes.  The sums over spin index
$s_k$ are performed in Appendix \ref{appD}. The results are the
following for the three MEC, seagull, pionic and $\Delta$ currents
\begin{eqnarray}
\nj_s(p,h)
&=&
4t_h\delta_{t_pt_h} 
\frac{f^2}{m_\pi^2}F_1^V
\int \frac{d^3k}{(2\pi)^3}
\left(
\frac{\delta_{s_ps_h}\nk_1+i\nsigma_{ph}\times\nk_1}{\nk_1^2+m_{\pi}^2}
-\frac{\delta_{s_ps_h}\nk_2+i\nk_2\times\nsigma_{ph}}{\nk_2^2+m_{\pi}^2}
\right)
\label{seagullph} 
\\
\nj_\pi(p,h)
&=&
4t_h\delta_{t_pt_h} 
\frac{f^2}{m_\pi^2}F_1^V
\int \frac{d^3k}{(2\pi)^3}
\frac{\delta_{s_ps_h}\nk_1\cdot\nk_2
+i(\nk_1\times\nk_2)\cdot\nsigma_{ph}}
{(\nk_1^2+m_{\pi}^2)(\nk_1^2+m_{\pi}^2)}(\nk_1-\nk_2)  
\label{pionicph}
\\
\nj_\Delta(p,h)
&=&
4it_h\delta_{t_pt_h}
C_\Delta\nq\times 
\int \frac{d^3k}{(2\pi)^3}
\left(
\frac{\nk_1^2\nsigma_{ph}+(\nsigma_{ph}\cdot\nk_1)\nk_1}
     {\nk_1^2+m_\pi^2}
+\frac{\nk_2^2\nsigma_{ph}+(\nsigma_{ph}\cdot\nk_2)\nk_2}
     {\nk_2^2+m_\pi^2}
\right)
\label{deltaph} \nonumber \\
\end{eqnarray}
with
$\nk_1=\np-\nk$ 
and 
$\nk_2=\nk-\nh$. 

\subsection{Interference between one-body and MEC in the transverse response}

In this section, we give the MEC contribution to the effective
single-nucleon transverse response, focusing exclusively on the
interference between the MEC and OB currents. The pure MEC responses
have been previously computed in chapter 3 and have been found
negligible, allowing them to be safely disregarded. Here, we describe
the interference terms separately for the different MEC components:
Seagull, pionic, and \(\Delta\), in combination with the magnetization
and convection terms of the OB currents. This separation is essential
to analyze the relative contributions of each component to the overall
response.

It should be clarified that here we are
computing the single-nucleon response corresponding to either a proton
or a neutron, with the requirement that the isospin of \( p \) and \(
h \) must be the same $t_p=t_h$. At the end of the calculation, the
contributions from protons and neutrons must be summed to obtain the
total response.

We recall the effective single-nucleon tensor defined in chapter 3 that incorporates the
square of the sum of the 1b and 2b currents and can be expanded as
\begin{eqnarray}
w^{\mu\mu}(\np,\nh)
&=& \frac{1}{2}\sum_{s_ps_h} |j^\mu_{1b}(\np,\nh)+j^\mu_{2b}(\np,\nh)|^2
\nonumber\\
&=&
  w^{\mu\mu}_{1b}+ w^{\mu\mu}_{1b2b}+w^{\mu\mu}_{2b}.
\label{single-nucleon}
\end{eqnarray}
where 
\begin{eqnarray}
w^{\mu\mu}_{1b}   & = & \frac{1}{2}\sum_{s_ps_h} |j^\mu_{1b}|^2, \\
w^{\mu\mu}_{1b2b} & = & \mbox{Re}\sum_{s_ps_h} (j^\mu_{1b})^*j^\mu_{2b}, \label{interrr}\\
w^{\mu\mu}_{2b}   & = & \frac{1}{2}\sum_{s_ps_h} |j^\mu_{2b}|^2.
\end{eqnarray}
The first term, \(w^{\mu\mu}_{1b}\),
is the tensor corresponding to the one-body current alone, \(w^{\mu\mu}_{1b2b}\)
is the interference between 1b and 2b
currents, and \(w^{\mu\mu}_{2b}\) represents the contribution of the two-body
current alone that is neglected.

The magnetization-seagull (ms) interference is given by Eq. (\ref{interrr})
\begin{equation}
w^T_{ms}=w^{11}_{ms}+w^{22}_{ms}= {\rm Re}\sum_{s_ps_h}
\nj_m(p,h)^*\cdot\nj_s(p,h),
\end{equation}
where we use that the magnetization current
is perpendicular to $\nq$ and it has only $x,y$ components. 
Using Eq. (\ref{magnetizationv}) for the magnetization current we can write
\begin{eqnarray}
w^T_{ms}
&=&
\delta_{t_pt_h}{\rm Re} \sum_{s_ps_h}
(-\frac{G_M^h}{2m_N}i\nq\times\nsigma_{s_ps_h})^*
\cdot \nj_s(p,h)
\nonumber\\
&=&
\sum_{s_ps_h}
\frac{G_M^h}{2m_N}i(\nq\times\nsigma_{s_hs_p})
\cdot \nj_s(\np,\nh)_{s_ps_h}.
\label{mssum}
\end{eqnarray}
We have utilized the fact that the spin sum 
already yields a real number, as will be shown later, making it unnecessary to
explicitly take the real part. 

By following the same procedure, we express the various interferences
required between the convection and magnetization currents with the
seagull, pionic, and $\Delta$ operators, as follows:
\begin{eqnarray}
w^T_{cs} & = & {\rm Re}\sum_{s_ps_h} \nj^T_c(p,h)^*\cdot\nj_s(p,h) =
\sum_{s_ps_h}
\frac{G_E^h}{m_N} \delta_{s_hs_p} \nh_T \cdot \nj_s(\np,\nh)_{s_ps_h},
\label{cssum}
\\
w^T_{m\pi}&  = & {\rm Re}\sum_{s_ps_h} \nj_m(p,h)^*\cdot\nj_\pi(p,h) =
\sum_{s_ps_h}
\frac{G_M^h}{2m_N}i(\nq\times\nsigma_{s_hs_p})
\cdot \nj_\pi(\np,\nh)_{s_ps_h},
\label{mpisum}
\\
w^T_{c\pi} & = & {\rm Re}\sum_{s_ps_h} \nj^T_c(p,h)^*\cdot\nj_\pi(p,h) =
\sum_{s_ps_h}
\frac{G_E^h}{m_N} \delta_{s_hs_p} \nh_T \cdot \nj_\pi(\np,\nh)_{s_ps_h},
\label{cpisum}
\\
w^T_{m\Delta}&  = & {\rm Re}\sum_{s_ps_h} \nj_m(p,h)^*\cdot\nj_\Delta(p,h) =
\sum_{s_ps_h}
\frac{G_M^h}{2m_N}i(\nq\times\nsigma_{s_hs_p})
\cdot \nj_\Delta(\np,\nh)_{s_ps_h},
\label{mdsum}
\\
w^T_{c\Delta}&  = & 0.
\end{eqnarray}
Note that only the transverse component of the convection current
appears that is proportional to the transverse nucleon momentum,
$\nh_T=\nh-\frac{\nh\cdot\nq}{q^2}\nq$,
thereby selecting the \(x\) and \(y\) components in the
transverse response. The convection-\(\Delta\)
interference is zero because the convection current is
spin-independent, while the \(\Delta\) current is linear in the
\(\sigma\) operators.

The sums over spin in Eqs. (\ref{mssum}--\ref{mdsum}) 
are performed in Appendix \ref{appE}. The result is
\begin{eqnarray}
w^T_{ms}(p,h)
&=&
4t_h\frac{f^2}{m_\pi^2}F_1^V\frac{G_M^h}{2m_N}
\int \frac{d^3k}{(2\pi)^3}
\left(
\frac{4\nq\cdot\nk_1}{\nk_1^2+m_{\pi}^2}+
\frac{4\nq\cdot\nk_2}{\nk_2^2+m_{\pi}^2}
\right) \nonumber \\
&&
\equiv
4t_h\frac{f^2}{m_\pi^2}F_1^V\frac{G_M^h}{2m_N}\  {\cal I}_{ms}(\np,\nh)
\label{wmsv} 
\\
w^T_{cs}(p,h)
&=&
4t_h\frac{f^2}{m_\pi^2}F_1^V\frac{G_E^h}{m_N}
\int \frac{d^3k}{(2\pi)^3}
\left(
\frac{2\nh_T\cdot\nk_1}{\nk_1^2+m_{\pi}^2}-
\frac{2\nh_T\cdot\nk_2}{\nk_2^2+m_{\pi}^2}
\right) \nonumber \\
&&
\equiv
4t_h\frac{f^2}{m_\pi^2}F_1^V\frac{G_E^h}{m_N}\ {\cal I}_{cs}(\np,\nh) 
\\
w^T_{m\pi}(p,h)
&=&
-4t_h\frac{f^2}{m_\pi^2}F_1^V\frac{G_M^h}{2m_N}
\int \frac{d^3k}{(2\pi)^3}
\frac{4(\nq\times\nk_2)^2}{(\nk_1^2+m_{\pi}^2)(\nk_2^2+m_{\pi}^2)}
\nonumber \\
&&
\equiv
-4t_h\frac{f^2}{m_\pi^2}F_1^V\frac{G_M^h}{2m_N}\  {\cal I}_{m\pi}(\np,\nh)
\label{mpiv} 
\\
w^T_{c\pi}(p,h)
&=&
-4t_h\frac{f^2}{m_\pi^2}F_1^V\frac{G_E^h}{m_N}
\int \frac{d^3k}{(2\pi)^3}
\frac{4(\nq\cdot\nk_2-\nk_2^2)\nh_T\cdot\nk_2}{(\nk_1^2+m_{\pi}^2)(\nk_2^2+m_{\pi}^2)} \nonumber \\
&&
\equiv
-4t_h\frac{f^2}{m_\pi^2}F_1^V\frac{G_E^h}{m_N}\
 {\cal I}_{c\pi}(\np,\nh) 
\\
w^T_{m\Delta}(p,h)
&=&
-4t_h C_\Delta \frac{G_M^h}{2m_N}
\int \frac{d^3k}{(2\pi)^3}
2\left(
\frac{3q^2k_1^2-(\nq\cdot\nk_1)^2}{\nk_1^2+m_{\pi}^2}+
\frac{3q^2k_2^2-(\nq\cdot\nk_2)^2}{\nk_2^2+m_{\pi}^2}
\right) \nonumber \\
&&
\equiv
-4t_h C_\Delta \frac{G_M^h}{2m_N}\
 {\cal I}_{m\Delta}(\np,\nh),
\label{wmdv}
\end{eqnarray}
where $C_\Delta$ is defined in Eq. (\ref{cdelta}), $\nk_1=\np-\nk$.
and $\nk_2=\nk-\nh$.  In Eqs. (\ref{wmsv}--\ref{wmdv}) we have defined
the integrals ${\cal I}_{ab}(\np,\nh)$, that are spin independent.

Finally, the total interference between the one-body and two-body
currents is given by the sum of the individual interferences between
the different terms of the currents, namely the seagull, pionic, and
$\Delta$ contributions with magnetization and convection currents.
\begin{equation}
w^T_{1b2b}=
w^T_{ms}+
w^T_{cs}+
w^T_{m\pi}+
w^T_{c\pi}+
w^T_{m\Delta}.
\end{equation}

\subsection{Low-momentum proposition}

\newtheorem{theorem}{Proposition}

\begin{theorem}
 The transverse interference response between the $\Delta$ current and
 the OB current is negative in the Fermi gas model:
 $w^T_{m\Delta}<0$.
\end{theorem}

The low-momentum proposition is applicable, specifically at moderate
momentum and energy transfers. Moderate in this context refers to
values small compared to the nucleon mass.

To demonstrate the proposition, we first need to express the total
single-nucleon interference responses as the sum of the contributions
from protons and neutrons.
\begin{eqnarray}
w^T_{ms}(\np,\nh)
&=&
\frac{f^2}{m_\pi^2}F_1^V\frac{G_M^p-G_M^n}{m_N}\  {\cal I}_{ms}(\np,\nh)
\label{wmstot}
\\
w^T_{cs}(\np,\nh)
&=&
2\frac{f^2}{m_\pi^2}F_1^V\frac{G_E^p-G_E^n}{m_N}\ {\cal I}_{cs}(\np,\nh)
\\
w^T_{m\pi}(\np,\nh)
&=&
-\frac{f^2}{m_\pi^2}F_1^V\frac{G_M^p-G_M^n}{m_N}\  {\cal I}_{m\pi}(\np,\nh)
\label{wmpitot}
\\
w^T_{c\pi}(\np,\nh)
&=&
-2\frac{f^2}{m_\pi^2}F_1^V\frac{G_E^p-G_E^n}{m_N}\
 {\cal I}_{c\pi}(\np,\nh)
\\
w^T_{m\Delta}(\np,\nh)
&=&
-  \sqrt{ \frac32 } \frac29 
\frac{ff^*}{m_\pi^2}
\frac{C_3^V}{m_N^2}\frac{G_M^p-G_M^n}{m_\Delta-m_N}
 {\cal I}_{m\Delta}(\np,\nh).
\label{wmdtot}
\end{eqnarray}

It suffices to verify that the
single-nucleon interference response \( w^{T}_{m\Delta} < 0 \)
in Eq (\ref{wmdtot}). 
On the one hand, \( w^T_{m\Delta} \) is proportional to
the integral \( {\cal I}_{m\Delta}\), 
which contains the pion propagator multiplied by a factor 
that is
always positive.  
The term in question can be seen inside the integral of Eq. (\ref{wmdv}),
 given by 
\begin{equation}
3q^2k_i^2-(\nq\cdot\nk_i)^2 \geq 0,
\end{equation}
with $i=1,2$.
This ensures that the integral \( {\cal I}_{m\Delta} \geq 0\). On the other
hand, \( w^T_{m\Delta} \) is also proportional to \( G_M^p - G_M^n \),
which is positive as well.  Therefore, since \( w^T_{m\Delta} \) includes an
overall negative sign, the final result is negative, completing the
proof.

Typically, the proposition is valid for momentum transfers below
approximately 500 MeV, where the interference response $m\Delta$ is
explicitly negative.  For momentum transfers above this threshold,
relativistic effects become increasingly significant. In this regime,
the explicit determination of the sign is no longer straightforward
due to the complexity of the spin sum in the relativistic case. The
analysis requires numerical computations to verify the sign of the
interference, as the NR proposition no longer applies
directly.

From Eq. (\ref{mpiv}), we can also establish the following proposition for
the magnetization-pionic response:

\begin{theorem}
The transverse interference response between the pionic current and
the magnetization current is negative in the Fermi gas model:
$w^T_{m\pi}<0$

\end{theorem}

The proof of this proposition follows similarly to proposition 1, by noticing
that the integral \(\mathcal{I}_{m\pi}\) is positive, as it contains
the square of \(\mathbf{q} \times \mathbf{k}_2\), as seen in
Eq. (\ref{mpiv}). Then, according to Eq. (\ref{wmpitot}), we conclude that \(
w^T_{m\pi} < 0 \).

The convection-pionic interference does not generally have a
well-defined sign, but its contribution is much smaller than the
magnetization interference. Therefore, proposition 2 can be approximately
extended to the total pionic-OB interference.

For the seagull-magnetization interference, a rigorous result is also
difficult to establish. However, certain particular cases suggest a
trend. For instance, in the case \( \nh = 0 \), it can be demonstrated
that \( w^T_{ms} > 0 \). Additionally, by analyzing the integrand of
\( \mathcal{I}_{ms} \) for \( \mathbf{k} = 0 \), we observe that it
remains positive below the quasielastic peak and changes sign for \(
\omega > (q^2/2m_N) (1+(2m_\pi/q)^2)^{1/2} \). This suggests a general
tendency: the interference starts positive at small \(\omega\) and
eventually changes sign at some point beyond the quasielastic peak,
though the precise location cannot be determined analytically.

The integrands and signs in the equations for the \( ms \), \( m\pi
\), and \( m\Delta \) transverse responses are consistent with those
in the pioneering work of Kohno and Ohtsuka \cite{Koh81}, which was
among the first to compute 1b-2b interferences using Riska’s
currents. Similar expressions were also obtained in
\cite{Ama94a,Ama94b}, although written in a different form. One of the
key contributions and novelties of the present chapter is the observation
that the signs of the \( m\Delta \) and \( m\pi \) contributions are
evidently negative, which follows trivially from Eqs. (98) and (100),
as established in our propositions.

\section{Results}

Here we present results for the transverse response functions due to
the interference between the MEC and OB current in the 1p1h
channel. As discussed in the previous sections, these interferences
are expressed as an integral of an effective single-nucleon
interference. In the NR Fermi gas, which we have
examined in great detail, the single-nucleon interferences are
represented through relatively complex integrals after analytically
computing the spin traces. In the case of the seagull and $\Delta$
currents, these integrals are analytical. For the $\Delta$ current, it
has been proven that the associated interference response is always
negative for all values of \(q\) and \(\omega\) (Proposition 1). In
this results section, we calculate the interference transverse
responses for various values of \(q = 300, 400, 550 \, \text{MeV/c}\),
and show the results as a function of \(\omega\) for
\(^{12}\text{C}\).  We employ several nuclear models to compare the
responses, primarily aiming to observe if the results deviate or not
from the Fermi gas significantly.  The nuclear models we use include:
non-relativistic Fermi gas, relativistic Fermi gas, mean-field models,
semi-relativistic models (both Fermi gas and mean field), and the
spectral function model. Relativistic mean field and superscaling
models with effective nucleon mass are also considered. The mean-field
models include the Woods-Saxon potential, Dirac-equation-based
potential, and the plane wave approximation. Many of these models have
been previously applied in calculations for the study of electron
scattering. Our results show approximate agreement in both magnitude
and sign of the different MEC contributions. In particular, all
examined models verify the proposition that the effect of the $\Delta$
current is negative for these values of momentum transfer, and the
total MEC effect is small and predominantly negative. This supports
the consensus that models without NN correlations, such as
single-particle models or models based on one-hole spectral function
do not produce an enhancement of the transverse response.

\begin{figure}
  \centering
\includegraphics[width=8.5cm,bb=160 380 380 810]{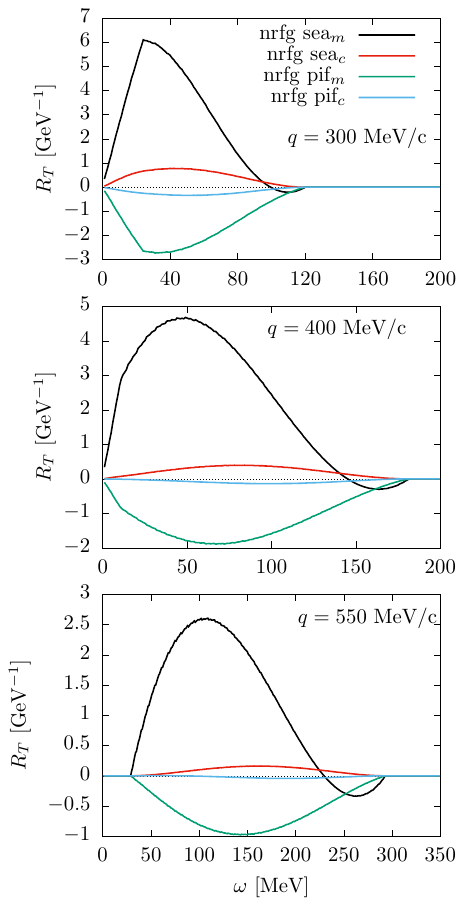}
\caption{ Interferences between the components of the one-body current
  and the MEC in the 1p1h transverse response. Specifically, we represent
  the magnetization-seagull (ms), convection-seagull (cs),
  magnetization-pionic (m$\pi$), and convection-pionic (c$\pi$) interferences
  for different values of $q$.}
\label{norel14}
\end{figure}

\subsection{Fermi gas}

We begin by presenting in Fig. \ref{norel14} results for the
non-relativistic Fermi gas, with $k_F=225$ MeV/c.  We show the
interference of the seagull and pionic currents with the magnetization
and convection currents. This is done to demonstrate that the
contribution of the convection current in the MEC-OB interference is
very small, particularly in the case of the pionic current. As a
result, the magnetization current is dominant in the interference for
these low to intermediate momentum transfer values. Specifically, we
can conclude that, according to proposition 2, the contribution of the
pionic current is negative if the small convection contribution is
disregarded.

Taking this into account, it is worth mentioning the calculation
performed by Alberico et al. \cite{Alb90}. In that reference, a
positive result was obtained for the pionic-OB interference, which
clearly points to an error in the calculation, as it also considered a
Fermi gas model. The proposition establishes that this interference is
negative when convection is neglected. In fact, by inspecting
Eq. (2.41) of Ref. \cite{Alb90}, it can be observed that the
contribution of the pionic current is positive in that reference,
indicating a possible error in performing the spin summation.

\begin{figure}
  \centering
\includegraphics[width=8.5cm,bb=160 380 380 810]{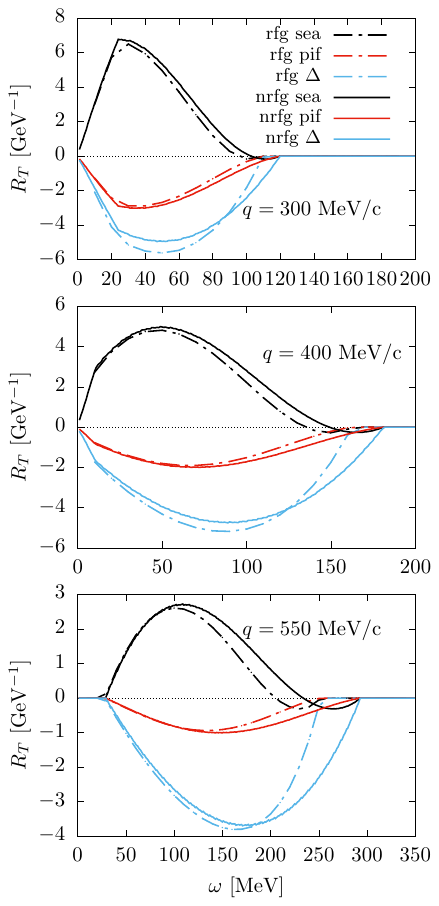}
\caption{ 
Interference between one-body and two-body currents
in the transverse response,
  separated into seagull, pion-in-flight, and $\Delta$ currents.
We  compare the non-relativistic Fermi Gas
  (NRFG) with the Relativistic Fermi Gas (RFG) for three values of the
  momentum transfer in the C12 nucleus. The Fermi momentum is \( k_F =
  225 \) MeV/c.  Diagrams for the 1p1h MEC matrix elements}
\label{norel10}
\end{figure}

In Fig. \ref{norel10}, we present the interferences of the separate
MEC components—seagull, pionic, and $\Delta$—with the OB current in
the transverse response. Here, we compare the non-relativistic Fermi
gas to the relativistic Fermi gas. Both models yield similar results,
with increasing differences as the momentum transfer increases,
primarily due to the different kinematics. In fact, in chapter 3 we
have checked that the relativistic result converges numerically to the
NR one when both the momentum transfer \(q\) and the Fermi momentum
approach zero \cite{Cas23}. NR responses extend to higher values of
\(\omega\), due to the kinematics.  We observe that both propositions
1 and 2 remains valid in the relativistic case, even though it was
proven in the NR limit.

\begin{figure}
  \centering
\includegraphics[width=8.5cm,bb=160 380 380 810]{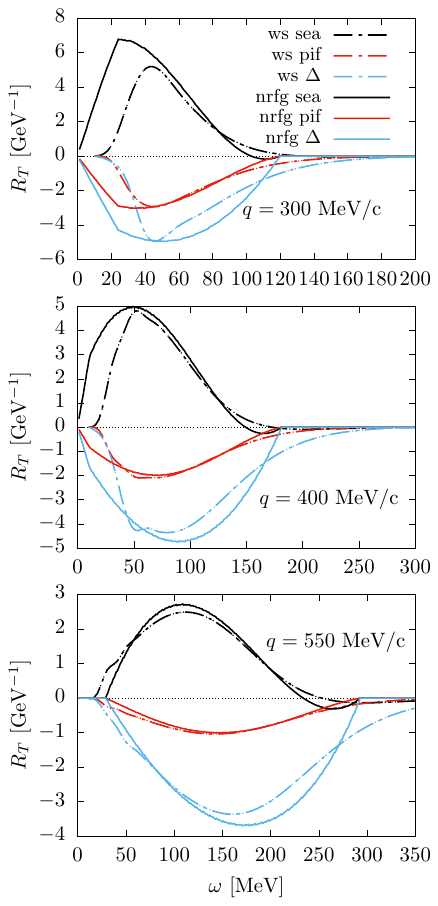}
\caption{ 
The same as Fig. \ref{norel10},
comparing between two models: the mean
  field with Woods-Saxon potential (ws) and the non-relativistic Fermi
  Gas (nrfg) for different values of momentum transfer \( q \) with \(
  k_F = 225 \) MeV/c.
}
\label{norel1}
\end{figure}

\subsection{Mean field with Woods-Saxon potential}

In Fig. \ref{norel1}, we compare the Fermi gas results to those of the
mean field model for finite nuclei, using a Woods-Saxon (WS) potential
\cite{Ama94a,Ama94b}. In this model, the initial and final nucleon
states are solutions to the Schrödinger equation
\begin{equation}
\left[ -\frac{1}{2m_N}\nabla^2+V(r)\right]\psi(\nr)=\epsilon \psi(\nr),
\end{equation}
for positive and negative values of the energy $\epsilon$. 
The WS potential is given by 
\begin{equation}
V(r)= -V_0f(r)
+\left(\frac{\hbar}{m_\pi c}\right)^2\frac{\nl\cdot\nsigma}{r}
\frac{df}{dr}+V_C(r),
\end{equation}
where the function $f(r)$ is the standard Woods-Saxon shape function
\begin{equation}
f(r)= \frac{1}{1+e^{(r-R)/a}},
\end{equation}
and  $V_C(r)$ in the Coulomb potential of a homogeneously charged sphere
with radius $R$. For $^{12}$C 
we use the  WS parameters for $^{12}$C given in Table \ref{cuadro1}.

\begin{table}
  \centering
\caption{Parameters of the Woods-Saxon potential used in this work for
  the nucleus $^{12}$C.}
\vspace{0.5cm}
\begin{tabular}{ccccc}
  \hline
  \hline
         &$V_0$ [MeV] & $V_{ls}$ [MeV]& $a$ [fm] & $R$ [fm] \\ \hline
protons  &$62$         & $3.2$            & 0.57   &  2.86 \\
neutrons &$60$         & $3.15$            & 0.57   &  2.86 \\
\hline
\hline
\label{cuadro1}
\end{tabular}
\end{table}

For $^{12}$C, the initial states include nucleons in the occupied shells
$1s_{1/2}$ and
$1p_{3/2}$. More details can be found in
Refs. \cite{Ama94a,Ama94b}. Note that there is a
typographic error in Ref. \cite{Ama94a} regarding the relative sign
between the central and spin-orbit potentials. This is merely a
mistake in the written formula and does not affect the results. The
energy of the \(1p_{3/2}\) state is lower than that of the
\(1p_{1/2}\) state because the spin-orbit potential is negative for
the \(1p_{3/2}\) state and positive for the \(1p_{1/2}\) state.

The mean-field approach accounts for some effects of the final-state
interaction (FSI) in the response functions.  Additionally, unlike the
Fermi gas model, it incorporates finite-size effects along with
surface effects of the nucleus.  In Fig. \ref{norel1} we observe some
differences between the WS model and the Fermi gas, particularly at
low momentum transfer and low energy, where Pauli blocking affects the
Fermi gas more significantly. The WS response shows a slight tail
extending to higher energies, unlike the Fermi gas. However, at \(q =
550\) MeV/c, the two models become more similar, except for the
high-energy tail seen in the WS model. A possible explanation for this
similarity at intermediate momentum is that the wavelength of the
exchanged photon is small compared to both the nuclear surface and the
extent of the nucleon orbits or wave functions in the occupied shells.

\begin{figure}
  \centering
\includegraphics[width=8.5cm,bb=160 380 380 810]{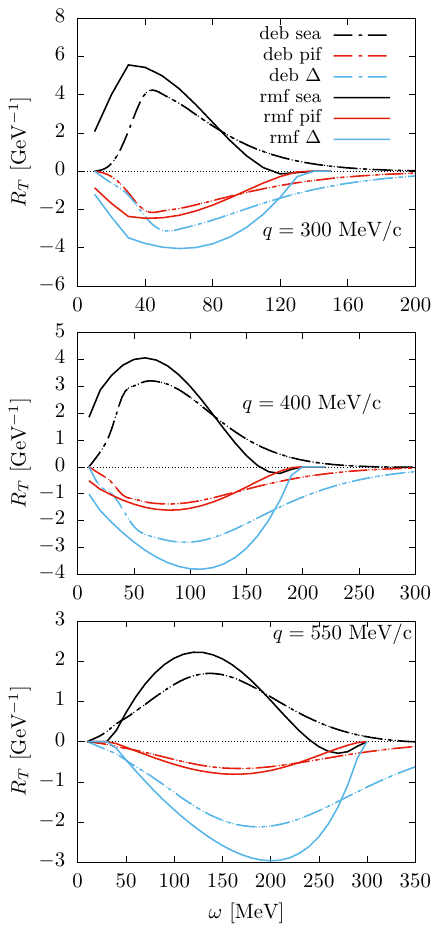}
\caption{ 
The same as Fig. \ref{norel10}.
Results are compared between two models: the
  relativistic mean field with DEB potential (deb) and the
  relativistic mean field of nuclear matter (rmf) 
  with effective mass \( M^* = 0.8 \), for different values of
  momentum transfer \( q \).
}
\label{norel2}
\end{figure}

\subsection{Relativistic mean field}

In Fig. \ref{norel2} we present the interference responses
for  two relativistic mean field models:
  the  Dirac-equation based (DEB) model  and the
  relativistic mean field of nuclear matter 
  with effective mass.

In the  RMF model nucleons
move in the presence of scalar \( U_S(\mathbf{r}) \) and vector \(
U_V(\mathbf{r}) \) potentials, and satisfy a Dirac-like equation for
the four-component nucleon wave function \(\Psi(\mathbf{r})\):

\begin{equation} \label{dirac}
\left[\gamma^0\left(E - U_V(\mathbf{r})\right) 
- \boldsymbol{\gamma} \cdot \mathbf{p} 
- \left(M + U_S(\mathbf{r})\right)\right] \Psi(\mathbf{r}) = 0.
\end{equation}
where $\Psi$ has up and down components
\begin{equation}
\Psi(\nr)= \left(
\begin{array}{c}
\psi_{u}(\nr)
\\
\psi_{d}(\nr)
\end{array}
\right)
\end{equation}
The DEB potential is obtained by rewriting the Dirac equation (\ref{dirac})
as a  Klein-Gordon equation for the upper
component of the wave function. In this reduction, the upper component
is written in the form
\begin{equation}
\Psi_{u}(\mathbf{r}) = A^{1/2}(r,E)\, \phi(\mathbf{r}),
\end{equation}
where \(E\) is the energy in the final state and \( A(r,E)\)
is the Darwin term
\begin{equation}
A(r,E)=1+ \frac{U_S(r)-U_V(r)}{E+M}.
\end{equation}
With this definition the function $\phi(\nr)$ verifies 
the equation
\begin{equation}
\left[ -\frac{1}{2m_N}\nabla^2+ U_{DEB}(r,E)\right] \phi(\nr)=
\frac{E^2-m_N^2}{2m_N}\phi(\nr).
\end{equation}
The DEB potential is given by \cite{Hor91,Udi95}
\begin{equation}
V_{DEB}=V_C + V_{so}\nsigma\cdot\nl+V_D+V_{coul}
\end{equation}
where the central, spin-orbit and Darwin potentials are given by
\begin{eqnarray}
V_C(r,E) &=& 2m_NU_S(r)+2EU_V(r)+U_S(r)^2-U_V(r)^2 
\nonumber\\
V_{so}(r,E) &=& -\frac{1}{rA}\frac{\partial A}{\partial r}
\nonumber \\
V_D(r,E) &=& 
         \frac{3}{4A^2}\left(\frac{\partial A}{\partial r}\right)^2
        -\frac{1}{rA}\frac{\partial A}{\partial r}
        -\frac{1}{2A}\frac{\partial^2 A}{\partial r^2},
\nonumber
\end{eqnarray}
and $V_{coul}$ is the Coulomb potential of a homogeneously charged sphere 
with nuclear radius $R$.

In Fig. \ref{norel2} we present the results of the interference 1b-2b
transverse response using the DEB potential within the
semi-relativistic model of Refs. \cite{Ama05b,Ama10b}. It is observed
that the contribution from the $\Delta$ current is negative, as is the
contribution from the pion-in-flight current. Consequently, this model
verifies the low momentum propositions.

In the same Figure \ref{norel2}, the results using the DEB potential
are compared with those obtained from the RMF of nuclear matter
computed in chapter 3. In this model, the scalar and vector potentials are
constant, making it similar to the RFG but with the nucleon mass
replaced by an effective mass \(m_N^*=m_N+U_S\) and the energy
increased by a constant vector energy \(E_V=U_V\).  For the
\(^{12}\text{C}\) case shown in Fig. \ref{norel2}, the values used are
\(m_N^* = 0.8\,m_N\) and \(E_V = 141\) MeV. More details of the RMF
nuclear matter model with MEC can be found in
Ref. \cite{Mar21}.  As seen in Figure \ref{norel2}, both the
DEB model and the RMF model with effective mass yield qualitatively
similar results, with the peaks of the interference responses largely
coinciding. This similarity arises because both models incorporate
final-state interaction effects. However, the absolute values obtained
with the DEB model are smaller. This is a consequence of the fact
that, in the DEB model, the effective mass depends on \(r\), leading
to responses that exhibit a tail extending much further than those in
the shell model or nuclear matter. Essentially, it appears as if the
strength is spread over a wider energy interval.  In any case, it is
remarkable that the low momentum propositions remain verified in the
models presented in Figure \ref{norel2}: the 1b-$\Delta$ interference is negative
and the 1b-pionic one is negative.

\begin{figure}
  \centering
\includegraphics[width=8.5cm,bb=180 380 430 810]{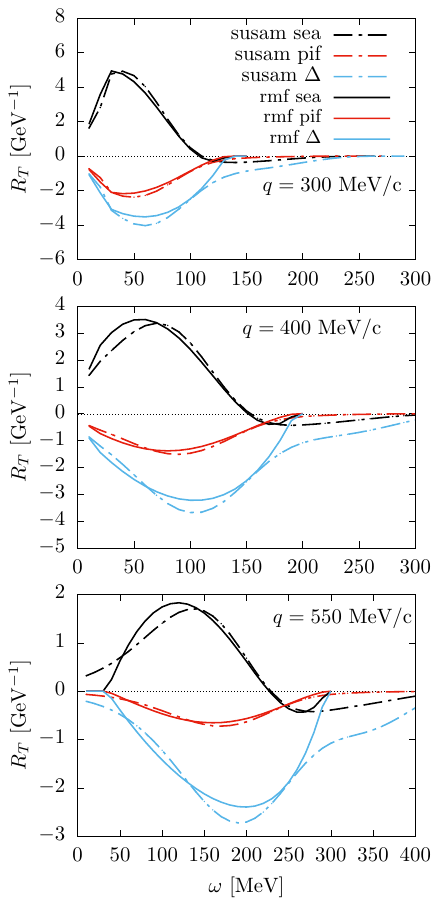}
\caption{ 
The same as Fig. \ref{norel10}.
Results are compared between two models: the
  superscaling model with relativistic effective mass
and the relativistic mean field of nuclear matter (rmf) 
  with effective mass \( M^* = 0.8 \), for different values of
  momentum transfer \( q \).
}
\label{norel15}
\end{figure}

\subsection{SuSAM* model}
In this section we present the results obtained with the SuSAM* model
for low momentum transfer. This model was introduced in chapter~2 and
later extended in chapter 3. In the first step, the single--nucleon
prefactor of the superscaling model was modified by computing an
averaged value with a Fermi momentum distribution, instead of using
the direct extrapolation of the RFG result. In chapter 3, this
averaging procedure was further employed to include MEC consistently
within the same framework. For a detailed description of the
formalism and its derivation, the reader is referred to those
chapters.

In Fig.~\ref{norel15} we show the interference transverse responses of
${}^{12}\mathrm{C}$ between the one--body current and the separate MEC
contributions (seagull, pionic, and $\Delta$) for momentum transfers
$q = 300$, $400$, and $550~\text{MeV}/c$. This figure complements
Figs. 3.7--3.10, where the individual interferences were displayed for
higher momentum transfers. Here, the SuSAM* results are compared with
those of the RMF model with an effective mass $M^* = 0.8$.

As in chapter 3, both models produce very similar results except for
the high--$\omega$ tail, where the SuSAM* responses extend slightly
beyond the RMF ones due to the phenomenological scaling function
inherent to the scaling approach. The overall behavior of the
different interference terms remains consistent with that found in the
RMF: the seagull contribution is positive, while the pionic and
$\Delta$ terms are negative. These results confirm that the SuSAM*
model also verifies the low--momentum propositions, reproducing the
same qualitative features of the RMF calculation. Both frameworks are
relativistic and include the effects of an effective nucleon mass.
\subsection{Strong form factor and relativistic effects}

\begin{figure}
  \centering
\includegraphics[width=8.5cm,bb=160 380 380 810]{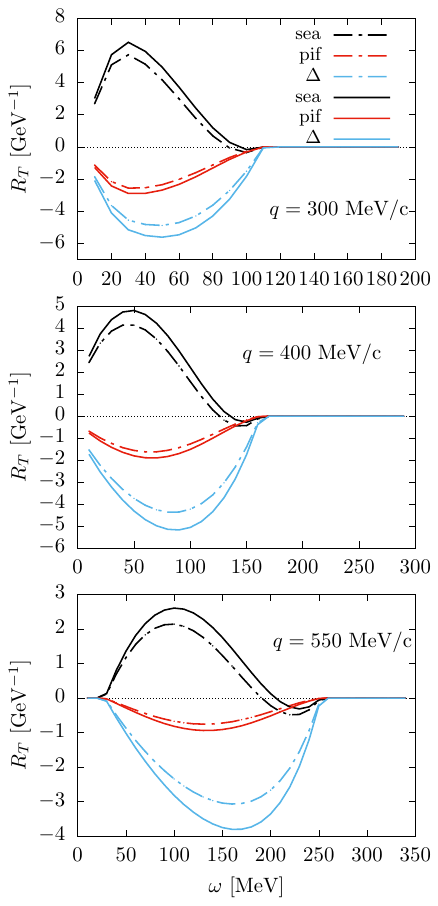}
\caption{  The same as Fig. \ref{norel10}
 but calculated with RFG, with and without the strong $\pi NN$ form factor. The dashed line represents the results with the form factor, 
while the solid line represents the results without it.
}
\label{norel5}
\end{figure}

In Fig. \ref{norel5}, we show the effect of including the \(\pi NN\)
and \(\pi N\Delta\) form factors. In our non-relativistic Fermi gas
equations and in the low-\(q\) propositions, we have omitted these
form factors for simplicity. These form factors are multiplicative
factors that would be included inside the internal integrals over the
intermediate nucleon momentum \(\mathbf{k}\). Their inclusion does not
affect the low momentum propositions since these form factors are
positive and do not alter the sign of the interference. Moreover, in
the NR calculation, some integrals can be evaluated analytically
without the form factors, which further simplifies the
computation. Given that we are considering small momentum transfers,
the effect of the form factors is minimal, as demonstrated in
Fig. \ref{norel5}, where the relativistic Fermi gas results are
compared with and without the strong form factor. As the form factor
is less than one, the inclusion produces a reduction of the maximum in
absolute value.

The relativistic Fermi gas can be compared with the semirelativistic
Fermi gas model (SRFG) developed in \cite{Ama03}. The SRFG model
starts from the non-relativistic Fermi gas, incorporating relativistic
kinematics and replacing the NR current with a
semirelativistic expansion in powers of the initial nucleon momentum
divided by the nucleon mass $(h/m)$, while preserving the exact
dependence on the final momentum. This approach was extended to
include MEC \cite{Ama02} and also applied to the $\Delta$ current,
although in the latter case the semirelativistic correction is not
exact due to the use of a static $\Delta$ propagator.  The
semirelativistic current is obtained from the relativistic one by
multiplying by a factor \(1/\sqrt{1+\tau}\).

\begin{figure}
  \centering
\includegraphics[width=8.5cm,bb=160 380 380 810]{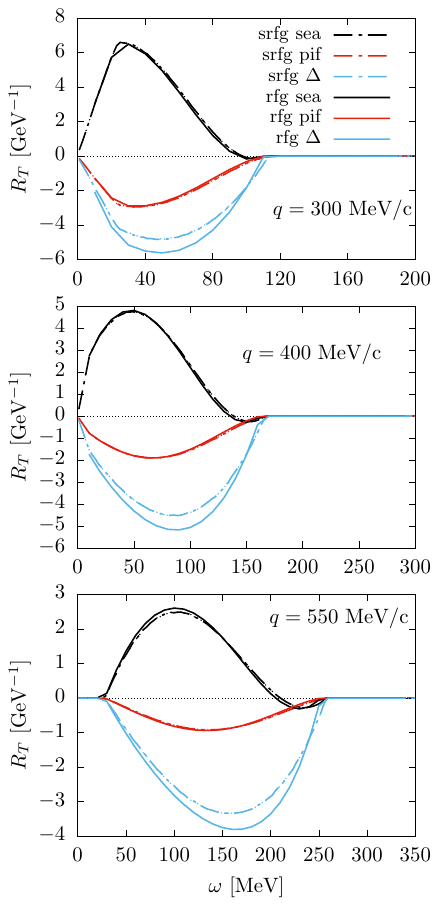}
\caption{ 
The same as Fig. \ref{norel10} but showing two models: RFG
  and semirelativistic Fermi gas (SRFG). The comparison illustrates
  the differences between the relativistic and semirelativistic
  approaches in the transverse response, for different values of $q$.  }
\label{norel6}
\end{figure}

\begin{figure}
  \centering
\includegraphics[width=8.5cm,bb=160 380 380 810]{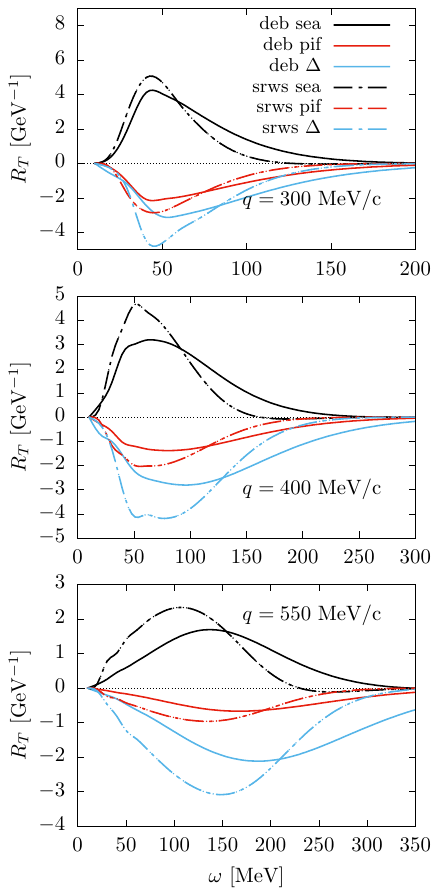}
\caption{ The same as Figure \ref{norel10} , but now comparing the
  models: relativistic mean field with DEB potential (deb) and
  semi-relativistic mean field with Woods-Saxon potential (srws) for
  different values of momentum transfer \( q \).
}
\label{norel3}
\end{figure}

In Fig. \ref{norel6},
the SRFG model is compared with the exact RFG for the interference
between the MEC and the one-body current. For the seagull and pionic
contributions, the SRFG model agrees very well with the relativistic
one. However, for the $\Delta$ contribution, the
approximation is less accurate because the static $\Delta$ propagator
limits the effectiveness of the semirelativistic factor.
In addition, the semirelativistic model was extended to be applied in
conjunction with the Woods-Saxon mean field model \cite{Ama10b}. This
extended model can be directly compared with the DEB model. In fact,
the DEB model also incorporates semirelativistic MEC currents, but
these currents are further modified because the pion propagator in the
DEB model is made dynamic by including the pion energy as the
difference in energy between the nuclear states of the mean field
model. The comparison between these two models, DEB and SRWS, as shown
in Fig. \ref{norel3}, reveals significant differences in both the width and the
height of the interference response peak. Specifically, the DEB model
extends to higher energies and exhibits a broader peak, which is
attributed to the fact that the DEB potential is much stronger than
the Woods-Saxon potential.
\subsection{Plane wave approximation}
In the shell model with a Woods-Saxon potential, the plane wave
approximation (PWA) assumes that the final nucleon with momentum
\(\np\) is described by a plane wave, meaning it is a solution of the
Schrödinger equation without final-state interactions. Therefore, in
this approach, the plane wave approximation is applied only to the
final outgoing particle state, while the initial state nucleons remain
described by the bound shell model wave functions.

\begin{figure}
  \centering
\includegraphics[width=8.5cm,bb=160 380 380 810]{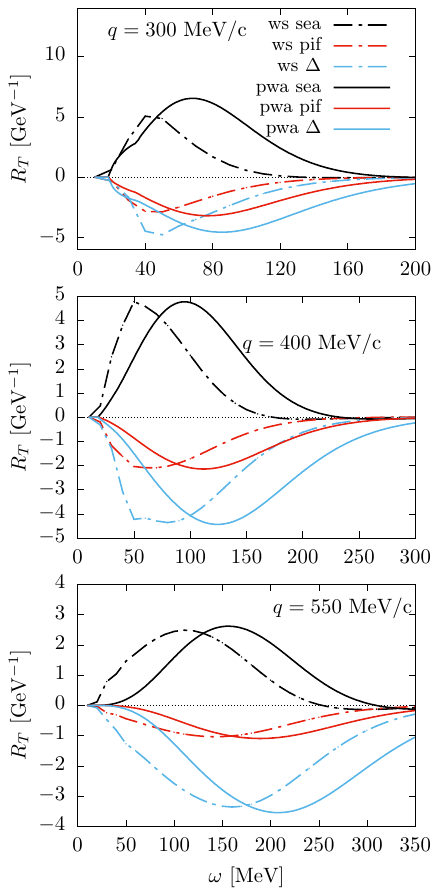}
\caption{ The same as Figure \ref{norel10}, but now comparing the models: mean
  field with Woods-Saxon potential (ws) and mean field in plane-wave
  approximation for the final state (pwa) for different values of
  momentum transfer \( q \).
}
\label{norel12}
\end{figure}

Results using this model are presented in Fig. \ref{norel12}, where they are
compared with the Woods-Saxon mean-field calculations for the 1b-2b
interference responses. The observed effect is similar to that seen in
the 1b response within the Plane Wave Impulse Approximation (PWIA)
\cite{Ama05b}. The impact of final-state interactions appears as a
shift in the response. This shift can be understood as a consequence
of the energy imbalance between the initial and final states. In the
initial state, the nucleon has both kinetic and potential energy,
whereas in the final state, only kinetic energy remains, since the
potential is neglected. This energy mismatch propagates to the
energy-conserving delta function, altering the position of the
response peak.
The shift in the response function can be qualitatively understood
using a back-of-the-envelope estimate. First we assume that the matrix
element of the current in PWA is
approximately equal to the matrix element in the Woods-Saxon model,
$\langle J^\mu \rangle_{PW} \simeq \langle J^\mu \rangle_{WS}$.
Second, we approximate the potential energy of the final-state
nucleon as a constant, $V_p \simeq -V < 0$.  Thus, the total energy of
the outgoing particle can be written as the sum of its kinetic and
potential energy: $\epsilon_p = t_p - V.$ Using this, the transverse response
function in PWA can be expressed as
\begin{eqnarray}
R^T_{PW}(q,\omega) &=& \sum_{ph} |\langle J_T \rangle_{PW}|^2
\delta(t_p - \epsilon_h - \omega) \nonumber
\\ 
&\simeq& \sum_{ph}
|\langle J_T \rangle_{WS}|^2 \delta(\epsilon_p+V- \epsilon_h - \omega)
\nonumber
\\ &=& R^T_{WS}(q,\omega-V)
\end{eqnarray}
This expression shows that the response function is effectively
shifted due to the neglect of the potential in the final state.

From Figure \ref{norel12}, we observe again that the low-momentum
proposition for the 1b-MEC interference response remains valid in both the
plane-wave approximation and the Woods-Saxon potential.

\subsection{Spectral function model}

In this subsection, we present results using the spectral function
(SF) model, which employs the one-hole spectral function, $S(\np,E)$,
that depends on the missing momentum and missing energy.  In the SF
model, the transverse response is computed assuming factorization of
the single-nucleon response and the one-hole spectral function for
one-particle emission.
\begin{equation} \label{spectral}
R_T(q,\omega)= \int d^3 p \, w_T(\np,\np-\nq) S(\np-\nq,\omega-T_p)
\end{equation}
where the single nucleon response is $w^T=w^{11}+w^{22}$, while
 $w^{\mu\mu}$ is defined in Eq. (\ref{single-nucleon}).

 Unlike the single-particle model that assumes holes with definite
 energy, the SF approach accounts for a continuous distribution of
 hole energies. It provides the probability that the system contains a
 hole state with momentum \(\mathbf{h} =
 \mathbf{p} - \mathbf{q}\) and a missing energy \(E = \omega - T_p\),
 where \(T_p = \mathbf{p}^2/(2m_N)\).  The basic theory of the SF
 approach to QE electron scattering is summarized in Appendix \ref{appF}.

\begin{figure}[t]
  \centering
\includegraphics[width=8.5cm,bb=145 80 520 620]{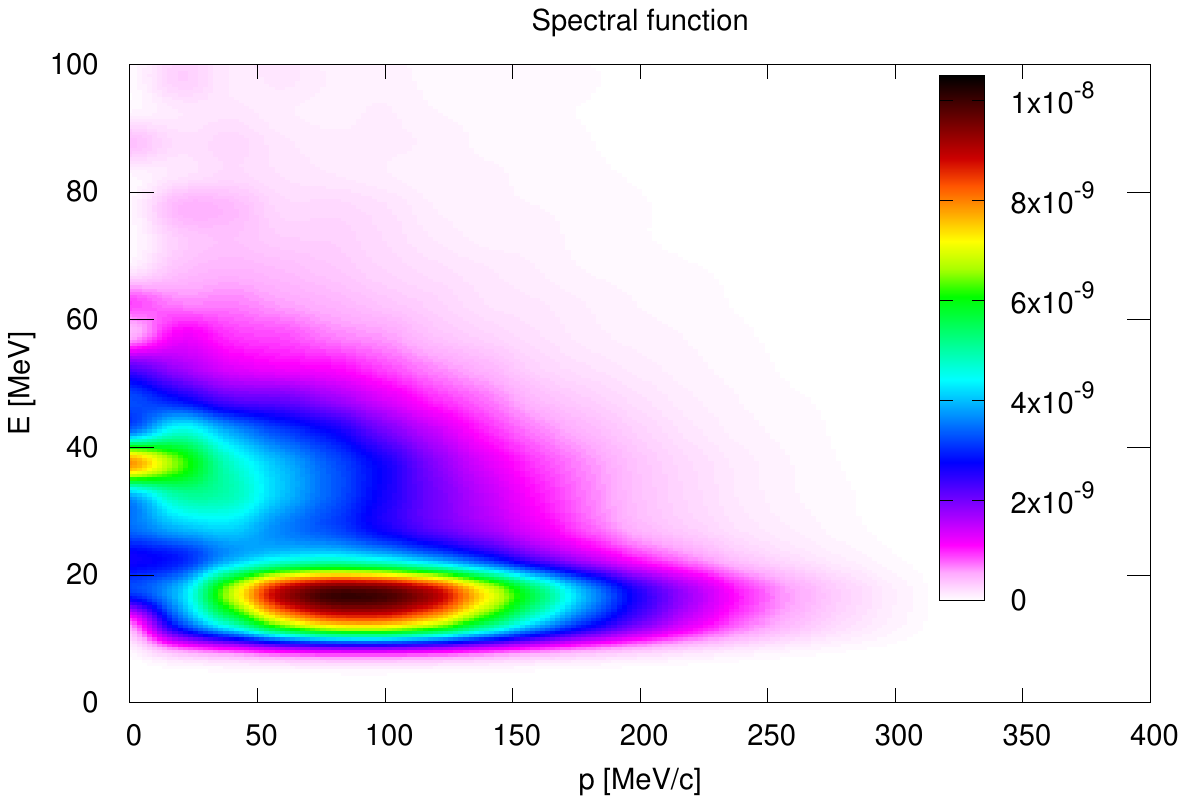}
\caption{Spectral function of $^{12}$C in units of MeV$^{-4} c^3$
}
\label{sf1c}
\end{figure}

We use the spectral function, $S(p,E)$, for \(^{12}\)C taken from
Ref. \cite{Ben94} for both protons and neutrons, as shown in
Fig. \ref{sf1c}. This spectral function exhibits peaks around \(E
\simeq 19\) MeV and \(E \simeq 39\) MeV as a function of energy. These
values are close to the binding energies of the \(1p_{3/2}\) and
\(1s_{1/2}\) shells in the extreme shell model, where the nuclear
wave function is described by a Slater determinant.

In the shell model, the spectral function is given by  
\begin{equation}
S(p,E) = \sum_{nlj} (2j+1) |\tilde{R}_{nlj}(p)|^2 \delta(E + \epsilon_{nlj})
\end{equation}
where the sum runs over occupied shells, and \( \tilde{R}_{nlj}(p) \)
are the shell radial wave functions in momentum space, with
single-particle energy $\epsilon_{nlj}$. In the more realistic
spectral function of Fig. \ref{sf1c}, the energy dependence is smeared around
the shell binding energies, resulting in a continuous energy
distribution instead of discrete shell levels.

\begin{figure}
  \centering
\includegraphics[width=8.5cm,bb=100 180 410 770]{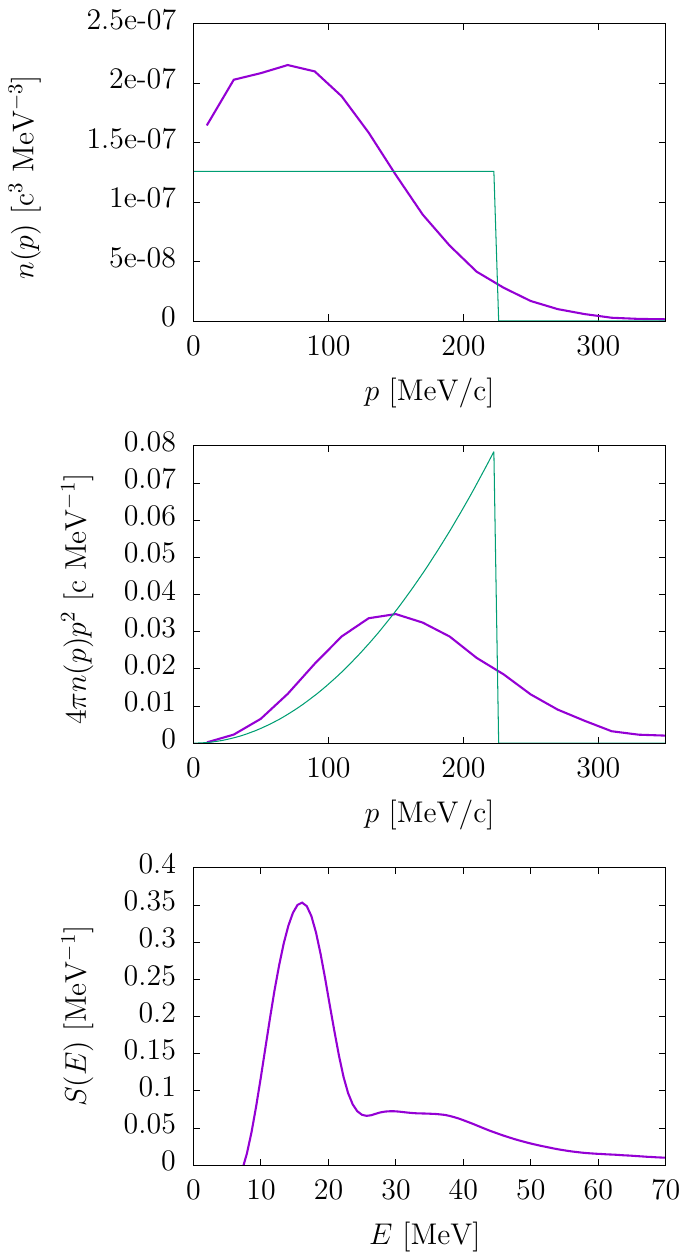}
\caption{ Proton momentum distribution of $^{12}$C, the radial
  momentum distribution, and missing energy distribution, obtained
  from the spectral function by integration.  }
\label{fig6}
\end{figure}

In Fig. \ref{fig6}, we show the proton momentum distribution $n(p)$ 
obtained by
integrating the spectral function over the missing energy. This
distribution is compared with the constant momentum distribution of
the Fermi gas model. Additionally, we present the radial momentum
distribution, \( 4\pi n(p) p^2 \), which highlights the probability
density of nucleons as a function of momentum. The missing energy
distribution, obtained by integrating the spectral function over
momentum, is also displayed. The normalization follows \( \int d^3p \,
n(p) = 6 \) for \( ^{12}C \), reflecting that the proton and neutron
distributions are identical in this model.

It is worth noting that the response function in the SF model,
Eq. (\ref{spectral}), is expressed as an integral over the final
nucleon momentum \( p \). To evaluate this integral, it is convenient
to first integrate over the missing energy and missing momentum. The
missing energy is given by
\begin{equation}
E = \omega - T_p = \omega - \frac{p^2}{2m_N}.
\end{equation}
Differentiating, we obtain \( dE = -p \, dp / m_N \), and the volume element in spherical coordinates is  
\[
d^3p = m_N \, p \, dE \, d\Omega
\]
where \( \theta \) and \( \phi \) are the nucleon emission angles. The
response function can then be rewritten as
\begin{equation}
R_T(q, \omega) 
= m_N \int dE d\Omega \, p\,   
w_T(\np, \np - \nq) S(\np - \nq, E) 
\label{integral121}
\end{equation}
Next, we define \(\nh = \np - \nq\), leading to the relation $h^2 =
p^2 + q^2 - 2pq\cos\theta$, where \(\theta\) is the angle between
\(\np\) and \(\nq\), with \(\nq\) chosen along the
\(z\)-axis. Differentiating with respect to \(\theta\), we obtain $h
\, dh = -pq \, d\cos\theta$.

Substituting this into the integral (\ref{integral121}), we can
express the transverse response as
\begin{equation}
R_T(q,\omega) = 2\pi \frac{m}{q} \int_0^\omega dE \int_{|p-q|}^{p+q} dh S(h,E) w_T(\nh+\nq, \nh),
\end{equation}
where $p=\sqrt{2m_N(\omega-E)}$. Note that $E<\omega$ ensures that $p$
is well defined.  The factor \(2\pi\) arises from the integration over
\(\phi\), and the integration limits in \(h\) correspond to nucleon
emission in the direction of \(\pm \nq\).

The effect of MEC is estimated by treating the spectator nucleon as an
on-shell plane-wave with momentum $\nk$, therefore we replace the
single-nucleon response by the effective single nucleon including MEC,
in Eq. (\ref{single-nucleon}).  This approximation has been done in
the past in previous calculations by the Pavia group for $(e,e'p)$
reactions \cite{Bof90}, and in recent RMF-based calculations
\cite{Fra23}, where the spectator nucleon is described using an
effective mass and vector energy. A similar approach to MEC was also
adopted in the spectral function model of Ref. \cite{Lov23}. Thus the
transverse response is evaluated using the effective single nucleon,
Eq. (\ref{single-nucleon}), which includes the MEC contribution, effectively decoupling
it from the spectral function.

\begin{figure}
  \centering
  \includegraphics[width=8.5cm,bb=160 380 380 810]{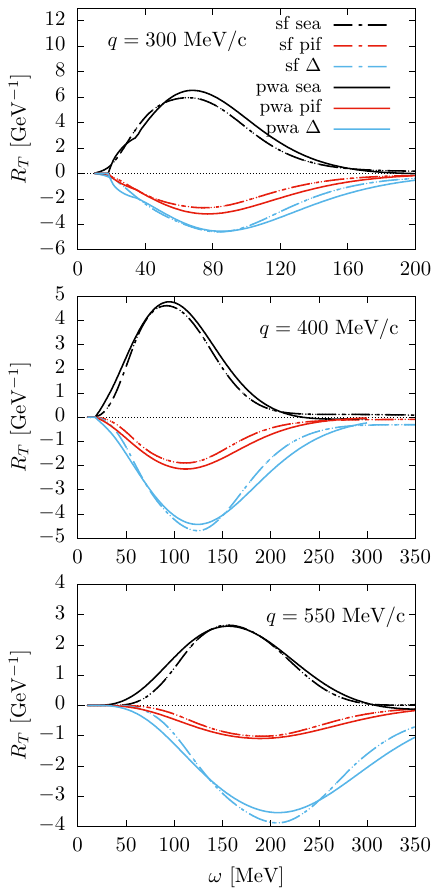}
\caption{ The same as Figure \ref{norel10}, but now comparing the
  models: spectral function (SF) and mean field with plane-wave
  approximation (PWA) for different values of momentum transfer \( q
  \).}
\label{sf}
\end{figure}

In the spectral function model, the interference MEC-OB responses are
presented in Fig. \ref{sf} for the separate contributions from the seagull,
pion-in-flight, and \(\Delta\) currents. The figure compares the SF
results with those obtained using the PW model from the previous
subsection. Both models yield quite similar results. This similarity
arises from the fact that both models assume plane waves for the
final-state nucleon. In the PW model, the response is obtained by
summing the contributions from each shell separately, while in the SF
model, the shell contributions are smeared according to the spectral
function's energy distribution. However, this smearing effect is
barely noticeable in the inclusive response, as the information about
the hole energy is lost.
Furthermore, the agreement between the SF and PWIA models reinforces
the validity of the approximation that treats the spectator nucleon as
a plane wave. While this approximation is not explicitly made in the
PW model, it is assumed in the SF model. In conclusion, the SF
model, as applied here, fully adheres to the low-momentum proposition,
consistent with all the models analyzed in this work.

\begin{figure}
  \centering
\includegraphics[width=8.5cm,bb=180 380 430 810]{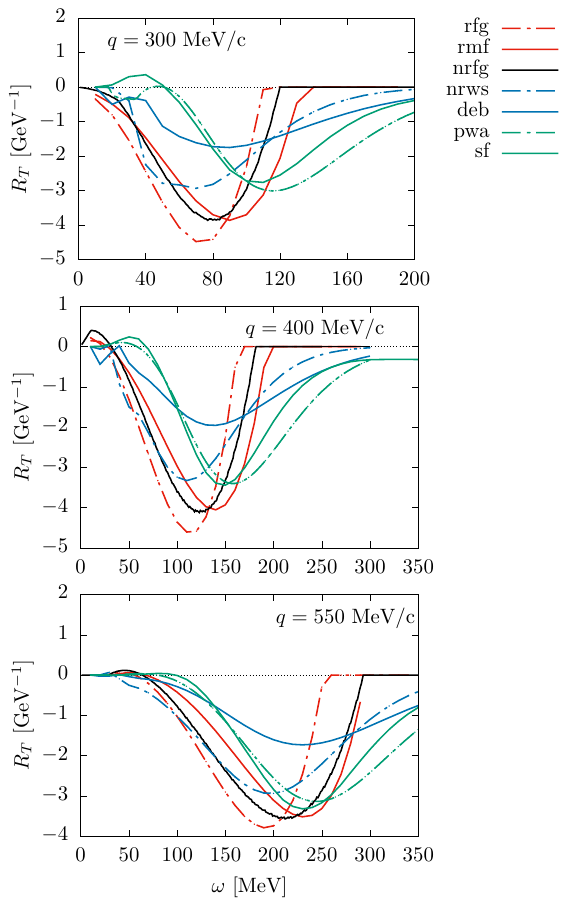}
\caption{
 Total interference OB-MEC compared across all different models considered in this work for all values of $q$. The $\pi NN$ form factor is not included.}
\label{norel13}
\end{figure}

\subsection{Total interference response.}
To conclude the results section, we present a comprehensive comparison
of most of the models discussed in this chapter to assess the overall
impact of MEC and the theoretical uncertainties. In
Fig. \ref{norel13}, we display the total 1b-MEC interference for a
selection of seven models. All models consistently predict a negative
interference, although there are significant quantitative differences
in the position of the peak and the width of the distribution. Despite
these variations, the overall magnitude remains comparable, with
differences up to a factor of two.  Importantly, the key takeaway from
this comparison, and one of the main objectives of this chapter, is that
the low-momentum proposition holds across all models analyzed. None of
the models considered in this chapter exhibit a qualitative deviation,
such as a sign change in the interference term.

To end we list here the models considered in this chapter:
\begin{enumerate}

\item Non relativistic Fermi gas (NRFG)

\item Relativistic Fermi gas  (RFG)

\item Mean field with Woods-Saxon potential (WS)

\item Mean field with Dirac-equation based potential (DEB)

\item Relativistic mean field of nuclear matter with effective mass (RMF)

\item Mean field with plane wave approximation (PWA)

\item Semirelativistic mean field with Woods-Saxon (SRWS)

\item Semirelativistic Fermi gas (SRFG)

\item Superscaling model with effective mass (SuSAM*).

\item Spectral function (SF)

\end{enumerate}

\section{Final remarks}

In this chapter, we have conducted a detailed reexamination of the
OB-MEC interference in the one-particle emission transverse response,
focusing on the low-momentum proposition. We compared various models,
obtaining qualitatively consistent results. A key aspect of our
analysis was the derivation of the low-momentum proposition within the
non-relativistic Fermi gas framework, ensuring full transparency and
reproducibility in our approach. The proposition clearly establishes
that the sign of the interference with the pionic and $\Delta$
currents is negative, which refers to the Riska current, considered
here as the standard. Once this current is fixed, its sign does not
present ambiguities. Our results show that all models considered
adhere to the low-momentum proposition. The common feature among these
models is that they are based on independent-particle descriptions,
either relativistic or non-relativistic, or extensions thereof, such
as the one-hole spectral function for one-particle emission in
electron scattering. Crucially, these models do not include explicit
two-body correlations beyond mean-field approximations. Consequently,
in all cases, the MEC particle-hole matrix element only retains the
exchange contribution involving the spectator nucleon, without
additional contributions.  Given these results, it does not seem
possible to explain any enhancement in the transverse response in
one-particle emission using models that do not include fundamentally
different ingredients that would violate the low-momentum
proposition. A clear candidate for producing such an enhancement, as
suggested by previous studies, is the inclusion of SRC correlations,
as in the microscopic calculation of \cite{Fab97}. These correlations
could introduce contributions that do not adhere to the low-momentum
proposition, thus altering the dynamics beyond what is captured by the
models analyzed here. We will study the SRC in chapters 6 and 7.

\chapter{CCQE neutrino-nucleus scattering with  Meson exchange currents }

In the preceding chapters, we have discussed the electromagnetic
scattering processes. The present chapter focuses on neutrino-nucleus
scattering. The effect of meson-exchange currents on CCQE neutrino
scattering with single-nucleon emission is computed and analyzed
within the simplest possible framework: the RFG model. This serves as
a first step towards more elaborate approaches, such as the RMF model
for nuclear matter and the SuSAM* model, both of which yield
qualitatively similar results. The calculations obtained show a
reduction of the vector, axial and vector-axial transverse response
functions and, consequently, a decrease in the total neutrino cross
section. A comparison with the NR limit is also presented. More
details can be found in Ref. \cite{Cas25b}.

\section{Charge-changing quasielastic neutrino scattering}
We will begin by summarizing the CCQE formalism in neutrino scattering
($\nu_\mu,\mu^-$) and ($\bar{\nu}_\mu,\mu^+$). Our approach is based
on previous works that have addressed charged-current (CC)
reactions in nuclei \cite{Rui17, Ama20}. We assume that the
four-momentum transfer is given by $k^\mu-k'^\mu=Q^\mu = (\omega,
\mathbf{q})$ with $Q^2=\omega^2-q^2<0$. The energy transfer is \(
\omega \), $\nq$ is the momentum transfer along the z-axis, and $\nk$
and $\nk'$ are the momenta of the incoming neutrino and the outgoing
muon, respectively. Their corresponding energies are $\epsilon=E_\nu$
for the neutrino and $\epsilon'=m_\mu+T_\mu$ for the muon. In this
way, assuming that \(\theta\) is the scattering angle between \(\nk\)
and \(\nk'\), the double-differential cross section is written as
follows,
\begin{equation}
  \frac{d^2\sigma}{dT_\mu d\Omega_\mu}= \frac{G^2cos^2\theta_c}{4\pi^2}\frac{k'}{\epsilon}
  \frac{v_0}{2} \left (V_{CC}R_{CC}+2V_{CL}R_{CL}+V_{LL}R_{LL}+V_{T} R_{T}\pm2V_{T'} R_{T'}\right )
\end{equation}
where $G=1.666 \times 10^{-11} \; MeV^{-2}$ is the Fermi weak constant
and $\theta_c$ is the Cabibbo angle. We define the factor $v_0 =
(\epsilon+\epsilon')^2-q^2$. The coefficients \(V_K\) are obtained
from the components of the leptonic tensor
\begin{eqnarray}
   V_{CC}&=&1+\delta^2\frac{Q^2}{v_0},\nonumber \\  
   V_{CL}&=&\frac{\omega}{q}-\frac{\delta^2}{\rho'}\frac{Q^2}{v_0}, \nonumber \\
   V_{LL}&=&\frac{\omega}{q^2}+\left (1+\frac{2\omega}{q \rho'}+\rho \delta^2 \right )
   \delta^2\frac{Q^2}{v_0}, \nonumber \\
   V_{T}&=&\frac{Q^2}{v_0}+\frac{\rho}{2}-\frac{\delta^2}{\rho'}
   \left (\frac{\omega}{q}+\frac{1}{2}\rho \rho' \delta^2 \right ) \frac{Q^2}{v_0}, \nonumber \\
   V_{T'}&=&\frac{1}{\rho'}\left (1-\frac{\omega \rho'}{q}\delta^2 \right )
   \frac{Q^2}{v_0},
\end{eqnarray}
with the dimensionless factors 
\begin{equation}
\delta=\frac{m_\mu}{ \sqrt{|Q^2|}},
\kern 1cm
\rho=\frac{|Q^2|}{q^2}, 
\kern 1cm 
\rho'= \frac{q}{\epsilon+\epsilon'}. 
\end{equation}
The five
nuclear responses \( R_K \), which only depend on $(q,\omega)$,
arise from various combinations of the components of the hadronic
tensor,
\begin{eqnarray}
  R_{CC}&=&W^{00},\nonumber \\  
  R_{CL}&=&-\frac{1}{2}(W^{03}+W^{30}), \nonumber \\
  R_{LL}&=&W^{33},\nonumber \\ 
  R_{T}&=&W^{11}+W^{22},\nonumber \\
  R_{T'}&=&-\frac{i}{2}(W^{12}-W^{21}).
\end{eqnarray}

In the case of charged-current weak interactions, the nuclear current
operator is the sum of a vector and an axial-vector component. As a
result, the response functions CC, CL, LL, and T, 
can each
be written as the sum of two separate contributions: one arising from
the vector-vector (VV) part of the current, and the other from the
axial-axial (AA) part:
\begin{equation}
  R_K = R_K^{VV} + R_K^{AA}, \;\;\;\;\;\;\;\; K=CC,\, CL,\, LL,\, T \, .
\end{equation}
On the other hand, the $T'$ response originates from the interference
between the vector and axial components of the current, 
and can be written as
\begin{equation}
  R_{T'} = R_{T'}^{VA} + R_{T'}^{AV} \, .
\end{equation}

The hadronic tensor for the 1p1h channel is
calculated for a RFG and is constructed from
the formalism that follows in parallel with the case of electron
scattering of previous chapters. The main differences are the presence
of an axial component in the charged current and the modified
structure of the isospin operators. The inclusive hadronic tensor for
the 1p1h channel is given by
\begin{eqnarray}
W^{\mu\nu}&=& \sum_{ph}
\left\langle
ph^{-1} \right|\hat{J}^{\mu} |\left. F \right\rangle^{*}
\left\langle
ph^{-1} \right|\hat{J}^{\nu} |\left. F \right\rangle 
\delta(E_{p}-E_{h}-\omega)
\theta(p-k_F)\theta(k_F-h),
\label{hadronic}
\end{eqnarray}
where $|F\rangle$ is the Fermi ground
state. Additionally, the theta functions ensure that the initial
nucleons have a momentum below the Fermi level \(h< k_F \) and that the
final nucleons, after the interaction, have a momentum \( p > k_F \).

The nuclear current  is taken as the sum of 
one-body and  two-body operators
\begin{equation}
\hat{J}^\mu = 
\hat{J}^\mu_{1b} 
+\hat{J}^\mu_{2b}.
\end{equation}
The 1p1h  matrix element of these operators in the RFG is given by 
\begin{equation}
\left\langle ph^{-1} \right|\hat{J}_{1b}^{\mu} |\left. F \right\rangle
=
\left\langle p \right|\hat{J}_{1b}^{\mu} |\left. h \right\rangle,
\label{j1b}
\end{equation}
\begin{equation}
\left\langle ph^{-1} \right|\hat{J}_{2b}^{\mu} |\left. F \right\rangle 
=
\sum_{k<k_F}\left[
\left\langle pk \right|\hat{J}_{2b}^{\mu} |\left. hk \right\rangle 
- \left\langle pk \right|\hat{J}_{2b}^{\mu} |\left. kh \right\rangle
\right].
\label{j2b}
\end{equation}
Note that the matrix element of the MEC, being a two-body operator,
involves a transition between pairs of nucleons; however, in a 1p1h
excitation, one of the nucleons,  $|k\rangle= |\nk,s_k,t_k\rangle $,
remains in its initial state and acts merely as a spectator.

The next step is to write the elementary matrix elements of the
one-body and two-body current operators between plane-wave states, using
momentum conservation 
 \begin{eqnarray}
 \langle p |\hat{J}_{1b}^{\mu} | h\rangle 
&=&
  \frac{(2\pi)^{3}}{V}\delta^{3}(\nq+\nh-\np)
j_{1b}^{\mu}(\np,\nh), 
\label{OBmatrix}
\\
\langle p'_{1}p'_{2}|\hat{J}_{2b}^{\mu}|p_{1}p_{2}\rangle
&=&
\frac{(2\pi)^{3}}{V^{2}}\delta^{3}(\np_1+\np_2+\nq-\np'_1-\np'_2)
j_{2b}^{\mu}(\np'_1,\np'_2,\np_1,\np_2).
\label{TBmatrix}
\end{eqnarray}
The current functions $j^\mu_{1b}(\np,\nh)$ and \(j_{2b}^{\mu}(\np'_1,
\np'_2, \np_1, \np_2)\) implicitly depend on spin and isospin indices.
Inserting Eqs. (\ref{OBmatrix}) and (\ref{TBmatrix}) in
Eqs. (\ref{j1b}) and (\ref{j2b}), respectively, we can write the total
current matrix element in the form
 \begin{equation}
\left\langle ph^{-1} \right|\hat{J}^{\mu} |\left. F \right\rangle 
=
  \frac{(2\pi)^{3}}{V}\delta^{3}(\nq+\nh-\np)
j^{\mu}(\np,\nh), 
\label{total}
\end{equation}
 where
\begin{equation}
j^{\mu}(\np,\nh) \equiv j_{1b}^{\mu}(\np,\nh)+ j_{2b}^{\mu}(\np,\nh), 
\end{equation}
and
\begin{eqnarray}
j_{2b}^{\mu}(\np,\nh) 
\equiv 
\frac{1}{V}
\sum_{k<k_F}
\left[ j_{2b}^{\mu}(\np,\nk,\nh,\nk)-j_{2b}^{\mu}(\np,\nk,\nk,\nh)\right].
\label{effectiveOB}
\end{eqnarray}
This effective current, $j_{2b}^\mu(\np,\nh)$, accounts for the fact that the
two-body current operator, when acting on a nucleon pair in
the Fermi sea, can lead to a 1p1h excitation if one of the nucleons
remains a spectator. This mechanism is responsible for the
interference contribution in the 1p1h channel.
Then the total current function $j^\mu(\np,\nh)$
incorporates the contribution of
the 1b and the 2b currents.

To evaluate the hadronic tensor (\ref{hadronic}) in the RFG
we insert the matrix element (\ref{total}), 
and take the thermodynamic limit by replacing the discrete sum over hole
states with an integral over momentum space,
\begin{equation}
\sum_{h} \longrightarrow \frac{V}{(2\pi)^3} \int d^3h \sum_{s_h t_h}
\end{equation}
The integration over the
final particle state can then be performed using the
momentum-conserving delta function,
which fixes the particle momentum to $\mathbf{p} =
\mathbf{h} + \mathbf{q}$
\begin{eqnarray}
W^{\mu\nu}
&=&
\frac{V}{(2\pi)^{3}}
\int d^3h\delta(E_{p}-E_{h}-\omega)
2 w^{\mu\nu}(\np,\nh)
\theta(p-k_{F})\theta(k_{F}-h),
\label{integralw2}
\end{eqnarray}  
where
\begin{equation}
w^{\mu\nu}(\np,\nh)\equiv\frac{1}{2}\sum_{s_ps_h}
j^\mu(\np,\nh)^*j^\nu(\np,\nh) 
\end{equation}
is the single-nucleon hadronic tensor.  The sums over isospin $t_p,t_h$ no
longer appear  because we have already imposed the condition
that, in the case of neutrino scattering, the hole state \( h \) corresponds
to a neutron and the particle \( p \) to a proton, while the opposite
holds for antineutrino scattering.

The single-nucleon tensor contains the square of the sum of the
one-body and two-body currents. By expanding this square, one obtains
\begin{equation}
w^{\mu\nu} = w^{\mu\nu}_{1b} + w^{\mu\nu}_{2b} + w^{\mu\nu}_{1b2b},
\end{equation}
where $w^{\mu\nu}_{1b}$ and $w^{\mu\nu}_{2b}$ are the pure one-body
and two-body contributions, respectively, and $w^{\mu\nu}_{1b2b}$ is
the interference term.
\begin{eqnarray}
w^{\mu\nu}_{1b}   & = & \frac{1}{2}\sum_{s_ps_h} (j^\mu_{1b})^* j^\nu_{1b} \\
w^{\mu\nu}_{2b}   & = & \frac{1}{2}\sum_{s_ps_h} (j^\mu_{2b})^* j^\nu_{2b} \\
w^{\mu\nu}_{1b2b} & = &
\frac12\sum_{s_ps_h} [(j^\mu_{1b})^*j^\nu_{2b}+(j^\mu_{2b})^*j^\nu_{1b}]
\label{w1b2b}
\end{eqnarray}
And the response functions also decompose accordingly:
\begin{equation}
R^K = R^K_{1b} + R^K_{2b} + R^K_{1b2b}
\end{equation}
From calculations in electron scattering in chapter 3, it has been observed that
the pure two-body MEC contribution, $R^K_{2b}$, is generally small and
can often be neglected when compared to the interference term
$R^K_{1b2b}$, which tends to dominate the MEC effects in the 1p1h
channel \cite{Ama03}.  This justifies our focus on the
interference term, which captures the leading MEC effect
in the 1p1h sector.

The one-body current operator consist of two
terms: $j_{1b}^{\mu}(\np,\nh)=j_{1bV}^{\mu}(\np,\nh)+j_{1bA}^{\mu}(\np,\nh)$.
The vector current is
\begin{equation}
  j^{\mu}_{1bV}(\np,\nh)
=\bar{u}(\np)
\left(F^V_{1}\gamma^{\mu}+i\frac{F^V_{2}}{2m_{N}}\sigma^{\mu\nu}Q_{\nu}
\right)u(\nh),
\end{equation}
where the isovector nucleon form factors are defined as 
$F_{1,2}^V=F_{1,2}^p-F_{1,2}^n$.
The axial current is
\begin{equation}
  j^{\mu}_{1bA}(\np,\nh)
=-\bar{u}(\np)
\left(G_A\gamma^{\mu}\gamma_5 + G_P\frac{Q^\mu}{2m_{N}}\gamma_5
\right)u(\nh),
\label{axial}
\end{equation}
where $G_A$ is the nucleon axial-vector form factor and $G_P$ is the
pseudo-scalar form factor, given by
\begin{eqnarray}
  G_A&=&\frac{g_A}{1-\frac{Q^2}{M_A^2}}
  \\
  G_P&=&\frac{4m_N^2}{m^2_\pi-Q^2}G_A
\end{eqnarray}
with $g_A=1.26$ and $M_A=1032$ MeV. 

Note that the minus sign in the axial one-body current (\ref{axial})
arises from our
convention of defining the total current as the sum of the vector and
axial parts, \( J^\mu = J_V^\mu + J_A^\mu \), whereas it is often
written in the literature as \( V - A \). We adopt this convention for
consistency with the meson-exchange currents, which are also defined
as the sum \( V + A \). Of course, physical results are independent of
this choice.

\section{Weak CC Meson exchange currents}
\begin{figure}
\centering
\includegraphics[width=10cm,bb=120 310 495 700]{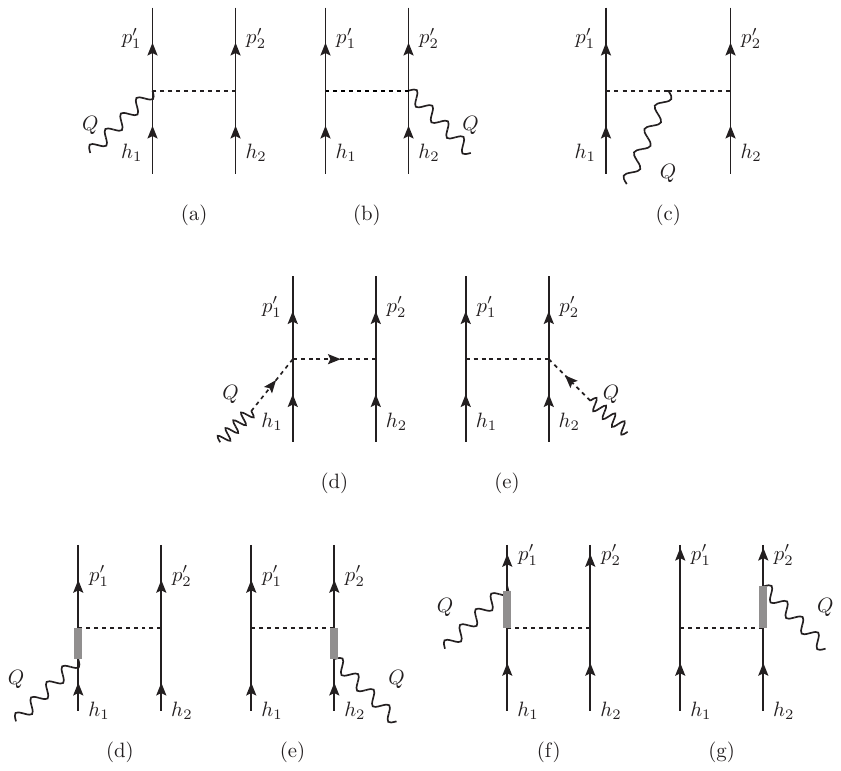}
\caption{Feynman diagrams for the electroweak MEC model used in 
this work.}\label{neudiag}
\end{figure}

We present the MEC for CC neutrino scattering
considered in this work, corresponding to the Feynman diagrams shown
in Figure \ref{neudiag}.  They were derived in \cite{Rui17} from the
pion weak production model of ref. \cite{Her07}.  Similarly to the 1b
current, the weak MEC are the sum of vector and axial currents.  The
different current operators are: the seagull current (diagrams a and
b), the pion-in-flight current (diagram c), the pion-pole current
(diagrams d and e), and the $\Delta$ forward (f,g) and backward
currents (diagrams h-i):
\begin{equation}
j_2^\mu(\np'_1,\np'_2,\np_1,\np_2) 
=  j^{\mu}_{sea}+  j^{\mu}_{\pi}+  j^{\mu}_{pole}+  j^{\mu}_{\Delta},
\end{equation}
The weak MEC operators exhibit a structure closely related to that of the
electromagnetic case of chapters 3 and 4, reflecting their common
origin in the underlying meson-nucleon dynamics. In particular, they
involve the same V-function defined in Eq. (\ref{Vfun})
\begin{equation}
  V_{s'_1s_1}(p'_1,p_1) \equiv 
 F_{\pi NN}(k_{1}^{2})
\frac{\bar{u}_{s'_1}(p'_1)\gamma^{5}\kbar_{1}u_{s_1}(p_1)}{k_1^2-m_{\pi}^2},
\label{V-function}
\end{equation}
where $k_1= p'_1-p_1$, is the momentum transfer to the individual
nucleon (and the momentum carried by the pion),
the spinors \(u_s(p)\) are the solutions of the Dirac equation with
momentum $\np$,
\( m_\pi \) is the pion mass,
and \( F_{\pi  NN}(k^2) \) 
is a strong form factor  \cite{Alb84,Som78,Mac87} 
\begin{equation}
  F_{\pi NN}(k^2) = 
  \frac{\Lambda^{2}-m_{\pi}^{2}}{\Lambda^{2}-k^{2}},
  \kern 1cm \Lambda_\pi=1300\;\rm MeV
\end{equation}

\subsubsection*{Seagull current}
The seagull current is given as the sum of the vector and axial operators,
\begin{eqnarray}
  j^{\mu}_{sea} &=& (j^{\mu}_{sea})_{V} + (j^{\mu}_{sea})_{A},  
\\
  (j^{\mu}_{sea})_V
&=&
i[\ntau^{(1)} \times \ntau^{(2)}]_{\pm}
\frac{f^2}{m_{\pi}^2}
F_{1}^{V} 
F_{\pi NN}(k_{1}^{2}) V_{s'_{1}s_{1}}(p'_{1},p_{1})
\bar{u}_{s'_2}(p'_{2}) \gamma^{5}\gamma^{\mu} u_{s_2}(p_{2}) 
 + (1 \leftrightarrow 2) \nonumber \\
\label{seaV}
\\
(j^{\mu}_{sea})_A
&=&
i[\ntau^{(1)} \times \ntau^{(2)}]_{\pm}
\frac{f^2}{m_{\pi}^2} \frac{F_{\rho}(k_{2}^{2})}{g_A}
F_{\pi NN}(k_{1}^{2}) V_{s'_{1}s_{1}}(p'_{1},p_{1})
 \bar{u}_{s'_2}(p'_{2})\gamma^{\mu}u_{s_2}(p_{2}) 
 + (1 \leftrightarrow 2),  \nonumber \\ 
\end{eqnarray}
where $f^2=1$ is the $\pi NN$ coupling constant, $\ntau^{(i)}$ is the
isospin operator of nucleon $i$, $F_1^V(Q^2) =F_1^p-F_1^n$ is the
isovector form factor of the nucleon, and $F_\rho$ is the $\rho$ meson
form factor. Note that the current has been written as proportional to
an isospin raising operator.  The sign $\pm$ in the isospin matrix
elements of Eqs. (\ref{seaV}--\ref{deltaB}) refers to neutrino $(+)$
or antineutrino $(-)$ scattering. The isospin matrix elements are
computed in Appendix \ref{appG}.

\subsubsection*{Pion-in-flight current}
The pion-in-flight has only vector part
\begin{eqnarray}
(j^{\mu}_{\pi})_V &=&
i[\ntau^{(1)} \times \ntau^{(2)}]_{\pm}
F_{1}^{V}
\frac{f^2}{m_{\pi}^2}
V_{s'_{1}s_{1}}(p'_{1},p_{1})
V_{s'_{2}s_{2}}(p'_{2},p_{2})(k_{1}^{\mu}-k_{2}^{\mu}) 
\\
(j^{\mu}_{\pi})_A&=&0  
\end{eqnarray}
\subsubsection*{Pion-Pole current}
The pion-pole current is purely axial 
\begin{eqnarray}
  (j^{\mu}_{pole})_V&=&0
  \\
  (j^{\mu}_{pole})_{A}&=&
 i[\ntau^{(1)} \times \ntau^{(2)}]_{\pm}
\frac{f^{2}}{m_{\pi}^{2}}\frac{F_{\rho}(k_{1}^{2})}{g_{A}}
F_{\pi NN}(k_{2}^{2})
\frac{Q^{\mu}\bar{u}_{s'_1}(p'_{1})\slashed{Q}u_{s_1}(p_{1})}{Q^2-m_{\pi}^2}
V_{s'_{2}s_{2}}(p'_{2},p_{2})
+ (1 \leftrightarrow 2) \nonumber \\
\end{eqnarray}
Note that since this current contains the term $Q^\mu$,
only contributes to the longitudinal and time components
of the hadronic tensor.

\subsubsection*{Delta ($\Delta$) current}

\begin{figure}
\centering
\includegraphics[width=10cm,bb=110 310 500 690]{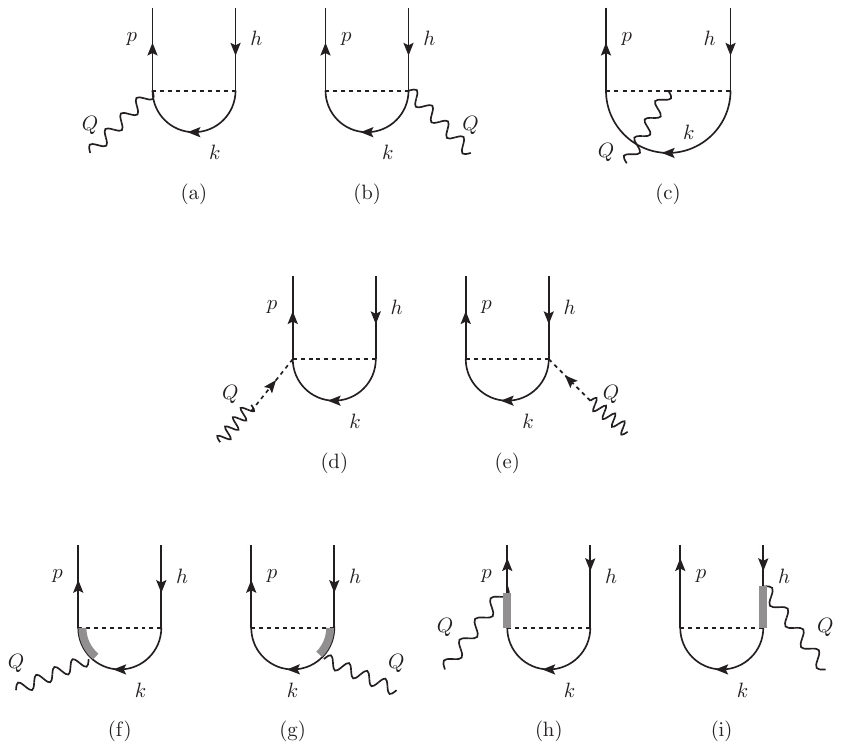}
\caption{Diagrams for the 1p1h MEC matrix elements}
\label{feyneu}
\end{figure}

The $\Delta$ excitation current operator has both
vector and axial parts,
\begin{equation}
 j^{\mu}_{\Delta}= (j^{\mu}_{\Delta})_V + (j^{\mu}_{\Delta})_A.
\end{equation}
corresponding to the vertices $\Gamma_V^{\beta\mu}$ and
$\Gamma_A^{\beta\mu}$, respectively,  in the 
\( N \rightarrow \Delta \) transition, given in  Eq. (\ref{tver}) below. 
The $\Delta$ current is further
divided into forward and backward operators
\begin{eqnarray}
  j^{\mu}_{\Delta F}
&=&
[U_{F}(1,2)_{\pm}]
\frac{f^{*}f}{m_{\pi}^2}
F_{\pi N \Delta}(k_{2}^{2})
V_{s'_{2}s_{2}}(p'_{2},p_{2})
\bar{u}_{s'_1}(p'_{1})k_{2}^{\alpha}G_{\alpha\beta}(p_{1}+Q)
\Gamma^{\beta\mu}(Q)u_{s_1}(p_{1}) \nonumber \\ 
&+& (1 \leftrightarrow 2), 
\label{deltaF}\\
  j^{\mu}_{\Delta B}
&=&
[U_{B}(1,2)_{\pm}]
\frac{f^{*}f}{m_{\pi}^2}
F_{\pi N \Delta}(k_{2}^{2})
V_{s'_{2}s_{2}}(p'_{2},p_{2})
\bar{u}_{s'_1}(p'_{1})k_{2}^{\beta}
\hat{\Gamma}^{\mu\alpha}(Q)G_{\alpha\beta}(p'_{1}-Q)u_{s_1}(p_{1}) \nonumber \\  
&+& (1 \leftrightarrow 2) 
\label{deltaB}
\end{eqnarray}
where the $\pi N \Delta$
coupling constant is $f^*=2.13 f$.
The $\gamma N\Delta$ vertices are 
\begin{eqnarray}
  \Gamma^{\beta\mu}(Q)&=&
  \Gamma_V^{\beta\mu}(Q) + \Gamma_A^{\beta\mu}(Q),
\\
\Gamma_V^{\beta\mu}(Q) &=& \frac{C_3^V}{m_N} 
(g^{\beta\mu}\slashed{Q}-Q^{\beta}\gamma^{\mu})\gamma_5
 \label{tver}\\
\Gamma_A^{\beta\mu}(Q)&=& C_5^Ag^{\beta\mu}.
\\
 \hat{ \Gamma}^{\beta\mu}(Q)&=&
  \hat{\Gamma}_V^{\beta\mu}(Q) + \hat{\Gamma}_A^{\beta\mu}(Q),
\\
\hat{\Gamma}_V^{\beta\mu}(Q) &=& -\Gamma_V^{\mu\beta}(Q),
\kern 1cm
\hat{\Gamma}_A^{\beta\mu}(Q) = \Gamma_A^{\beta\mu}(Q). 
\end{eqnarray}
The vector and axial form factors are taken from \cite{Her07}:
\begin{eqnarray}
  C_{3}^{V}(Q^{2})
&=&\frac{2.13}{(1-\frac{Q^{2}}{M_{V}^{2}})^{2}}
\frac{1}{1-\frac{Q^{2}}{4M_{V}^{2}}},\\
C_{5}^{A}(Q^{2})
&=&\frac{1.2}{(1-\frac{Q^{2}}{M_{A\Delta}^{2}})^{2}}
\frac{1}{1-\frac{Q^{2}}{4M_{A\Delta}^{2}}},
\end{eqnarray}
with $M_V=0.84$ GeV and $M_{A \Delta}=1.05$ GeV.
The $\pi$N$\Delta$ form factor
is taken as \cite{Alb84,Som78} 
\begin{equation}
  F_{\pi NN}(k^2)=   F_{\pi N\Delta}(k^2)
\label{pinnff}
\end{equation}
The forward $\Delta$ current corresponds to processes where the
$\Delta$ resonance is produced and then decays back to a nucleon,
while the backward $\Delta$ current involves the exchange of a pion,
leading to the creation of a $\Delta$ resonance in the intermediate
state. The charge dependence of these processes is embedded in the
isospin operators
$U_F(1,2)_{\pm} = U_{F}(1,2)_{x} \pm iU_{F}(1,2)_{y}$ for the forward term and
\( U_{B}(1,2)_{\pm} = U_{B}(1,2)_{x} \pm iU_{B}(1,2)_{y} \) for the backward
term, where
\begin{eqnarray}
 U_{F}(1,2)_i 
&=&
\sqrt{\frac{3}{2}}
\sum_{j=1}^{3}
T_{j}^{(1)}T_{i}^{(1)\dagger}\tau_{j}^{(2)},
\label{uf}\\ 
 U_{B}(1,2)_i 
&=&
\sqrt{\frac{3}{2}}
\sum_{j=1}^{3}
T_{i}^{(1)}T_{j}^{(1)\dagger}\tau_{j}^{(2)}.
\label{ub}
\end{eqnarray}
The operator $T_i^\dagger$ is the isospin raising operator that
connects isospin-$1/2$ states to isospin-$3/2$ states and satisfies
the condition
$T_i T_j^\dagger = \frac{2}{3} \delta_{ij} - \frac{i}{3} \epsilon_{ijk} \tau_k.$
Using this property it can be written
\begin{eqnarray}
  U_F(1,2)_\pm  &=&
   \frac{2}{\sqrt{6}} \tau_\pm^{(2)}
  -\frac{i}{\sqrt{6}} [\ntau^{(1)} \times \ntau^{(2)}]_{\pm}
\label{UF}  \\
    U_B(1,2)_\pm  &=&
   \frac{2}{\sqrt{6}} \tau_\pm^{(2)}
  +\frac{i}{\sqrt{6}} [\ntau^{(1)} \times \ntau^{(2)}]_{\pm}.
\label{UB}
\end{eqnarray}
Finally, the $\Delta$ propagator is as in chapter 3
\begin{equation}
  G_{\alpha\beta}(P)=
\frac{{\cal P}_{\alpha\beta}(P)}{
P^{2}-m_\Delta^2+im_\Delta\Gamma(P^{2})+\frac{\Gamma(P^{2})^{2}}{4}}
\end{equation} 
where $m_\Delta$ and $\Gamma$ are the $\Delta$ mass and width
respectively. The projector ${\cal P}_{\alpha\beta}(P)$ over spin-3/2  is
\begin{eqnarray}
{\cal  P}_{\alpha\beta}(P)
&=& -(\Pbar+M_{\Delta})
\nonumber\\
&& \kern -1cm \times
\left[
  g_{\alpha\beta}-\frac{\gamma_{\alpha}\gamma_{\beta}}{3}
  -\frac{2P_{\alpha}P_{\beta}}{3m_\Delta^2}
  +\frac{P_{\alpha}\gamma_{\beta}-P_{\beta}\gamma_{\alpha}}{3m_\Delta}
\right] .
\end{eqnarray}  
The vector part of the weak meson-exchange currents reduces to the
electromagnetic MEC of chapter 3 when the isospin-raising operators
are replaced by their third components \cite{Cas23}. The isospin sums
over $t_k$ that appear in the two-body current are provided in
Appendix \ref{appG}. The resulting effective one-body current consist
of a direct term minus an exchange term,
\begin{equation}
  j^\mu_{2b}(\np,\nh)= j^\mu_{2b}(\np,\nh)_{dir}- j^\mu_{2b}(\np,\nh)_{exch}
\end{equation}
The exchange contribution corresponds to the diagrams shown in
Fig. \ref{feyneu}. In symmetric nuclear matter, the direct term of the
vector current vanishes, similarly to the case of the electromagnetic
exchange current, because it arises solely from the $\Delta$ current,
which is transverse, and therefore vanishes upon contraction with
$Q^\mu$. When summing over spin in the direct matrix element, this
contraction appears explicitly, leading to a vanishing contribution.
However, for the axial $\Delta$ current, the direct term does not
vanish, and in principle should be included when computing the
neutrino response. Nevertheless, we have checked that this term
contributes significantly only to the longitudinal response $R_{LL}$,
which is known to be small. This behavior can be seen explicitly in
the non-relativistic limit, where the structure of the current becomes
more transparent. In the results section we show through explicit
calculations that the LL response gives a negligible contribution to
the cross section for quasielastic neutrino scattering. Since the
direct axial MEC term contributes significantly only to the LL
channel, and this channel plays a minor role in the kinematics of
interest, it is not included in the present work.

\section{Non-relativistic interference T-response}
In this section we present the expressions for the interference
responses in the NR limit, which will be compared
with the corresponding relativistic results in the results section.
This comparison serves several purposes: (i) to verify that
relativistic and non-relativistic results agree at low momentum and
energy transfer, as expected; (ii) to compare our results with known
expressions for axial MEC currents in the NR literature \cite{Ris89};
and (iii) to test the accuracy of our numerical implementation, since
the NR Fermi gas allows partial analytical evaluation of the response
functions \cite{Ama94a,Ama94b}. Furthermore, this limit helps identify
the dominant responses and current components at moderate momentum
transfer.
\subsection{Non-relativistic weak CC one-body current}

The vector part of the 1b current is the sum of magnetization and convection
currents:
\begin{eqnarray}
  \nj_{1b}(\np,\nh)_V
&=& 
  \nj_{mV}(\np,\nh)+  \nj_{cV}(\np,\nh), 
\\
  \nj_{mV}(\np,\nh)
&=&
-\frac{2G_M^V}{2m_N}i\nq\times\nsigma_{s_ps_h},
\label{magnetization}
\\
  \nj_{cV}(\np,\nh)
&=&
\delta_{s_ps_h}
\frac{2G_E^V}{m_N}(\nh+\frac{\nq}{2}).
\label{convection}
\end{eqnarray}
with $\nq=\np-\nh$ by momentum conservation.  Here $G_M^V$ ($G_E^V$)
is the isovector magnetic (electric) form factor of the nucleon.
$G_E^V=(G_E^p-G_E^n)/2.$ 

The transverse one-body axial current at leading order is \cite{Ama05}
\begin{equation}
\nj^\perp_{1b}(\np,\nh)_{A}=-G_A \nsigma^\perp_{s_ps_h}.
\end{equation}
The time component of the axial current, $j^0_{1b}$, or axial
charge-density is typically not included in non-relativistic
calculations because it is small at leading order; however, this
suppression is not a strict consequence of the NR expansion itself.
To obtain its static limit, we have considered the semi-relativistic
expansion of the electroweak current introduced in
Ref.~\cite{Ama96,Ama05}.  It can be written as the sum of a convective
term and a magnetization term, in analogy with the structure of the
transverse vector current \cite{Ama05}
\begin{eqnarray}
  j^0_{1b}(\np,\nh)_{A} & = & j^0_{mA}(\np,\nh)+j^0_{cA}(\np,\nh) \\
 j^0_{mA}(\np,\nh)     & = & - \frac{G_A'}{2m_N}\nq \cdot \nsigma_{s_ps_h}\\
 j^0_{cA}(\np,\nh)     & = &
-G_A\frac{\nh^\perp}{m_N} \cdot \nsigma_{s_ps_h}
\label{obaxial}
\end{eqnarray}
where $\nh^\perp$ is the transverse component of $\nh$, perpendicular
to the momentum transfer. The auxiliary form factor \( G_A' \) is given
by
\begin{equation}
  G_A'= G_A - \tau G_P=\left( 1- \frac{Q^2}{Q^2-m_\pi^2}\right)G_A.
\end{equation}

\subsection{Non-relativistic weak MEC}

The weak MEC operators are the sum of vector plus axial
components. The vector MEC are also isovectors and closely related to
the electromagnetic MEC, which are also isovectors. The key difference
lies in the isospin structure: while the electromagnetic current
corresponds to the third component of the isospin operator, the
charged weak current involves the raising and lowering \( \pm \)
operators.  Consequently, the NR expressions for
electromagnetic MEC derived in chapter 4 can be directly adapted to
the weak case by by substituting \( \tau_z \rightarrow \tau_\pm \),
and $[\ntau^{(1)} \times \ntau^{(2)}]_z\rightarrow [\ntau^{(1)} \times
\ntau^{(2)}]_\pm$, depending on the specific charge-changing
process.

The NR expansion of the MEC is obtained by applying again
standard reduction rules of chapter 4 to matrix elements involving products of
gamma matrices between Dirac spinors, retaining only leading-order
terms in $1/m_N$, summarized here as:
\begin{eqnarray}
\gamma^0 \longrightarrow 1, &
\gamma^i \longrightarrow 0, &
\gamma_5\gamma^0 \longrightarrow 0, 
\label{gammas1}\\
\gamma_5\gamma^i \longrightarrow -\sigma_i, &
\gamma^i\gamma^j \longrightarrow -\sigma_i\sigma_j, &
\gamma^0\gamma^j \longrightarrow 0. 
\label{gammas2}
\end{eqnarray}
For a nucleon momentum $ p^\mu \longrightarrow (m_N,p^i)$ and $\pbar
\longrightarrow p_0$. For the momentum transfer to nucleon $i$, $k^\mu
\longrightarrow (0,k^i)$ and $\gamma_5\kbar \longrightarrow
\nk\cdot\nsigma$. As a starting point, we recall the NR form of the
$V$-function, that was already derived in chapter4,
\begin{equation}
  V_{s'_1s_1}(p'_1,p_1) \longrightarrow 
-\frac{\nk_1\cdot\nsigma^{(1)}}{\nk_1^2+m_{\pi}^2}.
\label{Vnorel}
\end{equation}
The NR vector MEC are identical to the electromagnetic
ones, except for the isospin operators. As a result, only the spatial
components of the vector MEC survive at leading order in the
non-relativistic expansion. This feature is also confirmed with the
fully relativistic calculation. Then
\begin{eqnarray}
j^{\mu}_{s V} 
&\overset{\text{nr}}{\longrightarrow}& 
(0,\nj_{s V}) \nonumber
\\
j^{\mu}_{\pi V} 
&\overset{\text{nr}}{\longrightarrow}& 
(0,\nj_{\pi V}) \nonumber
\\
j^{\mu}_{\Delta V} 
&\overset{\text{nr}}{\longrightarrow}& 
(0,\nj_{\Delta V}) \nonumber
\end{eqnarray}
The axial seagull current is proportional to the matrix element of \(
\gamma^\mu \), and in the non-relativistic limit, only its time
component (proportional to \( \gamma^0 \)) survives.  On the other
hand, only the spatial components of the axial \(\Delta\) current
remain non-zero, as shown in Appendix \ref{appH}.  The pion-pole current,
being proportional to $q_\mu \gamma^\mu$,
vanishes at leading order.
The surviving axial currents are
\begin{eqnarray}
j^{\mu}_{sA} 
&\overset{\text{nr}}{\longrightarrow}& 
(j^{0}_{sA},\vec{0}) \nonumber 
\\
j^{\mu}_{\Delta A} 
&\overset{\text{nr}}{\longrightarrow}& 
(0,\nj_{\Delta A}) \nonumber
\end{eqnarray}
The corresponding NR operators are
\begin{eqnarray}
  \nj_{sV}(p'_1,p'_2,p_1,p_2)
  &=& 
i[\ntau^{(1)} \times \ntau^{(2)}]_+
\frac{f^{2}}{m_{\pi}^{2}}F_1^V
\left(
\frac{\nk_1\cdot\nsigma^{(1)}}{\nk_1^2+m_{\pi}^2}
\nsigma^{(2)}
-\frac{\nk_2\cdot\nsigma^{(2)}}{\nk_2^2+m_{\pi}^2}
\nsigma^{(1)}
\right).
\label{seagull}
\\
\nj_{\pi V}(p'_1,p'_2,p_1,p_2) &=&
  i[\ntau^{(1)} \times \ntau^{(2)}]_+
\frac{f^{2}}{m_{\pi}^{2}}F_1^V
\frac{\nk_1\cdot\nsigma^{(1)}}{\nk_1^2+m_{\pi}^2}
\frac{\nk_2\cdot\nsigma^{(2)}}{\nk_2^2+m_{\pi}^2}
(\nk_1-\nk_2).
\label{pionic}
\\
 \nj_{\Delta V}(p'_1,p'_2,p_1,p_2) &=&
i C_{\Delta}^V
\left\{
\frac{\nk_2\cdot\nsigma^{(2)}}{\nk_2^2+m_{\pi}^2}
\left[
4\tau_{+}^{(2)}\nk_2+
[\ntau^{(1)}\times\ntau^{(2)}]_{+}
\nk_2\times\nsigma^{(1)}
\right]
\right.
\nonumber\\
&&
\kern -0.3cm 
\left. \mbox{}+
\frac{\nk_1\cdot\nsigma^{(1)}}{\nk_1^2+m_{\pi}^2}
\left[
4\tau_{+}^{(1)}\nk_1-
[\ntau^{(1)}\times\ntau^{(2)}]_{+}
\nk_1\times\nsigma^{(2)}
\right]
\right\}
\times\nq.
\label{deltafinal}
\\
  j^{0}_{sA}(p'_1,p'_2,p_1,p_2) &=&
  -i[\ntau^{(1)} \times \ntau^{(2)}]_+
  \frac{f^2}{ m_\pi^2 }\frac{1}{g_A}
  \left[ \frac{ \nk_{1}\cdot\nsigma^{(1)} }{ \nk_{1}^{2}+m^{2}_{\pi} }
        -\frac{ \nk_{2}\cdot\nsigma^{(2)} }{ \nk_{2}^{2}+m^{2}_{\pi} }
  \right]
 \\
\nj_{\Delta A}(p'_1,p'_2,p_1,p_2) &=&
-C_{\Delta}^A
\Biggl\{
4\tau_{+}^{(1)}\frac{ ( \nk_{1} \cdot \nsigma^{(1)} )\nk_{1} }{ \nk_1^2+m^{2}_\pi }
+ 4\tau_{+}^{(2)}\frac{ ( \nk_{2} \cdot \nsigma^{(2)} )\nk_{2} }{ \nk_2^2+m^{2}_\pi }
  \nonumber \\
  &&
  \kern -1.7cm
  +[\ntau^{(1)}\times\ntau^{(2)}]_{+}
  \biggl[
  \frac{(\nk_{2}\cdot\nsigma^{(2)})(\nk_{2}\times\nsigma^{(1)})}{\nk_2^2+m^{2}_\pi}
  -\frac{(\nk_{1}\cdot\nsigma^{(1)})(\nk_{1}\times\nsigma^{(2)})}{\nk_1^2+m^{2}_\pi}
  \biggl]
  \Biggl\} 
\label{deltaAxial}
\end{eqnarray}
with
\begin{eqnarray}
  C_{\Delta}^V &=& \sqrt{ \frac32 } \frac29  \frac{ff^*}{m_\pi^2}\frac{C_3^V}{m_N}\frac{1}{m_\Delta-m_N} \\
  C_{\Delta}^A &=& \sqrt{ \frac32 } \frac29  \frac{ff^*}{m_\pi^2}C_{5}^{A}\frac{1}{m_\Delta-m_N} 
\end{eqnarray} 
These operators match the standard NR MEC in the
literature \cite{Ris89} modulo differences in coupling constants and
form factors.
\subsection{Effective one-body MEC}
The 1p1h matrix elements of the vector and axial MEC are
\begin{equation}
j^\mu_{2b}(p,h)=
-\int \frac{d^3k}{(2\pi)^3}
\sum_{t_ks_k}j^\mu_{2b}(p,k,k,h)
=
j^\mu_{s}(p,h)+j^\mu_{\pi}(p,h)+j^\mu_{\Delta}(p,h),
\end{equation}
where we have neglected the direct part in the axial $\Delta$ current,
as previously mentioned.  The results are the following for the three
MEC, seagull, pionic and $\Delta$ currents
\begin{eqnarray}
\nj_{sV}(p,h)
&=&
4
\frac{f^2}{m_\pi^2}F_1^V
\int \frac{d^3k}{(2\pi)^3}
\left(
\frac{\delta_{s_ps_h}\nk_1+i\nsigma_{ph}\times\nk_1}{\nk_1^2+m_{\pi}^2}
-\frac{\delta_{s_ps_h}\nk_2+i\nk_2\times\nsigma_{ph}}{\nk_2^2+m_{\pi}^2}
\right)
\label{seagullph5}
\\
\nj_{\pi V}(p,h)
&=&
4
\frac{f^2}{m_\pi^2}F_1^V
\int \frac{d^3k}{(2\pi)^3}
\frac{\delta_{s_ps_h}\nk_1\cdot\nk_2
+i(\nk_1\times\nk_2)\cdot\nsigma_{ph}}
{(\nk_1^2+m_{\pi}^2)(\nk_1^2+m_{\pi}^2)}(\nk_1-\nk_2),
\label{pionicph5}
\\
\nj_{\Delta V}(p,h)
&=&
 \sqrt{ \frac32 } \frac89  \frac{ff^*}{m_\pi^2}
\frac{iC_3^V}{m_\Delta-m_N}
\frac{\nq}{m_N} \nonumber \\
&&
\times 
\int \frac{d^3k}{(2\pi)^3}
\left(
\frac{\nk_1^2\nsigma_{ph}+(\nsigma_{ph}\cdot\nk_1)\nk_1}
     {\nk_1^2+m_\pi^2}
+\frac{\nk_2^2\nsigma_{ph}+(\nsigma_{ph}\cdot\nk_2)\nk_2}
     {\nk_2^2+m_\pi^2}
\right),
\label{deltaph5} \nonumber \\
\\
j_{s A}^0(p,h) 
&=&
-4\frac{f^{2}}{m_{\pi}^{2}}\frac{1}{g_A}
\int \frac{d^3k}{(2\pi)^3}
\left[ 
     \frac{ \nk_{1} \cdot \nsigma_{ph}  }{  \nk_{1}^{2}+m^{2}_{\pi} }
    -\frac{ \nk_{2} \cdot \nsigma_{ph}  }{  \nk_{2}^{2}+m^{2}_{\pi} } 
\right] 
\\
\nj_{\Delta A}(p,h) 
&=& 
 \sqrt{ \frac32 } \frac89  \frac{ff^*}{m_\pi^2}
 \frac{C_5^A}{m_\Delta-m_N} \nonumber \\
 &&
 \times
\int \frac{d^3k}{(2\pi)^3}
\left( \frac{\nk_1^2\nsigma_{ph}+(\nsigma_{ph}\cdot\nk_1)\nk_1}{\nk_1^2+m_\pi^2} 
+\frac{\nk_2^2\nsigma_{ph}+(\nsigma_{ph}\cdot\nk_2)\nk_2}{\nk_2^2+m_\pi^2} \right), 
\end{eqnarray}
with $\nk_1=\np-\nk$  and  $\nk_2=\nk-\nh$. 

\subsection{Non relativistic 1b2b interference single-nucleon responses}
Here we provide separate expressions for the
different contributions to the interference terms, which arise from
the cross products between the various components of the one-body
current and the meson exchange currents.
\begin{eqnarray}
w_{CC,1b2b}^{AA}&=& w_{CC,ms}^{AA}+ w_{CC,cs}^{AA}\\ 
\nonumber\\
  w_{CC,ms}^{AA}  &=&  
{\rm Re}\sum_{s_ps_h}j^0_{mA}(p,h)^*\; j_{sA}^0(p,h) \\
   w_{CC,cs}^{AA} 
 &=&  {\rm Re}\sum_{s_ps_h}j^0_{cA}(p,h)^* \; j_{sA}^0(p,h)
\\
w_{T,1b2b}^{VV}&=& 
w_{T,ms}^{VV} +
w_{T,cs}^{VV} +
w_{T,m\pi}^{VV}+
w_{T,c\pi}^{VV}+ 
w_{T,m\Delta}^{VV}
\end{eqnarray}
\begin{eqnarray}
w_{T,ms}^{VV} &=& {\rm Re}\sum_{s_ps_h} \nj_{mV}(p,h)^*\cdot\nj_{sV}(p,h)
\\
w_{T,cs}^{VV} & = & {\rm Re}\sum_{s_ps_h} \nj^T_{cV}(p,h)^*\cdot\nj_{sV}(p,h) 
\\
w_{T,m\pi}^{VV}&  = & {\rm Re}\sum_{s_ps_h} \nj_{mV}(p,h)^*\cdot\nj_\pi(p,h) 
\\
w_{T,c\pi}^{VV} & = & {\rm Re}\sum_{s_ps_h} \nj^T_{cV}(p,h)^*\cdot\nj_\pi(p,h) 
\\
w_{T,m\Delta}^{VV}&  = & {\rm Re}\sum_{s_ps_h} \nj_{mV}(p,h)^*\cdot\nj_{\Delta V}(p,h) 
\\
w_{T,1b2b}^{AA} & =&   w_{T,1b\Delta}^{AA} = 
 {\rm Re}\sum_{s_ps_h}
\nj^\perp_{1b A}(p,h)^*\cdot \nj_{\Delta A}(p,h) 
\end{eqnarray}
In the case of the $T'$ response we separate the VA and AV contributions
\begin{eqnarray}
  w_{T',1b2b} &=& w_{T',1b2b}^{VA}+ w_{T',1b2b}^{AV}
\\
  w_{T',1b2b}^{VA}& = &
\frac12 \mbox{Im} \sum_{s_ps_h}
\Big[  j_{1bV}^{1*}(p,h) j_{2bA}^{2}(p,h)+  j_{2bA}^{1*}(p,h) j_{1bV}^{2}(p,h) \Big]
\\
  w_{T',1b2b}^{AV}& = &
\frac12 \mbox{Im} \sum_{s_ps_h}
\Big[ j_{1bA}^{1*}(p,h) j_{2bV}^{2}(p,h)+ j_{2bV}^{1*}(p,h) j_{1bA}^{2}(p,h) \Big]
\end{eqnarray}
Then
\begin{eqnarray}
  w_{T',1b2b}^{VA} &=& w_{T',m\Delta}^{VA}
\\
  w_{T',1b2b}^{AV} &=&
  w_{T',1b s}^{AV}+w_{T',1b \pi}^{AV}+w_{T',1b \Delta}^{AV}
\end{eqnarray}
Note that in the case of the $\Delta$ current, both the vector and
axial parts contribute to the response $T'$, whereas in the case of the
seagull and pionic current, only the vector part is considered, since the axial
part is longitudinal in the NR limit.  

The explicit expressions for the different contributions to the single
nucleon interference responses are the following
(with  $\nk_1=\np-\nk$ and $\nk_2=\nk-\nh$):
\begin{eqnarray}
w^{VV}_{T,ms}
&=&
2\frac{f^2}{m_\pi^2}F_1^V\frac{G_M^V}{m_N}
\int \frac{d^3k}{(2\pi)^3}
\left(
\frac{4\nq\cdot\nk_1}{\nk_1^2+m_{\pi}^2}+
\frac{4\nq\cdot\nk_2}{\nk_2^2+m_{\pi}^2}
\right)
\label{wms}
\\
w^{VV}_{T,cs}
&=&
4\frac{f^2}{m_\pi^2}F_1^V\frac{G_E^V}{m_N}
\int \frac{d^3k}{(2\pi)^3}
\left(
\frac{2\nh_T\cdot\nk_1}{\nk_1^2+m_{\pi}^2}-
\frac{2\nh_T\cdot\nk_2}{\nk_2^2+m_{\pi}^2}
\right)
\\
w^{VV}_{T,m\pi}
&=&
-2\frac{f^2}{m_\pi^2}F_1^V\frac{G_M^V}{m_N}
\int \frac{d^3k}{(2\pi)^3}
\frac{4(\nq\times\nk_2)^2}{(\nk_1^2+m_{\pi}^2)(\nk_2^2+m_{\pi}^2)}
\label{mpi}
\\
w^{VV}_{T,c\pi}
&=&
-4\frac{f^2}{m_\pi^2}F_1^V\frac{G_E^V}{m_N}
\int \frac{d^3k}{(2\pi)^3}
\frac{4(\nq\cdot\nk_2-\nk_2^2)\nh_T\cdot\nk_2}{(\nk_1^2+m_{\pi}^2)(\nk_2^2+m_{\pi}^2)}
\\
w^{VV}_{T,m\Delta}
&=&
-2  \sqrt{ \frac32 } \frac29  \frac{ff^*}{m_\pi^2}
\frac{C_3^V}{m_N}\frac{1}{m_\Delta-m_N} \frac{G_M^V}{m_N} \nonumber \\
&&
\times
\int \frac{d^3k}{(2\pi)^3}
2\left(
\frac{3q^2k_1^2-(\nq\cdot\nk_1)^2}{\nk_1^2+m_{\pi}^2}+
\frac{3q^2k_2^2-(\nq\cdot\nk_2)^2}{\nk_2^2+m_{\pi}^2}
\right)
\label{wdelta1} 
\end{eqnarray}

\begin{eqnarray}
w_{CC,ms}^{AA} 
&=&
\frac{f^2}{m_{\pi}^{2}}\frac{1}{g_A}\frac{G_A'}{m_N}
\int \frac{d^3k}{(2\pi)^3}
\left(
\frac{4\nq\cdot\nk_1}{\nk_1^2+m_{\pi}^2}+
\frac{4\nq\cdot\nk_2}{\nk_2^2+m_{\pi}^2}
\right)
\\
w_{CC,cs}^{AA}
&=&
4 \frac{f^2}{m_{\pi}^{2}}\frac{1}{g_A}
\frac{(G_A-\xi G_A')}{m_N}
\int \frac{d^3k}{(2\pi)^3}
\left(
\frac{2\nh_T\cdot\nk_1}{\nk_1^2+m_{\pi}^2}-
\frac{2\nh_T\cdot\nk_2}{\nk_2^2+m_{\pi}^2}
\right)
\\
w_{T,1b\Delta}^{AA}
&=&
-\sqrt{ \frac32 } \frac{16}9  \frac{ff^*}{m_\pi^2}
\frac1{m_\Delta-m_N}
\frac{C_5^A G_A}{q^2} \nonumber \\
&&
\times
\int \frac{d^3k}{(2\pi)^3}
2\left(
\frac{3q^2k_1^2-(\nq\cdot\nk_1)^2}{\nk_1^2+m_{\pi}^2}+
\frac{3q^2k_2^2-(\nq\cdot\nk_2)^2}{\nk_2^2+m_{\pi}^2}
\right)
\label{wdelta2} 
\\
w_{T',1bs}^{AV}
&=&
2\frac{f^2}{m_{\pi}^{2}}
\frac{G_{A}F_1^V}{q} 
\int \frac{d^3k}{(2\pi)^3}
\left(
\frac{4\nq\cdot\nk_1}{\nk_1^2+m_{\pi}^2}+
\frac{4\nq\cdot\nk_2}{\nk_2^2+m_{\pi}^2}
\right)
\\
w^{AV}_{T',1b\pi}
&=&
-4\frac{f^2}{m_\pi^2}F_1^V\frac{G_A}{q}
\int \frac{d^3k}{(2\pi)^3}
\frac{4(\nq\times\nk_2)^2}{(\nk_1^2+m_{\pi}^2)(\nk_2^2+m_{\pi}^2)}
\label{mpi}
\\
w_{T',1b\Delta}^{AV}
&=&
-   \sqrt{ \frac32 } \frac49  \frac{ff^*}{m_\pi^2}
\frac{G_{A}   C_3^V}{m_N q}
\frac{1}{m_\Delta-m_N} \nonumber \\
&&
\times
\int \frac{d^3k}{(2\pi)^3}
2\left(
\frac{3q^2k_1^2-(\nq\cdot\nk_1)^2}{\nk_1^2+m_{\pi}^2}+
\frac{3q^2k_2^2-(\nq\cdot\nk_2)^2}{\nk_2^2+m_{\pi}^2}
\right)
\label{wdelta3}  \\
w_{T',m\Delta}^{VA}
&=&
-\sqrt{ \frac32 } \frac49  \frac{ff^*}{m_\pi^2}
\frac{G_M^VC_5^A}{m_N q}
\frac{1}{m_\Delta-m_N} \nonumber \\
&&
\times
\int \frac{d^3k}{(2\pi)^3}
2\left(
\frac{3q^2k_1^2-(\nq\cdot\nk_1)^2}{\nk_1^2+m_{\pi}^2}+
\frac{3q^2k_2^2-(\nq\cdot\nk_2)^2}{\nk_2^2+m_{\pi}^2}
\right)
\label{wdelta4}
\end{eqnarray}

\section{Results}
In this section, we present numerical results for the effect of
two-body MEC on the CC neutrino response functions
in the 1p1h channel.  We consider three nuclear models: the RFG, the
RMF in nuclear matter, and the Superscaling Approach with a
Relativistic Effective Mass. These results allow us to assess
the model dependence of the MEC contributions. We illustrate
the impact of these effects on neutrino cross sections with selected
examples. To this end, we will first examine in detail the
interference terms between the one-body and two-body currents, which
provide the dominant MEC contribution, and later compare them with the
pure one-body responses.

\begin{figure}
  \centering
 \includegraphics[width=13cm,bb=60 560 500 810]{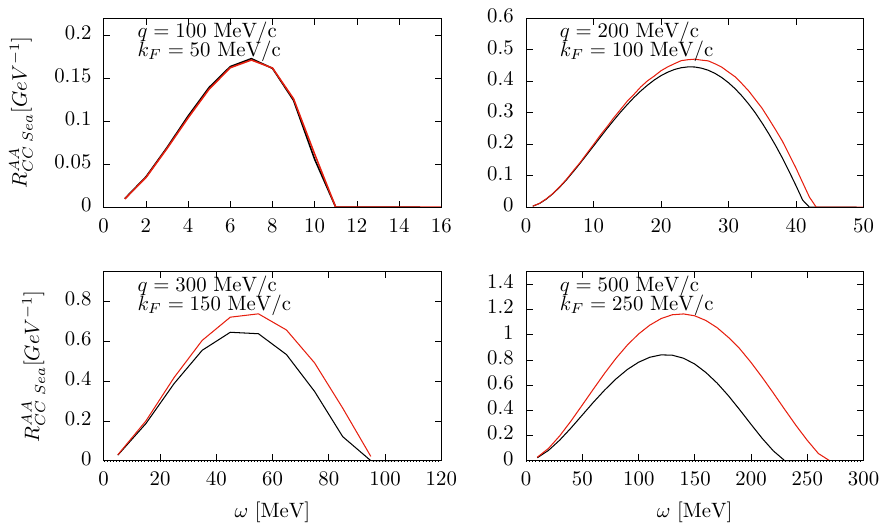}
 \caption{Interference response $R_{CC}^{AA}$ between the one-body
  axial current and the seagull current, for increasing values of the
  momentum transfer $q$, with $k_F = q/2$. In each panel, the
  non-relativistic results (red lines) are compared with the
  relativistic ones (black lines).}
 \label{relneu7}
\end{figure}

\begin{figure}
  \centering
\includegraphics[width=13cm,bb=65 520 490 810]{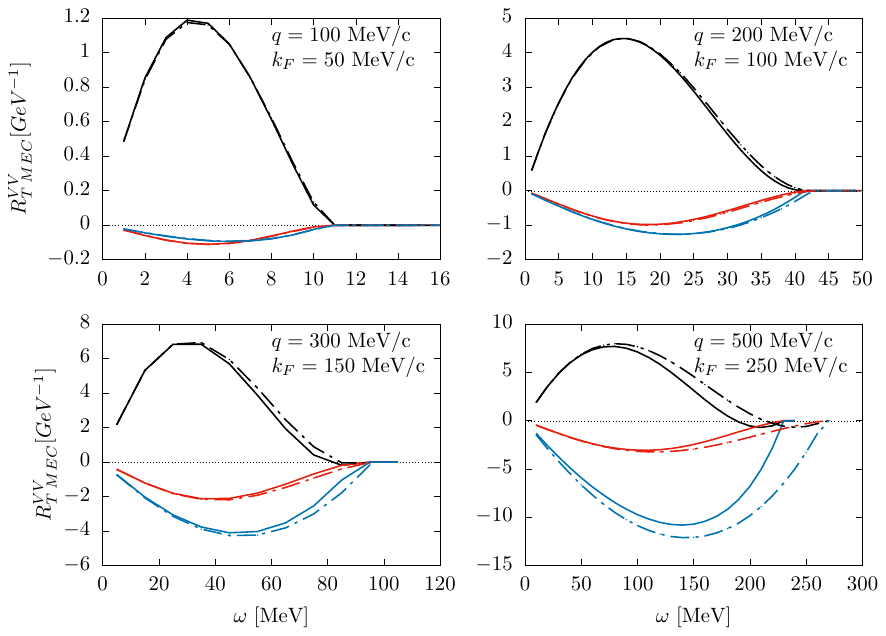}
  \caption{ Non-relativistic (dot-dashed lines) versus relativistic
    (solid lines) transverse interference response $R_T^{VV}$
    between the 1b current and the seagull (black),
    pion-in-flight (red), and $\Delta$-excitation (blue) two-body
    currents, for increasing values of $q$ and with $k_F = q/2$.  }
\label{relneu8}
\end{figure}

\begin{figure}
  \centering
\includegraphics[width=13cm,bb=65 550 490 810]{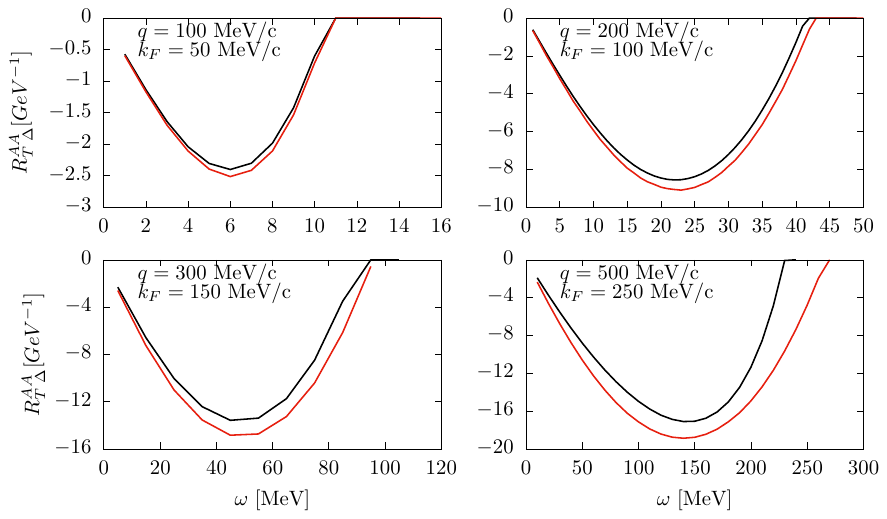}
\caption{The same as Fig. \ref{relneu7} for the
  Interference response $R_T^{AA}$ between the axial 1b and $\Delta$
  currents}
  \label{relneu9}
\end{figure}

\begin{figure}
  \centering
\includegraphics[width=13cm,bb=65 520 490 810]{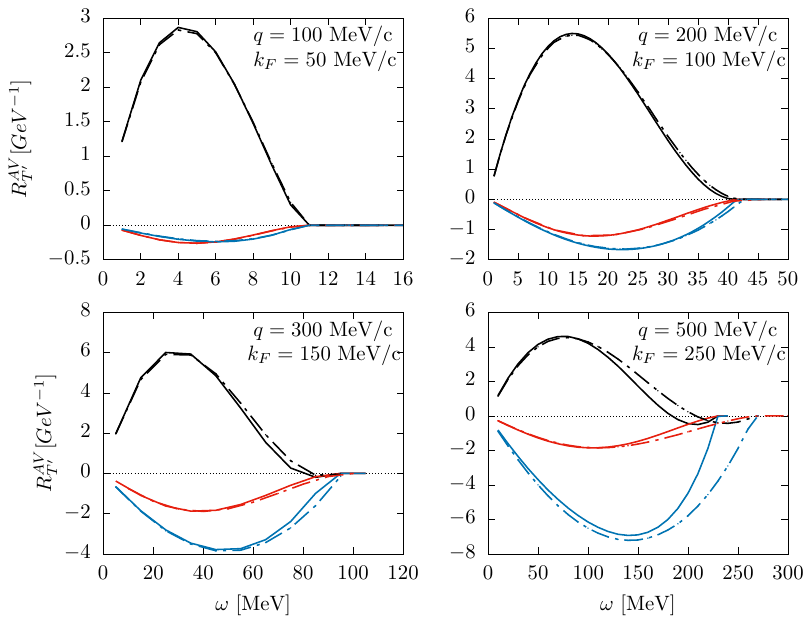}
\caption{The same as fig. \ref{relneu8} for the interference response
  $R_{T'1b2b}^{AV}$ between the axial 1b current and the
  vector seagull (black), pionic (red) and $\Delta$ (blue) currents.}
\label{relneu10}
\end{figure}

\begin{figure}
  \centering
\includegraphics[width=13cm,bb=65 520 490 810]{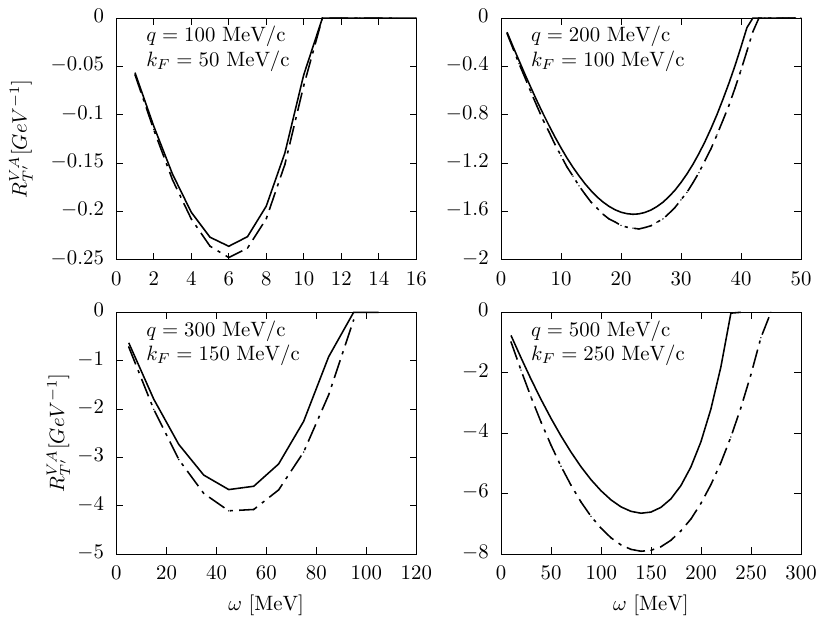}
\caption{The same as fig. \ref{relneu7}
  for the $R_{T' 1b \Delta}^{VA}$ interference response.}
\label{relneu11}
\end{figure}

\begin{figure}
  \centering
 \includegraphics[width=14cm,bb=20 300 550 810]{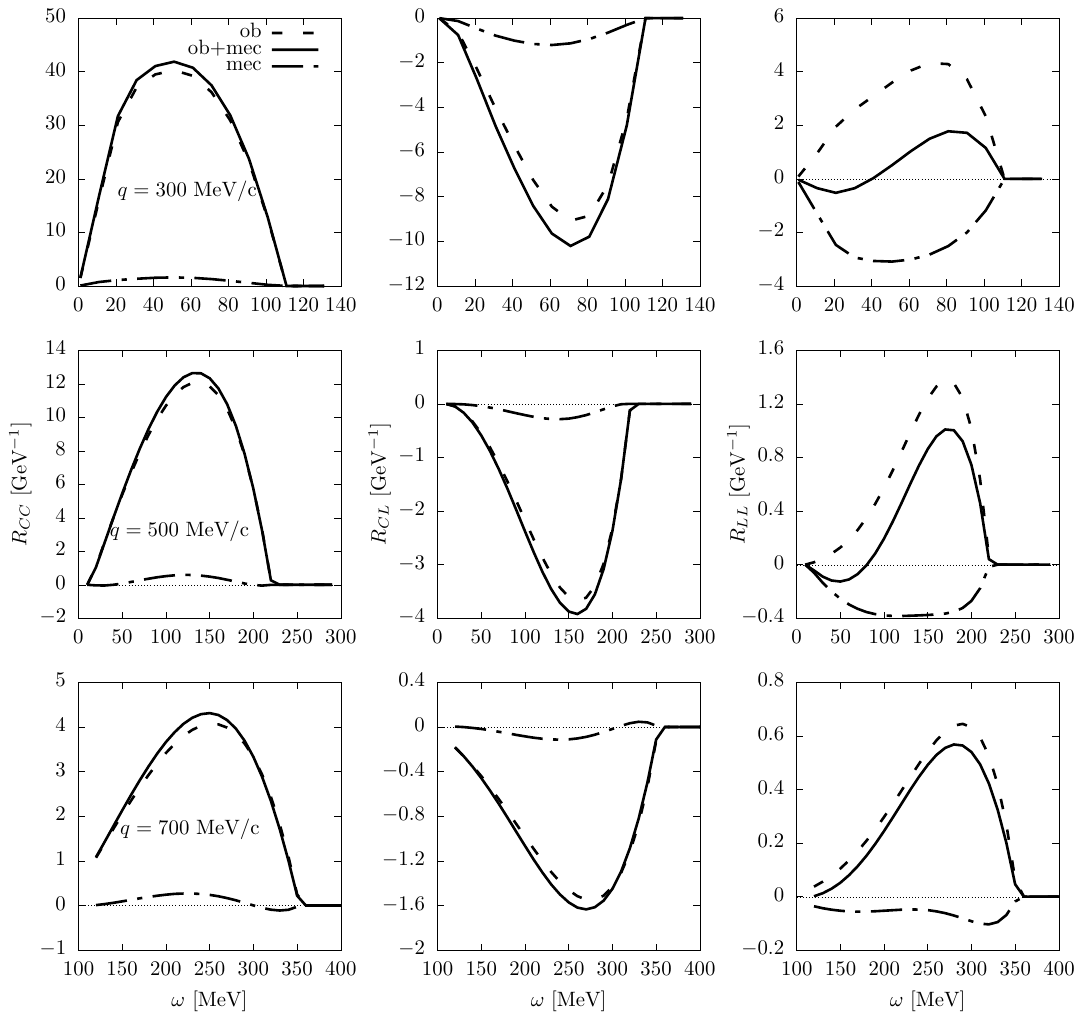}
 \caption{Response functions \(R_{CC}\), \(R_{CL}\), and \(R_{LL}\)
   computed in the RFG model for different values of $q$. We compare
   one-body responses with the interference MEC contribution and the total
   result (full lines).}
 \label{neufig7}
\end{figure}

We begin by analyzing the dominant response functions to leading order
in the non-relativistic Fermi gas (NRFG), and comparing them with the
fully relativistic results obtained within the RFG model. As
previously discussed, this comparison serves as a consistency test of
the calculation, since the NRFG and RFG models are implemented in
different ways. In the non-relativistic case, we integrate the
effective single-nucleon responses derived in the previous
section. Following the method of Refs.~\cite{Ama94a,Ama94b}, the
integrals over the momentum of the intermediate nucleon are performed
analytically except for the pion-in-flight current, where they reduce
to one-dimensional integrals. In addition, the spin traces have been
computed explicitly in the NR case. On the other hand, the RFG
responses are evaluated fully numerically. As expected, the RFG
results approach those of the NRFG in the low-$q$, low-$\omega$
region, as we explicitly demonstrate below.

In Fig.~\ref{relneu7} we show the interference response
$R_{CC,1b\,s}^{AA}$, which corresponds to the interference between the
axial one-body current and the axial seagull two-body current in the
CC channel. This is the dominant MEC contribution to the CC responses
at low momentum transfer. To examine how the relativistic response
approaches the non-relativistic one as $q$ becomes small, we present
results for several values of the momentum transfer: $q=100$, 200,
300, and 500 MeV$/c$. In each panel we set the Fermi momentum to
$k_F= q/2$, so that the momenta of the initial nucleons are also small
when $q$ is small. This choice also minimizes the effects of Pauli
blocking in the comparison. As expected, we observe that the
relativistic and non-relativistic responses are nearly identical for
$q = 100$ MeV$/c$, and begin to differ progressively as $q$ increases.

In Fig.~\ref{relneu8} we show the interference contributions to the
$R_T^{VV}$ response, separating the effects of the seagull,
pion-in-flight, and $\Delta$ currents. These vector-vector
responses are exactly twice the corresponding electromagnetic
responses, as proven in Appendix \ref{appG}. We observe that the seagull
contribution is positive, while the pion-in-flight and $\Delta$
contributions are negative, leading to a partial cancellation. The
$\Delta$ term increases with $q$ more rapidly than the seagull term,
and becomes the dominant contribution at $q = 500$~MeV$/c$. This
behavior is consistent with the electromagnetic response results.

In Fig.~\ref{relneu9} we present the axial transverse interference
response $R_T^{AA}$. At $q = 100$~MeV$/c$, the relativistic and
non-relativistic results are nearly identical, serving as a triple
consistency check: first, of the non-relativistic reduction of the
$\Delta$ current; second, of the numerical implementation of the RFG
and NRFG frameworks; and third, of the analytic and numerical
procedures used in each model. The response is negative and its
absolute value increases with $q$, becoming slightly larger than the
corresponding vector $\Delta$ contribution.

In Fig.~\ref{relneu10} we show the interference responses
$R_{T',1b\text{--}2b}^{AV}$ between the axial one-body current and the
vector MEC, separating the contributions of the vector seagull,
pionic, and $\Delta$ currents. These responses exhibit a behavior very
similar to the corresponding $R_{T,1b\text{--}2b}^{VV}$ ones. In fact,
by inspecting the expressions for the non-relativistic single-nucleon
responses, one finds that $w_{T,m\text{--}2b}^{VV}$ (interference with
magnetization current) and $w_{T',1b\text{--}2b}^{AV}$ involve exactly
the same integrals over the intermediate nucleon momentum $\nk$. The
difference lies in the coupling factors between the axial current and
the magnetization current. Since the convection current has little
impact on the transverse responses, the resulting behavior is nearly
identical in both cases.

Finally, in Fig.~~\ref{relneu11} we show the
$R_{T',1b\text{--}\Delta}^{VA}$ interference response. Once again, we
observe a close similarity between this response and both
$R_{T',1b\text{--}\Delta}^{AV}$ and $R_{T,1b\text{--}\Delta}^{VV}$,
for the same reasons discussed previously. This similarity originates
from the fact that the structure of the $\Delta$ current and the
magnetization current is nearly the same in the axial and vector
sectors. The main difference lies in a vector product with $\vec{q}$,
while the spin operator is identical in both cases, except for
different coupling constants associated with the axial and vector
currents. This can also be seen by inspecting the corresponding
single-nucleon tensors, which involve the same integrals in all three
cases.

The negative sign of the one-body--\(\Delta\) interference responses
can be clearly understood by inspecting the expressions of the
single-nucleon response functions, Eqs. (\ref{wdelta1}),
(\ref{wdelta2}), (\ref{wdelta3}) and (\ref{wdelta4}). All of them
carry an overall minus sign and are proportional to the same integral,
which is positive since the integrand contains terms of the form
$3q^2k_1^2 -(\mathbf{q} \cdot \mathbf{k}_1)^2 > 0$. This result for
the sign of the \(1b\)--\(\Delta\) interference was already discussed
in chapter 4 for the electromagnetic transverse response. Therefore
here the low-$q$ momentum theorem is extended to the vector-axial CC
$\Delta$ responses. A similar argument applies to the \(
R^{VV}_{T,m\pi} \) and \( R^{AV}_{T',1b\pi} \) responses, which are
also negative. Although these sign arguments are strictly valid in the
non-relativistic limit, numerical calculations show that they remain
valid in the relativistic case, except at very large \( q \), where
sign changes can occur for high \( \omega \).

\begin{figure}[t]
  \centering
\includegraphics[width=10cm,bb=85 320 450 810]{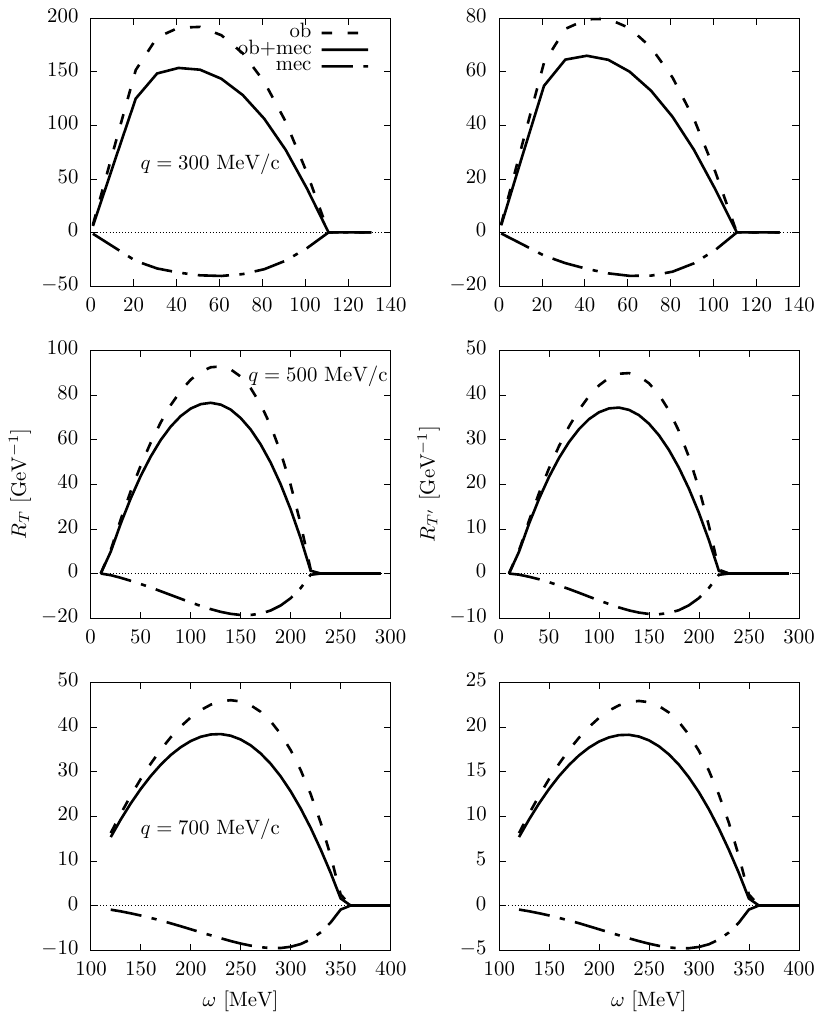}
 \caption{The same as in Fig. \ref{neufig7} for the transverse responses
   $R_T$ and $R_{T'}$}
 \label{neufig11}
\end{figure}

\begin{figure}
  \centering
 \includegraphics[width=10cm,bb=85 320 450 810]{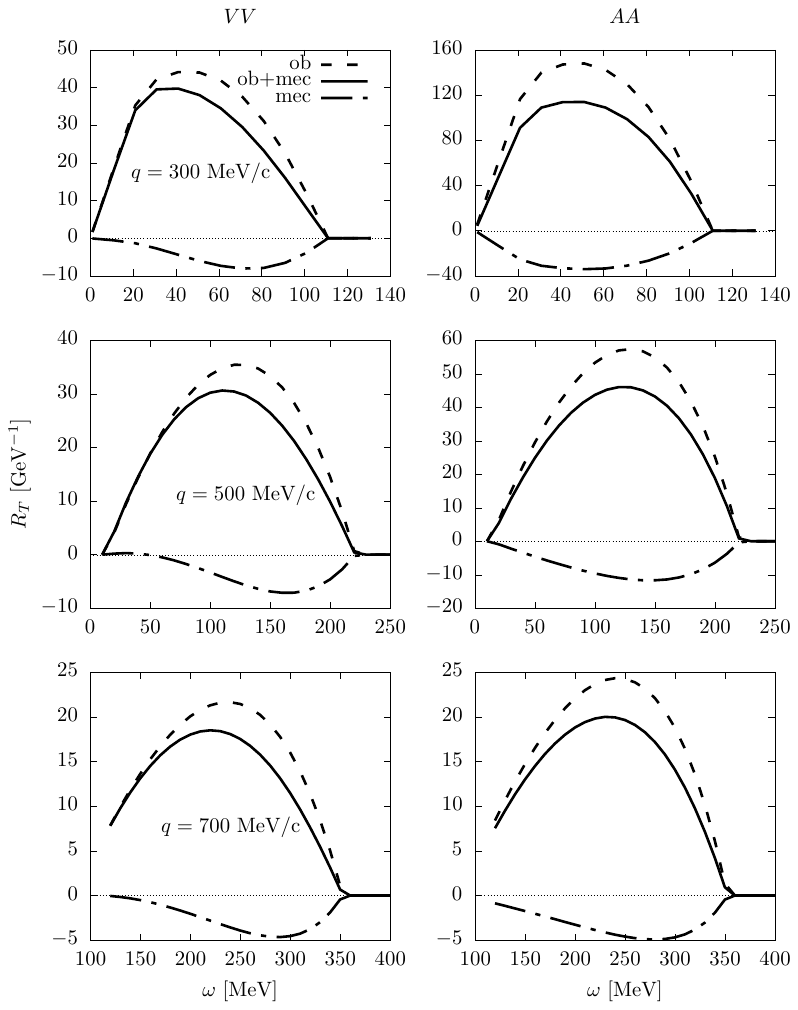}
 \caption{ The same as Fig. \ref{neufig11} for the $R_T^{VV}$ and $R_T^{AA}$
   response functions.}
 \label{neufig9}
\end{figure}

\begin{figure}
  \centering
 \includegraphics[width=14cm,bb=40 350 520 810]{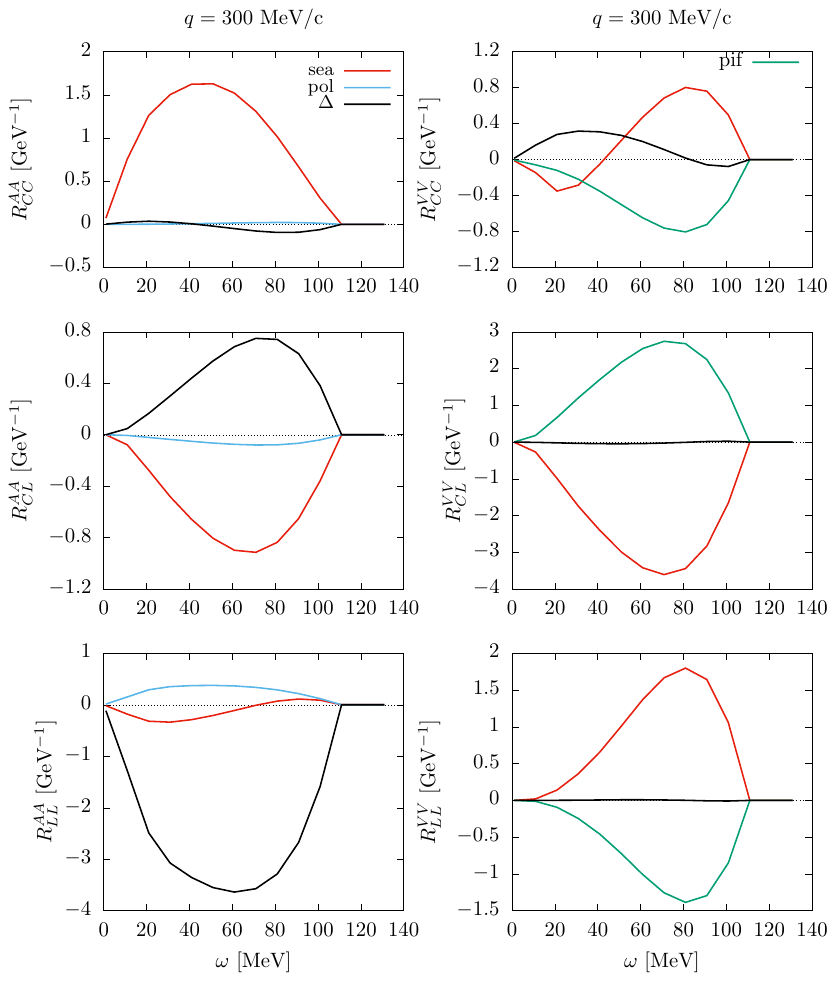}
 \caption{Interference longitudinal responses between the 1b current
   and the different MEC seagull (sea), pionic (pif), pion pole (pole) and $\Delta$,
   separated in vector and axial contributions, for $q=300$ MeV/c.  }
 \label{neufig12}
\end{figure}

\begin{figure}
  \centering
 \includegraphics[width=14cm,bb=40 350 520 810]{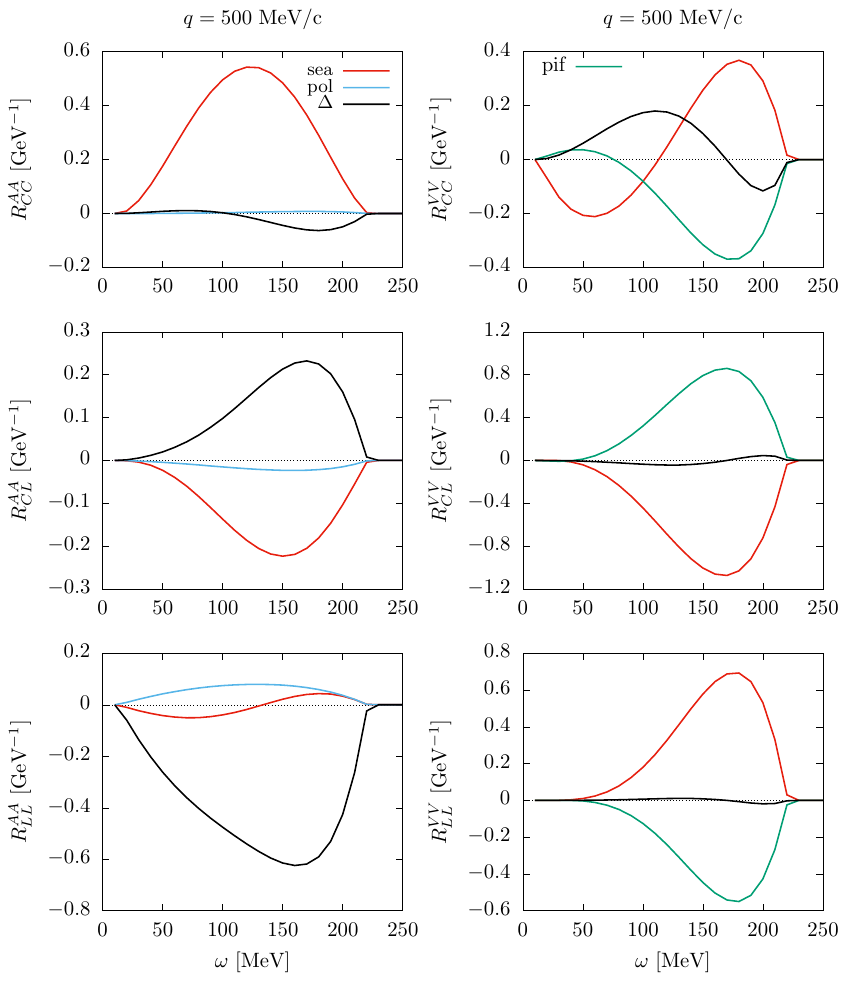}
 \caption{The same as in Fig. \ref{neufig12} for $q=500$ MeV/c. }
 \label{neufig13}
\end{figure}

\begin{figure}
  \centering
 \includegraphics[width=14cm,bb=40 350 520 810]{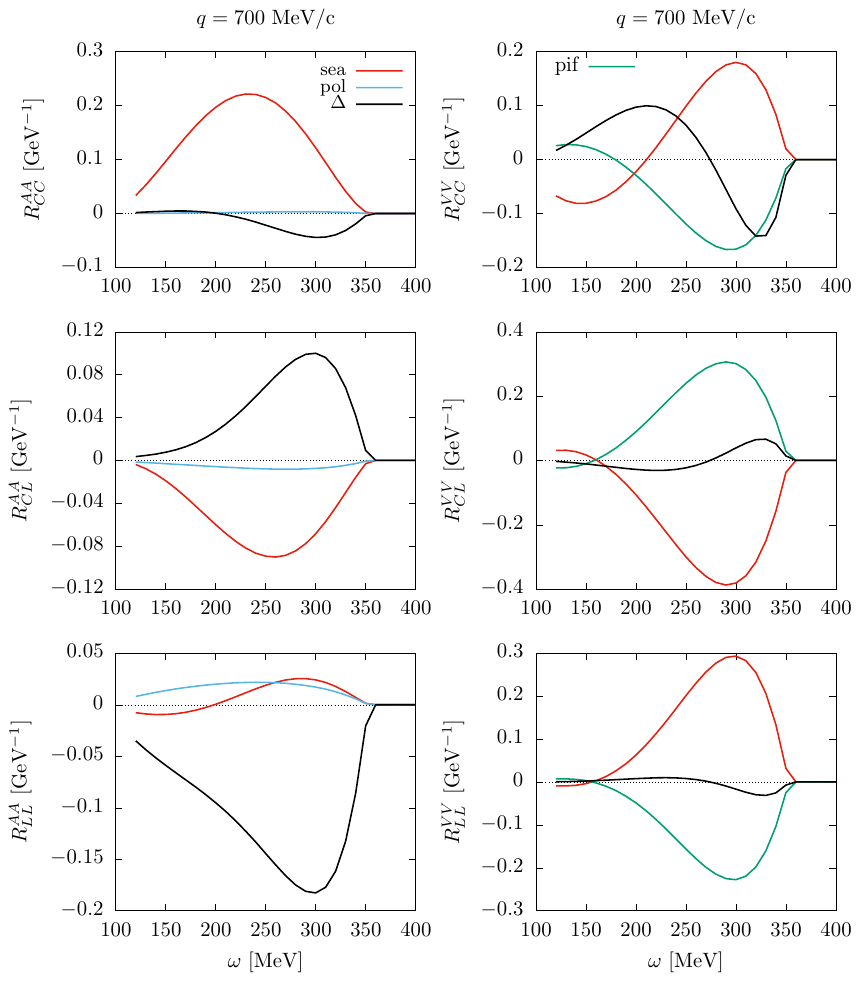}
 \caption{The same as in Fig. \ref{neufig12} for $q=700$ MeV/c. }
 \label{neufig14}
\end{figure}

\begin{figure}
  \centering
 \includegraphics[width=14cm,bb=60 480 500 810]{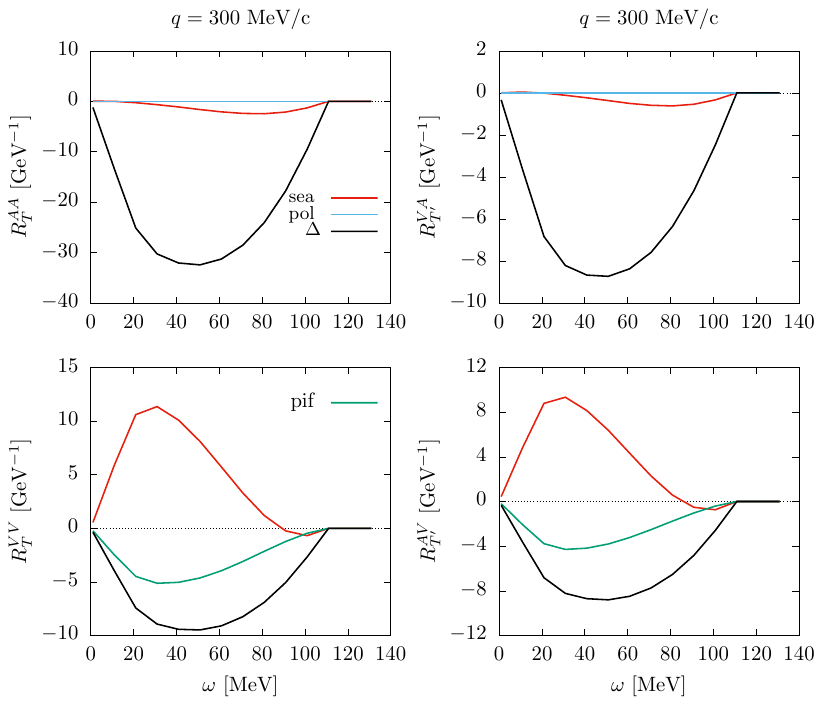}
 \caption{The same as Fig. \ref{neufig12} for the transverse
   responses for $q=300$ MeV/c.}
\label{neufig15}
\end{figure}

\begin{figure}
  \centering
 \includegraphics[width=14cm,bb=60 480 500 810]{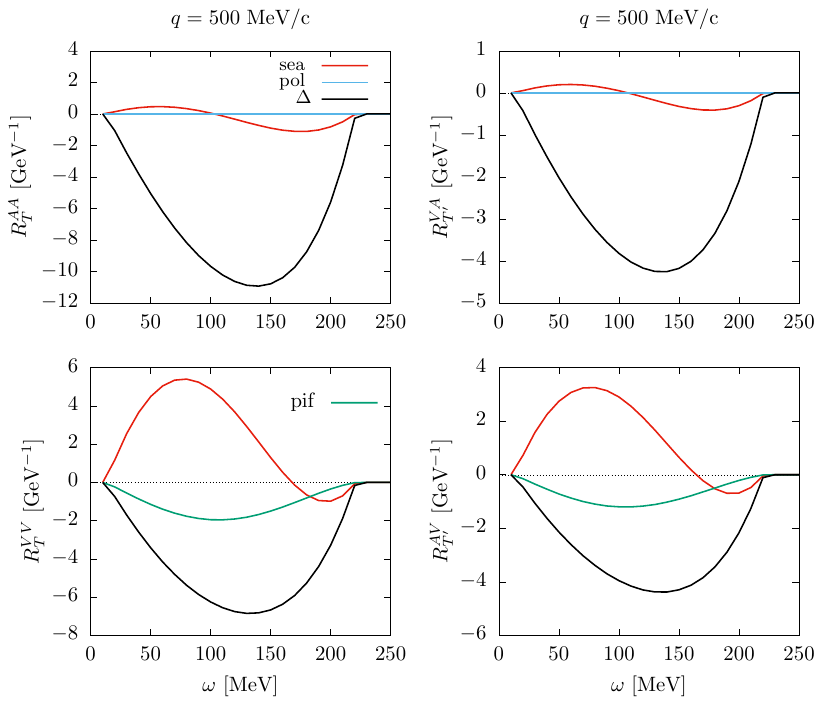}
 \caption{The same as Fig. \ref{neufig12} for the transverse
   responses for $q=500$ MeV/c.} 
 \label{neufig16}
\end{figure}

\begin{figure}
  \centering
 \includegraphics[width=14cm,bb=60 480 500 810]{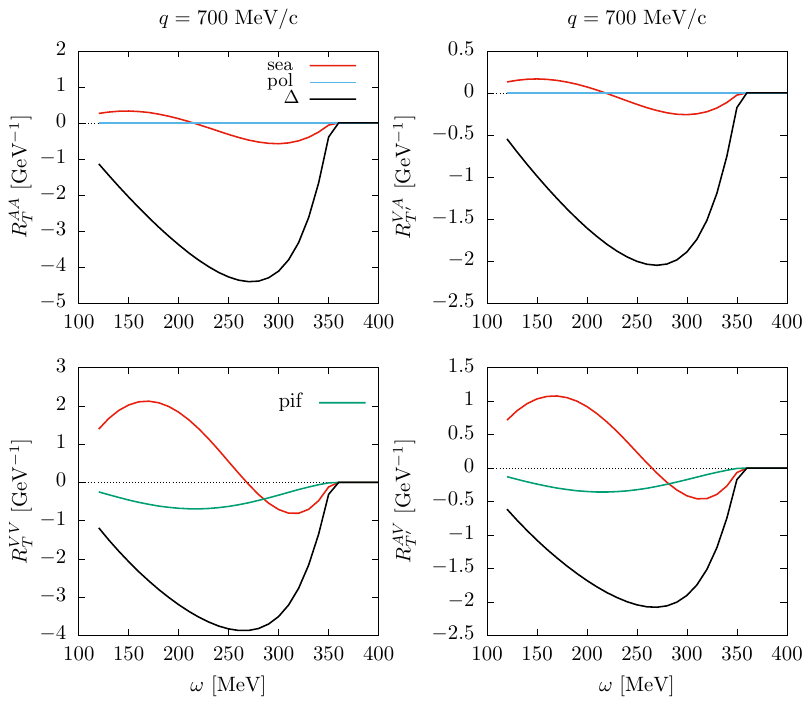}
 \caption{The same as Fig. \ref{neufig12} for the transverse
   responses for $q=700$ MeV/c.}
 \label{neufig17}
\end{figure}

We now introduce the results obtained within the RFG model. In this
section, we apply the formalism to the case of $^{12}$C, using a Fermi
momentum \( k_F = 225 \) MeV/c.

In Fig.~\ref{neufig7} we show the longitudinal responses
\( R_{CC} \), \( R_{CL} \), and \( R_{LL} \), while in
Fig.~\ref{neufig11} we display the transverse responses \(
R_T \) and \( R_{T'} \). For each response, we present three curves:
the pure one-body (1b) contribution, the full result including MEC
(1b+2b), and the interference term between one-body and two-body
currents.

It is observed that the effect of MEC in the \( R_{CC} \) response is
negligible. In the \( R_{CL} \) response, MEC introduce a visible
modification, and in the \( R_{LL} \) response they are of the same
order of magnitude as the one-body part. Nevertheless, this will have
little or no impact on neutrino cross sections, since \( R_{LL} \) is
very small and the longitudinal responses in general contribute less
significantly to the total cross section.

Figure~\ref{neufig11} shows the transverse responses \( R_T \) and \(
R_{T'} \) in the relativistic Fermi gas model for
\(^{12}\mathrm{C}\). The results include the one-body (1b)
contribution, the full result with meson exchange currents (1b+2b),
and the interference term between 1b and 2b currents. The interference
is negative and leads to a significant reduction of both
responses. This reduction is also found in the vector
transverse response \( R_T^{VV} \), and its size is comparable to that
observed in the axial \( R_T^{AA} \) and the axial-vector \(
R_{T'}^{VA} \) responses. This behavior is due to the interference
with the \(\Delta\) current, whose structure is similar in both the
vector and axial channels and gives a dominant contribution in the
relevant kinematics. As a result, the MEC-induced suppression of the
transverse neutrino responses is of similar magnitude in all three
cases, although slightly stronger in the axial-containing channels.
To illustrate this point, in Fig.~\ref{neufig9} we show separately the
vector-vector (\( VV \)) and axial-axial (\( AA \)) contributions to
the transverse response \( R_T \). It can be seen that the reduction
induced by MEC, relative to the one-body current, is similar in both
channels. In both cases, the dominant effect is the interference
between the one-body current and the \(\Delta\) current, which leads
to a negative contribution.

A more detailed scrutiny of the relevance of each MEC contribution to
the different response functions is provided in Figs. \ref{neufig12}
to \ref{neufig17} for $q=300,500,700$ MeV/c, where all the
interference terms are shown separately for the seagull, pionic,
pion-pole, and \(\Delta\) currents. The effect of MEC on the
longitudinal responses is diverse, see
Figs. \ref{neufig12}, \ref{neufig13}, \ref{neufig14}, although these
interferences have a limited impact on the total responses and an even
smaller one on the neutrino cross section. For example, the pionic and
seagull contributions tend to cancel each other in the VV-type
\(R_{CC}\), \(R_{CL}\), and \(R_{LL}\) responses. In the axial
\(R_{CC}\), the seagull clearly dominates, as anticipated in the
non-relativistic development, while in the axial \(R_{CL}\) and
\(R_{LL}\) responses a non-negligible contribution from the \(\Delta\)
current is observed, since the axial \(\Delta\) is no longer purely
transverse. The pion pole is negligible.  Nevertheless, we reiterate
that the overall impact of these interferences on observables is
minimal. The MEC effect that plays a significant role appears in the
transverse responses of
Figs. \ref{neufig15}, \ref{neufig16}, \ref{neufig17}, where the
\(\Delta\) current is clearly dominant and produces a sizable
reduction of the response, partially compensated by the seagull
contribution.

An important outcome of this analysis is the identification of a
sizable negative interference between the one-body and
$\Delta$-current contributions in the axial channel, particularly in
the AA transverse and AV,VA transverse responses. This effect, which to
our knowledge has not been previously highlighted in the literature on
quasielastic neutrino scattering, is a novel result of this work. It
is especially relevant for neutrino interactions, where the axial
current plays a central role. Since current neutrino event generators
and models often neglect such interference terms and treat MEC
contributions only a 2p2h additive term, this finding suggests that
existing models may require revision in order to properly account for
interference effects, especially those involving the $\Delta$ current
in axial channels.

\begin{figure}
  \centering
\includegraphics[width=12cm,bb=60 280 490 810]{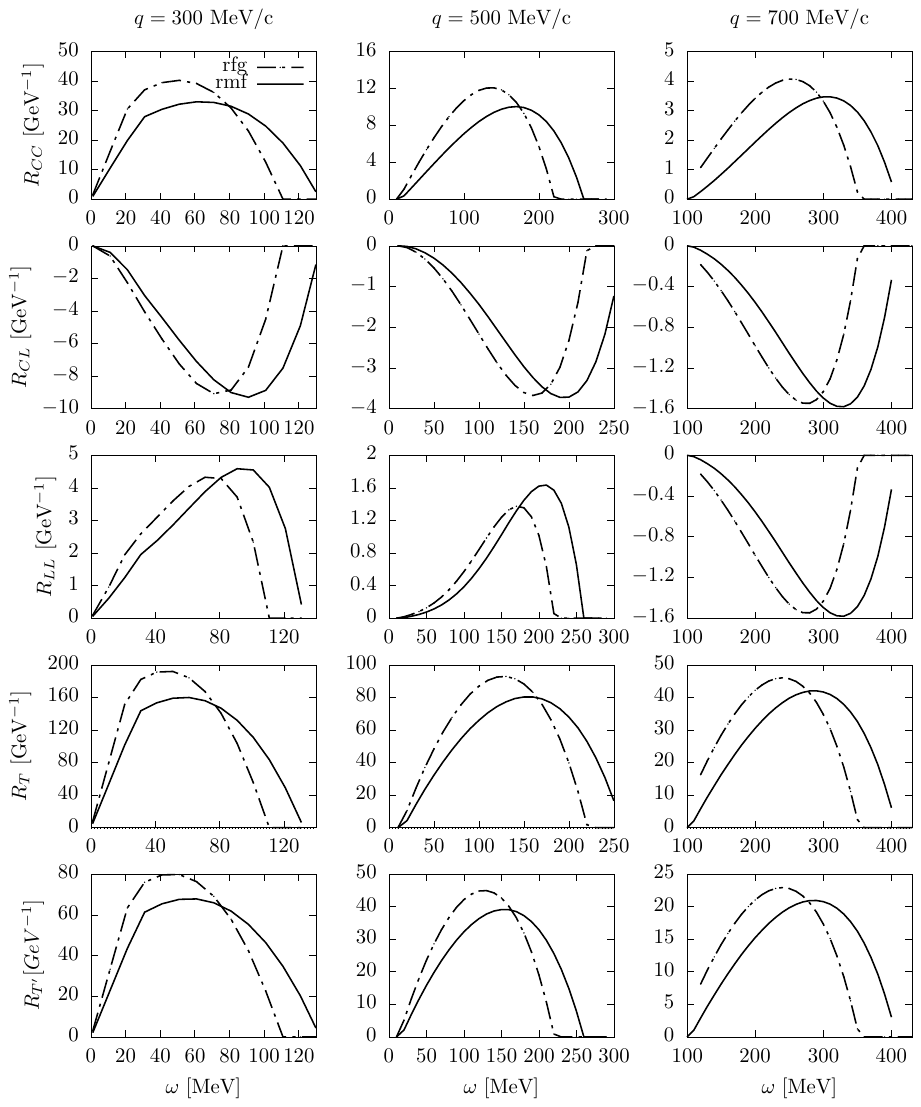}
\caption{Comparison of the five total response functions $R_{CC}$,
  $R_{CL}$, $R_{LL}$, $R_T$, and $R_{T'}$ computed within the
  RFG and the RMF models for momentum transfers $q = 300$, 500, and 700 MeV$/c$.}
 \label{neufig8}
\end{figure}

\begin{figure}
  \centering
 \includegraphics[width=14cm,bb=70 330 500 810]{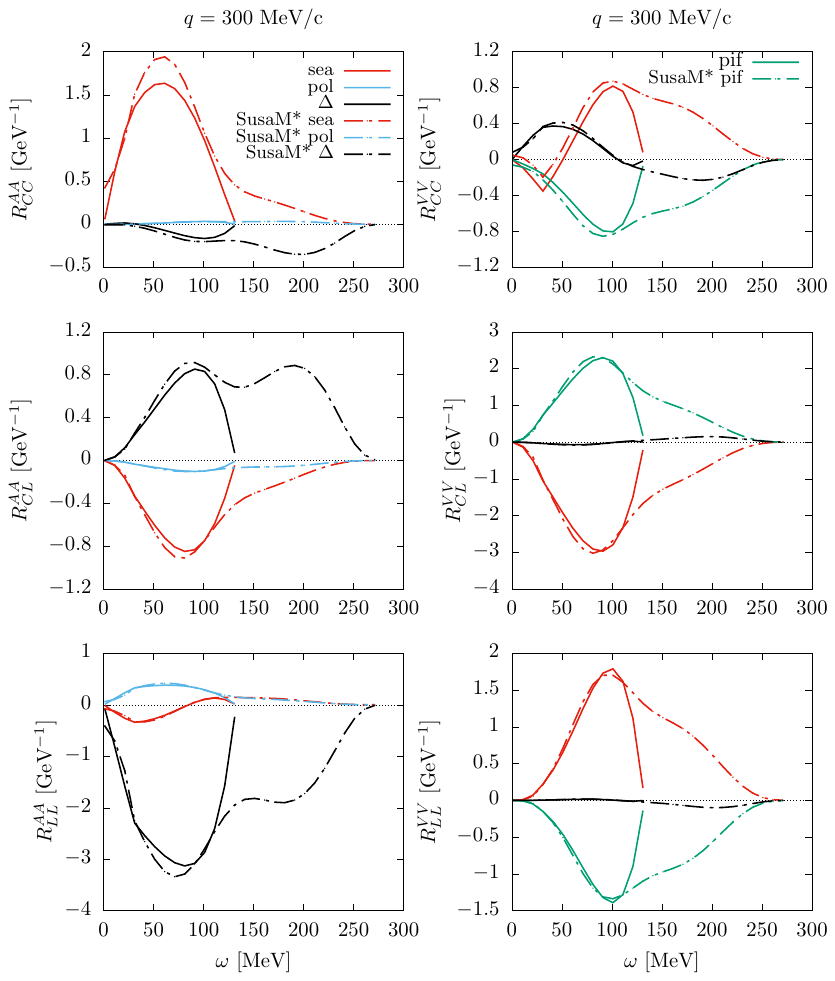}
 \caption{Longitudinal 1b-MEC interference responses in the SuSAM*
   approach compared to the RMF model for $q=300$ MeV/c.}
 \label{neufig1}
\end{figure}

\begin{figure}
  \centering
 \includegraphics[width=14cm,bb=70 330 500 810]{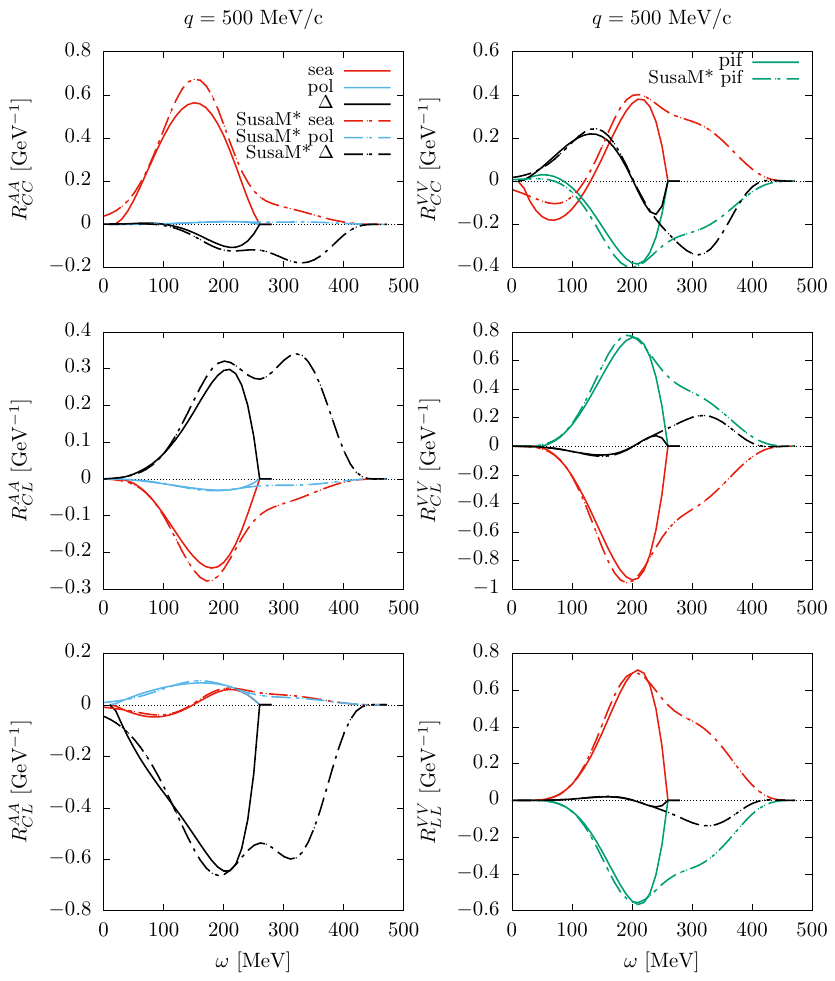}
 \caption{The same as in Fig. \ref{neufig1} for $q=500$ MeV/c.}
 \label{neufig2}
\end{figure}

\begin{figure}
  \centering
 \includegraphics[width=14cm,bb=70 330 500 810]{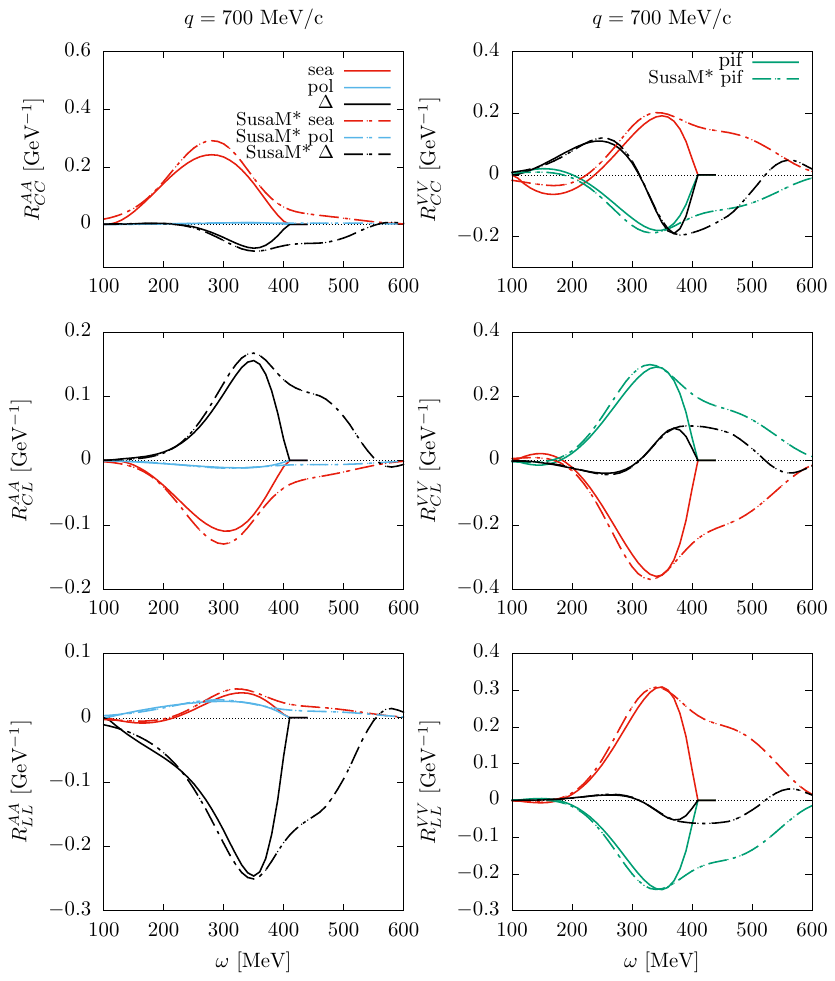}
 \caption{The same as in Fig. \ref{neufig1} for $q=700$ MeV/c.}
 \label{neufig6}
\end{figure}

\begin{figure}
  \centering
  \includegraphics[width=12cm,bb=60 470 500 810]{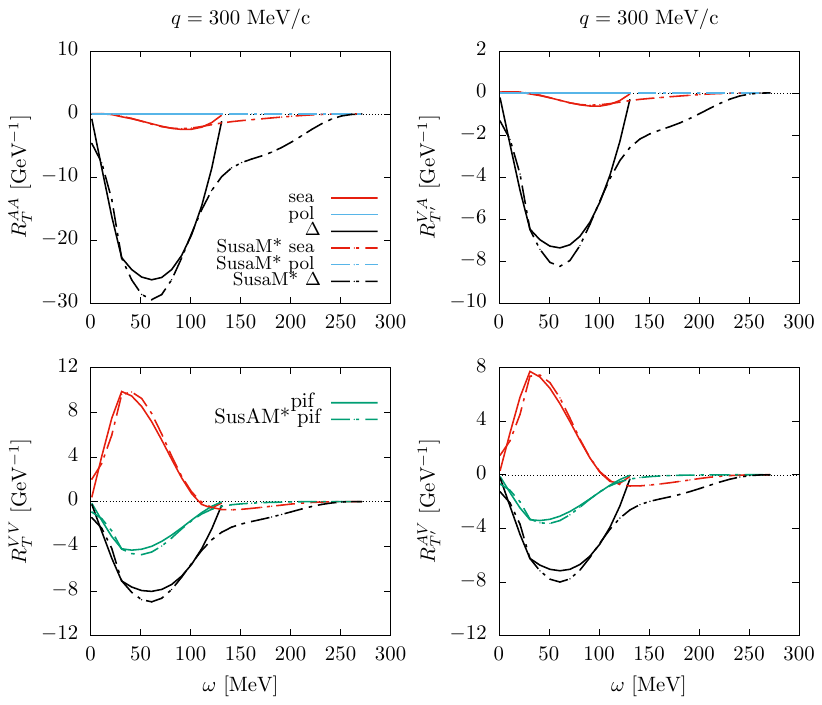}
  \caption{The same as in Fig. \ref{neufig1} for
    the transverse interference response for $q=300$ MeV/c.}
 \label{relneu4}
\end{figure}

\begin{figure}
  \centering
 \includegraphics[width=12cm,bb=60 470 500 810]{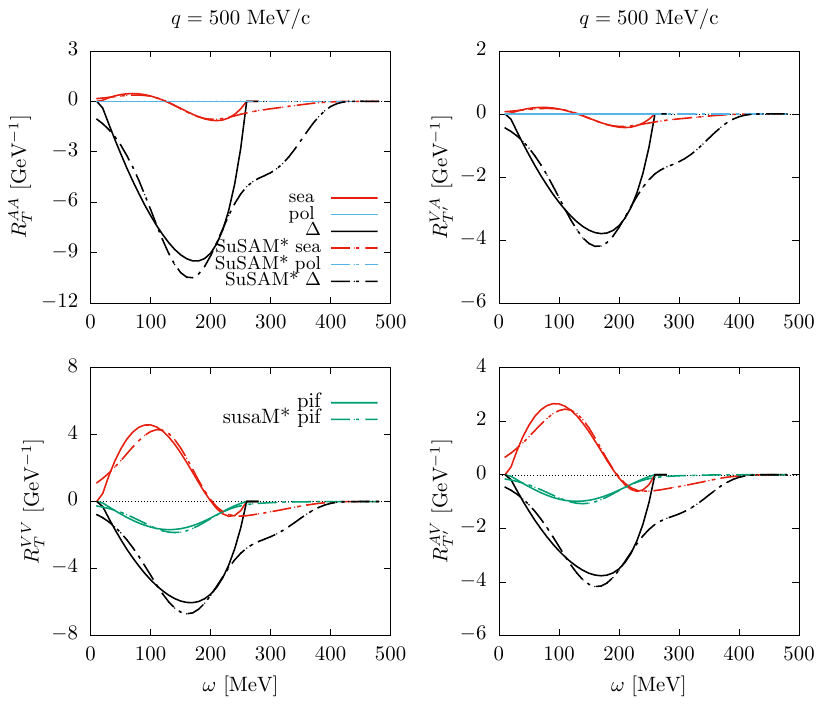}
 \caption{The same as in Fig. \ref{relneu4} for $q=500$ MeV/c. }
 \label{relneu6}
\end{figure}

\begin{figure}
  \centering
 \includegraphics[width=12cm,bb=60 470 500 810]{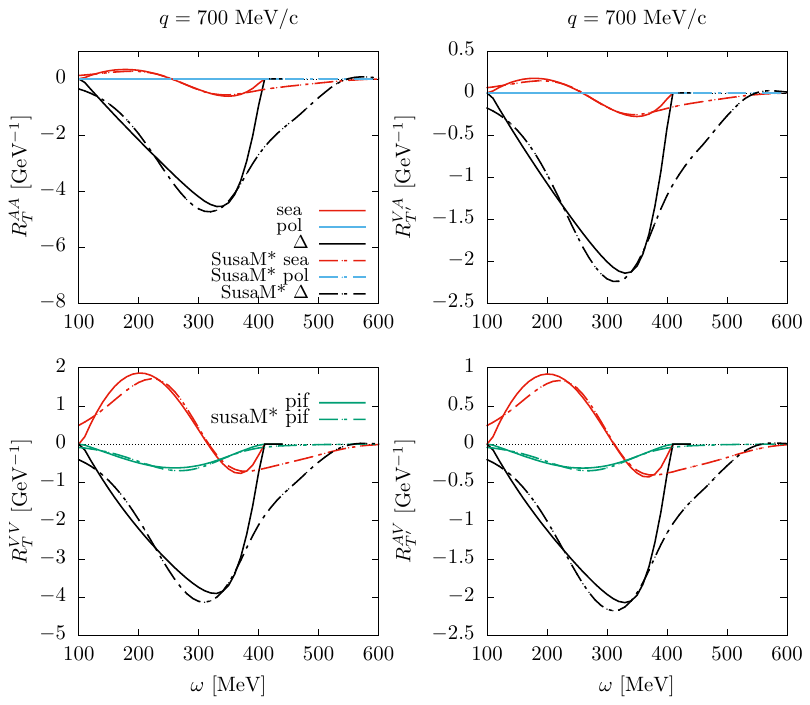}
 \caption{The same as in Fig. \ref{relneu4} for $q=700$ MeV/c.}
 \label{relneu5}
\end{figure}

\begin{figure}
  \centering
 \includegraphics[width=14cm,bb=60 280 500 810]{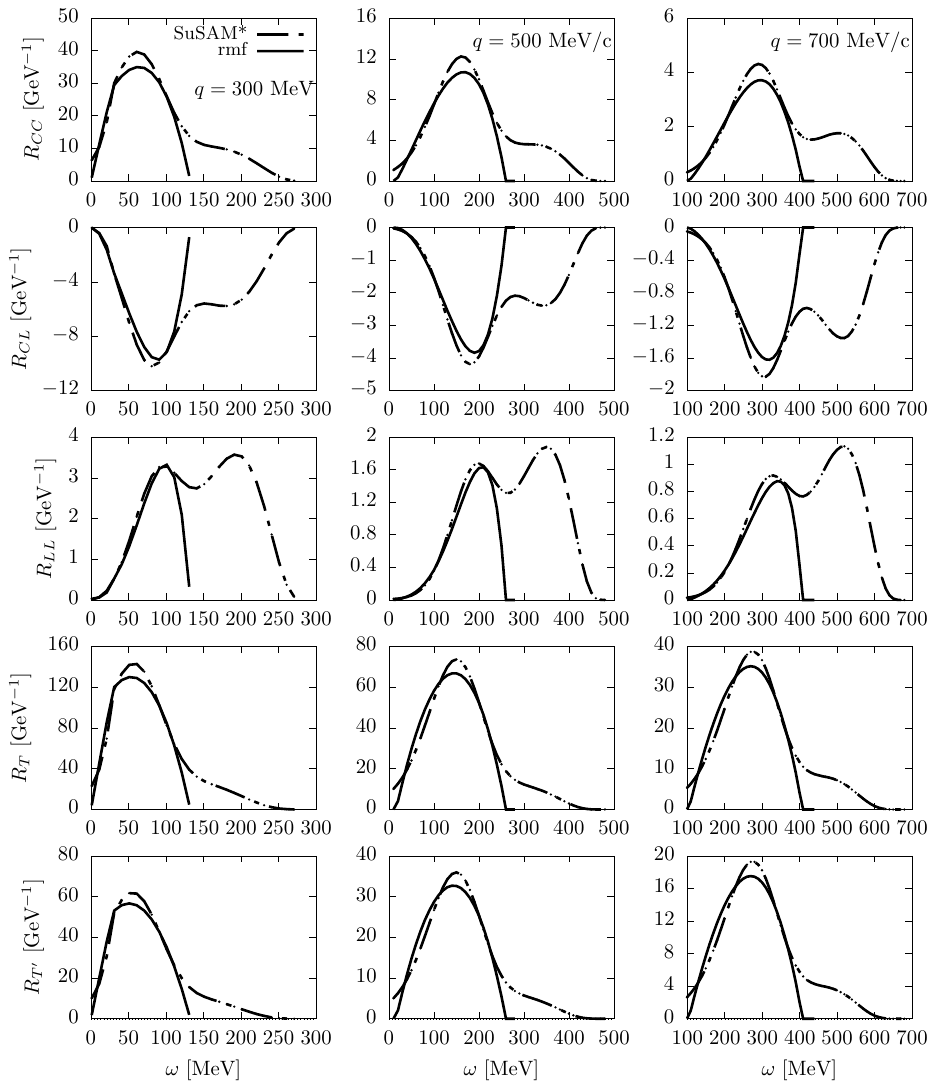}
 \caption{Total responses in the RMF and SuSAM* models for momentum
   transfers $q=300,500,700$ MeV/c.}
 \label{relneu12}
\end{figure}

Now we present results obtained with two additional nuclear models
beyond the RFG: the RMF model of nuclear matter and the SuSAM*. In
Fig.~\ref{neufig8}, we compare the five response functions calculated
in the RFG and RMF models for three values of the momentum
transfer. The main effect of the relativistic mean field is to produce
a shift in the responses due to the use of an effective mass for the
nucleons. This shift effectively incorporates, through the mean field,
part of the dynamical effects related to the binding and interaction
energy of the nucleon in the final state.

On the other hand, in the SuSAM* approach, we use Eq. (\ref{susam})
together with the phenomenological scaling function extracted in
chapter 3, Eq. (3.38). The results for the interference $1b$--MEC
response functions are shown in Figs.~\ref{neufig1}, \ref{neufig2},
\ref{neufig6} and \ref{relneu4}, \ref{relneu6} and \ref{relneu5} for
$q=300,500,700$ MeV/c, where we compare RMF and SuSAM* models. As we
can see, the interference SuSAM* responses extend well beyond the
allowed region of the RMF, enabling the estimation of MEC effects at
large $\omega$ values. The phenomenological scaling function was
parametrized as a sum of two Gaussians, which explains why some of the
responses display two peaks. It is also apparent that in the
transverse responses the total interference remains negative due to the
$\Delta$ current.

\begin{figure}
  \centering
\includegraphics[width=14cm,bb=15 298 530 810]{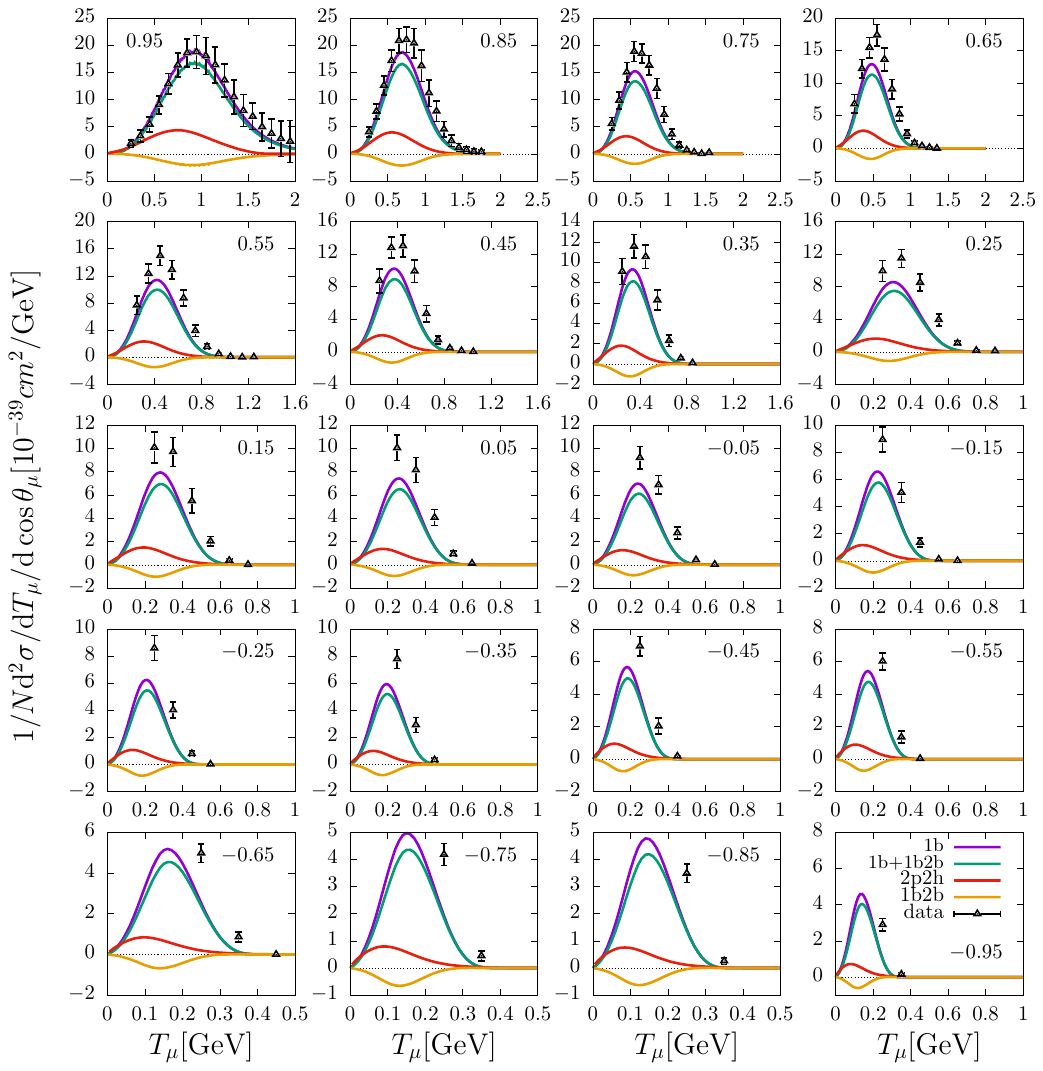}
\caption{Flux-integrated double-differential cross section per neutron
  for CC neutrino scattering on $^{12}$C in the RMF model. The
  experimental points are the inclusive CCQE measurements from
  MiniBooNE \cite{Agu10}. Shown are the one-body (1b) results, the
  one-body two-body interference (1b2b), their sum (1b+1b2b), and the
  2p2h contribution from Ref. \cite{Mar23b} computed within the same
  RMF model.}
 \label{neutrinocs}
\end{figure}

In Fig.~\ref{relneu12} we present the total response functions
computed in RMF and SuSAM* models. The second peak is only visible in the
longitudinal responses, which are small and contribute little to the
neutrino cross section. For the dominant responses---the transverse
($T$) and charge-charge ($CC$) ones---the effect of the SuSAM* model
is to introduce a high-energy tail in the responses. The relative size
of the responses can be clearly seen in this figure for $q=500$ MeV/c:
the $T$ response reaches a maximum of approximately $\sim
70$~GeV$^{-1}$, $T'$ is about half of that ($\sim 35$ ~GeV$^{-1}$),
$CC$ peaks around $\sim 12$, $CL \sim -4$, and $LL \sim 1.6$. This
shows that the $CC$ contribution is relatively small, $CL$ even
smaller, and $LL$ is almost negligible. Note, however, that each
response is weighted by a different kinematic factor $v_K$ in the
cross section. In particular, the $LL$ response is so small that it
becomes extremely sensitive to fine details of the model, but such
differences are likely to be un-observable in the total cross section,
which is largely dominated by the transverse responses and, to a
lesser extent, by the $CC$ component.

\subsection{Cross section}
\begin{figure}
  \centering
\includegraphics[width=15cm,bb=15 470 530 810]{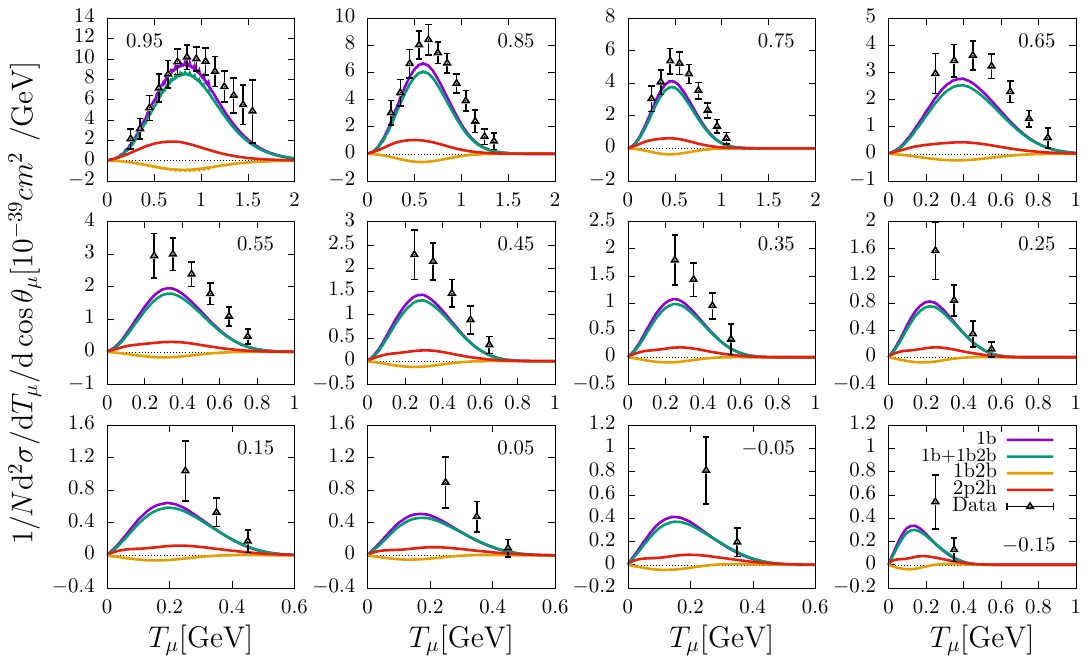}
\caption{The same as in Fig. \ref{neutrinocs} for the antineutrino
  cross section. Experimental data are the CCQE antineutrino
  measurements from MiniBoone \cite{Agu13}.}
 \label{antineutrinocs}
\end{figure}
To finish this section we present results for the neutrino
$(\nu_\mu,\mu^-)$ and antineutrino $(\overline{\nu}_\mu,\mu^+)$
inclusive cross sections.  To compare theoretical predictions with
experimental data, the double differential cross section, expressed as
a function of the muon kinetic energy and the scattering angle, must
be integrated over the neutrino flux. The flux-averaged cross section
is defined as:
\begin{equation}
  \frac{d^2\sigma}{dT_\mu dcos\theta}= \frac{1}{\Phi_{tot}}\int dE_\nu \Phi(E_\nu) \frac{d^2\sigma}{dT_\mu dcos\theta}(E_\nu)
\end{equation}
where $\Phi(E_\nu)$ is the neutrino flux, $\frac{d^2\sigma}{dT_\mu dcos\theta}(E_\nu)$ is the cross section evaluated at a fixed neutrino energy $E_\nu$ and $\Phi_{tot}$  is the total integrated flux,
\begin{equation}
  \Phi_{tot} = \int dE_\nu \phi(E_\nu).
\end{equation}
The experimental data are typically provided in bins of
$\cos\theta$, where $\theta$ is the scattering angle of the outgoing
muon. For each bin, what is actually given is the cross section
averaged over the bin width, which implies an integration over
$\cos\theta$.
\begin{equation}
  \left\langle \frac{d^2\sigma}{dT_\mu d\cos \theta} \right \rangle_{Bin} = \frac{1}{\Delta \cos \theta}
  \int_{\cos \theta_i}^{\cos \theta_f} \frac{d^2\sigma}{dT_\mu d\cos\theta} (\cos \theta) d \theta
\end{equation}
However, for large angles (i.e., low values of $\cos\theta$), the
variation of the cross section within the bin is generally small, and
one can approximate the averaged cross section by evaluating it at the
midpoint of the bin.

In Figs.~\ref{neutrinocs} and \ref{antineutrinocs} we present results
for the double-differential charged-current neutrino and antineutrino
cross sections, respectively, corresponding to the kinematics and flux
of the MiniBooNE
experiment~\cite{Agu10,Agu13}. Each
panel shows the cross section for a given $\cos\theta$ bin, with bin
width $\Delta\cos\theta = 0.1$, as a function of the kinetic energy of
the outgoing muon. A broad peak is observed, which arises from an
average over many cross sections corresponding to different values of
the incident neutrino energy $E_\nu$, weighted with the flux. This
averaging produces a much broader shape than what would be expected
from a quasielastic cross section at fixed $E_\nu$.

The theoretical calculations have been performed using the RMF model
with effective mass $M^* = 0.8$ and Fermi momentum $k_F = 225$ MeV/c,
corresponding to $^{12}$C.

In Figs.~\ref{neutrinocs} and \ref{antineutrinocs} 
we show the one-body (1b) results
together with the sum of 1b and the 1b–2b interference.
One observes that the interference with MEC produces a
reduction of the cross section of about 10\%. For antineutrinos
the effect is smaller due to the partial cancellation
between the $T$ and $T'$ responses. In the same figures we
also show separately the 1b–2b interference and the 2p2h
contributions for comparison. The 2p2h responses, computed
in Ref. \cite{Mar23b}, are positive, somewhat larger than
the interference, and display their maximum at lower
energy transfer. As a consequence, the two contributions
partially cancel each other, although the 2p2h still
dominates. Therefore, one expects that when both are
summed the net effect would be an enhancement smaller
than that produced by the 2p2h contribution alone. We
have not included this sum in the figure, since the focus
of the present work is on the interference, but it is
clear that both effects, 2p2h and 1b–2b interference, are
of comparable size.

\begin{figure}[t]
  \centering
\includegraphics[width=10cm,bb=15 280 530 750]{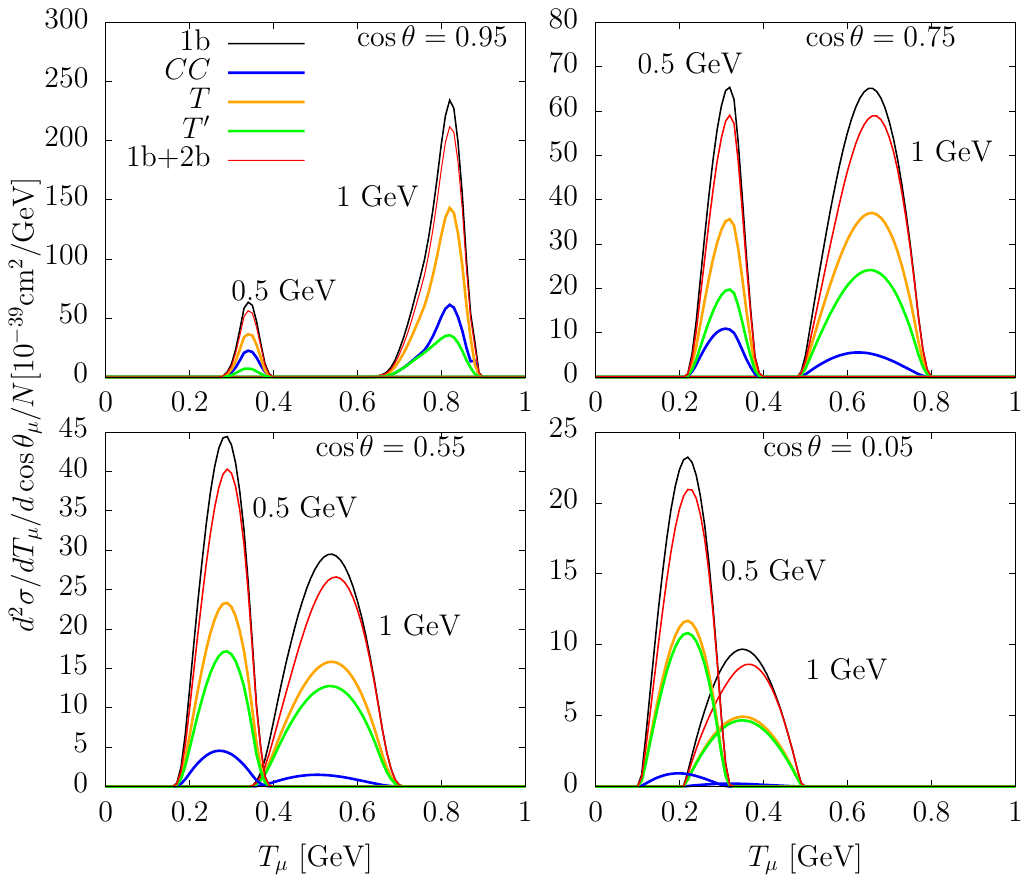}
 \caption{Double-differential cross sections for CC neutrino
   scattering on $^{12}\mathrm{C}$ in the RMF model at fixed neutrino
   energies $E_\nu = 0.5$ and $1~\mathrm{GeV}$. Results are shown for
   different scattering angles corresponding to $\cos \theta = 0.95$,
   $0.75$, $0.55$, and $0.05$.For the most forward bin, $\cos \theta =
   0.95$, the cross section is averaged over $\cos \theta \in [0.9,
     1]$, corresponding to angles between $0^{\rm o}$ and $25^{\rm o}$. In
   each panel we display the results with the one-body current alone
   (1b) and including the interference contribution (1b+1b2b). The 1b
   contributions of the CC, T, and T$'$ responses are also shown
   separately.  }
 \label{fig1neu}
\end{figure}

Other models in the literature often neglect the interference between
one-body and two-body currents, and their predictions vary
significantly among them, contributing to the theoretical systematic
uncertainties in neutrino oscillation experiments. Our results
indicate that the 1b–2b interference constitutes an additional source
of uncertainty that should also be taken into account.

In order to gain deeper insight into the flux–folded results of
Figs.~\ref{neutrinocs} and \ref{antineutrinocs},
in Fig. \ref{fig1neu} we show the impact of the interference
term on double-differential cross sections at fixed neutrino energies
and different scattering angles. Specifically, results are presented
for $\cos\theta = 0.95,0.75,0.55$ and $0.05$. Because the cross section
varies rapidly at forward angles, the case $\cos\theta = 0.95$ is averaged
over the bin $\cos\theta\in[0.9, 1]$, corresponding to scattering angles
between $0^{\rm o}$ and $25^{\rm o}$. For larger angles the cross section varies
smoothly within the bin, and it is sufficient to show the result at
the midpoint, corresponding to $41^{\rm o}$, $57^{\rm o}$ and $87^{\rm o}$, respectively. Each
panel displays results for incident neutrino energies of 0.5 and 1
GeV, which approximately span the region where the flux is most
intense. For each kinematics we show the cross section obtained with
the 1b current and after including the interference (1b+2b). The
results indicate that the interference effect depends only weakly on
kinematics, consistent with the behavior observed in Figs. ~\ref{neutrinocs} and
\ref{antineutrinocs}.

One might expect that the transverse response becomes more
relevant at larger angles, thereby increasing the relative effect of
MEC. This is also illustrated in Fig. \ref{fig1neu}, where the separated contributions
of the CC, T, and T' responses are shown (the CL and LL components
are not displayed because they are smaller). For forward angles the CC
response is larger than T' , but the T response dominates. This
arises because in neutrino scattering the axial current contributes
significantly to T, while its role in CC is minor. With increasing
angle the CC contribution decreases relative to T and T' and the
T' component becomes more important. In summary, except for very
small angles, the neutrino cross section is dominated by the
transverse response, which explains why the relative impact of MEC
does not vary strongly with kinematics. 

\section{Final Remarks}
In this chapter we have presented a detailed study of meson exchange
currents in the 1p1h channel of CC neutrino and antineutrino
scattering on nuclei. Our analysis focused on the interference between
one-body and two-body currents, an effect neglected in neutrino
event generators but that, as it have been shown, can be sizable and may
contribute to the systematic uncertainties in oscillation experiments.

The formalism employed was developed starting from the RFG model,
including 1p1h excitations produced by one-body and two-body current
operators.  In this context, we computed the five nuclear response
functions ($CC$, $CL$, $LL$, $T$, and $T'$) including the interference
between the 1b current and the seagull, pion-in-flight, pion-pole and
$\Delta$ currents.  The non-relativistic limit of the response
functions has been carefully examined as a preliminary step to
validate the relativistic model; in this limit, the spin sums can be
calculated analytically, allowing us to understand the signs and
relative importance of the dominant terms.  In particular, it has been
found that the $\Delta$ current gives rise to a strong negative
interference in the transverse responses $T$ and $T'$. This is a new
and important result, which is absent in existing neutrino models and
calls for a careful reevaluation of the role of 1b2b interference
terms in neutrino-nucleus scattering. The analysis is extended beyond
the RFG by employing two additional nuclear models: the RMF in nuclear
matter and the SuSAM* approach.

Although this chapter employs some of the most elementary nuclear
models, this choice was necessary to formulate the problem rigorously
and within a fully reproducible framework. This strategy provides a
solid baseline that enables more realistic models to incorporate and
test the effects identified. It is expected that this study will serve
as a starting point for deeper investigations and help indicate the
direction in which further theoretical efforts should proceed.

\chapter{Analysis of the SRC with the Bethe-Goldstone equation}

In this chapter we examine an important phenomenon occurring in
nuclei: short-range correlations. These correlations arise when two
nucleons interact strongly in the presence of the surrounding medium,
generating high-momentum components in the nuclear wave function. The
ultimate goal of this analysis is to compute the wave function of a
correlated nucleon pair, which will then be incorporated into our
model of MEC responses. This provides the necessary foundation for
chapter 7, where we will investigate in detail the combined effect of
MEC and SRC on the transverse response of nuclei.

Short-range correlations are described in the independent pair
approximation by solving the Bethe--Goldstone (BG) equation
\cite{Bet56,Gol57,Dah69}, which may be regarded as the in–medium
two–body Schrödinger equation. Within Brueckner theory of nuclear
matter \cite{bruck,Brue54,Brue55} it defines the $G$–matrix, which
encodes the effective nucleon–nucleon interaction in the medium
\cite{Koh61,Haf70,Jeu74,Nak84,Hos85,Nak88,Boe94}. We employ the
coarse-grained Granada 2013 potential \cite{cor3}, written as a sum of
Dirac deltas in each partial wave, which reduces the integral equation
to a simple algebraic system. The resulting correlated wave functions
are analyzed in position and momentum space via a multipole expansion,
with particular emphasis on their dependence on the center-of-mass
(CM) motion of the pair. Unlike the traditional back-to-back picture,
correlated nucleons can carry arbitrary CM momenta, extending previous
studies \cite{cor1,cor2} by the Granada group. The novelty of this
work is the explicit inclusion of CM motion, which is required for the
following chapter.  See more details in Ref.~\cite{Cas23b}.

\section{The Bethe-Goldstone equation}
We consider two particles with momenta $\mathbf{p}_1$ and
$\mathbf{p}_2$ in the Fermi gas, with $p_1,p_2<k_F$. In the absence of
interaction they are described as independent plane waves,
\[
\langle \nr_1|\np_1\rangle=\frac{1}{(2\pi)^{3/2}}e^{i\np_1\cdot\nr_1},
\kern 2cm
\langle \nr_2|\np_2\rangle=\frac{1}{(2\pi)^{3/2}}e^{i\np_2\cdot\nr_2}
\]
If we allow them to interact through a
two–body potential $V$, the wave function of the pair
$|\Psi_0\rangle=|\mathbf{p}_1\mathbf{p}_2\rangle$ is modified. Since
the interaction takes place in the medium and not in the vacuum, the
particles cannot scatter into states already occupied in the Fermi
sea. Consequently, due to the effect of the medium, scattering is only
possible into momenta above the Fermi surface, and the pair acquires
virtual high–momentum components. We thus seek the solution of the
Schrödinger equation for a pair of correlated nucleons in the form
\begin{equation}
    \Psi=\Psi_0+Q\Psi \label{beteshort}
\end{equation}
where $\Psi_0=|\np_1 \np_2 \rangle$ is the uncorrelated two-particle
and the second term represents the high-momentum components induced by
correlations. We've imposed the Pauli-Blocking condition using the
operator $Q$, which ensures our pair scatters only into unoccupied states
above $k_F$. This operator is defined as,
\begin{eqnarray}
Q|\np'_1\np'_2\rangle = \left\{ \begin{array}{lcc}
             |\np'_1\np'_2 \rangle &  \mbox{ if } & |\np'_i| > k_F \\
               0   & \mbox{otherwise} \\
\end{array}\right.
\label{pb}
\end{eqnarray}

The equation that describes this process is the Bethe-Goldstone
equation, which is the Schrödinger equation for a pair of correlated
nucleons in the nuclear medium. To deduce it, we need to determine the
condition that the state $\Psi$ must satisfy in order to fulfill the
Schrödinger equation in the presence of Pauli blocking. Starting with
the correlated wave function in Eq. (\ref{beteshort}), we want this
function to satisfy the Schrödinger equation in the presence of an
interaction,
\begin{eqnarray}
 (T+V)\Psi&=&E\Psi.
\end{eqnarray}
At the same time, the free wave satisfies,
\begin{equation}
 T\Psi_0=E\Psi_0 
\end{equation}
Substituting the full wave function into the Schrödinger equation, we
isolate the term $Q\Psi$. After some algebra, this leads to:
 \begin{eqnarray}
  Q\Psi=\frac{Q}{E-T}V\Psi, 
\end{eqnarray}
so finally, the BG equation is
 \begin{equation}
   \Psi=\Psi_0+\frac{Q}{E-T}V\Psi.  \label{betef}
 \end{equation}

We now write the Bethe–Goldstone equation in the form of an integral
equation by substituting the Pauli projector $Q$ with its
representation as an integral over intermediate states, 
\begin{equation}
Q =\int_{\np'_1 \np'_2 >k_F }d^{3}\np'_1d^{3}\np'_2 \; | \np'_1 \np'_2
  \rangle \langle \np'_1 \np'_2 |.
\end{equation}
The BG equation then reads
\begin{equation}
|\Psi \rangle
 =|\np_1 \np_2 \rangle + 
 \int d^{3}p'_1d^{3}p'_2
 \frac{\theta(p'_1-k_F)\theta(p'_2-k_F)}{E-E'} | \np'_1 \np'_2
  \rangle \langle \np'_1 \np'_2 |V|
\Psi \rangle.
\label{bethe1}
\end{equation}
where 
 $E'=E'_1+E'_2=({p'_1}^2+{p'_2}^2)/(2m_N)$. 
Note that the energy 
   $E'$ in the denominator is always larger than the 
  energy $E$ of the correlated pair, 
since $\mathbf{p}_1'$ and $\mathbf{p}_2'$ correspond to
  unoccupied states above the Fermi surface. Thus one has $E'>2E_F$
  while $E<2E_F$. The only problematic case arises when $E=2E_F$ and
  $E'\to 2E_F$. To avoid this
  issue we introduce an infinitesimal energy gap so that $E$ never
  coincides exactly with $E_F$.

Since the NN potential conserves the total momentum of the
two--nucleon system, it is convenient to switch from single--particle
coordinates $\nr_1,\nr_2$ to relative, $\nr$, and
center--of--mass, $\nR$, coordinates. Correspondingly,
the momenta are expressed in terms of the total momentum, $\mathbf{P}$,
 and the relative momentum, $\np$:
\begin{align}
\textbf{R}&=\frac{1}{2}(\textbf{r}_{1}+\textbf{r}_{2}), &
\textbf{P}&=\textbf{p}_{1}+\textbf{p}_{2}, \nonumber
\\ \textbf{r}&=\textbf{r}_{1}-\textbf{r}_{2}, &
\textbf{p}&=\frac{1}{2}(\textbf{p}_{1}-\textbf{p}_{2}). 
\end{align}
Up to this point we have omitted the spin indices for simplicity, but
we now include them explicitly since the potential depends on the spin
degrees of freedom.  We do not introduce isospin at this stage, since
the interaction we employ is charge independent. Isospin will be
relevant when discussing the different symmetry properties of
$pp$/$nn$ and $np$ pairs.
The uncorrelated and correlated wave functions
are then written
\begin{eqnarray}
  \langle \nr_1\nr_2  |p_1 p_2\rangle
  &=& 
  \frac{1}{(2\pi)^{3}}e^{i\nP \cdot    \nR}
  e^{i\np \cdot \nr}
  \textstyle{|\frac{1}{2} s_1 \frac{1}{2} s_2 \rangle}
 \equiv |\nP\rangle\otimes|\np\, s_1 s_2\rangle
  \\
  \langle\nr_1\nr_2 |\Psi\rangle
  &=& 
  \frac{1}{(2\pi)^{3/2}}e^{i\nP \cdot\nR}
\psi_{\nP,\np}^{s_1s_2}(\nr)
 \equiv |\nP\rangle\otimes|\psi_{\nP,\np}^{s_1s_2}\rangle
  \end{eqnarray}
Here $|\mathbf{p}\, s_1s_2\rangle$ denotes the relative wave function of
the uncorrelated pair of nucleons with spin projections $s_1$ and
$s_2$, while $|\psi_{\nP,\np}^{\,s_1s_2}\rangle$ represents the
corresponding correlated relative wave function.
Note that the center--of--mass wave function $|\nP\rangle$
remains unaffected by the interaction because,
\begin{equation} \label{conservCM}
  \langle \nP',\np' | V | \Psi\rangle
  = \delta(\nP'-\nP) \langle \np' | V | \psi_{\nP,\np}^{s_1s_2}\rangle.
\end{equation}

Since the NN potential commutes with $S^2$ and therefore conserves the
total spin $S=0,1$ of the two nucleons, it is also convenient to express both
the uncorrelated and correlated relative states in the coupled spin
basis
\begin{eqnarray}
|\np\, s_1 s_2\rangle
&=& \sum_{SM_S}
\langle \tfrac{1}{2}s_1\tfrac{1}{2}s_2 | S M_S \rangle\;
|\np; S M_S\rangle ,
\\
|\psi_{\nP,\np}^{\,s_1 s_2}\rangle
&=& \sum_{SM_S}
\langle \tfrac{1}{2}\,s_1\;\tfrac{1}{2}\,s_2 \,|\, S M_S \rangle\;
|\psi^{SM_S}_{\nP,\np}\rangle .
\end{eqnarray}
The state $\psi_{\nP\np}^{SM_S}$ denotes the relative correlated wave
function obtained from the uncorrelated relative state
$|\np;SM_S\rangle$, where the two nucleons are coupled to total spin
$S$ with projection $M_S$.  Exploiting the conservation of both the
total momentum $\nP$ and the total spin $S$, the center–of–mass wave
function appears as a common factor $|\mathbf{P}\rangle$ on both sides
of the BG equation (\ref{bethe1}), and therefore cancels out.
In the energy denominator we write the energies of the pair as the sum
of center–of–mass and relative contributions,
$E=\mathbf{P}^2/4m_N+\mathbf{p}^2/m_N$ and
$E'=\mathbf{P}^2/4m_N+\mathbf{p}'^2/m_N$. The center–of–mass terms
also cancel, so that only the relative energies of the pair remain, leaving
an equation for the relative correlated wave function
$|\psi_{\nP,\np}^{SM_S}\rangle$:
\begin{equation}
  |\psi_{\nP,\np}^{SM_S}\rangle
  = |\np;SM_S\rangle
+\int d^3p'\frac{{Q}(\nP,\np')}{p^2-p'{}^2}
\sum_{M'_S}|\np';SM'_S\rangle
\langle \np';SM'_S| m_NV  |\psi_{\nP,\np}^{SM_S}\rangle,
\label{bg_rel}
\end{equation}
where now the Pauli-blocking function is
\begin{equation}
Q(\nP,\np)=
\theta\left(\left|\frac{\nP}{2}+\np\right|-k_F\right)
\theta\left(\left|\frac{\nP}{2}-\np\right|-k_F\right).
\end{equation}
Since the nucleon--nucleon potential does not conserve the value of
the spin projection $M_S$, the correlated wave function for $S=1$ in
general contains admixtures of all three components $M_S=-1,0,1$.  The
wave function $\psi_{\nP,\np}^{SM_S}(\nr)$ depends on the relative
coordinate $\mathbf{r}$, the initial relative momentum $\mathbf{p}$,
and also on the total momentum $\mathbf{P}$ of the pair, since the
Pauli–blocking operator $Q$ carries an explicit
($\mathbf{P},\np)$–dependence. Actually $Q$ depends on the moduli
$P$, $p$, and on the angle between them. 
This angular dependence  breaks rotational invariance in Eq.
(\ref{bg_rel}) even for a central N-N potential
\cite{werner1959solution}, causing a mixing among different angular momenta
in a partial-wave expansion.
In this work we adopt an approximation first proposed by Brueckner
\cite{bruck}, in which $Q(\mathbf{P},\mathbf{p}')$ is replaced by its
angular average around the direction of the center–of–mass
momentum. This amounts to the substitution
\begin{equation}\label{def_angular_average}
Q(\nP,\np^\prime)\longrightarrow\,
\overline{Q}(P,p^\prime)\equiv
\frac{1}{4\pi}\int d\Omega\;
Q(\nP,\np^\prime)
\end{equation}
in Eq. (\ref{bg_rel}).  With this replacement, the angle–averaged
Pauli–blocking function $\overline{Q}(P,p')$ depends only on the
magnitudes of the CM and relative momenta, and not on the angle
between them.  The functional form of $\overline{Q}(P,p')$ for
$P<2k_F$ has been well known since the early work of Brueckner
\cite{bruck}.
\begin{equation}\label{Qbar_function}
\overline{Q}(P,p')=
\left\{
\begin{array}{ccc}
  0
  & {\rm if} &  0 \leq p' \leq \sqrt{k^2_F - \frac{P^2}{4}}
  \\
\frac{1}{P\, p'}(\frac{P^2}{4}+p^{\prime\,2}-k^2_F)
  &{\rm if}  & \sqrt{k^2_F - \frac{P^2}{4}}< p^\prime \leqslant k_F+\frac{P}{2}
  \\
1 & {\rm if} & p' > k_F+\frac{P}{2}.
\end{array}
\right.
\end{equation}
The use of the angle–averaged projector ensures that the correlated
wave functions depend only on the magnitude of the total momentum,
which simplifies the solution of the BG equation in a partial–wave
expansion. Although a few works have treated the general case
\cite{cheon,schiller,suzuki,samma,setp,white}, the angle–averaging
approximation is standard and accurate in nuclear matter
calculations~\cite{Preston75}, and is essential in this work for numerical
feasibility. The resulting BG equation is: 
\begin{equation}\label{bethegfin}
  |\psi_{\nP,\np}^{SM_S}\rangle
  = |\np;SM_S\rangle
+\int d^3p'\frac{\overline{Q}(P,p')}{p^2-p'{}^2}
\sum_{M'_S}|\np';SM'_S\rangle
\langle \np';SM'_S| m_NV  |\psi_{\nP,\np}^{SM_S}\rangle.
\end{equation}

\section{Multipole expansion of the BG equation}

The solution of the BG equation is expanded in multipoles
with total angular momentum $J M$, obtained by coupling the orbital
angular momentum $l$ with the total spin $S$ of the pair.
 This reduces the original integral
equation to a set of coupled equations for the radial components,
As a motivation, we first write the multipole expansion of the
relative plane wave with total spin $S$ in terms of the coupled states
 $\mathcal{Y}_{lSJM}(\hat{\nr})$ 
\begin{eqnarray}
   \langle \nr|\np,S\;m_S \rangle =
   A\sum_{JM}{\sum_{lm_{l}}{i^{l}Y^{*}_{lm_{l}}(\hat{\np})}
     \langle lm_{l}Sm_{s}|JM \rangle j_{l}(pr) 
\mathcal{Y}_{lSJM}(\hat{\nr})}, 
\label{mul1}
 \end{eqnarray}
with $A=\frac{4\pi}{(2\pi)^{3/2}}=\sqrt{2/\pi}$, and
\begin{equation}
 \mathcal{Y}_{lSJM}(\hat{\nr})=\sum_{m\mu}\langle l m S \mu|JM \rangle Y_{lm}(\hat{\nr}) |S\mu \rangle
\end{equation}
The solution of the Bethe–Goldstone equation can be expressed in a
similar multipole expansion, but with coupling between different
orbital angular momenta, since the potential $V$ does not conserve
$l$. We therefore propose the following ansatz for the relative wave function
\begin{eqnarray}
   |\psi\rangle =
   A\sum_{JM}{\sum_{ll^{\prime}m}{i^{l^\prime}Y^{*}_{l^{\prime}m}(\hat{\np})}
     \langle l^{\prime}m S m_{s}|JM \rangle \phi_{ll^{\prime}}^{SJ}(r) \mathcal{Y}_{lSJM}(\hat{\nr})}, \label{mulde}
 \end{eqnarray}
In Appendix I it is shown that
this ansatz satisfies the Bethe–Goldstone equation if and only if the
radial functions $\phi_{ll'}^{S J}(r)$ 
fulfill the coupled system of equations
\begin{equation}\label{rad1}
  \phi^{SJ}_{l\,l^\prime}(r)
  = j_l(pr)\delta_{ll^\prime} + \int^{\infty}_0  dr'r'^2 \; G_{l}(r,r^\prime)
  \sum_{l_1} U^{SJ}_{l_1,l}(r^\prime)\; 
\phi^{SJ}_{l_1\,l'}(r^\prime).
\end{equation}
Here, $U_{l',l}^{S J}(r) = m_N\, V_{l',l}^{S J}(r)$ is the reduced
nucleon–nucleon potential connecting the corresponding partial waves,
and the radial Green's function is defined as
\begin{equation}
  G_{l}(r,r')= \frac{2}{\pi} \int dp' p'^2 
\frac{\overline{Q}(P,p')}{p^{2}-p'^{2}} j_{l}(p'r)j_{l}(p'r').
\end{equation}
Note that $G_l(r,r')$ also depends implicitly on the momenta $P$ and $p$.

The same multipole expansion also applies to the defect function, $\Delta\psi$,
which carries the high–momentum components and is defined as
\begin{equation}
  |\psi \rangle = |\np,S\;m_S \rangle + |\Delta\psi\rangle \label{BGcompleta},
\end{equation}
with
\begin{eqnarray}
   |\Delta\psi\rangle &=&
   A\sum_{JM}{\sum_{ll^{\prime}m}{i^{l^\prime}Y^{*}_{l^{\prime}m}(\hat{\np})}
     \langle l^{\prime}m S m_{s}|JM \rangle
     \Delta \phi_{ll^{\prime}}^{SJ}(r) \mathcal{Y}_{lSJM}(\hat{\nr})}, \label{mulde}
\\
     \label{rad1}
\Delta \phi^{SJ}_{l\,l^\prime}(r) &=& \int^{\infty}_0  dr'r'^2 \; G_{l}(r,r^\prime)
  \sum_{l_1} U^{SJ}_{l_1,l}(r^\prime)\; \phi^{SJ}_{l_1\,l'}(r^\prime), 
\end{eqnarray}

In Ref. \cite{cor2} the case $P=0$ was studied using the same
formalism. Therefore, the present work can be regarded as an extension
of that formalism to the general case $P\neq 0$. In the cited
reference, only two functions appeared in the coupled–channel case
with $S=1$ and $l=J-1,J+1$, whereas in the present formalism there are
four functions. This should not be viewed as a shortcoming of
Ref. \cite{cor2}, since there the coupled equations were derived from
the corresponding NN scattering equations, where the initial momentum
is usually chosen along the $z$–axis. It can be shown that if the
initial momentum $\mathbf{p}$ is along the $z$–axis, only two radial
functions are needed in the multipole expansion. Thus the results of
Ref. \cite{cor2} are correct for that particular case, while our
equations have general validity.

Finally, it is customary to express the partial–wave equations in
terms of the reduced radial functions and the reduced Green's
function, defined as
\begin{align} \label{reduced}
  \phi^{SJ}_{l\,l^\prime}(r)&=\frac{u^{SJ}_{l\,l^\prime}(r)}{pr},&
  \kern -0.8cm
  j_l(pr)&=\frac{\hat{j}_l(pr)}{pr},&
  \kern -0.8cm
  G_{l}(r,r^\prime)=\frac{\hat{G}_{l}(r,r^\prime)}{rr^\prime}
\end{align}
where the reduced Green's function is given by,
\begin{equation} \label{green}
\hat{G}_{l}(r,r^\prime)=
\frac{2}{\pi} \int^{\infty}_0 dp^\prime\; \hat{j}_l(p^\prime r)\,
\frac{\overline{Q}(P,p^\prime)}{p^2-p^{\prime\,2}}\,
\hat{j}_l(p^\prime r^\prime).
\end{equation}
The Bethe–Goldstone equations for the partial waves then take the form
\begin{equation}\label{radial_wf}
u^{SJ}_{l\,l^\prime}(r)=
\hat{j}_l(pr)\delta_{ll^\prime}+ \int^{\infty}_0 
dr^\prime \; 
\hat{G}_{l}(r,r^\prime)
\sum_{l_1} U^{SJ}_{l_1,l}(r^\prime)\;
u^{SJ}_{l_1\,l^\prime}(r^\prime).
\end{equation}
Additionally, note that
the functions $u_{l l'}^{S J}(r)$ also depend implicitly on the
center–of–mass momentum $P$ and the initial relative momentum $p$,
although this dependence is omitted in the notation for brevity.

\section{Application to the Granada 2013 potential}

In the previous section we have presented the general formalism for
solving the BG equation for any nucleon–nucleon potential
that can be expressed in a partial–wave expansion. We now specialize
to the case of the Granada 2013 potential, a coarse–grained
interaction that greatly simplifies the solution of the integral
equations. This potential was fitted in a partial–wave analysis of
$pp$ and $np$ scattering data below the pion–production threshold with
high precision. Here, we apply it directly to the study of short–range
correlations in nuclear matter.

The potential is defined as a sum of delta–shells for each channel,
specified by the total spin $S$ and total angular momentum $J$:
\begin{equation}\label{delta-shells}
U^{SJ}_{l,l^\prime}(r)= \sum^{N_\delta}_{i=1}
\left(\lambda_i \right)^{SJ}_{l,l^\prime}\; \delta(r-r_i)\;,
\end{equation}

In the potential the number of delta-shells is fixed to $N_\delta=5$,
with $r_i=i\,\Delta r$ and $\Delta r=0.6$ fm.
The values of the  delta-shells strengths $\left(\lambda_i
\right)^{SJ}_{l,l^\prime}$ are given in Table (\ref{tab:lambdas}),
The chosen value of $\Delta r$
provides the optimal resolution allowed by the data. As
shown in Ref.~\cite{cor3}, adding more delta-shells does not improve
the description of the elastic $NN$ scattering data. On the contrary,
from a statistical point of view, the quality of the fits does not
increase, while correlations among fitting parameters become stronger,
rendering the additional deltas largely redundant.

In this calculation we neglect the one-pion exchange (OPE)
contribution, which starts at distances larger than $3$ fm. While this
contribution is essential to describe the physical scattering data
with a high quality fit (particularly for the peripheral waves), its
influence becomes marginal for the study of short distance
correlations and makes the calculation unnecessarily more cumbersome.

\begin{table}[t]
  \centering
   \begin{tabular}{S S S S S c}
        \hline
        \hline
       { $\lambda_1 [\textrm{fm}^{-1}]$} 
      & {$\lambda_2  [\textrm{fm}^{-1}]$} 
      & {$\lambda_3 [\textrm{fm}^{-1}]$ }
      & {$\lambda_4  [\textrm{fm}^{-1}]$} 
      & {$\lambda_5  [\textrm{fm}^{-1}] $} 
      & Partial Waves \\ 
        \hline
        1.31  & -0.723  & -0.187  &  0.000  & -0.024  & $^1S_0$ \\
        
        0.0  &  1.19   &  0.000  &  0.076  &  0.000  & $^1P_1$ \\  
        0.0  & -0.23   & -0.199  &  0.000  & -0.0195 & $^1D_2$ \\  
        0.0  &  0.000  &  0.130  &  0.091  &  0.000  & $^1F_3$ \\  
        0.0  &  1.00   & -0.339  & -0.054  &  0.025  & $^3P_0$ \\  
        0.0  &  1.361  &  0.000  &  0.0579 &  0.000  & $^3P_1$ \\  
        0.0  & -1.06   & -0.140  & -0.243  & -0.019  & $^3D_2$ \\  
        0.0  &  0.52   &  0.000  &  0.000  &  0.000  & $^3D_3$ \\  
        1.58   & -0.44   &  0.000  & -0.073  &  0.000  & $^3S_1$ \\  
        0.0  & -1.65   & -0.34   & -0.233  & -0.020  & $\epsilon_1$  \\  
        0.0  &  0.000  &  0.35   &  0.104  &  0.014  & $^3D_1$ \\  
        0.0  & -0.483  &  0.000  & -0.0280 & -0.0041 & $^3P_2$ \\  
        0.0  &  0.28   &  0.200  &  0.046  &  0.0138 & $\epsilon_2$  \\  
        0.0  &  3.52   & -0.232  &  0.000  & -0.0139 & $^3F_2$ \\  
        \hline
        \hline
    \end{tabular}
   \caption{Parameters $\lambda_i$ used for each partial wave. For
     coupled channels, $\epsilon_1$ and $\epsilon_2$ refer to partial
     waves $^{3}D_{1}/^{3}S_{1},^{3}S_{1}/^{3}D_{1}$ for $J=1$ and
     $^{3}F_{2}/^{3}P_{2},^{3}P_{2}/^{3}F_{2}$ for $J=2$.}
    \label{tab:lambdas}
\end{table}

\subsection{Solutions in coordinate space}

With the potential given in Eq.~(\ref{delta-shells}), the integration
over the radial coordinate in Eq.~(\ref{radial_wf}) can be carried out
analytically. One then obtains the following algebraic equation:
\begin{equation}\label{rad2}
u^{SJ}_{l\,l^\prime}(r)=
\hat{j}_l(pr)\delta_{ll^\prime}+ \sum_i \hat{G}_{l}(r,r_i)
\sum_{l_1}(\lambda_i)^{SJ}_{l_1 l} u^{SJ}_{l_1 l^\prime}(r_i),
\end{equation}

This allows us to
transform the coupled integral equation for the radial wave functions,
Eq.~(\ref{radial_wf}), into a linear system of algebraic equations
for the values of radial wave functions at the \emph{grid points} $r_i$.
In fact, by evaluating
Eq.~(\ref{rad2}) at $r=r_j$ with $j=1,2,\dots,N_\delta$, one obtains
the coupled linear system:
\begin{equation}\label{rad3}
u^{SJ}_{l\,l^\prime}(r_j)=
\hat{j}_l(pr_j)\delta_{ll^\prime}+ \sum_i \hat{G}_{l}(r_j,r_i)
\sum_{l_1}(\lambda_i)^{SJ}_{l_1 l} u^{SJ}_{l_1 l^\prime}(r_i).
\end{equation}
By solving these equations for fixed $S$ and $J$, one obtains the
values of $u^{SJ}_{l l'}(r_i)$ at the grid points $r_i$. Then, by
inserting these values into Eq. (\ref{rad2}), the wave function
can be reconstructed at any arbitrary point $r$, thus providing the
complete solution of the problem.

We now proceed to write the explicit equations for each
channel, classified according to the value of
$S$ and $J$, taking into account that the interaction conserves
parity and therefore does not couple states with orbital angular
momenta of opposite parity, i.e. odd $l$ with even $l'$.

\paragraph{\underline{Case (a): $S=0$, $J=0,1,2,3$, and $l=l'=J$.}}

Partial waves: $^{1}S_{0}$, $^{1}P_{1}$,
$^{1}D_{2}$, and $^{1}F_{3}$.

Since $\nJ=\nL+\nS$, for $S=0$ the only allowed orbital angular
momentum is $l=J$, so there is a single radial function
$u^{0J}(r)\equiv u^{0J}_{J}(r)\equiv u^{0J}_{JJ}(r)$.
 Hence,
for each $J$ one obtains an independent set of $5$ algebraic
equations for the values of the radial functions $u^{0J}_{JJ}(r_i)$ at the
points $r_i$.
\begin{equation}
u^{0J}_{JJ}(r_j)=
\hat{j}_J(pr_j)+ \sum_i \hat{G}_{J}(r_j,r_i)
(\lambda_i)^{0J}_{JJ} u^{0J}_{JJ}(r_i).
\end{equation}

\paragraph{\underline{Case (b): $S=1$, $J=0$, and $l=l'=1$.}}

Partial wave: $^3P_0$.

Since $J=0$,  for $S=1$ the only allowed orbital angular
momentum is $l=1$,
so there is a single radial function
$u^{10}(r)\equiv u^{10}_{1}(r)\equiv u^{10}_{11}(r)$.
\begin{equation}
u^{10}_{11}(r_j)=
\hat{j}_1(pr_j)+ \sum_i \hat{G}_{1}(r_j,r_i)
(\lambda_i)^{10}_{11} u^{10}_{11}(r_i).
\end{equation}

\paragraph{\underline{Case (c): $S=1$, $J=1,2$, and $l=l'=J$.}}

Partial waves: $^3P_1$, $^3D_2$.

When $S=1$ and $J>0$, there are three possibilities,
$l=J-1,J,J+1$. But due to parity conservation, the partial waves with
angular momenta $l=l^\prime=J$ (parity $P=(-1)^J$) are decoupled
from those with $l,l^\prime=J\pm1$ (parity $P=(-1)^{J+1}$).  In the
former case, $l=l^\prime=J$, there is again a single radial function
$u^{1J}(r)\equiv u^{1J}_{J}(r)\equiv u^{1J}_{JJ}(r)$, and the
corresponding equations are
\begin{equation}
u^{1J}_{JJ}(r_j)=
\hat{j}_J(pr_j)+ \sum_i \hat{G}_{J}(r_j,r_i)
(\lambda_i)^{1J}_{JJ} u^{1J}_{JJ}(r_i).
\end{equation}

\paragraph{\underline{Case (d): $S=1$, $J=1,2$, and $l,l'=J\pm 1$.}}

Partial waves: 
   ${}^{3}S_{1},{}^{3}D_{1},{}^{3}D_{1}/{}^{3}S_{1},{}^{3}S_{1}/{}^{3}D_{1}$,
and  ${}^{3}P_{2},{}^{3}F_{2},{}^{3}F_{2}/{}^{3}P_{2},{}^{3}P_{2}/{}^{3}F_{2}$
  
  The partial waves  $u^{1J}_{J\pm1,J\pm 1}(r)$. 
  are  coupled
  due to the tensor part of the NN interaction, which has
  off-diagonal components in the orbital angular momentum basis.
  The corresponding equations are
  \begin{eqnarray}
  u^{1J}_{J+1,J+1}(r_j) &=& \hat{j}_{J+1}(pr_j)
 + \sum_i \hat{G}_{J+1}(r_j,r_i)
              [(\lambda_i)^{1J}_{J+1,J+1} u^{1J}_{J+1,J+1}(r_i)\nonumber\\
&&\kern 4.5cm   +(\lambda_i)^{1J}_{J-1,J+1} u^{1J}_{J-1,J+1}(r_i)]
\label{upp}\\
u^{1J}_{J-1,J+1}(r_j)  &=&
   \sum_i \hat{G}_{J-1}(r_j,r_i)
             [ (\lambda_i)^{1J}_{J+1,J-1} u^{1J}_{J+1,J+1}(r_i) \nonumber\\
&&\kern 2.5cm   +(\lambda_i)^{1J}_{J-1,J-1} u^{1J}_{J-1,J+1}(r_i)]
\label{ump}\\
  u^{1J}_{J-1,J-1}(r_j) &=& \hat{j}_{J-1}(pr_j)
 + \sum_i \hat{G}_{J-1}(r_j,r_i)
                [(\lambda_i)^{1J}_{J+1,J-1} u^{1J}_{J+1,J-1}(r_i)\nonumber\\
&&\kern 4.5cm   +(\lambda_i)^{1J}_{J-1,J-1} u^{1J}_{J-1,J-1}(r_i)]
\label{umm}\\
u^{1J}_{J+1,J-1}(r_j)  &=&
   \sum_i \hat{G}_{J+1}(r_j,r_i)
               [ (\lambda_i)^{1J}_{J+1,J+1} u^{1J}_{J+1,J-1}(r_i) \nonumber\\
&&\kern 2.5cm   +(\lambda_i)^{1J}_{J-1,J+1} u^{1J}_{J-1,J-1}(r_i)]
\label{upm}
  \end{eqnarray}
Upon closer inspection, we find that there are in fact two independent
systems of coupled equations: one for the pairs $u^{1J}_{J\pm1,J+1}$,
Eqs.~(\ref{upp},\ref{ump}), and another for the pairs $u^{1J}_{J\pm1,J-1}$,
Eqs.~(\ref{umm},\ref{upm}). Thus, the problem reduces to solving two
linear systems of 10 equations each, instead of a single system of
20 equations for the grid-point values.

\paragraph{\underline{Case (e): $S=1$, $J=3$, and $l=l'=2$.}}

Partial wave: $^3D_3$.

In fact, this is a particular case of (d) with $J=3$ and $l,l'=2,4$,
but where the $l=4$ components vanish, i.e., the $^3H_{3}$ wave is
absent. This happens because partial waves with higher orbital angular
momentum ($l \geq 4$) were not needed to reproduce the scattering data
in the partial-wave analysis. Consequently, the equations for this
wave reduce to those of the previous uncoupled case.
\begin{equation}
u^{13}_{22}(r_j)=
\hat{j}_2(pr_j)+ \sum_i \hat{G}_{2}(r_j,r_i)
(\lambda_i)^{13}_{22} u^{13}_{22}(r_i).
\end{equation}
To compute the correlated wave function of a pair with total momentum
(P) and relative momentum (p), one must solve the linear systems
described above: five equations for uncoupled waves and (10+10)
equations for coupled waves in all the (S,J) channels considered. This
requires only the $\lambda_i$ parameters of the potential from the table
and the Green’s function values $G_l(r_i,r_j)$, the latter being
computed with the method outlined in Appendix \ref{appJ}.

\section{Solutions in momentum space}

The probability amplitude to find the
correlated state with relative momentum
$\np'$ and spin projection $m'_S$ is
\begin{eqnarray}
  \langle \np'Sm'_s | \psi \rangle = \frac{2}{\pi}
  \Bigl\langle \sum_{J_1M_1}\sum_{l_1m_1}i^{l_1}Y^{*}_{l_1m_{1}}(\hat{\np'})
  \langle l_1m_{1}Sm'_{s}|J_1M_1 \rangle j_{l_1}(p'r) \mathcal{Y}_{l_1SJ_1M_1}(\hat{\nr}) \Big| \nonumber \\
  &&
  \kern -9.7cm
 \times  \Big|\sum_{JM}\sum_{ll'm}i^{l'}Y^{*}_{l'm}(\hat{\np})
     \langle l'mSm_{s}|JM \rangle \phi_{ll'}^{SJ}(r) \mathcal{Y}_{lSJM}(\hat{\nr}) \Bigr \rangle.
\end{eqnarray}
By exploiting the orthonormality of the spin-coupled spherical harmonics, 
\[
\langle \mathcal{Y}_{l_1SJ_1M_1}(\hat{\nr})
| \mathcal{Y}_{lSJM}(\hat{\nr}) \rangle=
\delta_{l_1l} \delta_{J_1J}\delta_{M_1M},
\]
we obtain
\begin{eqnarray}
  \langle \np'Sm'_s | \psi \rangle &=& \frac{2}{\pi}
   \sum_{JM}\sum_{m_1} \sum_{l,l',m}
  i^{l'-l}Y_{lm_{1}}(\hat{\np'})Y^{*}_{l'm}(\hat{\np})\langle lm_{1}Sm'_{s}|JM \rangle
  \langle l'mSm_{s}|JM \rangle \nonumber \\
  &&
  \times
  \int^{\infty}_0 dr r^2 j_l(p'r)\phi_{ll'}^{SJ}(r).
\end{eqnarray}
We identify the radial partial wave function in momentum representation as
\begin{equation}
  \tilde{\phi}_{ll'}^{SJ}(p')=\sqrt{\frac{2}{\pi}} \int^{\infty}_0 dr r^2 j_l(p'r)\phi_{ll'}^{SJ}(r)
\end{equation}
or, in terms of the reduced partial waves $u^{SJ}_{ll'}(r)$,
\begin{equation} \label{fredu}
  \tilde{\phi}_{ll'}^{SJ}(p')=\sqrt{\frac{2}{\pi}}
  \frac{1}{pp'}\int^{\infty}_0 dr r^2 \hat{j}_l(p'r)u_{ll'}^{SJ}(r).
\end{equation}
Using that
\begin{equation}
\langle Sm'_s |\mathcal{Y}_{lSJM}(\hat{\np'}) \rangle
=
\sum_{m_1} 
Y_{lm_{1}}(\hat{\np'})
\langle lm_{1}Sm'_{s}|JM \rangle
\end{equation}
we can write
\begin{equation}
  \langle \np'Sm'_s | \psi \rangle = 
 \sqrt{\frac{2}{\pi}}
 \sum_{JM}\sum_{l,l',m} i^{l'-l} Y^{*}_{l'm}(\hat{\np}) 
 \langle l'mSm_{s}|JM \rangle
 \tilde{\phi}_{ll'}^{SJ}(p')\langle Sm'_s |\mathcal{Y}_{lSJM}(\hat{\np'})\rangle.
 \label{perturbed_wf_mom_space}
\end{equation}
Then the 
correlated wave function in momentum space expands into the multipole
series
\begin{equation}
  \langle \np'| \psi \rangle = 
 \sqrt{\frac{2}{\pi}}
 \sum_{JM}\sum_{l,l',m} i^{l'-l} Y^{*}_{l'm}(\hat{\np}) \langle l'mSm_{s}|JM \rangle
 \tilde{\phi}_{ll'}^{SJ}(p')\mathcal{Y}_{lSJM}(\hat{\np'})
 \label{perturbed_wf_mom_space}
\end{equation}
To obtain an explicit form of the radial wave function 
$\tilde{\phi}_{ll'}^{SJ}(p')$, one substitutes the coordinate-space 
function $u^{SJ}_{ll'}(r)$ from Eq.~(\ref{rad3}) into 
Eq.~(\ref{fredu}), employs the Green’s function $\hat{G}_{l}(r,r')$ 
of Eq.~(\ref{green}), and performs the radial integration. 
The orthogonality of the reduced spherical Bessel functions 
must also be used:
\begin{equation}\label{orthogonality_bessel}
  \int^{\infty}_0 dr \; \hat{j}_l(pr) \; \hat{j}_l(p'r)
  = \frac{\pi}{2}\; \delta(p-p'),
\end{equation}
The final result is:
\begin{equation}\label{radialp}
  \tilde{\phi}_{l\,l'}^{SJ}(p')
  = \sqrt{\frac{\pi}{2}}\frac{1}{pp'}\delta_{ll'}\;\delta(p-p')
  + \Delta\tilde{\phi}^{SJ}_{l\,l'}(p'),
\end{equation}
where
\begin{equation}\label{radialp-high}
  \Delta\tilde{\phi}^{SJ}_{ll'}(p')
  =\sqrt{\frac{2}{\pi}}\,\frac{1}{p\,p'}\,
\frac{\overline{Q}(P,p')}{p^2-p'^2}\sum^{N_\delta}_{i=1}
\hat{j}_l(p'r_i)\sum_{l_1} (\lambda_i)^{SJ}_{l_1,l}\;u^{SJ}_{l_1l'}(r_i).
\end{equation}
The first term of Eq.~(\ref{radialp}) corresponds to
the unperturbed radial component of the state $\left|\psi
\right\rangle$ of Eq.~(\ref{bg_rel}), while the second term, given
explicitly in Eq.~(\ref{radialp-high}), corresponds genuinely
to the high momentum components induced in the perturbed relative wave
function by the N-N interaction and the medium.
The angle--averaged Pauli projector $\overline{Q}(P,p')$ ensures that 
low--momentum components are excluded, so that only high--momentum states contribute. 
The explicit dependence on $p'$ appears both in the denominator 
$1/[p'(p^2-p'{}^2)]$ and in the spherical Bessel functions $\hat{j}_l(p'r_i)$, 
which in the asymptotic limit behave as
\begin{equation}
\hat{j}_l(p'r) \;\xrightarrow[p'\to\infty]{}\; \sin\!\big(p'r-l\pi/2\big).
\end{equation}
Consequently, the overall asymptotic behavior of the high--momentum 
wave function is
\begin{equation} \label{1pp3}
\Delta\tilde{\phi}^{SJ}_{ll'}(p') \;\sim\; \frac{1}{p'^3}, 
\qquad p'\to\infty.
\end{equation}

\subsection{High-momentum distribution}

We consider a nucleon pair with total spin $S$, total momentum $\nP$,
and relative momentum $\np$. The high--momentum distribution is
computed by integrating over the angles
the square of the high-momentum wave function
and summing over the final spin projections $\mu'$,
while averaging over the initial projections $\mu$. This procedure
yields the angle--averaged radial high--momentum distribution of the
nucleon pair induced by SRC:
\begin{eqnarray}
  &&
  \rho^{S}_{P,p}(p')
  =\frac{1}{2S+1}\sum_{\mu\mu'} \int d\Omega'
  \left| \left\langle \np' S \mu' \right|\left. \psi
  \right\rangle  \right|^2
  \nonumber\\ &&
  =\sum_{\mu\mu'}\int d\Omega'
  \biggl[ \sqrt{\frac{2}{\pi}} \sum_{JM}\sum_{l,l',m} i^{l'-l}
    Y^{*}_{l'm}(\hat{\np}) \langle l'mS\mu|JM \rangle
    \tilde{\phi}_{ll'}^{SJ}(p')\langle S\mu' |\mathcal{Y}_{lSJM}(\hat{\np'})\rangle
    \biggr]^*
  \nonumber \\
  && \times \sqrt{\frac{2}{\pi}}
  \sum_{J_1M_1}\sum_{l_1,l'_1,m_1} i^{l'_1-l_1}
  Y^{*}_{l'_1m_1}(\hat{\np}) \langle l'_1m_1S\mu|J_1M_1 \rangle
  \tilde{\phi}_{l_1l'_1}^{SJ}(p')
  \langle S\mu' |\mathcal{Y}_{l_1SJ_1M_1}(\hat{\np'}) \rangle.
\label{high_dens_mom_dist_SM_S}\nonumber\\
\end{eqnarray}
Where $\Omega'$ refers to the angles of $\np'$.
We use the orthogonality property of the spin-coupled 
spherical harmonics $\mathcal{Y}_{l S J M}(\hat{p})$ to perform the integration:
\begin{equation}\label{orthogonality_spin_angular}
\sum_{\mu'}\int d\Omega'   
\langle \mathcal{Y}_{lSJM}(\hat{\np'}) | S\mu' \rangle
\langle S\mu' |\mathcal{Y}_{l_1SJ_1M_1}(\hat{\np'}) \rangle
=\delta_{l, l_1}\;\delta_{J, J_1}\; \delta_{M, M_1}\,.
\end{equation}
Upon summation over $l_1,J_1,M_1$ we arrive at
\begin{eqnarray}
  \rho^{S}_{P,p}(p')
  & =&
  \frac{2}{\pi}
  \frac{1}{2S+1}
  \sum_{\mu}\sum_{JM}\sum_{ll'm}\sum_{l'_1m_1} i^{l'_1-l'}
  Y_{l'm}(\hat{\np})
  Y^{*}_{l'_1m_1}(\hat{\np})
  \nonumber\\
  &&\times
  \langle l'm S \mu|JM \rangle
    \langle l'_1 m_1 S \mu|JM \rangle
    \tilde{\phi}_{ll'}^{SJ}(p')
    \tilde{\phi}_{ll'_1}^{SJ}(p')
    \label{rho2}
\end{eqnarray}
Using the symmetry properties of the coupling coefficients under the
interchange of angular momenta $l'$ and $J$,
\begin{eqnarray}\label{C-G_symmetry_property}
  \langle l'mS\mu | JM \rangle
  &=&
  (-1)^{S+m_s} \sqrt{\frac{2J+1}{2l'+1}}
  \langle J,-M S \mu | l',-m \rangle.
  \nonumber\\
  \langle l'_1m_1 S\mu | JM \rangle
  &=&
  (-1)^{S+m_s} \sqrt{\frac{2J+1}{2l'+1}}
  \langle J,-M S \mu | l'_1,-m_1 \rangle.
\end{eqnarray}
we can perform the sum over $M,\mu$ of the product of two CG coefficients:
\begin{equation}
\sum_{\mu M}  \langle l'   m  S \mu|JM \rangle
    \langle l'_1 m_1 S \mu |JM \rangle
    = \frac{2J+1}{2l'+1}\delta_{l'l'_1}\delta_{mm_1}
\end{equation}    
After carrying out the sums over $l'_1$ and $m_1$, the momentum
distribution (\ref{rho2}) takes the form
\begin{equation}
  \rho^{S}_{P,p}(p') =
  \frac{2}{\pi}
  \frac{1}{2S+1}
  \sum_{J}\sum_{ll'm}
  |Y_{l'm}(\hat{\np})|^2
  \frac{2J+1}{2l'+1}
    \tilde{\phi}_{ll'}^{SJ}(p')^2
    \label{rho3}
\end{equation}
In the last step, we make use of the property that the sum over $m$ of
the squared spherical harmonics, $\sum_m |Y_{lm}(\hat{p})|^2$, has a
simple analytical expression.
\begin{equation}
\sum_m\; |Y_{l' m}(\hat{\np})|^{2}= \frac{2l'+1}{4\pi}, 
\end{equation}
thus obtaining the final result
\begin{equation} 
  \rho^{S}_{P,p}(p')=
  \frac{1}{2S+1}\sum_{J}\frac{2J+1}{4\pi}\sum_{l,l'}
\frac{2}{\pi}\left|\tilde{\phi}^{SJ}_{l,l'}(p')\right|^2.
\label{rho4}
\end{equation}

The high--momentum components of the correlated nucleon pair wave
functions arise from short--range correlations in the nuclear medium.
From the defect function defined in Eq.~(\ref{radialp-high}), for
large relative momenta $p'$ one finds
$\tilde{\phi}^{SJ}_{ll'}(p')=
\Delta\tilde{\phi}^{SJ}_{ll'}(p')
\sim 1/p'^3$, so that the square of
the wave function scales as $|\Delta\tilde{\phi}^{SJ}_{ll'}(p')|^2
\sim 1/p'^6$ for large values of $p'$.
Including the phase--space factor $p'^2$, the
corresponding momentum distribution of the pair given by
Eq. (\ref{rho4}), exhibits a
characteristic  $1/p'^4$ tail \cite{Rio14,Bao16}.

\section{Results}

In this section we present the results obtained for SRC from the
solution of the BG equation with the Granada 2013 interaction. We
display the correlated radial wave functions both in coordinate and
momentum space, for the different channels arising in the multipole
expansion. In addition, we analyze the associated correlation
functions and momentum distributions, which provide direct insight
into the role of SRC in generating high--momentum components in
nuclear matter.  In particular, we investigate the dependence of the
results on the center-of-mass momentum $\nP$, in order to assess
whether the short-range correlations are significantly affected by
$\nP$, considering that correlated pairs are typically back-to-back
(i.e., $\nP=0$).
We show the results for a Fermi momentum of $k_F=250$
MeV/c and an initial relative momentum of the pair of $p=140$ MeV/c,
in order to compare with what was done in
the previous study of Ref. Ref.~\cite{cor2} for $P=0$.
For definiteness, we present results for a relative momentum of $p =140$ MeV/c,
roughly half the Fermi momentum; we have verified,
however, that our conclusions regarding the dependence on the
center-of-mass momentum $\nP$ are largely insensitive to the
particular choice of $p$.

\subsection{Uncoupled waves in coordinate space}

\begin{figure}
  \centering
\includegraphics[width=12cm,bb=30 170 530 770]{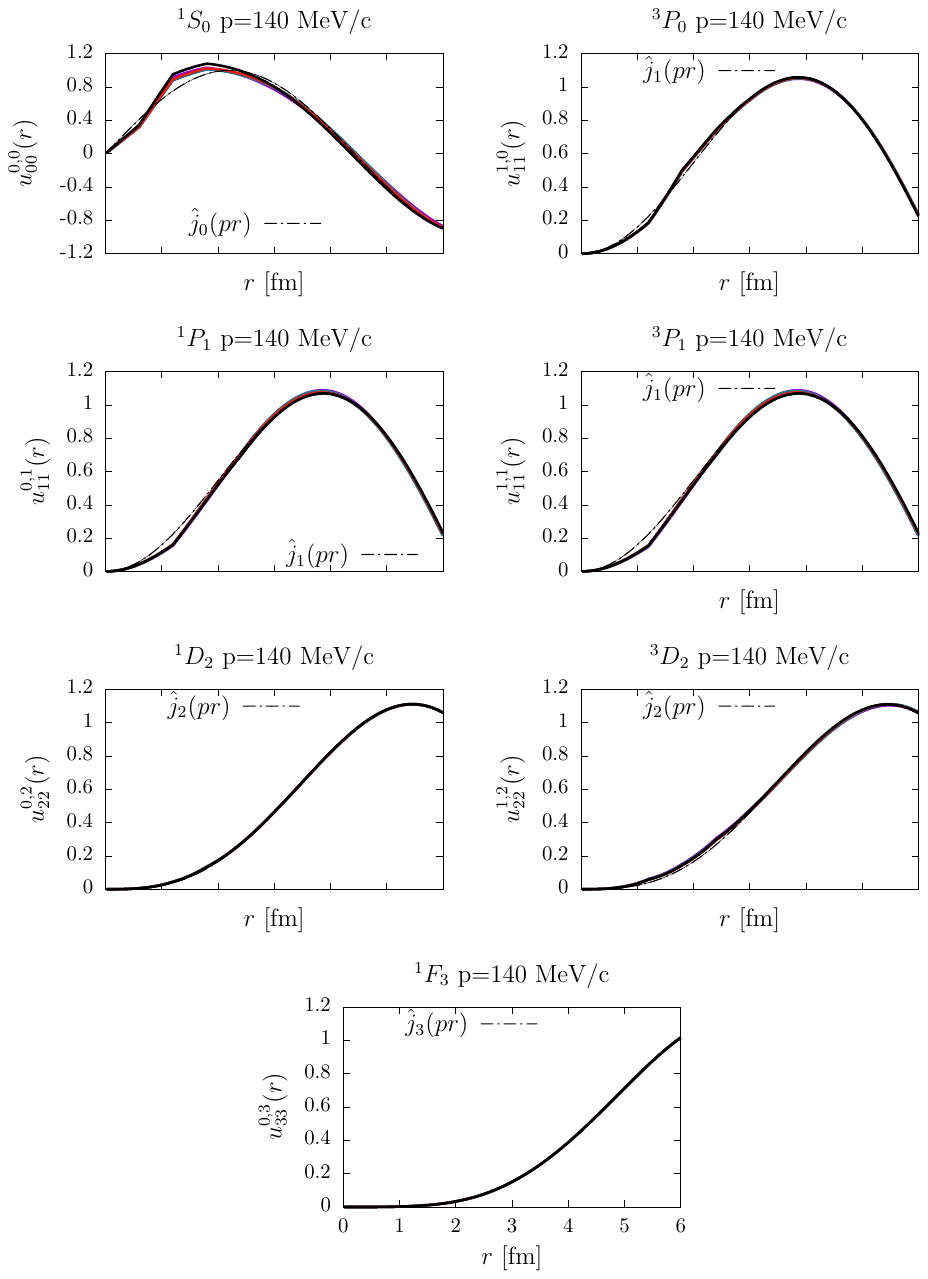}
\caption{
Reduced radial wave functions $u^{SJ}_{ll'}(r)$ for the uncoupled 
N-N partial waves ($l=l'$) at relative momentum $p=140$~MeV/c. 
For each partial wave, both the free solution $\hat{j}_l(pr)$ and the 
correlated waves for various CM momenta are shown.
}
\label{ondas3}
\end{figure}

\begin{figure}
\centering
\includegraphics[width=12cm,bb=30 170 500 770]{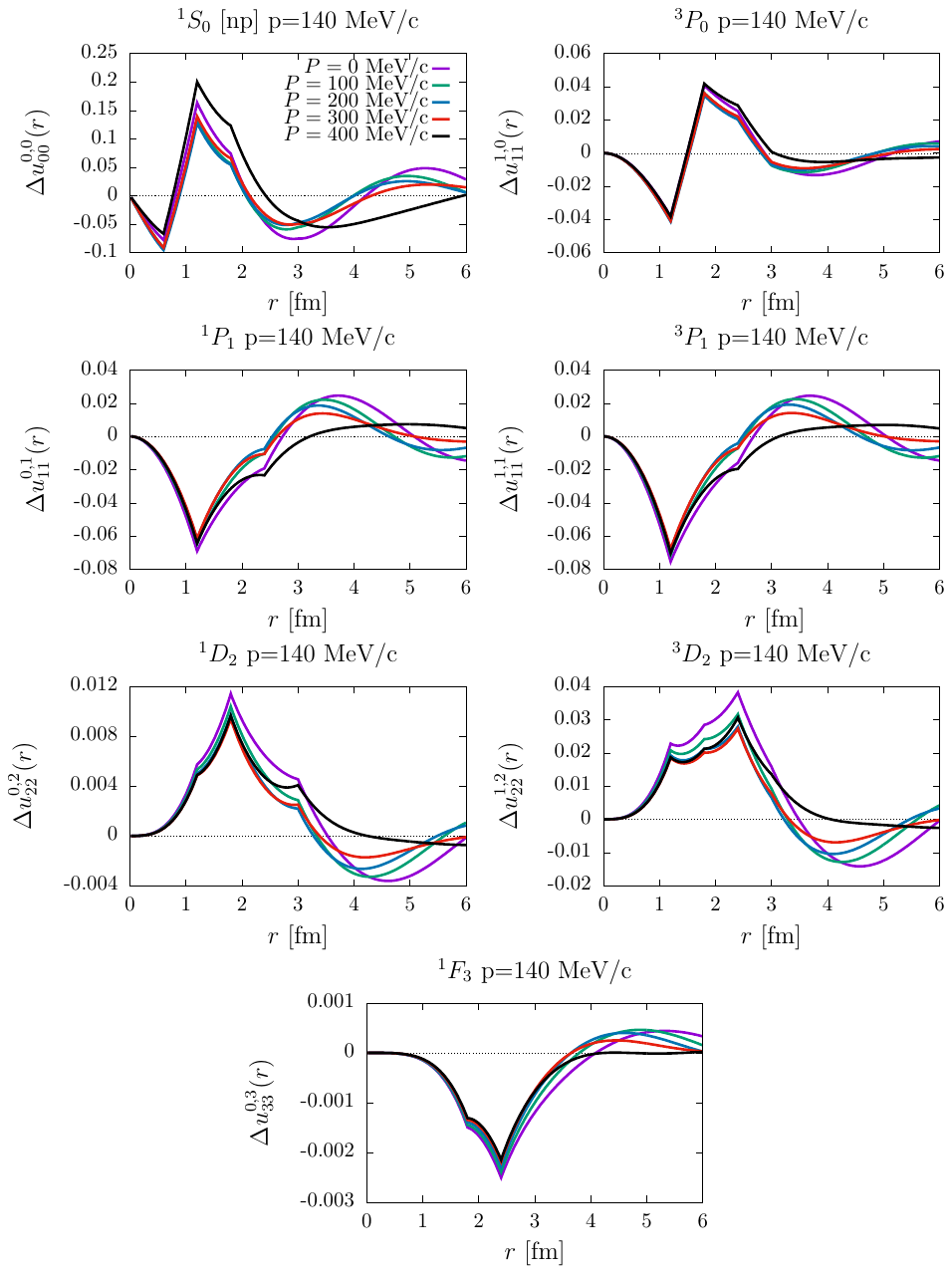}
\caption{
Defect wave functions 
$\Delta u^{SJ}_{ll'}(r) \equiv u^{SJ}_{ll'}(r)-\hat{j}_l(pr)$ 
for the uncoupled N-N partial waves ($l=l'$) at relative momentum 
$p=140$~MeV/c. Results are shown for several values of the  CM momentum.
}
\label{ondas4}
\end{figure}

\begin{figure}
  \centering
\includegraphics[width=12cm,bb=40 160 500 770]{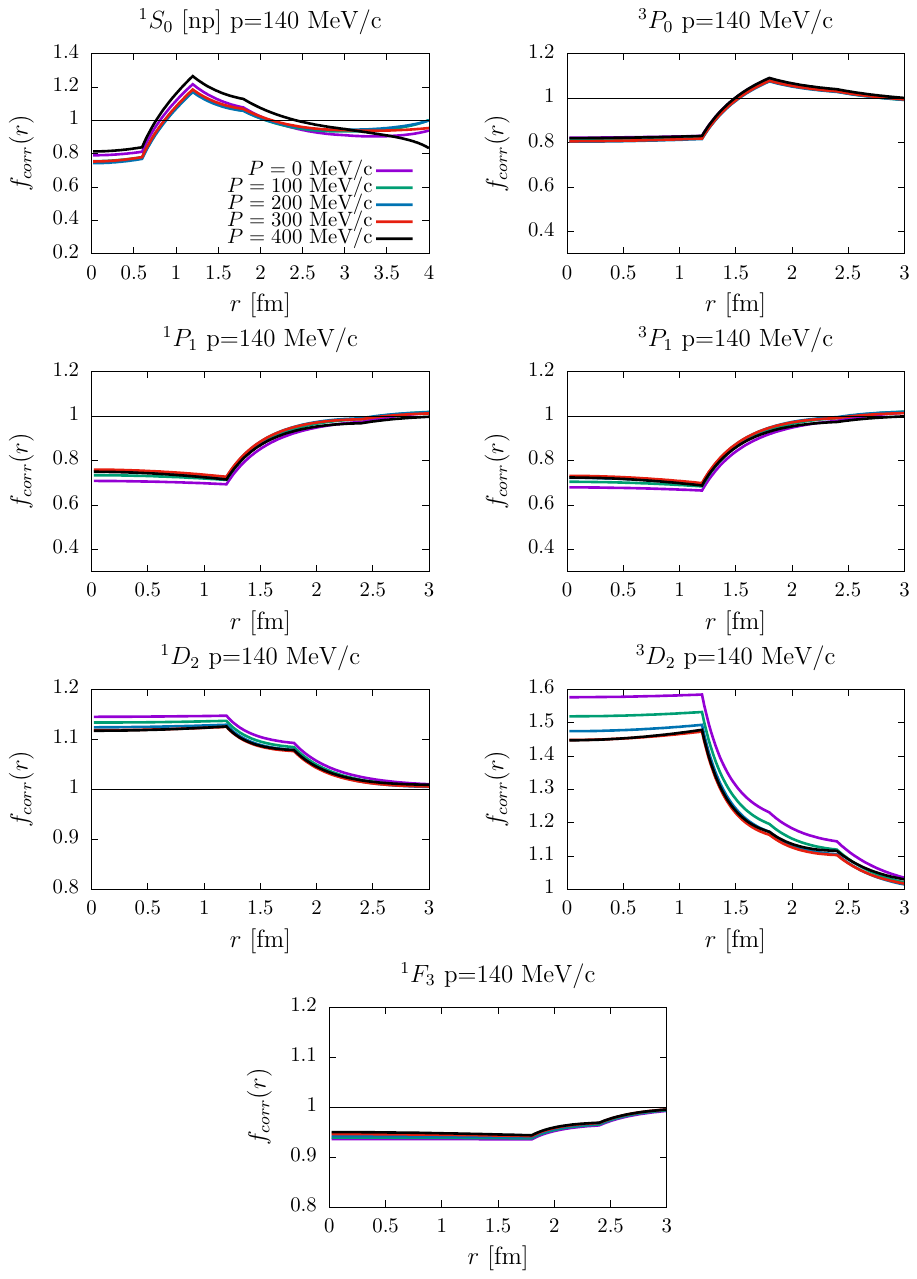}
\caption{
  Correlation functions 
$f_{\rm corr}(r) \equiv u^{SJ}_{ll'}(r)/\hat{j}_l(pr)$ 
for the uncoupled N-N partial waves ($l=l'$) at relative momentum 
$p=140$~MeV/c. Results are shown for several CM momenta.
    }
\label{ondas5}
\end{figure}

In Fig.~\ref{ondas3} we present the reduced radial wave functions 
$u^{SJ}_{ll}(r)$ for the uncoupled partial waves, obtained as solutions 
of Eq.~(\ref{rad2}). Each panel displays several curves corresponding 
to different total center-of-mass momenta of the nucleon pair, 
$P=0,\,100,\,200,\,300$, and $400$~MeV/c. These values are consistent 
with a fixed relative momentum $p=140$~MeV/c, while ensuring that both 
initial single-nucleon momenta remain below the Fermi momentum $k_F$.
The figure also includes the unperturbed wave function for comparison. 

We observe that the effect of SRC is to modify the wave function at 
short distances ($r<3$~fm), enhancing its curvature, whereas at large 
distances all correlated curves converge to the uncorrelated one.
The stronger impact of SRCs on low-$l$ partial waves arises both from 
the strength parameters of the delta-shell potential, 
$\left(\lambda_i \right)^{SJ}_{l,l^\prime}$, and from the centrifugal 
barrier, which increases with $l$ and hinders the nucleons from 
approaching each other. Consequently, SRC effects are most visible at 
short distances in S- and P-waves, while they are strongly suppressed 
in higher partial waves such as D or F.

Figure~\ref{ondas3} shows that the effect of the two-nucleon CM motion 
on the radial wave functions is generally small, with the correlated 
waves largely independent of the CM momentum. Small differences are 
slightly more visible at short distances in low-lying partial waves 
(S and P) and in the triplet ${}^{3}$D$_2$ compared to the singlet 
${}^{1}$D$_2$, reflecting the stronger attraction of the first 
delta-shell parameter.

To highlight the effect of SRCs, Fig.~\ref{ondas4} shows the defect 
wave functions, defined as
\begin{equation}
 \Delta u^{SJ}_{ll'}(r) \equiv u^{SJ}_{ll'}(r) - \hat{j}_l(pr),
\end{equation}
which quantify the distortion of the radial waves due to SRCs. 
The defect generally decreases with increasing orbital angular momentum 
$l$, reflecting the centrifugal barrier that limits the nucleons' approach 
to short distances, although the delta-shell strength parameters also 
contribute. In contrast, the magnitude of the distortion is largely 
insensitive to the CM momentum. We anticipate that the amplitudes of 
these distortions are directly related to the high-momentum components 
of the wave function, which will be reflected in the relative momentum 
distributions discussed later.

Figure~\ref{ondas5} shows the correlation functions for each uncoupled 
N-N partial wave,
\begin{equation}
f_{\rm corr}(r) \equiv \frac{u^{SJ}_{ll'}(r)}{\hat{j}_l(pr)},
\end{equation}
which deviate from unity only at short distances, illustrating the 
``healing'' property of the BG equation: correlations, or ``wounds'' in 
the unperturbed wave, are confined to short relative distances 
\cite{Preston75}. The correlation functions show little sensitivity to the 
CM momentum, with the largest variations near the origin in the 
${}^{3}$D$_2$ partial wave.  

In the literature, correlation functions are often defined such that 
$f(r)\rightarrow 1$ at large distances (beyond the healing distance). 
With the BG solutions, one might be concerned if the zeros of the 
correlated wave do not exactly coincide with those of the free wave 
when forming the quotient. However, this is not an issue here, since 
we compute the correlated wave function directly from the BG equation, 
rather than as a product of a correlation function and the free wave. 
Any differences at the nodes are infinitesimal and do not affect the 
physical content of the wave functions.
For example, the ${}^{1}$S$_0$ correlation function in Fig.~\ref{ondas5} 
is plotted only up to 4~fm, since the corresponding node for the chosen 
$p$ lies between 4 and 5~fm in Fig.~\ref{ondas3}.

\subsection{Coupled waves in coordinate space}

\begin{figure}
  \centering
\includegraphics[width=12cm,bb=30 420 520 770]{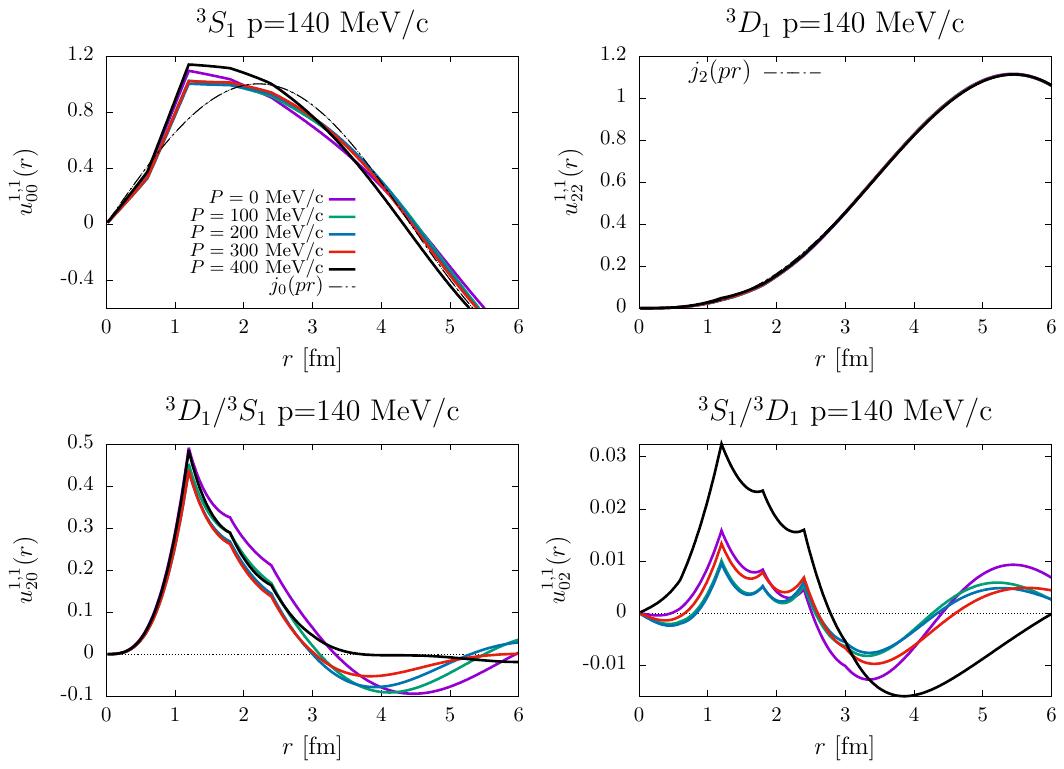}
\caption{
Radial wave functions $u^{SJ}_{ll'}(r)$ for the coupled partial waves 
with $S=J=1$ and $l,l'=0,2$, at relative momentum $p=140$~MeV/c. 
The corresponding uncorrelated waves $\hat{j}_l(pr)$ are also shown. 
Results are presented for CM momenta $P=0,100,200,300,400$~MeV/c. 
Top panels: diagonal components ($l=l'$); bottom panels: off-diagonal 
components ($l\neq l'$).
}
\label{ondas6}
\end{figure}

\begin{figure}
  \centering
\includegraphics[width=12cm,bb=30 420 520 770]{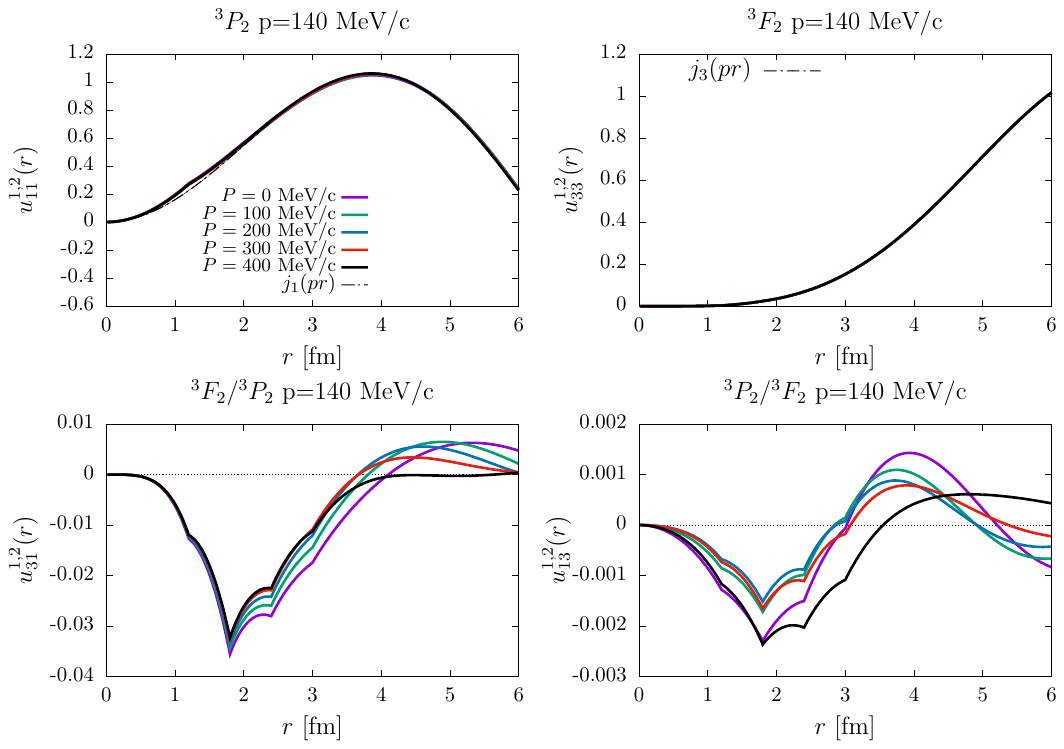}
\caption{The same as fig. \ref{ondas6} but for $l,l'=1,3$ and $J=2$}
\label{ondas7}
\end{figure}

Results for the radial wave functions of the coupled partial waves are
shown in Fig.~\ref{ondas6} for $J=1$ and in Fig.~\ref{ondas7} for
$J=2$.
We present results
for different values of the CM momentum of the two-nucleon system and
for a relative momentum of $p=140$ MeV/c.

In the case $S=J=1$ (Fig.~\ref{ondas6}), the four wave functions split
into two independent pairs, each arising from a distinct set of
coupled equations.  The first pair, $u^{11}_{00}(r)$ and
$u^{11}_{20}(r)$, obtained from Eqs.~(\ref{umm},\ref{upm}), correspond
to the ${}^3S_1$ (top left) and to the interference ${}^3D_1/{}^3S_1$
(bottom left) components.  The second pair, $u^{11}_{22}(r)$ and
$u^{11}_{02}(r)$, obtained from Eqs.~(\ref{upp},\ref{ump}), correspond
to the ${}^3D_1$ (top right) and the interference ${}^3S_1/{}^3D_1$
(bottom right) components.

Correlations strongly modify the ${}^{3}S_{1}$ wave, much more than in the 
${}^{1}S_{0}$ channel. They also generate a non-vanishing 
${}^{3}D_{1}/{}^{3}S_{1}$ interference, entirely produced by the tensor 
force, which couples partial waves with $l\pm1$ and drives the dynamics 
in the coupled channel.
The effect of the CM momentum is limited to small quantitative changes 
in the correlated wave functions. Overall, the dependence on $P$ is weak, 
with only minor deviations from the $P=0$ case of back-to-back nucleons.

For the second pair of solutions in the $J=1$ channel, namely 
${}^{3}D_{1}$ and ${}^{3}S_{1}/{}^{3}D_{1}$, the ${}^{3}D_{1}$ wave remains 
almost indistinguishable from the uncorrelated solution. This behavior 
is mainly due to the centrifugal barrier, which prevents the nucleons 
from approaching short distances at higher orbital angular momentum, 
together with the fact that in the ${}^{3}D_{1}$ channel the first two 
delta-shell strengths vanish. On the other hand, the mixing with the 
${}^{3}S_{1}$ component, shown in the bottom-right panel, is visible but 
comparatively small, especially when contrasted with the much stronger 
${}^{3}D_{1}/{}^{3}S_{1}$ interference discussed earlier.
Of course, in the case of the small ${}^{3}S_{1}/{}^{3}D_{1}$ interference, 
the relative effect of the CM momentum appears amplified. However, this 
is simply because one is examining a very small effect on a 
quantity that is itself already small.

In the coupled-channel case $J=2$, shown in Fig.~\ref{ondas7}, the 
effect of SRCs is extremely small. A slight bending of the radial 
functions can be seen between $1$ and $2$~fm, but the departure from 
the free Bessel solution is negligible, and in the ${}^{3}F_{2}$ wave 
no visible effect appears within the scale of the figure. Some 
tensor-induced mixing is observed in the ${}^{3}F_{2}/{}^{3}P_{2}$ and, 
to a much lesser extent, in the ${}^{3}P_{2}/{}^{3}F_{2}$ components, 
but overall the SRC effects in this channel are practically 
insignificant.

\begin{figure}
  \centering
\includegraphics[width=12cm,bb=70 470 500 770]{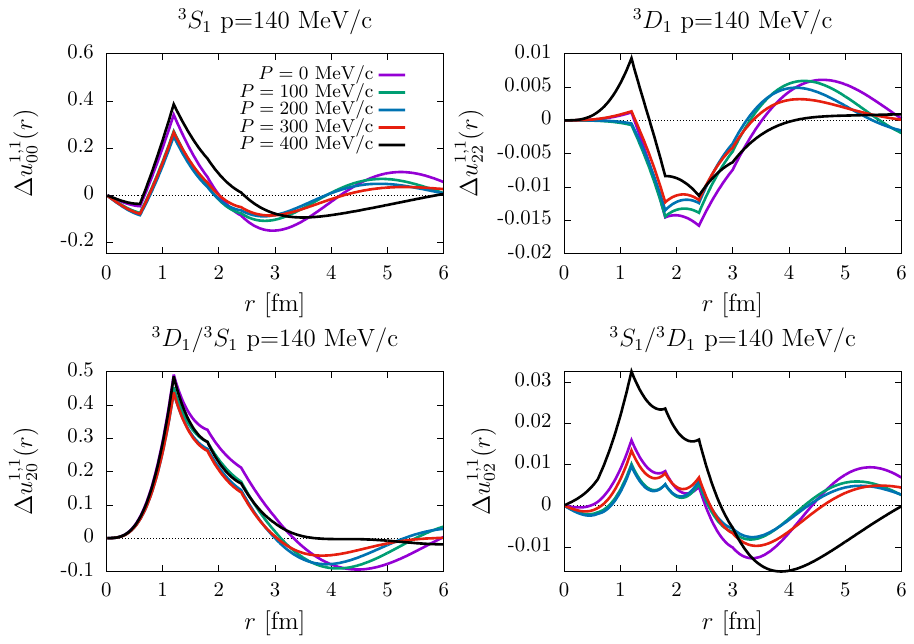}
\caption{
Defect wave functions for the coupled channel with $J=1$. 
The top panels display the diagonal components, 
$\Delta u^{SJ}_{ll}(r) \equiv u^{SJ}_{ll}(r) - \hat{j}_l(pr)$, 
while the bottom panels show the off-diagonal components, 
$\Delta u^{SJ}_{l\,l'}(r) \equiv u^{SJ}_{l\,l'}(r)$ for $l \neq l'$. 
Results are given for relative momentum $p=140$ MeV/c and for several 
values of the CM momentum.
}
\label{ondas8}
\end{figure}

\begin{figure}
  \centering
\includegraphics[width=12cm,bb=40 480 500 770]{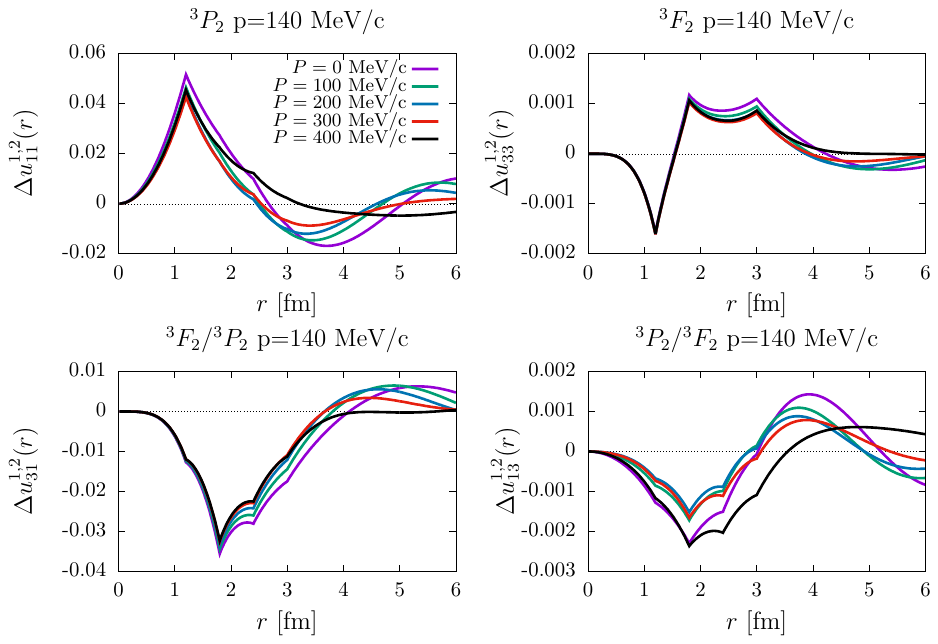}
\caption{The same as fig. \ref{ondas8} for $J=2$ and $l,l'=1,3$}
\label{ondas9}
\end{figure}

For completeness, we show in Figs.~\ref{ondas8} and \ref{ondas9} the 
defect wave functions corresponding to the coupled channels $J=1$ and 
$J=2$, respectively. In this case we define 
$\Delta u_{ll}(r) \equiv u_{ll}(r) - \hat{j}_{l}(pr)$ for the diagonal 
components, and $\Delta u_{ll'}(r) \equiv u_{ll'}(r)$ for the 
off-diagonal ones ($l \neq l'$). The results indicate that 
$\Delta u_{ll}$ is comparable in magnitude to its coupled partner, 
as can be appreciated by comparing the top and bottom panels of the 
figures.

\begin{figure}
  \centering
\includegraphics[width=12cm,bb=40 480 500 770]{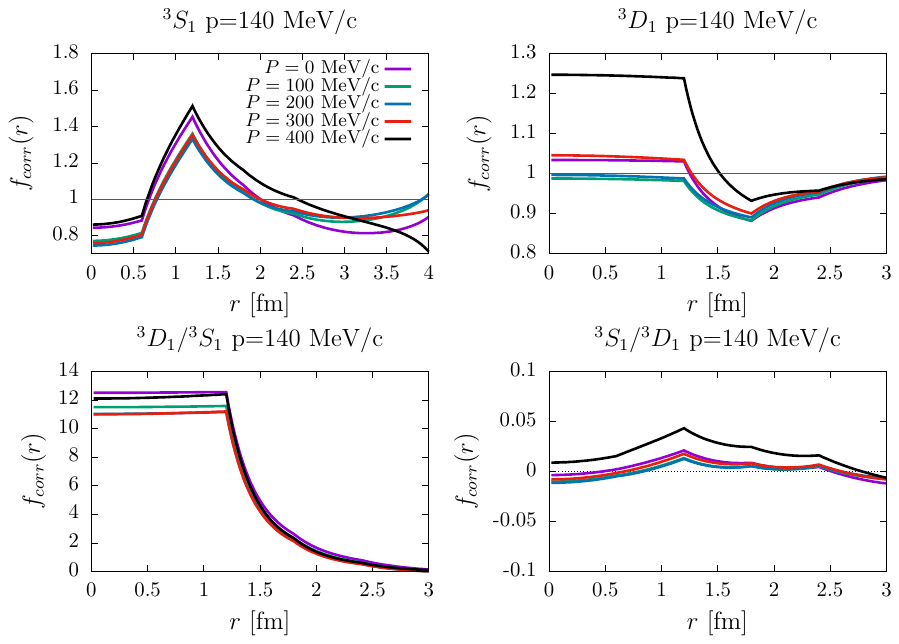}
\caption{
  Correlation functions $f_{\rm corr}(r)$ for the diagonal
   wave functions (top panels) and off-diagonal wave
  functions (bottom panels) 
  for the coupled  partial
  waves for $J=1$ and $l,l'=0,2$.
}
\label{ondas10}
\end{figure}

\begin{figure}
  \centering
\includegraphics[width=12cm,bb=40 450 500 770]{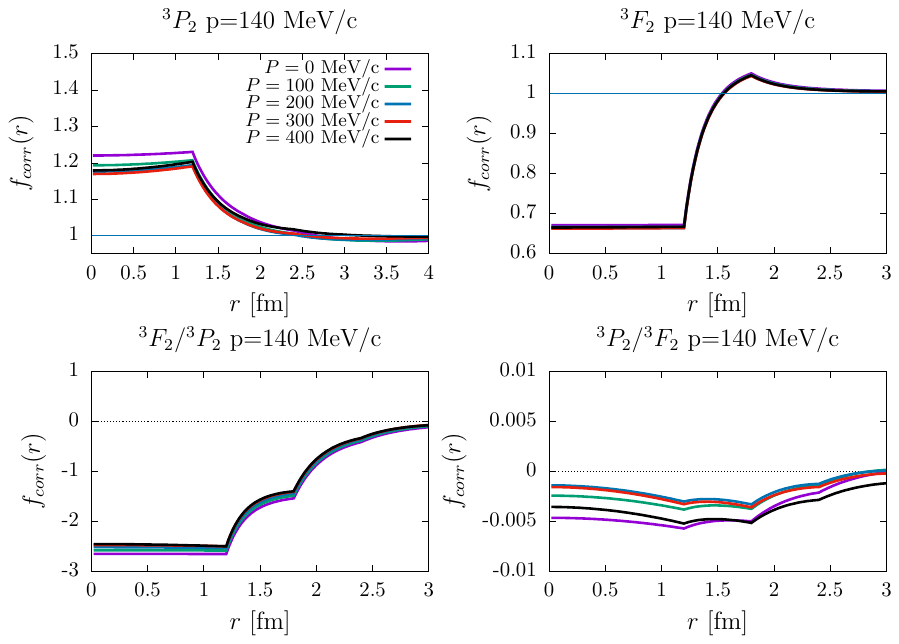}
\caption{The same as fig. \ref{ondas10} for $J=2$ and $l,l'=1,3$}
 \label{ondas11}
\end{figure}

In Figs.~\ref{ondas10}--\ref{ondas11} we present the correlation functions 
for the coupled partial waves. This definition only applies to the diagonal 
components, for which
\[
f_{\rm corr}(r) = \frac{u_{ll}(r)}{\hat j_l(pr)} \, ,
\]
so that $f_{\rm corr}(r)\to 1$ at large distances. As in the uncoupled
case, one should keep in mind the mismatch of the nodes between
numerator and denominator, although this does not affect the
interpretation at short distances ($r\lesssim 3\,$fm), where $f_{\rm
  corr}(r)$ provides a clear visualization of the wound and its
healing. In Fig.~\ref{ondas10}, the correlation function of the
${}^{3}S_{1}$ wave shows a behavior very similar to the ${}^{1}S_{0}$
case: a strong oscillation from $\sim 0.8$ to $\sim 1.6$ before
relaxing back to unity around $r\sim 3\,$fm. The effect of the CM
motion in this channel remains small. By contrast, the ${}^{3}D_{1}$
wave displays practically no wound, except for very large CM momenta,
but in any case this component is subdominant.  In the case of the
${}^{3}S_{1}$ channel, the correlation function shows a clear wound at
short distances: for $r \lesssim 0.6$ fm one finds $f_{\rm
  corr}(r)\approx 0.8$, indicating an effective repulsion, since the
correlated wave is suppressed with respect to the free one.  In the
intermediate region $0.6 \lesssim r \lesssim 1.8$ fm, the correlation
function rises above unity, reaching values around $f_{\rm
  corr}(r)\simeq 1.5$. This reflects the strong enhancement of the
wave amplitude due to short-range correlations, which in turn signal
the attractive part of the interaction in this channel.  Beyond
$r\gtrsim 3$ fm, $f_{\rm corr}(r)$ smoothly approaches unity, as
expected from the healing property.

There is no strict analogue of the correlation function for the 
off-diagonal coupled waves ($l\neq l'$), since these components 
vanish at large distances. Nevertheless, one can formally introduce the 
quotient
\[
f_{\rm corr}(r) = \frac{u_{ll'}(r)}{\hat j_l(pr)} \, ,
\]
which we denote as a ``non-diagonal correlation function.'' The results 
are displayed in the bottom panels of Fig.~\ref{ondas10} for $J=1$. 
For the $l l'=20$ component, corresponding to the 
${}^{3}D_{1}/{}^{3}S_{1}$ interference, $f_{\rm corr}(r)$ remains nearly 
constant for $r \lesssim 1.5$ fm before decreasing to zero at larger 
distances (with a small residual oscillation). This behavior shows that 
the interference acts effectively as a $D$ wave at short distances, but 
with a much larger amplitude than a free $D$ wave, reaching values of 
order $\sim 12$. In contrast, the $l l'=02$ component, associated with 
the ${}^{3}S_{1}/{}^{3}D_{1}$ interference, yields an almost vanishing 
$f_{\rm corr}(r)$, reflecting the negligible strength of this mixing.

The ${}^{3}P_{2}$ channel is attractive at short distances, whereas the 
${}^{3}F_{2}$ is strongly repulsive. As a consequence, the corresponding 
correlation functions shown in Fig.~\ref{ondas11} exhibit an enhancement 
($f_{\rm corr}>1$) for ${}^{3}P_{2}$ and a pronounced wound for 
${}^{3}F_{2}$. In both cases, the functions display a smooth and rapid 
healing toward unity at $r\simeq 3$ fm. The non-diagonal correlation 
function for the ${}^{3}F_{2}/{}^{3}P_{2}$ interference behaves in a 
similar way to the ${}^{3}D_{1}/{}^{3}S_{1}$ case: it remains nearly 
constant for $r<1.2$ fm, indicating that the mixing acts effectively as a 
pure $F$ wave at short distances. However, this clear behavior for the 
$J=2$ channel is mainly of theoretical interest, serving as a didactic 
example, since its contribution to the total wave function is negligible 
in practice.

\subsection{High-momentum distributions}

\begin{figure}
  \centering
\includegraphics[width=14cm,bb=30 170 470 770]{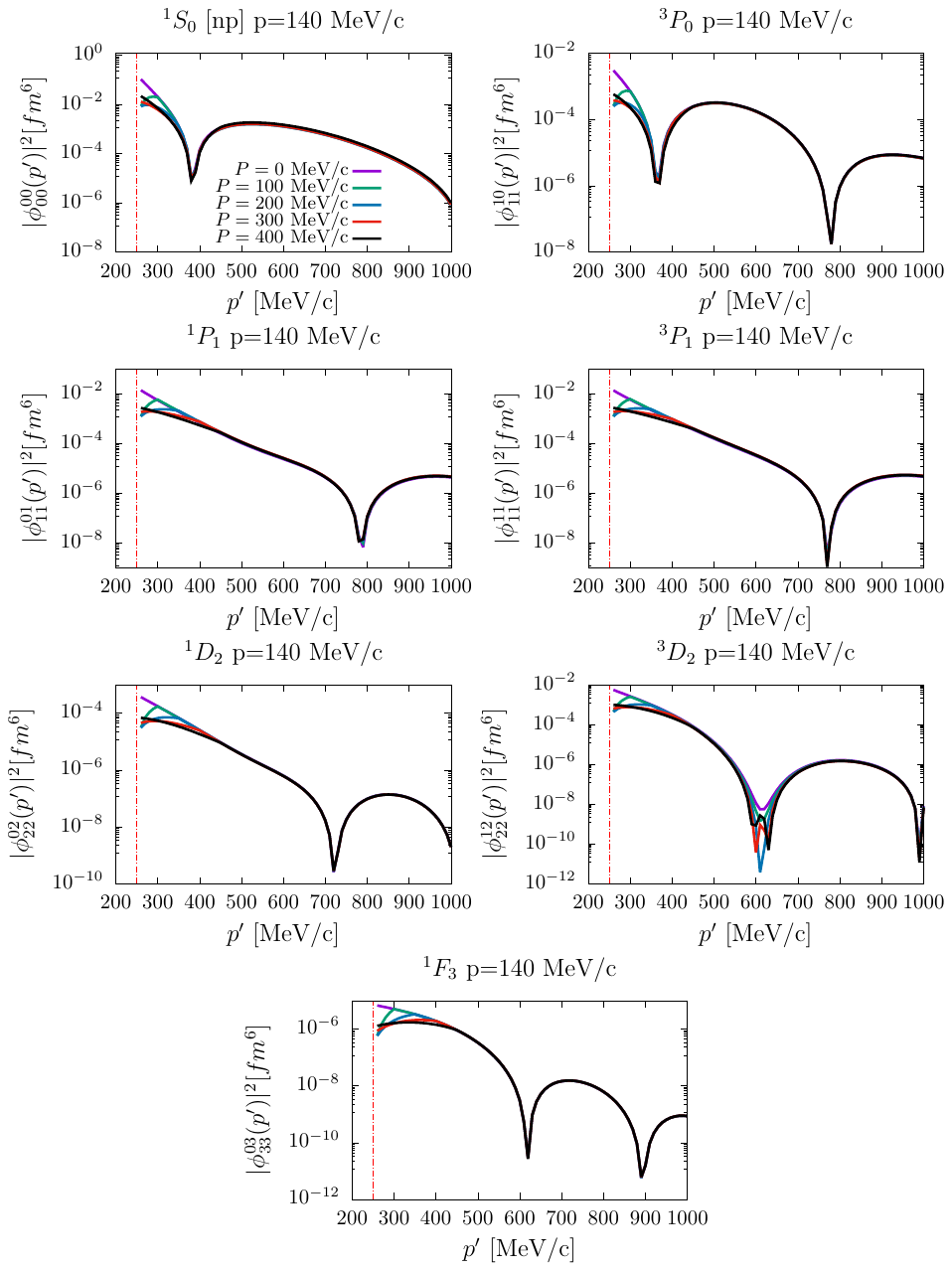}
\caption{
  Squared radial wave functions 
$\left|\tilde{\phi}^{SJ}_{l,l}(p')\right|^2$ for uncoupled partial waves, 
plotted as a function of $p'$ in the high-momentum region above $k_F$. 
Results are shown for a relative momentum $p=140$ MeV/c and several 
values of the CM momentum.
}
\label{ondas12}
\end{figure}

\begin{figure}
  \centering
\includegraphics[width=14cm,bb=40 470 470 770]{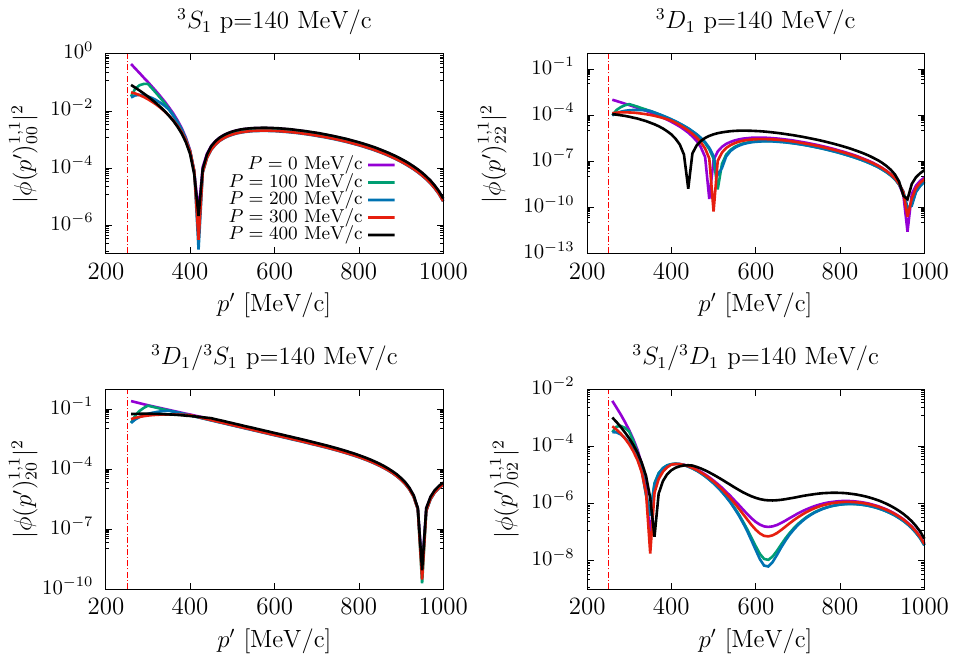}
\caption{
  Squared radial wave functions 
$\left|\tilde{\phi}^{SJ}_{l,l'}(p)\right|^2$ for the coupled 
partial waves with $J=1$, evaluated at relative momentum 
$p=140$ MeV/c for different CM momenta.
}
\label{ondas13}
\end{figure}

\begin{figure}
  \centering
\includegraphics[width=14cm,bb=10 450 490 770]{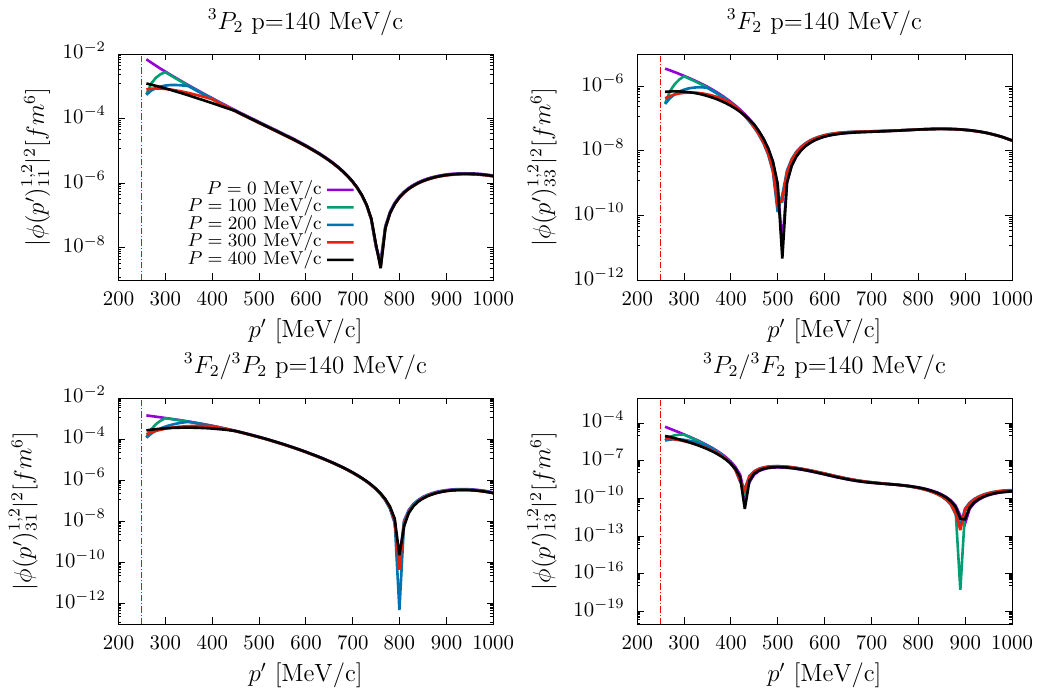}
\caption{The same as fig. \ref{ondas13} for $J=2$.}
 \label{ondas14}
\end{figure}

To end this chapter, we present the results for the high-momentum
distributions in momentum
space. Figures~(\ref{ondas12})--(\ref{ondas14}) display the squared
wave functions, $\left|\phi^{SJ}_{l\,l'}(p')\right|^{2}$, for
different CM momenta of the nucleon pair, computed for $p=140$ MeV/c
as a function of the relative momentum $p'$. We only show results for
$p'>k_{F}$, as indicated by the vertical lines in the plots. For very
large CM momentum, $P=400$ MeV/c, high-momentum components may also
appear at $p'<k_{F}$, since in this kinematic configuration the two
nucleons are nearly parallel. Nevertheless, our main interest lies in
the high-momentum tail characteristic of short-range correlations,
which emerges predominantly at large $p'$.

Now we turn to Fig.~\ref{ondas12}, which shows the results for the
uncoupled waves. Each panel displays the squared high-momentum wave
function, where characteristic oscillations appear. These oscillations
originate from the structure of the defect function
$\Delta\tilde{\phi}^{SJ}_{ll'}$, given in Eq.~\ref{radialp-high}. From
that expression it is clear that the momentum dependence of the wave
function arises from a linear combination of Bessel functions
$\hat{j}_l(p' r_i)$ evaluated at the grid points $r_i$. For large $p'$
this can be written schematically as
\begin{equation}
  \Delta\tilde{\phi}^{SJ}_{ll'}(p')=\sqrt{\frac{2}{\pi}}
  \frac{1}{pp'}\frac{1}{p^2-p'{}^2}
  \sum_i a_{ll'}(r_i)\,\hat{j}_l(p'r_i),
\end{equation}
where the coefficients are defined as  
\begin{equation}
  a_{ll'}(r_i) = \sum_{l_1}(\lambda_i)^{SJ}_{l_1l}\,
  u^{SJ}_{l_1l'}(r_i).
\end{equation}
The oscillatory behavior thus results from the superposition of Bessel
functions with different weights for each partial wave, leading to
distinct positions of the zeros. The coefficients $a_{ll'}(r_i)$
depend on the product of the $\lambda_i$ coefficients and the reduced
wave functions in coordinate space, i.e. on the local product of the
potential and the wave function at the point $r_i$.

From Fig.~\ref{ondas12} it is apparent that the $S$- and $P$-waves
tend to dominate the momentum distribution of the uncoupled channels,
while the $D$- and $F$-waves are much smaller. An important feature is
that, at high momentum, the dependence on the CM momentum is
weak. This suggests that setting $P=0$ provides a good approximation
for evaluating these distributions.

In the case of the coupled waves with $S=1$ and $J=1$, shown in
Fig.~\ref{ondas13}, the most relevant contributions come from the
${}^3S_1$ and the interference ${}^3D_1$/${}^3S_1$, while the other
two channels are much smaller. The ${}^3S_1$ component exhibits a zero
around $p' \simeq 425$ MeV/c, whereas the non-diagonal
${}^3D_1$/${}^3S_1$ wave develops a much longer tail, with its first
zero only appearing beyond $p' \simeq 900$ MeV/c. This indicates that
the interference term dominates the momentum distribution in the
tensor-force region between 400 and 550 MeV/c. The effect of the CM
momentum is negligible for the dominant waves. For the much smaller
${}^3D_1$ and ${}^3S_1$/${}^3D_1$ components, the CM dependence is
stronger but irrelevant due to their negligible strength.

Finally, in Fig.~\ref{ondas14} we show the momentum distributions for
the coupled waves with $S=1$, $J=2$. The same qualitative features as
in the $J=1$ case are observed: the dominant contributions are the
${}^3P_2$ and the interference ${}^3F_2$/${}^3P_2$, while the other
two components are negligible. The dependence on the CM momentum is
again irrelevant. However, these $J=2$ distributions are almost two
orders of magnitude smaller than the ${}^3S_1$ and ${}^3D_1$/${}^3S_1$
channels, making it clear that the high-momentum tail of a nucleon
pair is dominated by the $J=1$ tensor-driven components.

\begin{figure}
  \centering
\includegraphics[width=15cm,bb=30 599 480 770]{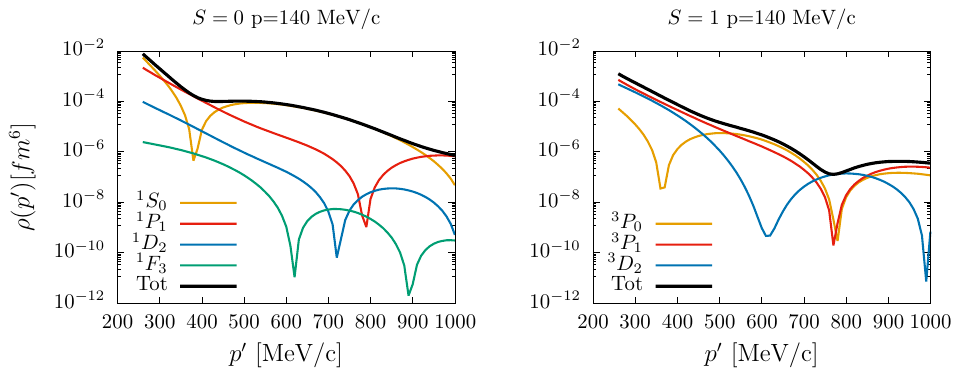}
\caption{ Contributions of the uncoupled partial waves to the total
  high-momentum distribution $\rho(p')$ for a nucleon pair with
  relative momentum $p=140$ MeV/c and for $P=0$. Left panel: $S=0$
  partial waves; right panel: $S=1$ partial waves.  The total
  distribution is obtained from
  Eq.~(\ref{rho-partial}).}
   \label{densi1}
\end{figure}

Figure~\ref{densi1} compares the contributions of the uncoupled
partial waves to the total momentum distribution, obtained from
Eq.~(\ref{rho4}):
\begin{equation}
  \rho^{S}(p') = \sum_{J} \sum_{l,l'} \rho^{SJ}_{ll'}
\end{equation}
where the partial momentum distributions, plotted in Fig \ref{densi1}, of
the different waves are
\begin{equation}
\rho^{SJ}_{ll'}=
  \frac{1}{2S+1} \frac{2J+1}{4\pi} \frac{2}{\pi} 
  \left| \tilde{\phi}^{SJ}_{l,l'}(p') \right|^2 \, ,
\label{rho-partial}
\end{equation}
where $2J+1$ counts the magnetic sub-states and $1/(2S+1)$ averages over spin.

\begin{figure}
  \centering
\includegraphics[width=15cm,bb=40 610 480 770]{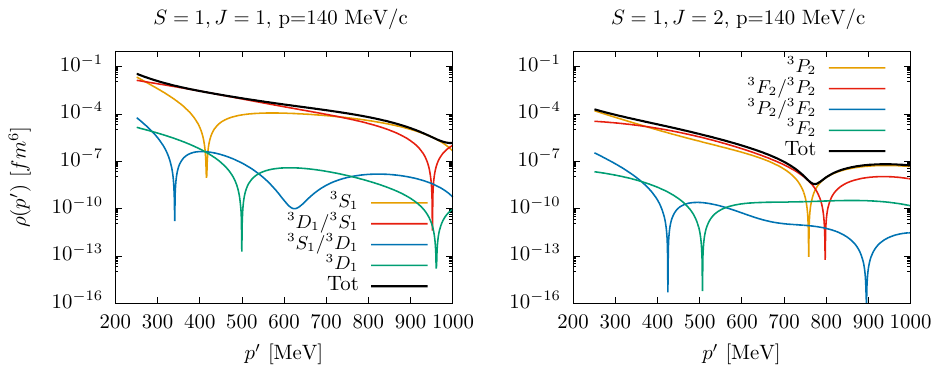}
\caption{ Partial high-momentum distributions $\rho^J_{ll'}(p')$ of
  the coupled waves for proton-neutron pairs with relative momentum
  $p=140$ MeV/c. Left panel: $S=1$, $J=1$ channel; right panel: $S=1$,
  $J=2$ channel. The figure highlights the dominant role of the
  ${}^3S_1$ and ${}^3D_1/{}^3S_1$ interference in the high-momentum
  region associated with the tensor force, while the other components
  are orders of magnitude smaller and negligible.  }
 \label{densi2}
\end{figure}

For $S=0$ (left panel), the ${}^1S_0$ wave dominates, with the
${}^1P_1$ partially filling the depression near its zero, while the
remaining waves are negligible. For $S=1$ (right panel), the
${}^3P_0$, ${}^3P_1$, and ${}^3D_2$ waves have comparable magnitudes
and collectively smooth the dips near the zeros of each component,
complementing each other in the total distribution.

The partial momentum distributions of the coupled waves are shown in
Fig.~\ref{densi2}. The left panel displays the $S=1$, $J=1$ channel,
which provides the dominant contribution to the high-momentum
distribution of proton-neutron pairs. It is evident that the
${}^3D_1/{}^3S_1$ interference fills the dip around the zero of the
${}^3S_1$ wave and dominates the high-momentum region associated with
the tensor force, roughly between 400 and 550 MeV/c \cite{Rio14}. The
individual ${}^3D_1$ and ${}^3S_1/{}^3D_1$ components are more than
two orders of magnitude smaller and can be considered negligible, as
the potential in the ${}^3D_1$ channel (i.e., the values of
$\lambda^{11}_{22}$) vanishes at short distances and becomes
appreciable only around $r \sim 1.8$ fm. The right panel shows the
$J=2$ channel, whose partial momentum distributions are several orders
of magnitude smaller than the $J=1$ case and therefore also
negligible.

\begin{figure}
  \centering
\includegraphics[width=12cm,bb=130 600 390 790]{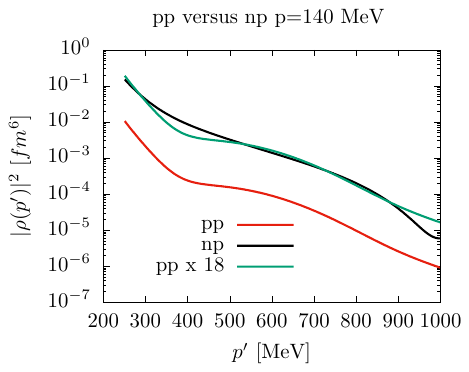}
\caption{Comparison of the $np$ and $pp$ momentum distributions.
 The $pp$ distribution multiplied by 18 is shown to agree
with the $np$ distribution, consistent with the experimental ratio of
high-momentum $np$ to $pp$ pairs reported in Ref.~\cite{sube}.
}\label{bete}
\end{figure}

Finally, in Fig.~\ref{bete} we present the total momentum
distributions for neutron--proton (np) and proton--proton (pp) pairs.
In the pp case, the two--nucleon wave function must be antisymmetric
under the exchange of coordinates and spins. This implies that for
$S=0$ the spatial wave function is symmetric, and therefore the
relative orbital angular momentum $l$ must be even, while for $S=1$
the spatial wave function is antisymmetric and $l$ must be odd. As a
consequence, many partial waves do not contribute to the high--momentum
distribution of pp pairs. In particular, the most relevant np channels,
namely the ${}^3S_1$ and the ${}^3D_1/{}^3S_1$ interference, are absent
in the pp case. In contrast, for np pairs all partial waves contribute.
This explains why the momentum distribution of pp pairs is
significantly suppressed compared to that of np pairs, especially in
the high--momentum region.

In Ref.~\cite{sube}, it was experimentally found that the ratio of
high-momentum $np$ to $pp$ pairs in $^{12}$C is $18 \pm 5$. This
agrees with the results of Fig.~\ref{bete}, where multiplying the $pp$
distribution by 18 brings it close to the $np$ distribution.

\section{Final remarks}
In this chapter we have examined the effect of the short--range
nucleon--nucleon interaction on the wave function of a nucleon pair in
the nuclear medium, with particular attention to the role of
Pauli--blocking. Within the independent pair approximation, we have
shown that the two--body Schrödinger equation reduces in these
conditions to the Bethe--Goldstone equation, where the Pauli principle
translates into the induction of high--momentum components in the
relative wave function.  

We have solved the Bethe--Goldstone equation in a partial--wave
expansion using the realistic Granada 2013 potential. The results show
that the radial wave functions are strongly modified at short
distances ($r < 3$ fm), where the short--range repulsion and tensor
components of the interaction produce a wound in the wave function and
new interference components. At larger distances the correlations die
out and the solutions heal to the uncorrelated wave function,
consistently with the physical picture of short--range correlations.

An essential feature of these correlations is the role of the tensor
force, which mixes the $^3S_1$ and $^3D_1$ partial waves. This mixing,
absent in the uncorrelated case, generates a new component of the wave
function that is fundamental for the appearance of a long tail in the
momentum distribution. In particular, the $^3D_1$–$^3S_1$ interference
dominates the so--called tensor--force region between $p=300 $ and 600
MeV/$c$ and extends well beyond, making it the leading contribution to
the high--momentum distribution of correlated $np$ pairs.  
We stress that this mechanism is absent in $pp$ pairs,
because the coupled $S=1$, $J=1$ channel is forbidden by the
Pauli principle. As a consequence, the dominant source of high--momentum
components in nuclei arises from $np$ pairs, while the $pp$ high-momentum
distribution remains comparatively suppressed.  

Finally, we note that the formalism developed here explicitly includes
the center--of--mass motion of the nucleon pairs under
angular-average approximation. Although we have shown that its effect
on the high--momentum components is small, this treatment allowed us
to compute the full correlated wave function of nucleon pairs in
nuclear matter. This result will be essential in the next chapter,
where we apply the formalism to evaluate the contribution of
short--range correlations to an observable in electron scattering: the
transverse response enhancement produced by meson--exchange currents
in the presence of correlations, within the independent pair
approximation.

 \chapter{Transverse enhancement from MEC and SRC}
 
The study of the electron and neutrino nuclear responses requires a
careful consideration and have indicated the need of including
mechanisms that enhance the transverse response
\cite{Ama94,Jou96,Bod22,Bod23,Bod23b,Bod23c}. Apart from meson exchange
currents, which we have seen play a significant role in the
transverse response, another possible candidate responsible for this
enhancement is the presence of nuclear short-range correlations.

In previous chapters, we have explored these two effects separately:
first, by analyzing the effects of MEC in a relativistic framework and
second, by solving the Bethe-Goldstone equation to obtain the
high-momentum wave function of correlated pairs, providing a realistic
description of SRC in nuclear matter. In this chapter, we focus on
investigating the impact of the high momentum components in the
interference transverse response including MEC. The core of this
theoretical work lies in the BG wave function, solved in the previous
chapter. The initial two-nucleon plane wave states are replaced with the
correlated wave functions for two nucleons. This replacement allowed us
to incorporate SRC explicitly into the calculation in order to compare
with previous uncorrelated approaches.

The motivation behind this work comes from earlier studies where SRC
played a central role. A key reference is the work of Fabrocini
\cite{Fab97}, who explored the effect of  correlations on the
transverse interference response between one-body and two-body
currents. 
Fabrocini found that
including tensor correlations led to an enhancement of the
interference responses, shifting them from negative values (as found
using Jastrow type correlations) to positive ones.

As we have seen in chapter 6, the SRC effect is also manifested
clearly in the high-momentum tails of the pp and np pair
distributions, which increase significantly as a result of these
nuclear correlations. This behavior is in agreement with in studies
such as the Green Function Montecarlo (GFM) \cite{Wir13} and the
Correlated Shell Model \cite{Col15}. According to these studies, the
short distance nucleon-nucleon interaction enhances the probability of
finding np pairs with high relative momentum, compared to pp or nn
pairs. This enhancement is often attributed to the tensor component of
the NN interaction. In chapter 6 we have seen that this effect is
mainly due to the dominance of the mixed $^3D_1/^3S_1$ wave, prevalent
in np pairs at high momentum, while it is suppressed in pp
configurations.

Electron scattering experiments have played a crucial role in probing
short-range correlations, helping to clarify properties such as the
np-pair dominance and the presence of high-momentum tails in the
nuclear wave function.  In the \((e,e'p)\) experiment of Subedi
\textit{et al.}  \cite{sube}, nucleon knockout measurements from
\(^{12}\mathrm{C}\) revealed that about 20\% of the nucleons belong to
short-range correlated pairs, while the remaining 80\% behave as
independent particles described by the shell model. The study
demonstrated that most of the correlated pairs are pn pairs, with a
much smaller fraction of pp and neutron--neutron nn pairs, providing
clear experimental evidence of the dominance of the tensor component
of the nucleon--nucleon interaction.

In this chapter, we show that the high-momentum components of the
nucleon pair wave functions also contribute to observable effects in
the transverse response induced by MEC, providing additional evidence
of SRC and, in particular, of the tensor force.  More details can be
found in Ref. \cite{Cas25c}.

\section{Formalism}
Our methodology is based on the matrix element of the two-body current
operator between the ground state of the Fermi gas and a particle–hole
state (1p1h). We start recalling the definition of the hadronic tensor
from chapter 3. 
In the non-relativistic Fermi gas model it reads
\begin{equation}
W^{\mu\nu}= \sum_{ph}
\left\langle
ph^{-1} \right|\hat{J}^{\mu} |\left. F \right\rangle^{*}
\left\langle
ph^{-1} \right|\hat{J}^{\nu} |\left. F \right\rangle 
 \delta(E_{p}-E_{h}-\omega)
\theta(p-k_F)\theta(k_F-h).
\label{eq1}
\end{equation}
Here, the index $p$ stands for the set of quantum numbers
$(p,s_p,t_p)$, namely momentum, spin, and isospin, and similarly for
$h$. The final and initial states are plane-waves states for particles
and holes, dependent also of spin and isospin. In our model, the total
current operator $\hat{J^\mu}$ is the sum of
one-body and two-body currents,
\begin{equation}
\hat{J}^\mu = 
\hat{J}^\mu_{1b} 
+\hat{J}^\mu_{2b},
\end{equation}
Substituting the total operator in Eq. (\ref{eq1}), we obtain
the many-body matrix elements of these operators:
\begin{eqnarray}
\left\langle ph^{-1} \right|\hat{J}^{\mu} |\left. F \right\rangle
&=&
\left\langle ph^{-1} \right|\hat{J}_{1b}^{\mu} |\left. F \right\rangle
+
\left\langle ph^{-1} \right|\hat{J}_{2b}^{\mu} |\left. F \right\rangle
\nonumber \\
&=&
\left\langle [p] \right|\hat{J}_{1b}^{\mu} |\left. [h] \right\rangle
+
\sum_{k<k_F}\left[
\left\langle [pk] \right|\hat{J}_{2b}^{\mu} |\left. [hk] \right\rangle 
- \left\langle [pk] \right|\hat{J}_{2b}^{\mu} |\left. [kh] \right\rangle
\right].
\label{eq2}
\end{eqnarray}
The first term corresponds to the one-body current matrix element,
where the operator acts on a single nucleon. The second term is the
matrix element of the two body current, where a sum over all occupied
states $|k\rangle=|\nk,s_k,t_k\rangle$ in the Fermi sea is
included.

To avoid confusion with previous chapters, inside the matrix
elements of Eq.(\ref{eq2}), we have introduced the notation
$|[p]\rangle$ (with brackets) to denote states normalized to unity in
a finite volume $V$:
\begin{equation}
  |[\np] s t \rangle = \frac{e^{i\np \cdot \nr}}{\sqrt{V}} 
\textstyle |\frac{1}{2}s \rangle \otimes |\frac{1}{2} t \rangle
\end{equation}
In contrast, states without brackets, $|p\rangle$, refer to continuum
states with spatial wave function normalized to a momentum delta
function in the whole space,
\begin{equation}
  |\np s t \rangle = \frac{e^{i\np \cdot \nr}}{(2\pi)^{3/2}}
\textstyle  |\frac{1}{2}s\rangle \otimes | \frac{1}{2} t \rangle
\end{equation}
This notation is included because in  chapters 2--5 the states
are normalized with $V$, while in chapter 6 they are normalized with
$(2\pi)^3$, and in this chapter we use both normalizations. 

The focus is placed exclusively on the transverse response $R_T$,
which is more sensitive to MEC. Therefore only the $\mu=1,2$
components of the electromagnetic current are considered. Due to MEC
contribution, an interference appears between one-body and two-body
currents. Then the diagonal components of the uncorrelated hadronic
tensor (\ref{eq1}) are
\begin{eqnarray}
  W^{\mu\mu}&=& \sum_{ph}\Big\{ |\langle [p]| \hat{J}_{1b}^\mu |[h] \rangle|^2 
  +2 \mbox{Re}\, \langle [p]| \hat{J}_{1b}^\mu |[h] \rangle^*
  \sum_{k<k_F} \langle [pk] |\hat{J}_{2b}^\mu | [hk-kh] \rangle \Big\} \nonumber \\
  &\times& \delta(E_{p}-E_{h}-\omega)\theta(p-k_F)\theta(k_F-h),
  \label{nfg}
\end{eqnarray}
where the pure MEC contribution is disregarded because its effect is
small, as we have seen in chapter 3. 

 To go beyond the uncorrelated
Fermi gas model, we include the effects of SRC 
between two nucleons in the initial state. These correlations
modify the two-nucleon wave function, introducing high-momentum
components,
\begin{eqnarray}
\Psi_{hk}&=& | hk \rangle + | \Delta \Psi_{hk} \rangle \label{bg1}  \\
\Psi_{kh}&=& | kh \rangle + | \Delta \Psi_{kh} \rangle \label{bg2},
\end{eqnarray}
$|hk\rangle$ and $|kh\rangle$ are the free wave functions and
$\Delta \Psi_{hk}$ and $\Delta \Psi_{kh}$ are the defect functions.
By replacing $|hk\rangle$ and $|kh\rangle$ with the correlated states
$\Psi_{hk}$ and $\Psi_{kh}$ obtained from the BG equation in chapter
6, 
\begin{eqnarray}
W^{\mu\mu}&=& \sum_{ph}\Big\{ |\langle [p]| \hat{J}_{1b}^\mu |[h] \rangle|^2 
  +2 \mbox{Re}\, \langle [p]| \hat{J}_{1b}^\mu |[h] \rangle^*
  \sum_{k<k_F} \langle [pk] | \hat{J}_{2b}^\mu | [\Psi_{hk}-\Psi_{kh}]\rangle \Big\} \nonumber \\
  &\times&  \delta(E_{p}-E_{h}-\omega)\theta(p-k_F)\theta(k_F-h)
  \label{nfg}
\end{eqnarray}
we obtain,
\begin{eqnarray}
W^{\mu\mu}&=& \sum_{ph} \Big\{ |\langle [p]| \hat{J}_{1b}^\mu |[h] \rangle|^2 
+2 \mbox{Re}\, \langle [p]| \hat{J}_{1b}^\mu |[h] \rangle^* 
\sum_{k<k_F} \langle [pk] | \hat{J}_{2b}^\mu | [hk-kh]\rangle \nonumber \\
&+& 2 \mbox{Re}\, \langle [p]| \hat{J}_{1b}^\mu |[h] \rangle^*
\sum_{k<k_F} \langle [pk] |\hat{J}_{2b}^\mu | [\Delta \Psi_{hk}-\Delta \Psi_{kh} ]\rangle \Big\}
\nonumber \\
&\times&\delta(E_{p}-E_{h}-\omega)\theta(p-k_F)\theta(k_F-h).
\label{eq3}
\end{eqnarray}
A new interference term appears, between the one-body current and
two-body current acting on high-momentum nucleon pairs. The sum in $k$
includes the spin and isospin $s_k$ and $t_k$, respectively. In the
thermodynamic limit, it is $1/V\sum_{s_kt_k}\int d^3k/(2\pi)^3$.

\subsection*{Sum of isospin}

We have seen in chapter 6 that the high-momentum components are
different for np and pp pairs because of antisymmetrization. Therefore,
we have to pay attention to the isospin summation that gives the
contribution of np and pp pairs.
It is convenient to separate explicitly
the two-body isospin structure from the current operator. As
in Eq. (\ref{descom}) in chapter 4, the full two body current can be decomposed
into isospin channels,
\begin{equation}
 \hat{J}_{2b}^\mu =\tau_3^{(1)} J_1^\mu + \tau_3^{(2)} J_2^\mu +
  i[\ntau^{(1)}\times\ntau^{(2)}]_3 \;J_3^\mu =\sum_{a=1}^{3}U_aJ_a^\mu,
\end{equation}
where $U_1=\tau_3^{(1)}$, $U_2=\tau_3^{(2)}$ and
$U_3=i[\ntau^{(1)}\times\ntau^{(2)}]_3$.
Then we can write the isospin summation as,
\begin{equation}
  \sum_{t_k} \langle [pk] |\hat{J}_{2b}^\mu | [\Delta \Psi_{hk}-\Delta \Psi_{kh} ]\rangle =
  \sum_{t_k}\sum_{a} \langle pk | U_a J_{a}^\mu | \Delta \Psi_{hk}- \Delta \Psi_{kh} \rangle. 
\end{equation}
We next perform the isospin sum for proton and neutron emission, separating the direct and exchange matrix elements.

\begin{table}[t]
\centering
\begin{tabular}{|c|c|c|}
\hline
$\langle p p | U_1 | p p \rangle=1$ & $\langle p n | U_1 | pn \rangle=1$ & $\langle p n | U_1 | n p \rangle=0$ \\
\hline
$\langle p p | U_2 | p p \rangle=1$ & $\langle p n | U_2 | pn \rangle=-1$ & $\langle p n | U_2 | n p \rangle=0$ \\
\hline
$\langle p p | U_3 | p p \rangle=0$ & $\langle p n | U_3 | pn \rangle=0$ & $\langle p n | U_3 | n p \rangle=-2$ \\
\hline
\end{tabular}
\caption{Isospin matrix elements between two-nucleon states for proton emission.} \label{cuadr}
\end{table}

\subsubsection*{Proton emission}
\begin{itemize}
\item $t_p=t_h=1/2$ and $t_k=\pm 1/2$.

Direct matrix element:
\begin{eqnarray}
  \sum_{a}\sum_{t_k}\langle t_pt_k | U_a| t_ht_k \rangle
  \langle pk| J_{a}^\mu | \Delta \Psi_{hk}\rangle  
&=&
  \sum_{a} 
\textstyle
\langle\frac{1}{2}\frac{1}{2}| U_a | \frac{1}{2}\frac{1}{2}\rangle
  \langle pk |J_{a}^\mu| \Delta \Psi_{hk}\rangle \nonumber \\
  &&
 \kern -2cm  +\sum_{a} 
\textstyle
\langle\frac{1}{2}{-}\frac{1}{2}| U_a | \frac{1}{2}{-}\frac{1}{2}\rangle
  \langle pk |J_{a}^\mu| \Delta \Psi_{hk}\rangle. 
\end{eqnarray}
Exchange matrix element:
\begin{eqnarray}
\sum_a  \sum_{t_k}\langle t_pt_k | U_a| t_kt_h \rangle \langle pk| J_{a}^\mu | \Delta \Psi_{kh}\rangle  &=&
  \sum_{a} 
\textstyle
\langle\frac{1}{2}\frac{1}{2}| U_a | \frac{1}{2}\frac{1}{2}\rangle
  \langle pk |J_{a}^\mu| \Delta \Psi_{kh}\rangle \nonumber \\
  &&
 \kern -2cm +\sum_{a} 
\textstyle
\langle\frac{1}{2}{-}\frac{1}{2}| U_a | {-}\frac{1}{2}\frac{1}{2}\rangle
  \langle pk |J_{a}^\mu| \Delta \Psi_{kh}\rangle.
\end{eqnarray}
The combination of these two contributions gives the full result:
\begin{eqnarray}
  \sum_{t_k}\sum_{a} \langle pk | U_a J_{a}^\mu | \Delta \Psi_{hk}- \Delta \Psi_{kh} \rangle 
&=&
  \sum_{a}
\textstyle
\langle\frac{1}{2}\frac{1}{2}| U_a | \frac{1}{2}\frac{1}{2}\rangle
  \langle pk |J_{a}^\mu| \Delta \Psi_{hk}- \Delta \Psi_{kh} \rangle \nonumber \\
  &+&
  \sum_{a}
  \textstyle
\langle\frac{1}{2}{-}\frac{1}{2}| U_a | {-}\frac{1}{2}\frac{1}{2}\rangle
  \langle pk |J_{a}^\mu| \Delta \Psi_{hk}\rangle \nonumber \\
  &+&
  \sum_{a}
  \textstyle
\langle\frac{1}{2}{-}\frac{1}{2}| U_a |{-}\frac{1}{2}\frac{1}{2}\rangle
  \langle pk |J_{a}^\mu| \Delta \Psi_{kh}\rangle
\end{eqnarray}
Using the explicit values of the isospin coefficients from Table
\ref{cuadr}, this becomes:
\begin{eqnarray}
  \sum_{t_k}\sum_{a} \langle pk | U_a J_{a}^\mu | \Delta \Psi_{hk}- \Delta \Psi_{kh} \rangle &=&
  \langle pk |J_{1}^\mu+J_{2}^\mu| \Delta \Psi_{hk}- \Delta \Psi_{kh} \rangle \nonumber \\
  &&
  \kern -0.8cm
  + \langle pk |J_{1}^\mu-J_{2}^\mu| \Delta \Psi_{hk}\rangle 
  +2\langle pk |J_{3}^\mu| \Delta \Psi_{kh}\rangle .
  \label{isospinp}
\end{eqnarray}
\end{itemize}
\subsubsection*{Neutron emission}
\begin{itemize}
\item $t_p=t_h=-1/2$  and $t_k=\pm 1/2$ \\
  The derivation for neutron emission follows analogously,
  \begin{eqnarray}
    \sum_{t_k}\sum_{a} \langle pk | U_a J_{a}^\mu | \Delta \Psi_{hk}- \Delta \Psi_{kh} \rangle &=&
 - \langle pk |J_{1}^\mu+J_{2}^\mu| \Delta \Psi_{hk}- \Delta \Psi_{kh} \rangle \nonumber \\
 &&
 \kern -0.8cm
  - \langle pk |J_{1}^\mu-J_{2}^\mu| \Delta \Psi_{hk}\rangle 
  -2\langle pk |J_{3}^\mu| \Delta \Psi_{kh}\rangle. 
  \label{isospinn}
\end{eqnarray}
The isospin matrix elements for neutron emission are shown in Table \ref{cuadr2} 
\end{itemize}
\begin{table}[t]
\centering
\begin{tabular}{|c|c|c|}
\hline
$\langle nn | U_1 |nn \rangle=1$ & $\langle np | U_1 | np \rangle=-1$ & $\langle np | U_1 | pn \rangle=0$ \\
\hline
$\langle nn | U_2 |nn \rangle=-1$ & $\langle np | U_2 | np \rangle=1$ & $\langle np | U_2 | pn \rangle=0$ \\
\hline
$\langle nn | U_3 | nn \rangle=0$ & $\langle np | U_3 | np \rangle=0$ & $\langle np | U_3 | pn \rangle=2$ \\
\hline
\end{tabular}
\caption{Isospin matrix elements between two-nucleon states for neutron emission.} \label{cuadr2}
\end{table}
\subsection{Correlated matrix elements}
We work at the non-relativistic regime as the SRC model is non
relativistic. We start with the one-body matrix element between
particle and hole plane wave states, which is given by
\begin{eqnarray}
  \langle [p] | \hat{J}_{1b}^\mu |[h] \rangle \equiv \frac{(2\pi)^3}{V}\delta^3(\nh+\nq-\np)
  j^\mu_{1b}(\np,\nh).
\end{eqnarray}
We consider the non-relativistic reduction of the transverse response,
i.e., only $j_{1b}^i$ for $i=1,2$ components are involved. Then the
one-body magnetization current is
\begin{equation}
  \nj_{1b}(\np,\nh)=-i\delta_{t_pt_h}\frac{G_M^h}{2m_N}\nq \times \nsigma_{ph},
\end{equation}
where $G_M^h$ is the magnetic form factor of a nucleon with isospin $t_h$.\\
\begin{figure}[t]
  \centering
  \includegraphics[width=14cm,bb=100 460 520 700]{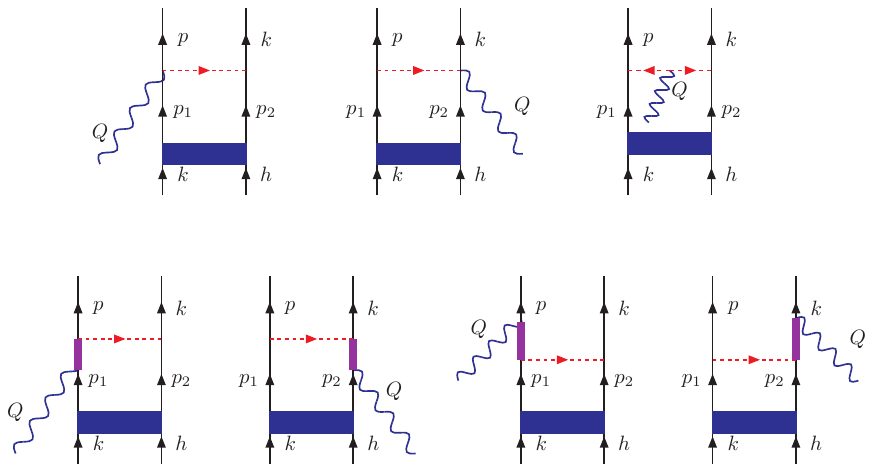}
  \caption{Correlation Feynman diagrams for the Seagull,
    Pion-in-flight and $\Delta$ resonance two body currents for the
    direct term. The blue square represent the correlation process and
    $p_1$ and $p_2$ are the high-momentum components.}
  \label{figsrc}
\end{figure}
The matrix element of the two body operator
$\hat{J}^\mu_{2b}=\hat{\nJ}_{2b}$ between correlated states and 1p1h
final states can be written as,
\begin{eqnarray}
  \langle [ \np s_pt_p \nk s_kt_k ]| \hat{\nJ}_{2b}| [\Delta\Psi_{\nh\nk}^{s_hs_k},t_ht_k ]\rangle &=&
  \int d^3p_1 d^3p_2 \sum_a\sum_{s_1s_2} \langle [\np s_pt_p \nk s_kt_k] |U_a\nJ_{a}
  |\np_1 s_1 t_h \np_2 s_2 t_k \rangle \nonumber \\
  &&
  \times
  \langle \np_1 s_1 t_h \np_2 s_2 t_k|[\Delta\Psi_{\nh\nk}^{s_hs_k},t_ht_k]\rangle, \nonumber \\
  \label{mesrc}
\end{eqnarray}
where $\np_1$ and $\np_2$ are the high momentum components of the
correlated particles with spin $s_1$ and $s_2$ appearing in
Fig. \ref{figsrc}. Moreover, in Eq. (\ref{mesrc}), we have inserted a
complete set of momentum-spin states $|\np_1 s_1,\np_2s_2\rangle$,
making the spin dependence explicit,
\begin{equation}
 \sum_{s_1} 
 \sum_{s_2} \int d^3 p_1  d^3 p_2 
|\np_1 s_1 \np_2 s_2\rangle \langle \np_1 s_1 \np_2 s_2| =1.
\end{equation}
By applying Eq. (\ref{two-body-matrix}), 
 non-relativistic version with bracket notation, the matrix element inside
the integral (\ref{mesrc}) can be expressed as,
\begin{eqnarray}
\sum_a  \langle [\np s_pt_p \nk s_kt_k]| U_a \nJ_a |\np_1 s_1 t_h \np_2 s_2 t_k \rangle
  &=&
  \frac{V}{{(2 \pi)^3}}
 \sum_a \langle [\np s_pt_p \nk s_kt_k]| U_a\nJ_a |[\np_1 s_1 t_h \np_2 s_2 t_k ]\rangle  \nonumber \\
  &&
\kern -6cm = \frac{V}{{(2 \pi)^3}} \frac{(2 \pi)^3}{V^2}
\sum_a  \langle t_pt_k| U_a | t_ht_k \rangle
  \delta(\np_1+\np_2+\nq-\np-\nk)\nj_a(\np,s_p,\nk,s_k,\np_1,s_1,\np_2,s_2). \nonumber \\
\end{eqnarray}
Recalling chapter 4, the currents $\nj_1$ and
$\nj_2$ correspond to the terms of the $\Delta$ current in
Eq. (\ref{deltafinalv}) that are accompanied by the isospin factors
$\tau_3^{(1)}$ and $\tau_3^{(2)}$ respectively,
\begin{eqnarray}
\nj_1(\np,\nk,\np_1,\np_2) &=&
iC_\Delta
\frac{\nk_1\cdot\nsigma^{(1)}}{\nk_1^2+m_{\pi}^2} 4\nk_1\times\nq,  \\
\nj_2(\np,\nk,\np_1,\np_2) &=&
iC_\Delta
\frac{\nk_2\cdot\nsigma^{(2)}}{\nk_2^2+m_{\pi}^2}
4\nk_2\times\nq,
\end{eqnarray} 
with 
\begin{equation}  
C_\Delta\equiv
 \sqrt{\frac{3}{2}} \frac{2}{9} 
\frac{ff^*}{m_\pi^2}
\frac{C_3^V}{m_N}\frac{1}{m_\Delta-m_N}.
\end{equation}
where now $\nk_1=\np-\np_1$ and $\nk_2=\nk-\np_2$.
The current $\nj_3$ is composed of those terms
where the isospin is $i[\ntau^{(1)}\times\ntau^{(2)}]_3$. Therefore,
it receives contributions from the seagull, pion in flight and $\Delta$
currents:
\begin{eqnarray}
  \nj_3(\np,\nk,\np_1\np_2)&=&\nj_{s3}(\np,\nk,\np_1\np_2)+\nj_{\pi3}(\np,\nk,\np_1\np_2)
  +\nj_{d3}(\np,\nk,\np_1\np_2), \\
\nj_{s3}(\np,\nk,\np_1\np_2)&=&
\frac{f^{2}}{m_{\pi}^{2}}F_1^V
\left(
\frac{\nk_1\cdot\nsigma^{(1)}}{\nk_1^2+m_{\pi}^2}
\nsigma^{(2)}
-\frac{\nk_2\cdot\nsigma^{(2)}}{\nk_2^2+m_{\pi}^2}
\nsigma^{(1)}
\right),  \\
\nj_{\pi 3}(\np,\nk,\np_1\np_2) &=&
\frac{f^{2}}{m_{\pi}^{2}}F_1^V
\frac{\nk_1\cdot\nsigma^{(1)}}{\nk_1^2+m_{\pi}^2}
\frac{\nk_2\cdot\nsigma^{(2)}}{\nk_2^2+m_{\pi}^2}
(\nk_1-\nk_2), \\
\nj_{d3}(\np,\nk,\np_1\np_2) &=&
iC_\Delta \left (
\frac{\nk_2\cdot\nsigma^{(2)}}{\nk_2^2+m_{\pi}^2} (\nk_2\times\nsigma^{(1)}) -
\frac{\nk_1\cdot\nsigma^{(1)}}{\nk_1^2+m_{\pi}^2} (\nk_1\times\nsigma^{(2)}) \right)
\times \nq.
\end{eqnarray}
On the other side, as in Eq. (\ref{conservCM}), the correlated wave
function of a nucleon pair carries the same CM momentum,
\begin{eqnarray}
  \langle \np_1 s_1 \np_2 s_2 |[\Delta\Psi_{\nh\nk}^{s_hs_k}] \rangle &=&
  \frac{(2 \pi)^3}{V} 
\langle \np_1 s_1 \np_2 s_2 |\Delta\Psi_{\nh\nk}^{s_hs_k} \rangle \nonumber \\
 &=&   \frac{(2 \pi)^3}{V} 
\delta(\np_1 + \np_2 - \nh - \nk )\Delta\psi_{\nh\nk}^{s_hs_k}(\np')_{s_1s_2}, 
\end{eqnarray}
where $\np'=(\np_1-\np_2)/2$ is the relative momentum of $\np_1$ and
$\np_2$. It is worth pointing out that the relative wave function
$\Delta\psi_{\nh\nk}^{s_hs_k}(\np')$ is a bi-spinor and therefore its
spinorial components are written as subscripts,
$\Delta\psi_{\nh\nk}^{s_hs_k}(\np')_{s_1s_2}$. The superscripts $s_h$ and
$s_k$ refer to the spins of the uncorrelated pair before the
interaction. Substituting this in Eq. (\ref{mesrc}), we finally obtain
the direct matrix element
\begin{eqnarray}
   \langle [\np s_pt_p \nk s_kt_k]|\hat{\nJ}_{2b}|
[\Delta\Psi_{\nh\nk}^{s_hs_k} t_ht_k ]\rangle
    &=&
   \frac{(2 \pi)^3}{V^2}\delta(\nh+\nq-\np)
\sum_a   \langle t_pt_k| U_a | t_ht_k \rangle \nonumber \\
   &&
   \kern -5cm
   \times
   \int d^3p_1 \sum_{s_1s_2}
    \nj_a(\np,s_p,\nk,s_k,\np_1,s_1,\np_2,s_2) 
\Delta\psi_{\nh\nk}^{s_hs_k}(\np_1-\frac{\nh+\nk}{2})_{s_1s_2}. 
\end{eqnarray}
By momentum conservation, the integral over $\np_2$ gives
$\np_2=\nh+\nk-\np_1$. The derivation of the exchange matrix element
is conducted in a similar way. As seen in Eqs. (\ref{isospinp}) and
(\ref{isospinn}), the isospin summation leads to three distinct matrix
elements:
\begin{eqnarray} \label{me1}
  \langle[\np s_pt_p\nk s_kt_k] | \nJ_1+\nJ_2 |
[ \Delta\Psi_{\nh\nk}^{s_hs_k} -\Delta\Psi_{\nk\nh}^{s_ks_h} ]\rangle
  &=& \frac{(2 \pi)^3}{V^2}\delta(\nh+\nq-\np)\nonumber \\
  &&
  \kern-6.8cm
   \times
  \int d^3p_1 \sum_{s_1s_2}
 \nj_+(\np,s_p,\nk,s_k,\np_1,s_1,\np_2,s_2) 
\left[\Delta\psi_{\nh\nk}^{s_hs_k}(\np')-\Delta\psi_{\nk\nh}^{s_ks_h}(\np')\right]_{s_1s_2}
\end{eqnarray}
\begin{eqnarray}\label{me2}
    \langle [\np s_pt_p\nk s_kt_k]| \nJ_1-\nJ_2 |[ \Delta\Psi_{\nh\nk}^{s_hs_k}] \rangle
    &=&
    \frac{(2 \pi)^3}{V^2}\delta(\nh+\nq-\np) \nonumber \\
    &&
    \kern -2.7cm
    \times
    \int d^3p_1 \sum_{s_1s_2}
    \nj_-(\np,s_p,\nk,s_k,\np_1,s_1,\np_2,s_2) 
    \Delta\psi_{\nh\nk}^{s_hs_k}(\np')_{s_1s_2} 
\end{eqnarray}
\begin{eqnarray}\label{me3}
    \langle [\np s_pt_p\nk s_kt_k] | \nJ_3 |
[ \Delta\Psi_{\nk\nh}^{s_ks_h}] \rangle
    &=&
    \frac{(2 \pi)^3}{V^2}\delta(\nh+\nq-\np) \nonumber \\
    &&
    \kern -2cm 
    \times
    \int d^3p_1 \sum_{s_1s_2}
    \nj_3(\np,s_p,\nk,s_k,\np_1,s_1,\np_2,s_2)
    \Delta\psi_{\nk\nh}^{s_ks_h}(\np')_{s_1s_2} 
\end{eqnarray}
where we denoted by $\nj_+$ and $\nj_-$,
\begin{eqnarray}
  \nj_+(\np,s_p,\nk,s_k,\np_1,s_1,\np_2,s_2)
  &=&\nj_1(\np,s_p,\nk,s_k,\np_1,s_1,\np_2,s_2)+\nj_2(\np,s_p,\nk,s_k,\np_1,s_1,\np_2,s_2), \nonumber \\
  \nj_-(\np,s_p,\nk,s_k,\np_1,s_1,\np_2,s_2)
  &=&\nj_1(\np,s_p,\nk,s_k,\np_1,s_1,\np_2,s_2)-\nj_2(\np,s_p,\nk,s_k,\np_1,s_1,\np_2,s_2). \nonumber \\
\end{eqnarray}
The components of the hadronic tensor defined in Eq.~(\ref{eq3})
determine the nuclear transverse response \(R_T\). The first two terms
correspond to the pure one-body response, \(R_T^{1b}\),
and to the uncorrelated interference between one-body and two-body
currents, \(R_T^{1b2b}\). We then focus on the third term, which
represents the correlated interference response between the one-body
and two-body currents.
\begin{equation}\label{Rcor}
  (R_T^{1b2b})_{cor}= \frac{V}{(2\pi)^3}\int d^3h \delta(E_{p}-E_{h}-\omega)
  w_{T}^{cor}(\np,\nh)
  \theta(p-k_F)\theta(k_F-h)
\end{equation}
with $\np=\nq+\nh$ after integrating over $\np$. The integrated
$\omega_{T}^{cor}(\np,\nh)$ is the correlated single-nucleon function
for the transition $p\rightarrow p$,
\begin{eqnarray}
  w_{T}^{cor}(\np,\nh)_{pp}&=& 2Re \; \sum_{s_ps_h}\nj_{1b,p}(\np,\nh)^*
  \sum_{s_k}\sum_{s_1s_2} 
  \int \frac{d^3k}{(2\pi)^3} \int d^3p_1 
  \nonumber \\
  &&
  \times
  \big [\nj_+(\np,s_p,\nk,s_k,\np_1,s_1,\np_2,s_2)
  \left[\Delta\psi_{\nh\nk}^{s_hs_k}(\np')-\Delta\psi_{\nk\nh}^{s_ks_h}(\np')\right]_{s_1s_2} \nonumber \\
  &&
  +
   \nj_-(\np,s_p,\nk,s_k,\np_1,s_1,\np_2,s_2)
  \Delta\psi_{\nh\nk}^{s_hs_k}(\np')_{s_1s_2} \nonumber \\
  &&
  +
  2\nj_3(\np,s_p,\nk,s_k,\np_1,s_1,\np_2,s_2) \Delta\psi_{\nk\nh}^{s_ks_h}(\np')_{s_1s_2} \big]
  \label{Rpp}
\end{eqnarray}
while for $n\rightarrow n$ channel is,
\begin{eqnarray}
  w_{T}^{cor}(\np,\nh)_{nn}&=& - 2Re \; \sum_{s_ps_h}\nj_{1b,n}(\np,\nh)^*
  \sum_{s_k}\sum_{s_1s_2} 
  \int \frac{d^3k}{(2\pi)^3} \int d^3p_1 
  \nonumber \\
  &&
  \times
  \big [\nj_+(\np,s_p,\nk,s_k,\np_1,s_1,\np_2,s_2)
  \left[\Delta\psi_{\nh\nk}^{s_hs_k}(\np')-\Delta\psi_{\nk\nh}^{s_ks_h}(\np')\right]_{s_1s_2} \nonumber \\
  &&
  +
   \nj_-(\np,s_p,\nk,s_k,\np_1,s_1,\np_2,s_2)
  \Delta\psi_{\nh\nk}^{s_hs_k}(\np')_{s_1s_2} \nonumber \\
  &&
  +
  2\nj_3(\np,s_p,\nk,s_k,\np_1,s_1,\np_2,s_2) \Delta\psi_{\nk\nh}^{s_ks_h}(\np')_{s_1s_2} \big]
  \label{Rnn}
\end{eqnarray}
with $\np'=(\np_1-\np_2)/2$ and $\np_2=\nh+\nk-\np_1$.  To compute the
interference $w_T^{cor}$, it is necessary to solve a 3-dimensional
integral over the spectator nucleon momentum $\nk$, restricted to
values below the Fermi surface, and  a second 3-dimensional integral over
the high-momentum component $\np_1$, which encodes the short-range
correlation effects through the defect function in momentum space. In
addition to these, the $T$-response (\ref{Rcor}) also contains the
integration over $d^3h$, which simplifies to a one–dimensional
integral in the hole momentum $h$, 
yielding a total of seven dimensional numerical integration that must
be carried out for each kinematical point $(q,\omega)$.

\subsection{High momentum wave functions}
The correlated wave function in momentum space, that was obtained in
previous chapter, expanded in multipoles is written as
\begin{equation}
  \langle \np' s_1s_2| \psi \rangle \equiv \Delta\psi_{\nh\nk}^{S\mu}(\np')_{s_1s_2} =
 \sqrt{\frac{2}{\pi}}
 \sum_{JM}\sum_{ll'm} i^{l'-l} Y^{*}_{l'm}(\hat{\np}) \langle l'mS \mu |JM \rangle
 \Delta\tilde{\phi}_{hk\; ll'}^{SJ}(p')
 \mathcal{Y}_{lSJM}(\hat{\np}')_{s_1s_2},
 \label{totalwfhk}
\end{equation}
for the direct term. The spin-coupled spherical harmonics
$\mathcal{Y}_{lSJM}(\hat{\np}')_{s_1s_2}$, were defined in chapter 6
and are bi-spinors with two indices. Explicitly, they are given
\begin{equation}
  \mathcal{Y}_{lSJM}(\hat{\np}')_{s_1s_2}= \sum_{m_lm_s}
  \langle lm_lSm_s|JM \rangle  Y_{lm_l}(\hat{\np}') \langle s_1 s_2 | S m_s\rangle.
\end{equation}
The radial wave functions in momentum space, which are functions of
the initial and final (high) relative momenta $p$, $p'$ of the pair, and
also of the CM momentum P, are
\begin{equation}\label{radwf}
\Delta\tilde{\phi}^{SJ}_{hk\;ll'}(p')=\sqrt{\frac{2}{\pi}}\,\frac{1}{p\,p'}\,
\frac{\overline{Q}(P,p')}{p^2-p'^2}\sum^{N_\delta}_{i=1}
\hat{j}_l(p'r_i)\sum_{l_1} (\lambda_i)^{SJ}_{l_1,l}\;u^{SJ}_{hk\;l_1l'}(r_i).
\end{equation}
with $\np=(\nk-\nh)/2$ and CM momentum $\nP=\nk+\nh$. In Equations
(\ref{me1})-(\ref{me3}), we need the correlated wave function
corresponding to the uncorrelated state $|\nh,s_h,\nk,s_k\rangle$ with
uncoupled spins third components. Using Eqs. (6.13)-(6.14),
\begin{equation}
  \Delta\psi_{\nh\nk}^{s_hs_k}(\np')= \sum_{S \mu} \langle \frac{1}{2}s_h\frac{1}{2}s_k|S\mu \rangle \Delta\psi_{\nh\nk}^{S\mu}(\np')
\end{equation}
To obtain the exchange term, we change $(\nh,s_h)\leftrightarrow
(\nk,s_k)$, and then the initial relative momentum $\hat{\np} \rightarrow
-\hat{\np}$. By doing this we'll have a sign of difference in the
spherical harmonic because
\begin{equation}
  Y_{l'm}(-\hat{\np})=(-)^{l'} Y_{l'm}(\hat{\np}).
\end{equation}
The radial wave function is the same for the exchange case
$\Delta\tilde{\phi}_{kh\; ll'}^{SJ}(p')=\Delta\tilde{\phi}_{hk\;
  ll'}^{SJ}(p')$. Then, the exchange wave function is
\begin{equation}
  \Delta\psi_{\nk\nh}^{s_ks_h}(\np')=
  \sum_{S \mu} \langle \frac{1}{2}s_k\frac{1}{2}s_h|S\mu \rangle \Delta\psi_{\nk\nh}^{S\mu}(\np')
\end{equation}
with
\begin{equation}
  \Delta\psi_{\nk\nh}^{S\mu}(\np')_{s_1s_2} =
 \sqrt{\frac{2}{\pi}}
 \sum_{JM}\sum_{ll'm} i^{l'-l} Y^{*}_{l'm}(-\hat{\np}) \langle l'mS \mu |JM \rangle
 \Delta\tilde{\phi}_{kh\; ll'}^{SJ}(p')
 \mathcal{Y}_{lSJM}(\hat{\np}')_{s_1s_2}
 \label{totalwfkh}
\end{equation}
Writing explicitly the singlet $S=0$ and triplet $S=1$ cases, we have
to compute the following sums involving the partial waves:
\begin{itemize}
\item $S$=0 and $\mu=0$
    \begin{equation}
      \Delta\psi_{\nh\nk}^{00}(\np')_{s_1s_2} =\sqrt{\frac{2}{\pi}}\sum_{JM}Y^{*}(\hat{\np})_{JM}
      \Delta\tilde{\phi}_{hk}^{0J}(p')\mathcal{Y}_{J0JM}(\hat{\np}')_{s_1s_2},
     \end{equation}
    \begin{equation}
      \Delta\psi_{\nk\nh}^{00}(\np')_{s_1s_2} =\sqrt{\frac{2}{\pi}}\sum_{JM} (-1)^J Y^{*}(\hat{\np})_{JM}
      \Delta\tilde{\phi}_{kh}^{0J}(p')\mathcal{Y}_{J0JM}(\hat{\np}')_{s_1s_2}.
    \end{equation}
\end{itemize}
In this sum, the partial waves are:  $^{1}S_{0}$, $^{1}P_{1}$, $^{1}D_{2}$, and $^{1}F_{3}$.
\begin{itemize}
  \item $S=1$ and $\mu=-1,0,1$
    \begin{eqnarray}
      \Delta\phi_{\nh\nk}^{1\mu} (\np')_{s_1s_2} &=&\sqrt{\frac{2}{\pi}}\sum_{m}Y^{*}(\hat{\np})_{1m}
      \langle 1 m 1 \mu |00 \rangle
      \Delta\tilde{\phi}_{hk \; 11}^{10}(p')
      \mathcal{Y}_{1100}(\hat{\np}')_{s_1s_2}
      \nonumber \\
      &&
      \kern -0.7cm
      +\sqrt{\frac{2}{\pi}}\sum_{J=1}^2\sum_{Mm}Y^{*}(\hat{\np})_{Jm}
      \langle J m 1 \mu |JM \rangle
      \Delta\tilde{\phi}_{hk\;JJ}^{1J}(p')\mathcal{Y}_{J1JM}(\hat{\np}')_{s_1s_2}
      \nonumber \\
      &&
      \kern -1.5cm 
      + \sqrt{\frac{2}{\pi}}\sum_{J=2}^2\sum_{l,l'=J\pm1}\sum_{Mm}i^{l'-l}Y^{*}(\hat{\np})_{l'm}
      \langle l' m S \mu |JM\rangle \Delta\tilde{\phi}_{hk\;ll'}^{1J}(p')\mathcal{Y}_{l1JM}(\hat{\np}')_{s_1s_2}, \nonumber \\
    \end{eqnarray}
    \begin{eqnarray}
      \Delta\phi_{\nk\nh}^{1\mu} (\np')_{s_1s_2} &=& - \sqrt{\frac{2}{\pi}}\sum_{m}Y^{*}(\hat{\np})_{1m}
      \langle 1 m 1 \mu |00 \rangle
      \Delta\tilde{\phi}_{kh \; 11}^{10}(p')
      \mathcal{Y}_{1100}(\hat{\np}')_{s_1s_2}
      \nonumber \\
      &&
      \kern -0.7cm
      +\sqrt{\frac{2}{\pi}}\sum_{J=1}^2\sum_{Mm} (-1)^J Y^{*}(\hat{\np})_{Jm}
      \langle J m 1 \mu |JM \rangle
      \Delta\tilde{\phi}_{kh\;JJ}^{1J}(p')\mathcal{Y}_{J1JM}(\hat{\np}')_{s_1s_2}
      \nonumber \\
      &&
      \kern -1.5cm 
      + \sqrt{\frac{2}{\pi}}\sum_{J=1}^2\sum_{l,l'J\pm1}\sum_{Mm}i^{l'-l} (-1)^{l'}Y^{*}(\hat{\np})_{l'm}\langle l' m S \mu |JM\rangle
      \Delta\tilde{\phi}_{kh\;ll'}^{1J}(p')\mathcal{Y}_{l1JM}(\hat{\np}')_{s_1s_2}. \nonumber \\
    \end{eqnarray}
\end{itemize}
These formulas account for the computation of the partial waves
$^{3}P_{0}$,$^{3}P_{1}$,$^{3}D_{2}$,$^{3}S_{1}$,
$^{3}D_{1}/^{3}S_{1}$, $^{3}D_{1}$, $^{3}S_{1}/^{3}D_{1}$,
$^{3}P_{2}$, $^{3}F_{2}/^{3}P_{2}$, $^{3}F_{2}$ and
$^{3}P_{2}/^{3}F_{2}$. These multipole expansions are inserted into
Eqs. (\ref{Rpp}) and (\ref{Rnn}) to compute the interference
transverse responses.

\section{Results}

Having established the theoretical framework showing how SRC modify
the 1p1h MEC matrix elements within the independent pair
approximation, through their coupling to the high--momentum components
of nucleon pairs in the medium, we now turn to the computation of
their contribution to the interference between one--body and two--body
currents in the transverse response. Specifically, the SRC correction
to the transverse response is added to the previously calculated
uncorrelated MEC results studied in previous chapters.

We present the results obtained for $^{12}C$ , using a Fermi momentum
of $k_F=250$ MeV/c. The calculation procedure requires solving the
Bethe--Goldstone (BG) equation for each nucleon pair $\nh,\nk$ in
order to obtain the corresponding defect function $\phi_{\nh\nk}$ in a
multipole expansion (see chapter 6). Rigorously, the integral over the
high momentum $\np_1$ extends from $k_F$ to infinity. In practice,
however, the high--momentum radial functions $\phi_{ll'}^{SJ}$
decrease as a power of the relative momentum $1/p^3$, Eq. (\ref{1pp3}),
so it is sufficient to integrate up to $|\np_1|\sim 800$~MeV or less
to obtain convergence.

In the following, the results are obtained using the frozen--$k$
approximation, which provides a numerically efficient procedure while
preserving the accuracy of the calculation. In this approach, within
the integral over $\nk$, in Eq. (\ref{Rpp}), (\ref{Rnn}), one simply
sets $\nk=0$ (frozen). With this assumption the integration reduces to
a factor $4\pi k_F^3/3$. We have found that this approximation is
quite accurate, introducing only a $\sim 6\%$ variation in the
correlated interference transverse response. This represents a major
computational advantage, since the problem is reduced from a
seven--dimensional to a four--dimensional integral, and it is no
longer necessary to solve the BG equation for all $\nh,\nk$ pairs but only
for $h$ and $k=0$, which are far fewer.

\begin{figure}
  \centering
  \includegraphics[width=12cm,bb=50 40 550 400]{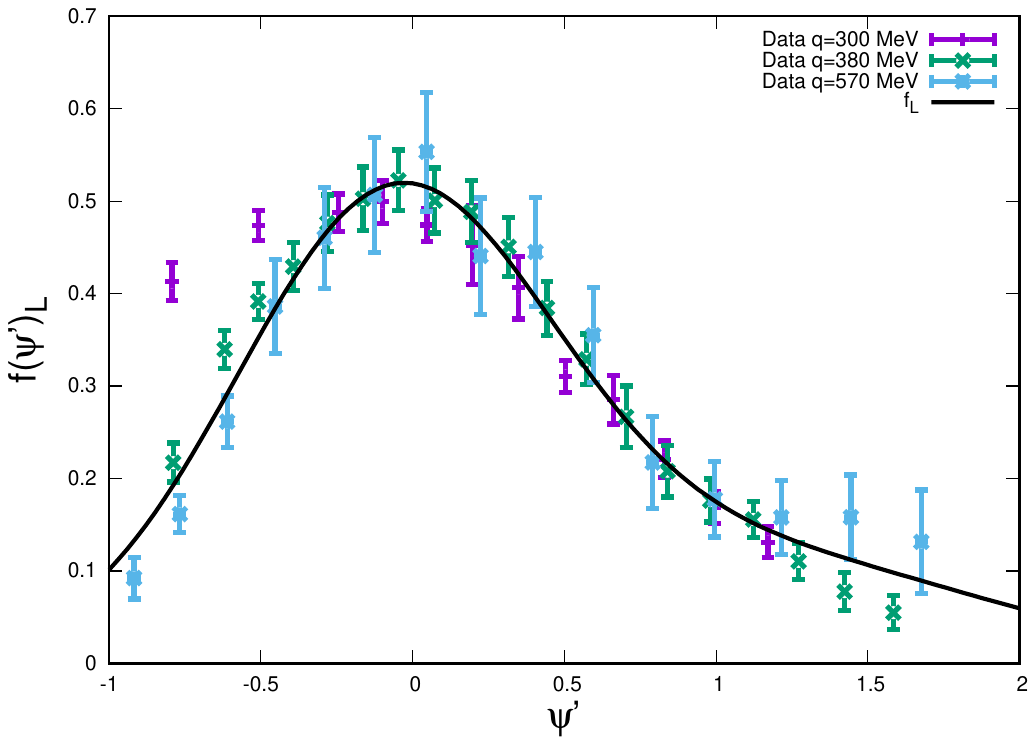}
  \caption{Longitudinal scaling function obtained phenomenologically
    for three values of $q=300,380,570$ MeV/c from \cite{Bod22}}
  \label{res1}
\end{figure}

We begin by performing a scaling analysis of the experimental data of
the longitudinal nuclear response $R_L$ for three values of the
momentum transfer: $q=300,380,570$ MeV/c.  To compare our results with
experimental data and investigate whether an enhancement is observed
in the transverse response $R_T$, we introduce a hybrid model, where
the one-body contribution is not calculated using the Fermi gas
approach, which overestimate the experimental data. Instead, we use
the SuSA model. This SuSA approximation is based on a phenomenological
scaling function extracted from longitudinal experimental data by
dividing them by the single-nucleon factor, as in Eq. (\ref{pheno}), but
for the response function instead of the cross section

\begin{equation} \label{fl}
  f_L(\psi')=\frac{R_L^{exp}}{r_L}
\end{equation}
where $r_L$ is given by Eq. (\ref{rsn}). In this chapter we do not
include the effective mass. Instead, we use $M^*=1$ and we introduce
an energy shift of 20 MeV so $\omega \rightarrow \omega-E_s$ with
$E_s=20$ MeV. The resulting longitudinal scaling function using the
definition described above is shown in Figure \ref{res1}. The black
line is parametrized as a sum of two gaussians. The figure illustrates
how the experimental data tend to collapse onto a single curve,
particularly near the quasielastic peak, corresponding to values of
the scaling variable within the interval $-1\le\psi'\le1$. This scaling
behavior supports the idea that the longitudinal response factorizes
and can be used to describe the QE kinematics.

Our goal is to compute the transverse response, so we assume that the
longitudinal and transverse scaling functions can be approximated as
equal, ie, $f_L=f_T$. This assumption assumes zeroth kind scaling
\cite{Don99a}. From Eq. (\ref{susam}) in chapter 2, we obtain:
\begin{equation}
  R_{T}^{1b} \equiv R_{T}^{SuSA}=r_T(q,\omega) f_T(\psi')
\end{equation}
with
\begin{equation}
  r_T(q,\omega)= \frac{\epsilon_F-1}{m_N \eta_F^3 \kappa}
   Z \overline{w}_T(q,\omega)
\end{equation}
and where the $\overline{w}_T(q,\omega)$ is the averaged single
nucleon tensor defined in Eq. (\ref{def_sn}).

\begin{figure}
  \centering
  \includegraphics[width=8cm,bb=170 250 400 830]{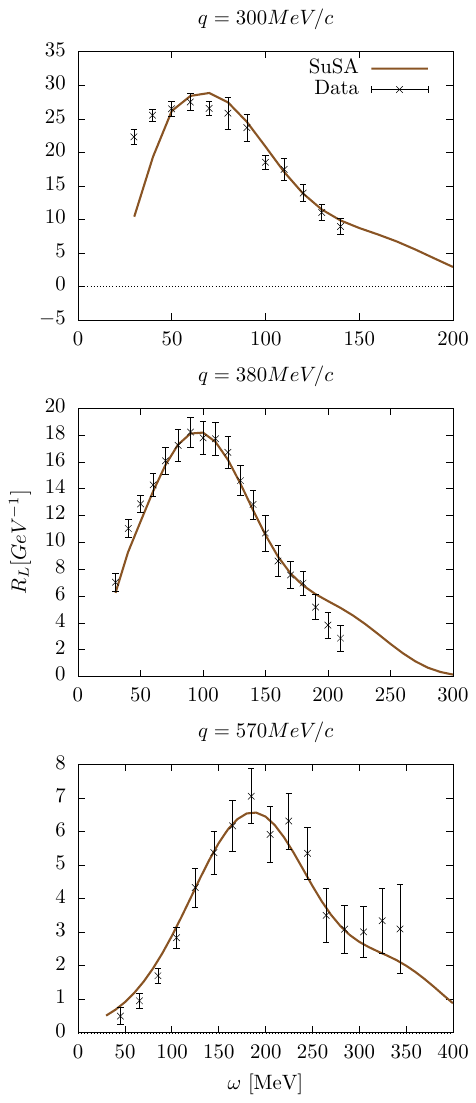}
  \includegraphics[width=8cm,bb=170 250 400 830]{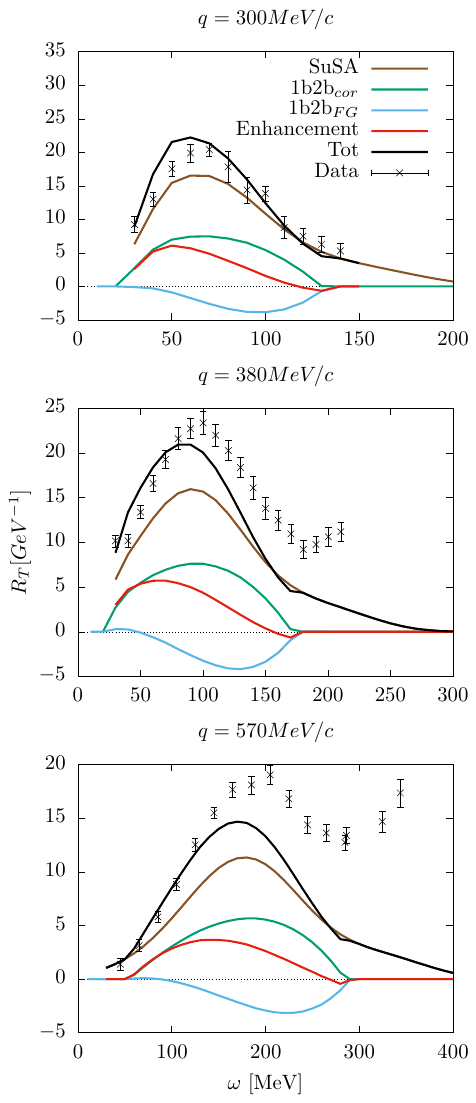}
  \caption{Left: Longitudinal one-body response $R_L$ computed with
    the $SuSA$ model and compared with experimental data for several
    momentum transfers $q=300,380,570$ MeV/c. Right: total transverse
    response function for the same $q$ values. The latter includes the
    $SuSA$ one-body contribution (brown line), the uncorrelated 1b2b
    interference $(R_T^{1b2b})_{FG}$ (blue line), the correlated 1b2b
    interference $(R_T^{1b2b})_{cor}$ (green line), and the net
    enhancement effect (red line). Experimental data are also included
    for comparison in both cases.}
  \label{corj}
\end{figure}

In Figure \ref{corj}, we present the longitudinal and transverse
responses for $^{12}C$ at three values of the momentum transfer:
$q=300,380,570$ MeV/c. The left panel shows the one-body longitudinal
response $R_L$ obtained with the $SuSA$ model, using the
phenomenological longitudinal scaling function of Fig. \ref{res1}. The
comparison with experimental data demonstrates that the $SuSA$ model
provides a good description of the longitudinal channel. The right
panel displays the corresponding transverse response $R_T$, including
all the contributions of our model. The one-body term is computed
within the $SuSA$ approach, assuming $f_L=f_T$, as discussed
earlier. Alongside this baseline, we show the uncorrelated 1b2b (MEC)
interference $(R_T^{1b2b})_{FG}$, which yields a negative
contribution, and the correlated 1b2b interference
$(R_T^{1b2b})_{cor}$ from the contribution from the high momentum
pairs, which is positive. The sum of these two interference terms
produces a net positive contribution that enhances the $SuSA$ one-body
response. As a result, the total transverse response, represented by
the solid black line, is given by
\begin{equation}
  R_T=R_T^{SuSA}+ (R_T^{1b2b})_{FG} + (R_T^{1b2b})_{cor}
\end{equation}

\begin{figure}
  \centering
  \includegraphics[width=8cm,bb=170 250 400 830]{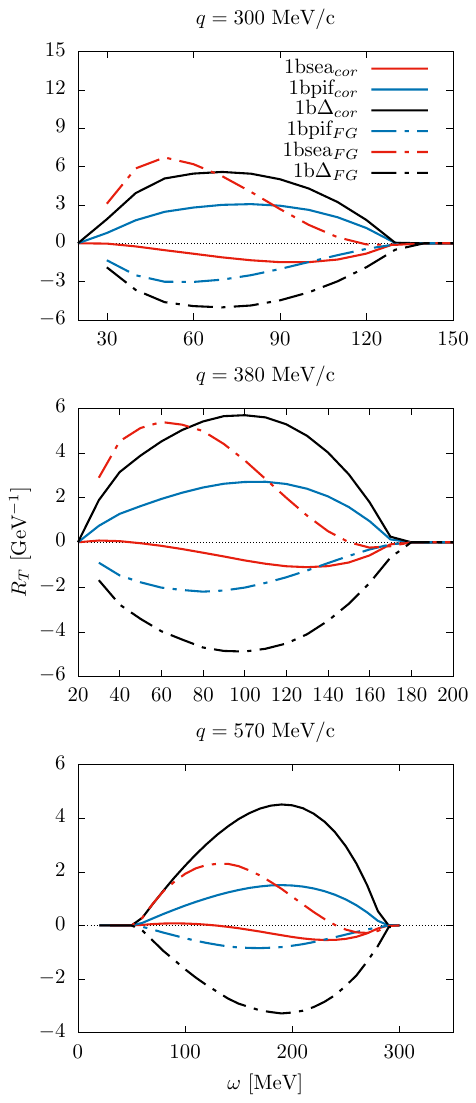}
  \includegraphics[width=8cm,bb=170 250 400 830]{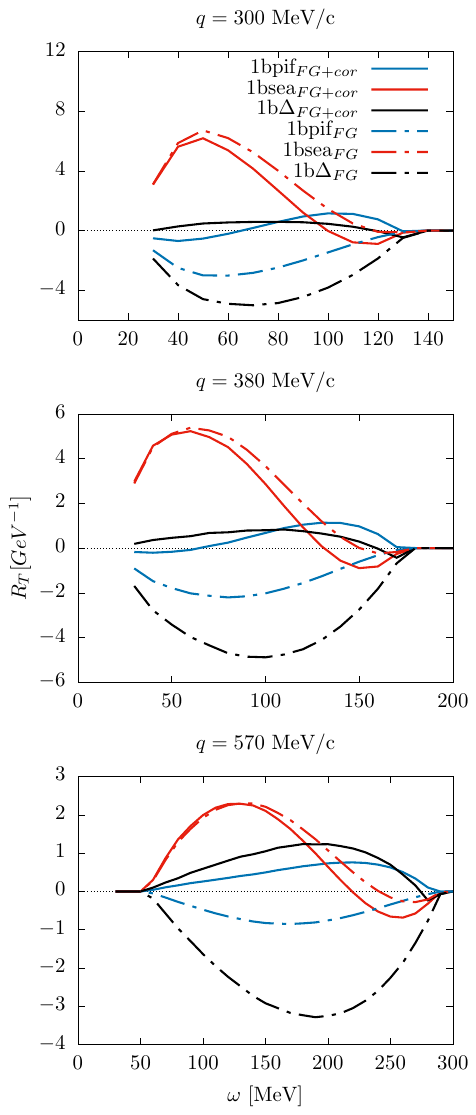}
  \caption{Left: Transverse 1b2b interference response at momentum
    transfers $q = 300, 380, 570$ MeV/c, shown separately for each MEC
    contribution: seagull (red), pion-in-flight (blue), and $\Delta$
    (black). Dashed lines represent the uncorrelated interference
    $(R_T^{1b2b})_{FG}$ while the solid lines correspond to
    $(R_T^{1b2b})_{cor}$ with the effect of short-range correlations.
    Right: Comparison between the uncorrelated interference
    $(R_T^{1b2b})_{FG}$ (dashed lines) and the total 1b2b contribution
    including correlations $(R_T^{1b2b})_{FG}+(R_T^{1b2b})_{cor}$
    (solid lines). Colors indicate the same MEC components as in the
    left panel: seagull (red), pion-in-flight (blue), and $\Delta$
    (black).}
  \label{corj2}
\end{figure}

The left panel of Fig. \ref{corj2} shows the comparison of the 1b2b
uncorrelated and correlation contribution interference transverse
responses $(R_T^{1b2b})_{FG}$ and $(R_T^{1b2b})_{cor}$ for seagull,
pionic and $\Delta$ currents. On the other hand, in the right panel,
Fig.\ref{corj2} shows the total effect of correlations with each MEC
current, given by $(R_T^{1b2b})_{FG}+(R_T^{1b2b})_{cor}$ in comparison
with the non-correlated response $(R_T^{1b2b})_{FG}$.  We can compare
our results from Figure \ref{corj2} with other calculations of the
1b--2b interference including correlations. The only available work of
this kind is that of Fabrocini~\cite{Fab97}, who employed the
correlated basis function (CBF) model in nuclear matter. A deeper
comparison is possible since Fabrocini published the separate
contributions of the seagull, pionic, and $\Delta$ currents.  The
first thing we conclude between both models is the qualitative
similarity of the results. In the independent-pair approximation (the
present calculation), the effect of correlations on the seagull
response is very small (See right panel of Fig. \ref{corj2}), so that
it practically does not change when correlations are included. The
seagull response is large and positive, becoming slightly negative
beyond the midpoint of the quasielastic region. Correlations have a
more noticeable effect on the pionic response, which is negative in
the absence of correlations and becomes positive when correlations are
included. The most significant effect is observed in the interference
with the $\Delta$ current: in the uncorrelated case it is negative and
dominates the three contributions, while correlations render it
positive. Similar results were obtained by Fabrocini using the CBF
model \cite{Fab97}. Although our approach is based on a microscopic
model, it is also worth noting the analysis by Bodek et al
\cite{Bod22}, who performed an empirical extraction of the transverse
enhancement using a global fit to inclusive electron scattering data
on $^{12}C$ and $^{16}O$ . Despite not being a theoretical
calculation, their result also shows to an enhancement of the same
size in the transverse channel, which aligns with the trend observed
in our work.

\begin{table}[t]
\centering
\renewcommand{\arraystretch}{1.4}
\begin{tabular}{llc}
\toprule
\multicolumn{3}{c}{\textbf{Partial-Wave Notation in Coupled Channels}} \\
\midrule
\multicolumn{3}{l}{\textit{Total Angular Momentum} \( J = 1 \)} \\
\cmidrule(r){1-3}
   & Partial Wave  & label \\
\midrule
       & \( ^3S_1 \) & SS \\
       & \( ^3S_1 \)/\( ^3D_1 \) & SD \\
     & \( ^3D_1 \)/\( ^3S_1 \) & DS \\
     & \( ^3D_1 \) & DD \\
\midrule
\multicolumn{3}{l}{\textit{Total Angular Momentum} \( J = 2 \)} \\
\cmidrule(r){1-3}
     & \( ^3P_2 \) & PP \\
     & \( ^3P_2 \)/\( ^3F_2 \) & PF \\
     & \( ^3F_2 \)/\( ^3P_2 \) & FP \\
     & \( ^3F_2 \) & FF \\
\bottomrule
\end{tabular}
\caption{Abbreviations used for each partial wave in the coupled
  channels with total angular momentum $J = 1$ and $J = 2$.}
\label{tabwaves}
\end{table}

\begin{figure}
  \centering
  \includegraphics[width=5.5cm,bb=170 250 400 830]{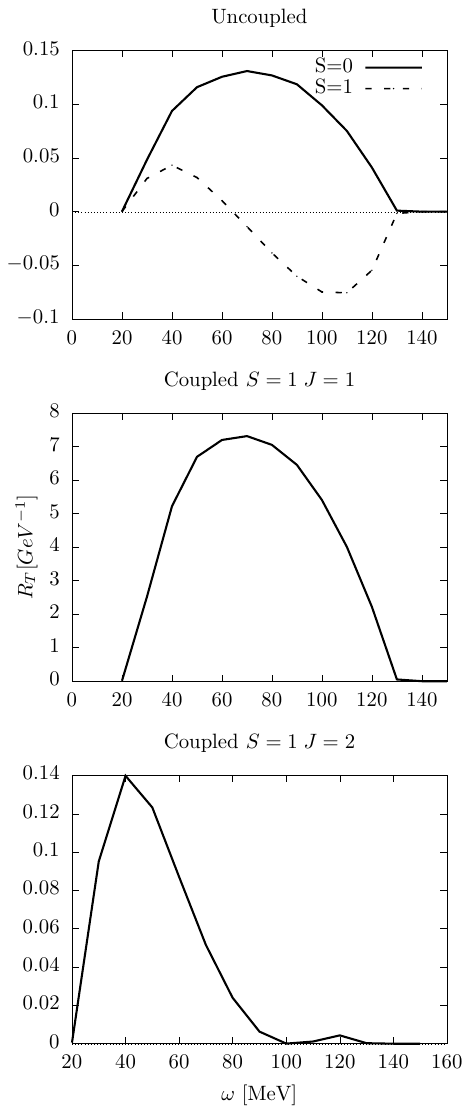}
  \hspace{-0.5cm}
  \includegraphics[width=5.5cm,bb=170 250 400 830]{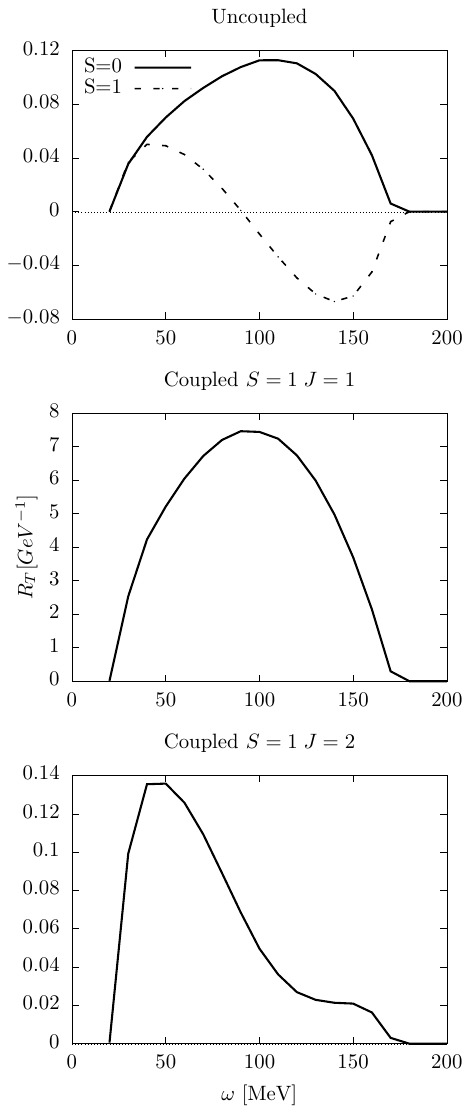}
  \hspace{-0.5cm}
   \includegraphics[width=5.5cm,bb=170 250 400 830]{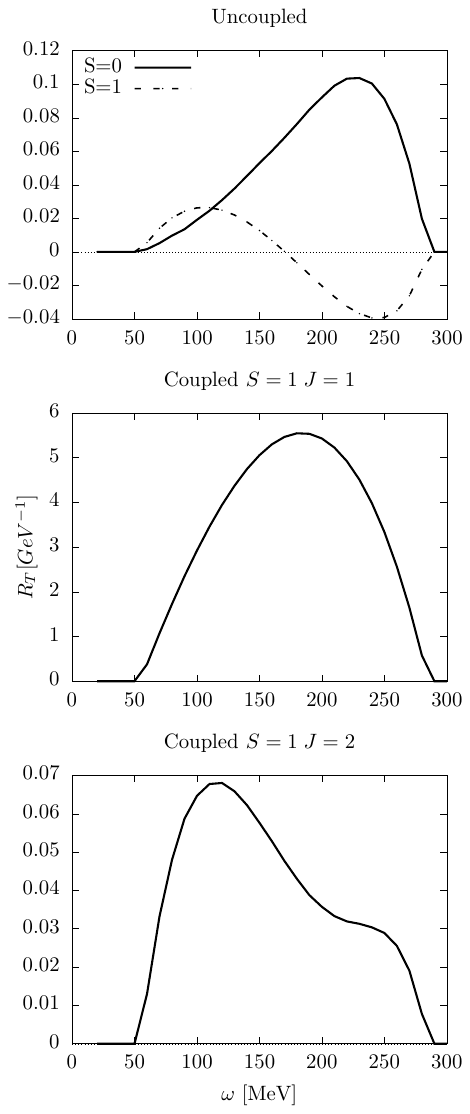}
  \caption{Partial wave decomposition of the correlation induced
    interference response $(R_T^{1b2b})_{cor}$ for $q = 300$ (left),
    $380$ (middle) and  $570$ (right) MeV/c.  The contributions are separated
    according to spin and total angular momentum: uncoupled channels
    with $S=0$ and $S=1$, and coupled channels with $S=1$, $J=1$ and
    $J=2$.}
  \label{cor7}
\end{figure}

A deeper insight into the effect of correlations on the 1b--2b
interference response is presented in Figure \ref{cor7} , where we show the
separate contributions in $^{12}$C, for $q=300,380,570$ MeV/c, from different
multipoles, $\phi_{ll'}^{SJ}$, of the correlated wave
function. Specifically, we display the contributions from the
uncoupled channels with $S=0$ and $S=1$ (top panels), the coupled
channels with $S=1$, $J=1$ (middle panels), and the coupled channels
with $S=1$, $J=2$ (bottom panels). It is evident that the dominant
contribution comes from the $S=1$, $J=1$ channels, i.e., the
$^3S_1$--$^3D_1$ partial waves, which exceed the other contributions by
roughly a factor of 50. This dominance arises primarily from the
$^3D_1$/$^3S_1$ interference, indicating that the main effect is
driven by the tensor force in the NN interaction, a feature also
reported by Fabrocini in Ref.~\cite{Fab97}. Fig.\ref{cor7} indicates that
correlations tend to generate a significant enhancement of the
transverse response. Since the dominant contribution arises from the
$^3S_1$--$^3D_1$ channel, i.e. the deuteron channel, this enhancement
is produced mainly by the interaction with correlated $np$ pairs in the
nuclear medium.

\begin{figure}
  \centering
  \includegraphics[width=10cm,bb=60 430 490 830]{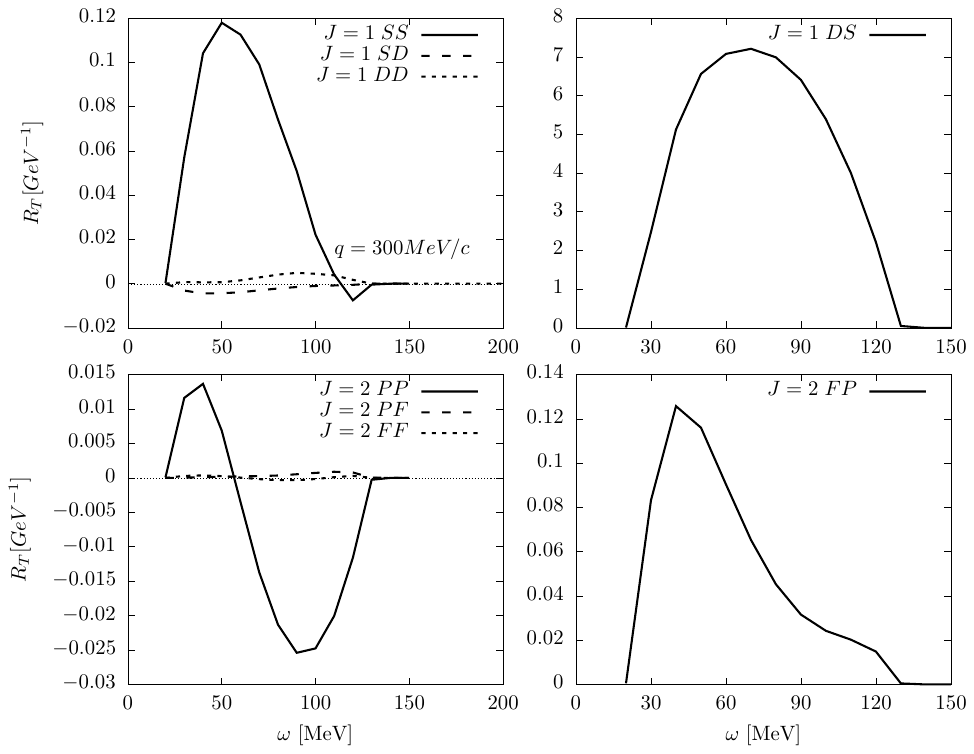}
  \caption{Correlated 1b2b interference response $(R_T^{1b2b})_{cor}$ for
    $q=300$ MeV/c, separated in partial waves, for each coupled angular
    momentum channel. The upper panels correspond to the $J=1$
    channel, and the lower panels to $J= 2$. The largest
    contributions appear in the right-hand panels, particularly in the
    DS component of the $J=1$ channel.}
    \label{cor4}
\end{figure}

\begin{figure}
  \centering
  \includegraphics[width=10cm,bb=60 430 490 830]{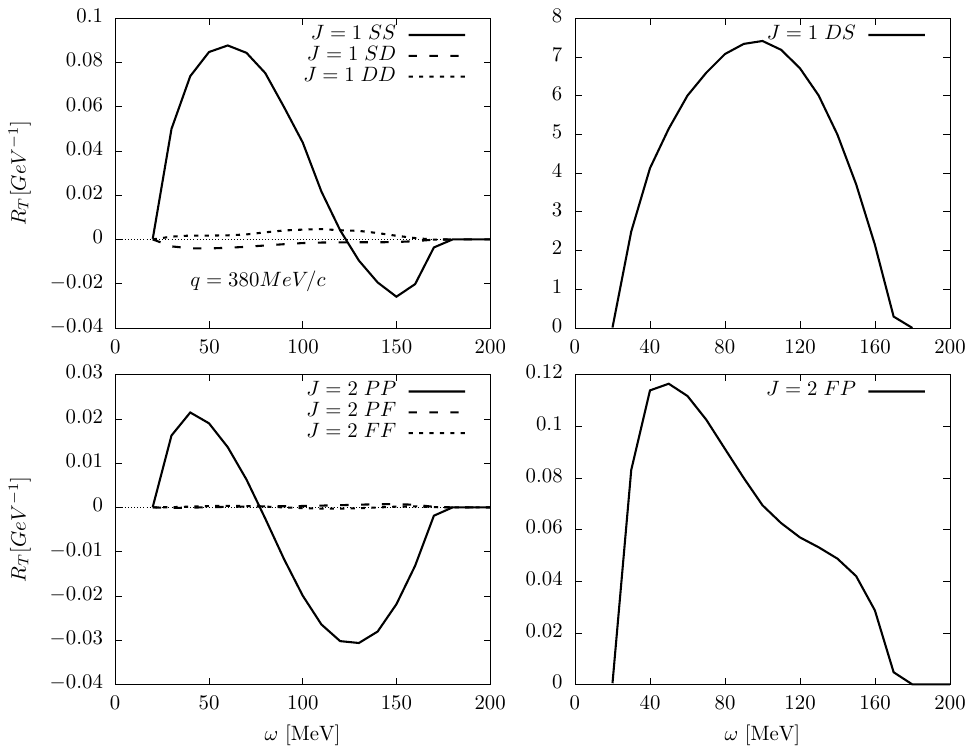}
  \caption{The same as Figure \ref{cor4} but for $q=380$ MeV/c}
  \label{cor5}
\end{figure}

\begin{figure}
  \centering
  \includegraphics[width=10cm,bb=60 430 490 830]{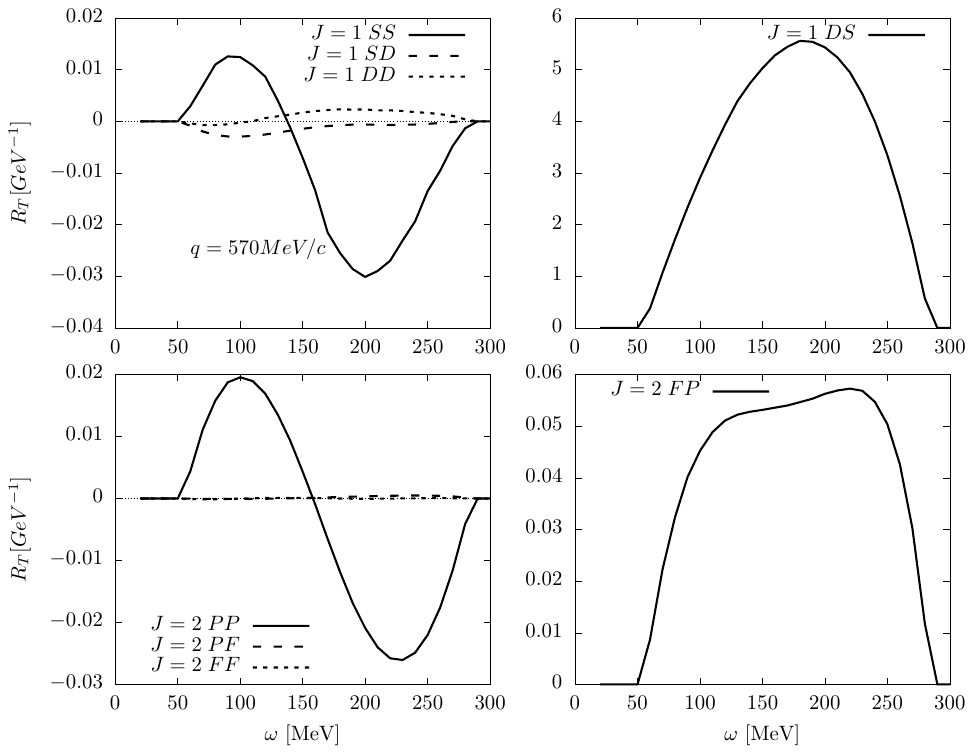}
  \caption{The same as Figure \ref{cor4} but for $q=570$ MeV/c}
  \label{cor6}
\end{figure}

In Figures \ref{cor4}, \ref{cor5} and \ref{cor6}, we present the correlated
interference response $(R_T^{1b2b})_{cor}$, separated again in its partial
wave contributions for each coupled channel. The notation used
throughout the figures is explained in Table \ref{tabwaves}.

Figure \ref{cor4} shows the results for $q=300$ MeV/c. The upper
panels display the contributions to the response from the coupled
$J=1$ channel, while the lower panels correspond to $J=2$. As we
already discussed and can be seen in the top right panel, the dominant
contribution clearly comes from the $DS$ component. This behavior is
consistent with what was observed in chapter 6, particularly in Figure
(6.14), where it is evident that the largest contribution corresponds
precisely to the partial wave mixing $^3D_1/^3S_1$ for $J=1$. Figures
\ref{cor5} and \ref{cor6} show the same partial wave decomposition of
the correlated interference response, evaluated for momentum transfers
$q=380$ and $q=570$ MeV/c, respectively. In both cases, the overall
behavior remains consistent with the results obtained at $q=300$
MeV/c.

\section{Final Remarks}

In this chapter we have investigated the effect of short-range
correlations on the 1p1h MEC contribution to the transverse response
of nuclei. Using the independent pair approximation, we have
incorporated SRC by replacing the uncorrelated pair wave functions with
the correlated waves obtained from the solution of the Bethe–Goldstone
equation of chapter 6 with a realistic NN interaction, namely the
Granada 2013 potential. This approach allows the two-nucleon wave
function to acquire high-momentum components above the Fermi sea,
which are absent in the uncorrelated Fermi gas.

The results show that SRC generate an enhancement of the transverse
1b2b interference response in the quasielastic peak region. The
multipole decomposition of the correlated wave function reveals that
the dominant contribution arises from the coupled $S=1$, $J=1$
channel, i.e., the $^3S_1$–$^3D_1$ (deuteron-like) components. This
indicates that the enhancement is primarily driven by the tensor
component of the NN interaction, in agreement with previous findings
from CBF calculations. Comparison with the CBF results shows a
qualitative agreement despite the very different treatments of SRC:
while the CBF approach applies a variational correlation operator to
uncorrelated Slater determinants, our method solves the BG equation.
We have combined our correlated 1b2b response with a SuSA
description of the quasielastic 1b response in a hybrid model to
compare with experimental transverse response data of $^{12}$C. We
find that the inclusion of SRC produces a net enhancement of the
transverse response that brings the calculation into the correct order
of magnitude to describe the data at low and intermediate momentum
transfer. While at higher $q$ additional contributions from 2p2h
excitations and pion emission are missing, the present calculation
highlights the essential role of SRC in generating the transverse
enhancement.

\chapter{Conclusions}

The present thesis has addressed the QE scattering of electrons and
neutrinos from nuclei, with particular emphasis on the role of
meson-exchange currents and short-range correlations in one-particle
emission channel. The work has combined theoretical developments,
model extensions, and detailed numerical analyses, aiming at a more
comprehensive description of nuclear responses in this regime.

Chapter 2 is devoted to the study of the superscaling framework,
revisiting the definition of single-nucleon responses within the
$SuSAM^*$ model for electron scattering. It was shown that the
conventional extrapolation procedure of the averaged on-shell single
nucleon responses leads to inconsistencies beyond the scaling region,
occasionally producing unphysical negative responses. To overcome this
limitation, a new definition is proposed, based on an average using
smeared momentum distribution around the Fermi surface. This approach
avoids the problem of the extrapolation while remaining consistent
with data in the scaling region.

The thesis has also explored in detail the interference between
one-body and two-body currents in the quasielastic regime. A method
that enables the consistent inclusion of meson-exchange currents
within the framework of the superscaling analysis with relativistic
effective mass is presented in chapter 3. The approach is rigorously
relativistic, drawing its foundation from the relativistic mean field
theory of nuclear matter. The generalized scaling analysis is
comprehensively illustrated and applied to the $(e,e')$ cross-section
of $^{12}C$. To achieve this, we apply the new definition of the
single-nucleon tensor studied in chapter 2. Calculations performed
within several independent particle models like RFG, RMF and SuSAM*
confirmed the general tendency of MEC to reduce the transverse
response, in line with previous findings \cite{Ama94},\cite{Ama10}.

In chapter 4, we have rigorously demonstrated, within the framework of
the non-relativistic Fermi gas, that the interference between one-body
(1b) and two-body (2b) currents—arising from $\Delta$-excitation and pionic
mechanisms—produces a negative contribution to the transverse
response. This finding has been critically evaluated through
comparisons with a wide range of independent single-particle
descriptions, including both relativistic and non-relativistic
mean-field models, as well as spectral function approaches, all of
which consistently support the conclusion of a negative interference
effect.

The formalism was then extended to charged-current quasielastic
neutrino and antineutrino scattering in chapter 5. The interference
between one-body and two-body operators was calculated for all nuclear
response functions (CC, CL, LL, T, and T'), showing that the
contribution of MEC generates strong negative interference effects in
the transverse responses and also in the neutrino and antineutrino
cross sections for 1p1h excitations in the relativistic Fermi gas. Our
results show that the 1b2b interference contributions are not
negligible and may be comparable in size to other nuclear effects
usually included in more sophisticated models. For instance, effects
such as long-range correlations modeled via the Random Phase
Approximation ~\cite{Nie04,Mar09}, final-state interactions
(FSI) in energy-dependent RMF
approaches~\cite{GonzalezJimenez:2017gcy}, finite-size shell effects
in Hartree–Fock based models ~\cite{Jachowicz:2002rr}, or GiBUU
transport theory~\cite{Mosel:2016cwa}, all modify the shape and size
of the nuclear response. The presence of a non-negligible 1b2b
interference is comparable to the 2p2h effect and implies that the
models may need to be revisited or extended to estimate at least the
order of magnitude of this contribution and its possible interplay
with the aforementioned mechanisms.

It is worth mentioning the existence of some calculations that
disagree with this result and suggest a different effect of MEC on the
transverse response. In particular, there are two notable model
calculations: the Green Function Monte Carlo (GFMC) model from
Reference \cite{Lov16} and the RFM and spectral function models from
\cite{Fra23,Lov23,Fra25}.  In those approaches, the effect of MEC is
positive in the quasielastic peak and quite significant, around 20\%,
in the transverse response. For GFMC model, this substantial effect is
attributed to the simultaneous effect of correlations in the
wave function and MEC. Conversely, the positive result of
Refs. \cite{Fra23,Fra25} is due to a $\Delta$ current with a sign
opposite to the standard operator, as was explicitly demonstrated in
the Appendix G of Ref. \cite{Cas25}. A similar feature can be
identified in Ref. \cite{Lov23} , where the propagator of the $\Delta$
resonance is taken with opposite sign. On the other hand,
Ref. \cite{Lov16a} employs a fully correlated model with the standard
operator, in which case the enhancement plausibly originates from the
role of correlations, consistent with the findings of Fabrocini from
Ref. \cite{Fab97}.

The persistence of negative response functions across different
theoretical frameworks strongly suggests that the mechanism
responsible for the experimentally observed transverse enhancement
\cite{Bod22} cannot be attributed to mean-field dynamics alone. This
observation naturally points to SRC as a
key ingredient that must be incorporated into the nuclear wave
function to capture the correct underlying physics. Consequently, the
study of SRC emerges as the natural next step, forming the focus of
chapter 6.

In chapter 6, the focus shifted toward the microscopic treatment
of SRC via the Bethe–Goldstone equation. The high-momentum components
of the nuclear wave function of two correlated nucleons were obtained
using the realistic 2013 Granada potential, which is a coarse grained
interaction represented as a sum of delta functions in each partial
wave. It has been observed that this coarse-grained potential reduces
the problem to an algebraic linear system of five (ten) equations for
uncoupled (coupled) partial waves that can be easily solved. The
analysis confirmed the dominance of the $^3S_1$ and $^3D_1/^3S_1$
channels, which implies that the pn pairs dominate the high-momentum
tail of the relative momentum distribution.

Finally, in chapter 7, these correlated wave functions were
incorporated into the calculation of MEC contributions within the
quasielastic transverse response. The results demonstrate that the
high momentum components generate an additional enhancement that
cannot be accounted for within uncorrelated frameworks, reinforcing
the necessity of their inclusion in any realistic description of
electron and neutrino scattering. In general, this study demonstrates
that the SRC mechanism incorporated via the BG equation and
independent pair approximation provides a physically transparent,
parameter-free, and essentially \emph{ab initio} way to account for
the observed enhancement in the 1p1h MEC transverse response. The
dominance of the $^3S_1$–$^3D_1$ channel underscores the importance of
the tensor force in the nuclear medium, and the results presented
suggest that SRC should be included in any realistic modeling of
quasielastic transverse responses. Future investigations stemming from
this work include extending the formalism to neutrino scattering,
since the transverse enhancement will appear in the vector response;
it is also important to determine whether the responses associated
with the axial current exhibit a similar enhancement due to SRC and
MEC, which would have direct implications for neutrino interaction
models.

\begin{appendix}
\inappendixtrue
  
  \chapter{Single-nucleon responses}
  \label{appA}
  
  The single-nucleon hadronic tensor 
is computed performing the spin traces 
(\ref{traza}) with the current matrix elements (\ref{corriente}),
and can be written as
\begin{equation}
w^{\mu\nu}= -w_1\left(g^{\mu\nu}-\frac{Q^\mu Q^\nu}{Q^2}\right)+
w_2V^\mu V^\nu,
\label{wmunu}
\end{equation}
where we have defined the four-vector $V^\mu= (H^\mu+Q^\mu/2)/m_N^*$,
and $H^\mu=(E,\nh)$ is the initial nucleon four-momentum with
effective mass $m_N^*$. The four-momentum of the final nucleon is
$P^\mu=H^\mu+Q^\mu$.
The nucleon structure functions are given by
\begin{eqnarray}
w_1(Q^2) &=& \tau (G_M^*)^2 > 0,  \label{w1}\\
w_2(Q^2) &=& \frac{(G_E^*)^2+\tau (G_M^*)^2}{1+\tau} > 0, \label{w2}
\end{eqnarray}
where the electric and magnetic form factors for nucleons with effective mass 
are
\begin{equation}
G_E^*  =  F_1-\tau \frac{m^*_N}{m_N} F_2, \kern 1cm
G_M^*  = F_1+\frac{m_N^*}{m_N} F_2.  \label{GM}
\end{equation}
For the $F_i$ form factors of the nucleon, we use the Galster
parametrizations \cite{Gal71}.  

Note that $w_1$ and $w_2$ are positive and depend only on $Q^2$.
Here we compute the longitudinal and transverse 
components of the hadronic tensor,
$w_L= w^{00}$ and $w_T=w^{11}+w^{22}$ respectively, 
 appearing in inclusive electron scattering.

\paragraph{\bf Longitudinal single-nucleon response}
We use the following results for the time components of the basic tensors
and vectors in terms of dimensionless variables, $\kappa,\lambda,\tau$
\begin{equation}
g^{00}-\frac{Q^0 Q^0}{Q^2}= -\frac{q^2}{Q^2}= \frac{\kappa^2}{\tau}, 
\end{equation}
\begin{equation}
V^0= \frac{E+\omega/2}{m_N^*}=\epsilon+\lambda.
\end{equation}
Substituting the values of these time components and of the structure
functions in the hadronic tensor (\ref{wmunu}), the longitudinal single-nucleon
response function becomes
\begin{equation}
w_L= 
-\kappa^2(G_M^*)^2
+\frac{(G_E^*)^2+\tau (G_M^*)^2}{1+\tau}(\epsilon+\lambda)^2.
\end{equation}
Rearranging terms containing $G_E^*$ and $G_M^*$ this becomes
\begin{equation} \label{wl}
w_L= 
\frac{(G_M^*)^2}{1+\tau}
[\tau(\epsilon+\lambda)^2-(1+\tau)\kappa^2]
+\frac{(G_E^*)^2}{1+\tau}
(\epsilon+\lambda)^2.
\end{equation}

\paragraph{\bf Transverse single-nucleon response}

In the case of the transverse response  
$g^{ii}=-1$ and $V^i= h_i/m_N^* = \eta_i$, for $i=1,2$, 
where we have defined the 
three-vector $\neta=\nh/m_N^*$. Then
 the T response is  
\begin{equation}\label{wt}
w_T= w^{11}+w^{22}= 2w_1+w_2(\eta_1^2+\eta_2^2).
\end{equation}
Note that
$\eta_1^2+\eta_2^2=\eta^2-\eta_3^2= \epsilon^2-1-\eta_3^2$. 
The value of $\eta_3^2$ is the projection
of the vector $\neta$  over the $\nq$ direction,  
which is determined by energy-momentum conservation. In fact,
using Eq (\ref{angulo}) 
\begin{equation}
\eta_3= \frac{h\cos\theta}{m_N^*}= \frac{E\omega+Q^2}{m_N^* q}
= \frac{\epsilon\lambda-\tau}{\kappa}.
\end{equation}
Then we have
\begin{equation}
\eta_1^2+\eta_2^2= \epsilon^2-1-
\left(\frac{\epsilon\lambda-\tau}{\kappa}\right)^2.
\end{equation}
Expanding the square and  using $\kappa^2-\lambda^2=\tau$, this gives
gives
\begin{equation}
\eta_1^2+\eta_2^2= \frac{\tau}{\kappa^2}
\left[\epsilon^2-\frac{\kappa^2}{\tau}-\tau+2\epsilon\lambda\right]=
\frac{\tau}{\kappa^2}
\left[(\epsilon+\lambda)^2-\kappa^2\frac{1+\tau}{\tau}\right].
\end{equation}
Inserting this result in Eq. (\ref{wt}) and using the values of 
$w_i$ from Eqs. (\ref{w1},\ref{w2}), the transverse response becomes
\begin{equation} \label{wtfinal}
w_T= 
2\tau (G_M^*)^2
+\frac{(G_E^*)^2+\tau (G_M^*)^2}{1+\tau}
\frac{\tau}{\kappa^2}
\left[(\epsilon+\lambda)^2-\kappa^2\frac{1+\tau}{\tau}\right].
\end{equation}

\chapter{Isospin Summations in the 1p1h MEC Matrix Element}
\label{appB}
Here we provide the sums over the isospin index \(t_k\) of the
spectator nucleon appearing in the 1p1h MEC matrix element. 
The isospin dependence of the MEC is of the form
\begin{equation}
\nj= \tau^{(1)}_z \nj_1+ \tau^{(2)}_z \nj_2 + 
i[ \ntau^{(1)} \times \ntau^{(2)}]_z \nj_3,
\end{equation}
where $\ntau^{(1)}$ and $\ntau^{(2)}$ are isospin operators of the
first and second particle, respectively.  We begin by referencing the
Pauli matrices, which also represent the isospin operators required.
\begin{equation}
\tau_1 = \begin{pmatrix}
0 & 1 \\
1 & 0
\end{pmatrix}, \quad
\tau_2 = \begin{pmatrix}
0 & -i \\
i & 0
\end{pmatrix}, \quad
\tau_3 = \begin{pmatrix}
1 & 0 \\
0 & -1
\end{pmatrix}.
\end{equation}

These matrices act on the isospin states of
nucleons, $|t\rangle$,  for protons (\(t =
+\frac{1}{2}\)) and neutrons (\(t = -\frac{1}{2}\)).
We need the basic result
\begin{eqnarray}
\tau_1|p\rangle = |n\rangle,
&&
\tau_1|n\rangle = |p\rangle 
\nonumber \\
i\tau_2|p\rangle = -|n\rangle
&&
i\tau_2|n\rangle = |p\rangle.
\nonumber
\end{eqnarray}
By expanding the vector product
\begin{equation}
i[\ntau^{(1)} \times \ntau^{(2)}]_z=
i\tau_1^{(1)}\tau_2^{(2)}
-i\tau_2^{(1)}\tau_1^{(2)},
\end{equation}
we obtain
\begin{eqnarray}
i[\ntau^{(1)} \times \ntau^{(2)}]_z  |pp\rangle &=& 0,\nonumber\\
i[\ntau^{(1)} \times \ntau^{(2)}]_z  |nn\rangle &=& 0,\nonumber\\
i[\ntau^{(1)} \times \ntau^{(2)}]_z  |pn\rangle &=& 
2|np\rangle = 4t_p |np\rangle,
\nonumber\\
i[\ntau^{(1)} \times \ntau^{(2)}]_z  |np\rangle &=&
-2|pn\rangle = 4t_n |pn\rangle.
\nonumber
\end{eqnarray}
These four equations can be written in unified form as
\begin{equation}
i[\ntau^{(1)} \times \ntau^{(2)}]_z|t_1t_2\rangle=
4t_1 (1-\delta_{t_1t_2})|t_2t_1\rangle.
\end{equation}
From these elementary results, we can compute the 
sums over $t_k$ appearing in the direct and exchange matrix elements of the current. 

\subsubsection{Direct terms.}

For the the direct terms we have:
\begin{equation} \label{iso1}
\sum_{t_k=\pm 1/2} 
\langle t_pt_k|  \tau^{(1)}_z  | t_ht_k\rangle
=
\sum_{t_k} 
     \delta_{t_pt_h}2t_h 
=     \delta_{t_pt_h}4t_h, 
\end{equation}
\begin{equation} \label{iso2}
\sum_{t_k} 
\langle t_pt_k| \tau^{(2)}_z  | t_ht_k\rangle
=
\sum_{t_k} 
 \delta_{t_pt_h}2t_k
= 0,
\end{equation}
\begin{eqnarray}
\sum_{t_k} 
\langle t_pt_k| i[\ntau^{(1)} \times \ntau^{(2)}]_z  | t_ht_k\rangle =
\sum_{t_k} 
\delta_{t_pt_k}\delta_{t_kt_h} 4t_k (1-\delta_{t_ht_k}) = 0.
\label{iso3}
\end{eqnarray}

\subsubsection{Exchange terms.}

For the the exchange matrix elements we have:
\begin{equation} \label{iso4}
\sum_{t_k} 
\langle t_pt_k|  \tau^{(1)}_z  | t_kt_h\rangle
=
\sum_{t_k} 
     \delta_{t_pt_k}
     \delta_{t_kt_h}2t_k 
=     \delta_{t_pt_h}2t_h, 
\end{equation}
\begin{equation} \label{iso5}
\sum_{t_k} 
\langle t_pt_k| \tau^{(2)}_z  | t_kt_h\rangle
=
\sum_{t_k} 
     \delta_{t_pt_k}
     \delta_{t_kt_h}2t_h 
=     \delta_{t_pt_h}2t_h, 
\end{equation}
\begin{eqnarray}
\sum_{t_k} 
\langle t_pt_k| i[\ntau^{(1)} \times \ntau^{(2)}]_z  | t_kt_h\rangle 
&=& \sum_{t_k} 
\langle t_pt_k| 4t_k (1-\delta_{t_kt_h})  | t_ht_k\rangle 
\nonumber\\
&=& \sum_{t_k} 
\delta_{t_pt_h} 4t_k (1-\delta_{t_kt_h}) = 
   - \delta_{t_pt_h}4t_h. 
 \label{iso6}
\end{eqnarray}

\subsubsection{Null $\Delta$ diagrams.}

Next, we will demonstrate that diagrams (f) and (g) corresponding to
the $\Delta$ current are zero after summing over isospin. To achieve
this, we must use the original form of the isospin operators,
Eqs. (\ref{ufc1} and \ref{ubc1}). The forward current involves the operators
\(U_F(1,2)\) and \(U_F(2,1)\), while the backward current contains the
isospin operators \(U_B(1,2)\) and \(U_B(2,1)\). By carefully
analyzing these operators, we can show that the specific contributions
from diagrams (f) and (g) cancel out, leading to a net zero result for
each.  First, from property (\ref{titj}) we can write the
following products
\begin{eqnarray}
T_1T_3^\dagger = \frac{i}{3}\tau_2
&& 
T_2T_3^\dagger = -\frac{i}{3}\tau_1
\\
T_3T_1^\dagger = -\frac{i}{3}\tau_2
&&T_3T_2^\dagger = \frac{i}{3}\tau_1
\\
T_3T_3^\dagger = \frac{2}{3}.
\end{eqnarray}
From here, using $\tau_1\tau_2=-\tau_2\tau_1=i\tau_3$, we have
\begin{eqnarray}
\sum_i  \tau_i T_iT_3^{\dagger} &=& 
\tau_1T_1T_3^\dagger
+\tau_2T_2T_3^\dagger
+\tau_3T_3T_3^\dagger
=
 \frac{i}{3}\tau_1\tau_2
- \frac{i}{3}\tau_2\tau_1
+ \frac{2}{3}\tau_3 \nonumber \\
&=&
-\frac13\tau_3-\frac13\tau_3+\frac23\tau_3=0. \nonumber \\
\end{eqnarray}
In the case of the forward current, the isospin sum of the exchange 
matrix element of the $(1\leftrightarrow 2)$ term is
\begin{eqnarray}
\sum_{t_k}
\langle pk| U_F(2,1) |kh\rangle 
&=&
\sum_{t_k}
\sqrt{\frac32}
\langle pk| \sum_i T_i^{(2)}T_3^{(2)\dagger}\tau_i^{(1)} |kh\rangle
\nonumber\\
&=&\sqrt{\frac32}
\sum_{t_k}\sum_i 
\langle p|\tau_i|k\rangle
\langle k| 
T_iT_3^{\dagger} |h\rangle =
\nonumber\\
&=&
\sqrt{\frac32}\sum_i 
\langle p|\tau_i
T_iT_3^{\dagger} |h\rangle = 0.
\end{eqnarray}
This demonstrates the result for the forward term, that diagram (f) of
Fig. \ref{feynman2} is zero.
 Analogously, the same steps can be applied to show
the result that diagram (g) for the backward term is zero. 
Fist we have
\begin{eqnarray}
\sum_i T_3T_i^\dagger \tau_i =T_3T_1^\dagger\tau_1
+T_3T_2^\dagger\tau_2
T_3T_3^\dagger\tau_3
=
 -\frac{i}{3}\tau_2\tau_1
+ \frac{i}{3}\tau_1\tau_2
+ \frac{2}{3}\tau_3
=
-\frac13\tau_3-\frac13\tau_3+\frac23\tau_3=0. \nonumber \\
\end{eqnarray}
The diagram (g) contain the isospin operator $U_B(1,2)$ and
the isospin sum of the exchange matrix element is
\begin{eqnarray}
\sum_{t_k}
\langle pk| U_B(1,2) |kh\rangle 
&=&
\sum_{t_k}
\sqrt{\frac32}
\langle pk| \sum_i T_3^{(1)}T_i^{(1)\dagger}\tau_i^{(2)} |kh\rangle
\nonumber\\
&=&\sqrt{\frac32}
\sum_{t_k}\sum_i 
\langle p| T_3T_i^{\dagger} |k\rangle
\langle k|\tau_i|h\rangle
\nonumber\\
&=&
\sqrt{\frac32}\sum_i 
\langle p|
T_3T_i^{\dagger} \tau_i|h\rangle = 0.
\end{eqnarray}

\chapter{Non relativistic reduction of the vector $\Delta$ current}
\label{appC}

Here we perform the non relativistic reduction of the four-vectors
$A^\mu$, and $B^\mu$, Eqs. (\ref{amu},\ref{bmu}), appearing in the
$\Delta$-current. Using the definition of the $\gamma N \Delta$
vertex, Eq. (\ref{gammabetamu}) we have
\begin{eqnarray}
A^\mu
&=&
\bar{u}(1')k_{2}^{\alpha}G_{\alpha\beta}(p_{1}+Q)
\frac{C_3^V}{m_N}
(g^{\beta\mu}\Qbar-Q^{\beta}\gamma^{\mu})\gamma_5
u(1) 
\nonumber\\
\label{amubis}\\
B^\mu
&=&
\bar{u}(1')k_{2}^{\beta}
\frac{C_3^V}{m_N}\gamma_5
(g^{\alpha\mu}\Qbar-Q^{\alpha}\gamma^{\mu})
G_{\alpha\beta}(p'_{1}-Q)u(1). 
\nonumber\\
\label{bmubis}
\end{eqnarray}
In the last equation
we have permuted the \(\gamma_5\) matrix, which introduces a minus sign that cancels with the negative sign from \(-Q\).
We only need to perform the non-relativistic reduction of the spatial
components (\(\mu = i\)) of the \(\Delta\) current, since we are
computing the transverse response. In the non relativistic limit we 
neglect the time components of $k^\mu$ and $Q^\mu$, i. e.
\begin{equation}
k_2^\mu \simeq (0,\nk_2), \kern 1cm Q^\mu\simeq (0,\nq).
\end{equation} 
Then for the $A^i$ components we have
\begin{eqnarray}
A^i
&\simeq& 
\bar{u}(1')k_{2}^{k}G_{kj}
\frac{C_3^V}{m_N}
(g^{j i}\Qbar-Q^{j}\gamma^{i})\gamma_5
u(1), 
\nonumber\\
&=&
\bar{u}(1')k_{2}^{k}G_{kj}(p_1+Q)\Gamma^{ji}(Q)u(1).
\label{amayusculai}
\end{eqnarray}
Hence at leading order in the non-relativistic limit, only the spatial
components of the \(\Delta\) propagator \( G_{kj} \) and the vertex \(
\Gamma_{ji} \) contribute, while the time components are suppressed.
Analogously, we obtain a similar result for the backward vector
components $B^i$,
\begin{eqnarray}
B^i
&\simeq& 
\bar{u}(1')k_{2}^{k}
\frac{C_3^V}{m_N}
\gamma_5(g^{j i}\Qbar-Q^{j}\gamma^{i})
G_{jk}
u(1), 
\nonumber\\
&=&
\bar{u}(1')  k_{2}^{k}\Gamma^{ji}(-Q) G_{jk}(p'_1-Q) u(1).
\label{bmayusculai}
\end{eqnarray}
Furthermore, the procedure we follow to compute the non-relativistic
reduction of a product of matrix operators is to perform the reduction
on each operator separately. This approach is valid at leading order.

\subsubsection{$\Delta$ propagator}

We begin with the $\Delta$ propagator. In the static limit, with
$p^\mu+Q^\mu \simeq (p^0,0) = (m_N,0)$, and neglecting the lower
components we have
\begin{equation}
\frac{\pbar+m_\Delta}{p^2-m_\Delta^2}
\rightarrow
\frac{p_0+m_\Delta}{p_0^2-m_\Delta^2}
=
\frac{1}{m_N-m_\Delta}
\end{equation}
Then the $\Delta$ propagator is written as
\begin{eqnarray}
G_{ij} 
&\simeq& -\frac{1}{m_N-m_\Delta}(g_{ij}-\frac13\gamma_i\gamma_j)
\nonumber\\
&\simeq & -\frac{1}{m_N-m_\Delta}(-\delta_{ij}+\frac13\sigma_i\sigma_j)
\nonumber\\
&=& \frac{1}{m_N-m_\Delta}(\frac23\delta_{ij}-i\frac13\epsilon_{ijk}\sigma_k)
\label{delta-prop}
\end{eqnarray}
where we have used the property
\begin{equation}
\sigma_i\sigma_j= i\epsilon_{ijk}\sigma_k
\end{equation}
and $\epsilon_{ijk}$ is the Levi-Civita tensor.

\subsubsection{$\gamma N \Delta$ vertex}

To obtain the non-relativistic reduction of the vertex
\begin{equation}
\Gamma^{ji}(Q)=
\frac{C_3^V}{m_N}
(g^{j i}\Qbar-Q^{j}\gamma^{i})\gamma_5,
\end{equation}
in the low energy limit, we have $Q^\mu\simeq (0,q^i)$. Then 
\begin{eqnarray}
(g^{j i}\Qbar-Q^{j}\gamma^{i})\gamma_5
&\simeq&
\delta_{j i}q^k\gamma^k\gamma_5-q^j\gamma^i\gamma_5
\nonumber\\
&\simeq&
\delta_{j i}q^k\sigma_k-q^j\sigma_i
\nonumber\\
&=&
q^k\sigma_l(\delta_{ij}\delta_{kl}-\delta_{il}\delta_{kj})
\end{eqnarray}
This expression can be rewritten using the contraction of two
Levi-Civita tensors
\begin{equation}   \label{contraccion}
\epsilon_{ikm}\epsilon_{jlm}=
\delta_{ij}\delta_{kl}-\delta_{il}\delta_{kj}.
\end{equation}
Therefore we have the non relativistic reduction
\begin{equation}
\Gamma^{ji}(Q) \simeq
\frac{C_3^V}{m_N}
(\epsilon_{ikm}q^k)(\epsilon_{jlm}\sigma_l)
\end{equation}

\subsubsection{Forward vector $A^i$}

 From Eq. (\ref{amayusculai}) we have
(we do not write the spinors, just the spin operators):
\begin{eqnarray}
A^i 
&\simeq&
k_2^kG_{kj} 
\frac{C_3^V}{m_N}
(\epsilon_{inm}q^n)(\epsilon_{jlm}\sigma_l)
\nonumber\\
&=&
\frac{C_3^V}{m_N}
\epsilon_{inm}q^n a_m,
\end{eqnarray}
where we have defined the vector
\begin{equation}
a_m \equiv  \epsilon_{jlm} k_2^kG_{kj} \sigma_l.
\end{equation}
(note that $G_{kj}$ and $\sigma_l$ do not commute).
Therefore we can write, in vector form
\begin{equation} \label{Avector}
\nA \simeq \frac{C_3^V}{m_N}(\nq\times\na).
\end{equation}
Hence $\nA$ is purely transverse.

\subsubsection{Backward vector $B^i$}

Similarly,  from Eq. (\ref{bmayusculai}),
\begin{eqnarray}
B^i 
&\simeq&
-k_2^k \frac{C_3^V}{m_N} (\epsilon_{inm}q^n)(\epsilon_{jlm}\sigma_l) G_{jk} 
\nonumber\\
&=&
\frac{C_3^V}{m_N} \epsilon_{inm}q^n b_m,
\end{eqnarray}
where we have defined the vector
\begin{equation}
b_m \equiv
-\epsilon_{jlm}k_2^k  \sigma_l G_{jk}. 
\end{equation}
In vector form we have
\begin{equation} \label{Bvector}
\nB \simeq \frac{C_3^V}{m_N}
(\nq\times\nb).
\end{equation}

\subsubsection{Vectors $a^i$ and $b^i$}

Next, we perform the necessary contractions to derive the explicit
expressions for the vectors \(\na\) and \(\nb\) in the non relativistic limit. 
Using Eq. (\ref{delta-prop}) for the static $\Delta$-propagator,
we have
\begin{eqnarray}
a_m &\simeq& 
\frac{\epsilon_{jlm} k_2^k}{m_N-m_\Delta}
\left(\frac23\delta_{kj}-i\frac13\epsilon_{kjn}\sigma_n\right)\sigma_l
\\
b_m &\simeq& 
\frac{-\epsilon_{jlm} k_2^k}{m_N-m_\Delta}
 \sigma_l \left(\frac23\delta_{kj}-i\frac13\epsilon_{jkn}\sigma_n\right).
\end{eqnarray}
Hence
\begin{eqnarray}
(m_N-m_\Delta) a_m 
&\simeq& 
\frac23  \epsilon_{jlm} k_2^j\sigma_l
\nonumber\\
&&
-\frac{i}3 k_2^k \epsilon_{kjn}\epsilon_{jlm} \sigma_n \sigma_l
\\
(m_N-m_\Delta) b_m 
&\simeq& 
-\frac23  \epsilon_{jlm} k_2^j\sigma_l
\nonumber\\
&&
+\frac{i}3 k_2^k \epsilon_{jkn}\epsilon_{jlm} \sigma_l \sigma_n
\end{eqnarray}
To compute the contractions in the second term, we employ 
again the property (\ref{contraccion}) of the Levi-Civita tensor
\begin{equation}
\epsilon_{jnk}\epsilon_{jlm}=\delta_{nl }\delta_{km }-\delta_{nm }\delta_{kl }
\end{equation}
Then we have
\begin{eqnarray}
k_2^k \epsilon_{kjn}\epsilon_{jlm} \sigma_n \sigma_l
&=&
k_2^k (\delta_{nl }\delta_{km }-\delta_{nm }\delta_{kl })\sigma_n\sigma_l
\nonumber\\
&=& k_2^m\sigma_n\sigma_n-k_2^l\sigma_m\sigma_l
\nonumber\\
&=& 3k_2^m -k_2^l(\delta_{ml}+i\epsilon_{mln}\sigma_n)
\nonumber\\
&=& 2k_2^m -i\epsilon_{mln}k_2^l\sigma_n
\nonumber\\
&=& (2\nk_2 -i\nk_2\times\nsigma)^m,
\end{eqnarray}
and
\begin{eqnarray}
k_2^k \epsilon_{jkn}\epsilon_{jlm} \sigma_l \sigma_n
&=&
k_2^k (\delta_{lk }\delta_{mn }-\delta_{ln }\delta_{mk })\sigma_l\sigma_n
\nonumber\\
&=& k_2^l\sigma_l\sigma_m-k_2^m\sigma_l\sigma_l
\nonumber\\
&=& -3k_2^m + k_2^l(\delta_{lm}+i\epsilon_{lmn}\sigma_n)
\nonumber\\
&=& -2k_2^m -i\epsilon_{mln}k_2^l\sigma_n
\nonumber\\
&=& (-2\nk_2 -i\nk_2\times\nsigma)^m.
\end{eqnarray}
Then we can write in vector form  $\na$ and $\nb$ as
\begin{eqnarray}
(m_N-m_\Delta) \na 
&\simeq& 
\frac23 (\nk_2\times\nsigma) 
-\frac{i}3 (2\nk_2 -i\nk_2\times\nsigma)
\nonumber \\
&=& 
-\frac{2}3 i\nk_2+
\frac13 \nk_2\times\nsigma 
\\
(m_N-m_\Delta) \nb 
&\simeq& 
-\frac23 (\nk_2\times\nsigma) 
+\frac{i}3 (-2\nk_2 -i\nk_2\times\nsigma)
\nonumber\\
&=& 
-\frac{2}3 i\nk_2-\frac13 \nk_2\times\nsigma .
\end{eqnarray}
Using this result in Eqs. (\ref{Avector},\ref{Bvector}) finally we find
\begin{eqnarray} 
\nA &\simeq& \frac{C_3^V}{m_N}
\frac{1}{m_N-m_\Delta}\nq\times
\left[-\frac{2}3 i\nk_2+ \frac13 \nk_2\times\nsigma \right] 
\\
\nB &\simeq& \frac{C_3^V}{m_N}
\frac{1}{m_N-m_\Delta}\nq\times
\left[-\frac{2}3 i\nk_2- \frac13 \nk_2\times\nsigma \right] 
\end{eqnarray}
from where Eqs. (\ref{Anonrel}) and (\ref{Bnonrel}) follow.

\chapter{Spin summations in the 1p1h MEC matrix elements}
\label{appD}

Here we perform the spin summations appearing in the exchange matrix
element, given by
\begin{eqnarray}
\sum_{t_ks_k}\nj_{2b}(p,k,k,h)
&=& \delta_{t_pt_h}2t_h \sum_{s_k}
[\nj_1(p,k,k,h)
+\nj_2(p,k,k,h)-2\nj_3(p,k,k,h)]
\end{eqnarray}

\subsubsection{Seagull current}

In the case of the seagull current only the current $\nj_3$ contribute, given by Eq. (\ref{seagull}). The sum over $t_k,s_k$ is
\begin{eqnarray}
\sum_{t_ks_k}\nj_{s}(p,k,k,h)
&=&
-4t_h\delta_{t_pt_h} 
\sum_{s_k}
\langle s_ps_k|
\frac{f^2}{m_\pi^2}F_1^V
\left(
\frac{\nk_1\cdot\nsigma^{(1)}}{\nk_1^2+m_{\pi}^2}\nsigma^{(2)}
-\frac{\nk_2\cdot\nsigma^{(2)}}{\nk_2^2+m_{\pi}^2}\nsigma^{(1)}
\right)
|s_ks_h\rangle
\nonumber\\
&=&
-4t_h\delta_{t_pt_h} 
\frac{f^2}{m_\pi^2}F_1^V
\sum_{s_k}
\left(
\frac{\nk_1\cdot\nsigma_{pk}}{\nk_1^2+m_{\pi}^2}\nsigma_{kh}
-\frac{\nk_2\cdot\nsigma_{kh}}{\nk_2^2+m_{\pi}^2}\nsigma_{pk}
\right).
\label{seagull2}
\end{eqnarray}
with $\nk_1=\np-\nk$ and $\nk_2=\nk-\nh$.
The separate spin sums are
\begin{equation}
\sum_{s_k}(\nk_1\cdot\nsigma_{pk})\nsigma_{kh}
= (\nk_1+i\nsigma\times\nk_1)_{ph},
\kern 2cm
\sum_{s_k}(\nk_2\cdot\nsigma_{kh})\nsigma_{pk}
= (\nk_2+i\nk_2\times\nsigma)_{ph}.
\end{equation}
We obtain
\begin{eqnarray}
\sum_{t_ks_k}\nj_{s}(p,k,k,h)
&=&
-4t_h\delta_{t_pt_h} 
\frac{f^2}{m_\pi^2}F_1^V
\langle s_p|
\left(
\frac{\nk_1+i\nsigma\times\nk_1}{\nk_1^2+m_{\pi}^2}
-\frac{\nk_2+i\nk_2\times\nsigma}{\nk_2^2+m_{\pi}^2}
\right)
|s_h\rangle.
\end{eqnarray}

\subsubsection{Pionic}

In the case of the pion in flight or pionic current the sum over
spin-isospin reads
\begin{eqnarray}
\sum_{t_ks_k}\nj_{\pi}(p,k,k,h)
&=&
-4t_h\delta_{t_pt_h} 
\sum_{s_k}
\langle s_ps_k|
\frac{f^2}{m_\pi^2}F_1^V
\frac{\nk_1\cdot\nsigma^{(1)}}{\nk_1^2+m_{\pi}^2}
\frac{\nk_2\cdot\nsigma^{(2)}}{\nk_2^2+m_{\pi}^2}(\nk_1-\nk_2)
|s_ks_h\rangle 
\nonumber\\
&=&
-4t_h\delta_{t_pt_h} 
\frac{f^2}{m_\pi^2}F_1^V
\sum_{s_k}
\frac{\nk_1\cdot\nsigma_{pk}}{\nk_1^2+m_{\pi}^2}
\frac{\nk_2\cdot\nsigma_{kh}}{\nk_2^2+m_{\pi}^2}(\nk_1-\nk_2)
\label{pionic2}
\end{eqnarray}
with $k_1=\np-\nk$ and $\nk_2=\nk-\nh$.
The sum over spin index $s_k$ is performed using
\begin{equation}
\sum_{s_k} (\nk_1\cdot\nsigma_{ph})(\nk_2\cdot\nsigma_{kh})=
\nk_1\cdot\nk_2\delta_{s_ps_h}+i(\nk_1\times\nk_2)\cdot\nsigma_{ph}
\end{equation}

\subsubsection{$\Delta$ current}

From the non-relativistic Eq. (\ref{deltafinal}) we can identify the
three contributions, $\nj_i$ to the $\Delta$ current. 
\begin{eqnarray}
\sum_{t_ks_k}\nj_{\Delta}(p,k,k,h)
&=&
-i\delta_{t_pt_h}2t_h 
C_\Delta\nq\times 
\sum_{s_k}
\langle s_ps_k|
\frac{\nk_1\cdot\nsigma^{(1)}}{\nk_1^2+m_{\pi}^2}4\nk_1
+\frac{\nk_2\cdot\nsigma^{(2)}}{\nk_2^2+m_{\pi}^2}4\nk_2
\nonumber\\
&&
+2i 
\frac{\nk_2\cdot\nsigma^{(2)}}{\nk_2^2+m_{\pi}^2}
(\nk_2\times\nsigma^{(1)})
-2i
\frac{\nk_1\cdot\nsigma^{(1)}}{\nk_1^2+m_{\pi}^2}
(\nk_1\times\nsigma^{(2)})
|s_ks_h\rangle.
\end{eqnarray}
with $\nk_1=\np-\nk$ and $\nk_2=\nk-\nh$.
Writing explicitly the spin indices in the Pauli matrices we have
\begin{eqnarray}
\sum_{t_ks_k}\nj_{\Delta}(p,k,k,h)
&=&
-i\delta_{t_pt_h}2t_h 
C_\Delta\nq\times 
\sum_{s_k}
\left\{
\frac{\nk_1\cdot\nsigma_{pk}}{\nk_1^2+m_{\pi}^2}4\nk_1\delta_{s_ks_h}
+\frac{\nk_2\cdot\nsigma_{kh}}{\nk_2^2+m_{\pi}^2}4\nk_2\delta_{s_ps_k}
\right.
\nonumber\\
&& +2i 
\left.
\frac{\nk_2\cdot\nsigma_{kh}}{\nk_2^2+m_{\pi}^2}
(\nk_2\times\nsigma_{pk})
-2i
\frac{\nk_1\cdot\nsigma_{pk}}{\nk_1^2+m_{\pi}^2}
(\nk_1\times\nsigma_{kh})
\right\}
\end{eqnarray}
with
$\nk_1=\np-\nk$, and $\nk_2=\nk-\nh$.
We need the following spin sums 
\begin{eqnarray}
\sum_{s_k}
(\nsigma_{kh}\cdot\nk_2)(\nk_2\times\nsigma_{pk})
&=&
-i\left[ 
k_2^2\nsigma_{ph}-
 (\nsigma_{ph}\cdot\nk_2)\nk_2
\right]
\nonumber\\
\sum_{s_k}
(\nsigma_{pk}\cdot\nk_1)(\nk_1\times\nsigma_{kh})
&=&
i\left[ 
k_1^2\nsigma_{ph}-
 (\nsigma_{ph}\cdot\nk_1)\nk_1
\right].
\nonumber\\
\end{eqnarray}
The result for the sum over $s_k$ is
\begin{eqnarray}
\sum_{t_ks_k}\nj_{\Delta}(p,k,k,h)
=-4it_h\delta_{t_pt_h} C_\Delta\nq\times 
\left\{
\frac{k_1^2\nsigma_{ph}+(\nsigma_{ph}\cdot\nk_1)\nk_1}
     {k_1^2+m_\pi^2}
+\frac{\nk_2^2\nsigma_{ph}+(\nsigma_{ph}\cdot\nk_2)\nk_2}
     {k_2^2+m_\pi^2}
\right\} \nonumber \\
\label{delta_sk}
\end{eqnarray}

\chapter{Spin summations in the interference responses}
\label{appE}

Here we compute the spin summations in the 1b-2b interference response
function.

In this appendix we use the notation $\nk_1=\np-\nk$  and  $\nk_2=\nk-\nh$. 
We use also the identities $\nk_1+\nk_2=\nq$, and $\nk_1=\nq-\nk_2$.

\subsubsection{Magnetization-seagull}

Inserting Eq. (\ref{seagullph}) into Eq. (\ref{mssum}) we have

\begin{eqnarray}
w^T_{ms}
&=&
4t_h\delta_{t_pt_h} 
\frac{G_M^h}{2m_N}
\frac{f^2}{m_\pi^2}F_1^V
\int \frac{d^3k}{(2\pi)^3}\nonumber \\
&&
\sum_{s_ps_h}
i(\nq\times\nsigma_{hp})
\cdot 
\left(
\frac{\delta_{s_ps_h}\nk_1+i\nsigma_{ph}\times\nk_1}{\nk_1^2+m_{\pi}^2}
-\frac{\delta_{s_ps_h}\nk_2+i\nk_2\times\nsigma_{ph}}{\nk_2^2+m_{\pi}^2}
\right).
\end{eqnarray}
The sums involved inside the integral are of the kind:
\begin{eqnarray}
\sum_{s_ps_h}
i(\nq\times\nsigma_{hp})\cdot
\delta_{s_ps_h}\nk
&=& 
\sum_{s_h} i(\nq\times\nsigma_{hh})\cdot\nk =0
\\
\sum_{s_ps_h}
i(\nq\times\nsigma_{hp})\cdot
(i\nsigma_{ph}\times\nk)
&=& 
-\sum_{s_ps_h}
(\nq\cdot\nsigma_{ph} \nsigma_{hp}\cdot \nk
-\nq\cdot\nk \nsigma_{hp}\cdot\nsigma_{ph}) \nonumber \\
&=&
-\mbox{Tr}(q^i\sigma_i\sigma_j k^j-\nq\cdot\nk \sigma_i\sigma_i)
\nonumber\\
&=&
-\mbox{Tr}(q^i\delta_{ij} k^j-3\nq\cdot\nk)=4\nq\cdot\nk
\label{qssk}
\end{eqnarray}
Therefore
\begin{eqnarray}
w^T_{ms}
&=&
4t_h\delta_{t_pt_h} 
\frac{G_M^h}{2m_N}
\frac{f^2}{m_\pi^2}F_1^V
\int \frac{d^3k}{(2\pi)^3}
\left(
\frac{4\nq\cdot\nk_1}{\nk_1^2+m_{\pi}^2}
+\frac{4\nq\cdot\nk_2}{\nk_2^2+m_{\pi}^2}
\right).
\end{eqnarray}

\subsubsection{Convection-seagull}

Inserting Eq. (\ref{seagullph}) into Eq. (\ref{cssum}) we have

\begin{eqnarray}
w^T_{cs} 
& = & 
4t_h\delta_{t_pt_h} 
\frac{G_E^h}{m_N} 
\frac{f^2}{m_\pi^2}F_1^V
\int \frac{d^3k}{(2\pi)^3}
\sum_{s_ps_h}
\delta_{s_hs_p} \nh_T \cdot 
\left(
\frac{\delta_{s_ps_h}\nk_1+i\nsigma_{ph}\times\nk_1}{\nk_1^2+m_{\pi}^2}
-\frac{\delta_{s_ps_h}\nk_2+i\nk_2\times\nsigma_{ph}}{\nk_2^2+m_{\pi}^2}
\right).
\nonumber\\
& = & 
4t_h\delta_{t_pt_h} 
\frac{G_E^h}{m_N} 
\frac{f^2}{m_\pi^2}F_1^V
\int \frac{d^3k}{(2\pi)^3}
\left(
\frac{\nh_T \cdot \nk_1}{\nk_1^2+m_{\pi}^2}
-\frac{\nh_T \cdot \nk_2}{\nk_2^2+m_{\pi}^2}
\right).
\end{eqnarray}

\subsubsection{Magnetization-pionic}

Inserting Eq. (\ref{pionicph}) into Eq. (\ref{mpisum}) we have

\begin{eqnarray}
w^T_{m\pi}&  = & 
4t_h\delta_{t_pt_h} 
\frac{G_M^h}{2m_N}
\frac{f^2}{m_\pi^2}F_1^V
\int \frac{d^3k}{(2\pi)^3}
\sum_{s_ps_h}
i(\nq\times\nsigma_{hp})
\cdot (\nk_1-\nk_2)
\frac{\delta_{s_ps_h}\nk_1\cdot\nk_2
+i(\nk_1\times\nk_2)\cdot\nsigma_{ph}}
{(\nk_1^2+m_{\pi}^2)(\nk_1^2+m_{\pi}^2)}. \nonumber \\
\end{eqnarray}
The sum over spin inside the integral is
\begin{eqnarray}
\sum_{s_ps_h} i(\nq\times\nsigma_{hp})\cdot (\nk_1-\nk_2)
\left[ \delta_{s_ps_h}\nk_1\cdot\nk_2 +i(\nk_1\times\nk_2)\cdot\nsigma_{ph} \right]=
\nonumber \\
&&
\kern -8.6cm
-\sum_{s_ps_h}(\nq\times\nsigma_{hp})\cdot (\nk_1-\nk_2)
(\nk_1\times\nk_2)\cdot\nsigma_{ph} =
\nonumber\\
&&
\kern -8.6cm
-\sum_{s_ps_h}
[(\nk_1-\nk_2)\times\nq]\cdot\nsigma_{hp}(\nk_1\times\nk_2)\cdot\nsigma_{ph} =
\nonumber \\
&&
\kern -8.6cm
-2
[(\nk_1-\nk_2)\times\nq]\cdot
(\nk_1\times\nk_2)=-4(\nq\times\nk_2)^2
 \label{trazampi}
\end{eqnarray}
where we have used that 
\begin{eqnarray}
\sum_{s_ps_h}(\na\cdot\nsigma_{hp})
(\nb\cdot\nsigma_{ph})=
{\mbox Tr}(a^i\sigma_i\sigma_jb^j)=2\na\cdot\nb
\label{asigbsig}
\\
\nk_1=\nq-\nk_2
\\
(\nk_1-\nk_2)\times\nq=
(\nq-2\nk_2)\times\nq=2\nq\times\nk_2
\\
\nk_1\times\nk_2=
(\nq-\nk_2)\times\nk_2=
\nq\times\nk_2
\end{eqnarray}
With the result of Eq (\ref{trazampi}), we  obtain
\begin{eqnarray}
w^T_{m\pi}&  = & 
4t_h\delta_{t_pt_h} 
\frac{G_M^h}{2m_N}
\frac{f^2}{m_\pi^2}F_1^V
\int \frac{d^3k}{(2\pi)^3}
\frac{-4(\nq\times\nk_2)^2}
{(\nk_1^2+m_{\pi}^2)(\nk_1^2+m_{\pi}^2)}.
\end{eqnarray}

\subsubsection{Convection-pionic}

Inserting Eq. (\ref{pionicph}) into Eq. (\ref{cpisum}) we have

\begin{eqnarray}
w^T_{c\pi} & = & 
4t_h\delta_{t_pt_h} 
\frac{G_E^h}{m_N} 
\frac{f^2}{m_\pi^2}F_1^V
\int \frac{d^3k}{(2\pi)^3}
\sum_{s_ps_h}
\delta_{s_hs_p} \nh_T \cdot (\nk_1-\nk_2)
\frac{\delta_{s_ps_h}\nk_1\cdot\nk_2
+i(\nk_1\times\nk_2)\cdot\nsigma_{ph}}
{(\nk_1^2+m_{\pi}^2)(\nk_1^2+m_{\pi}^2)}. \nonumber \\
\end{eqnarray}
Sum over spin inside the integral:
\begin{eqnarray}
\sum_{s_ps_h}
\delta_{s_hs_p} \nh_T \cdot (\nk_1-\nk_2)
[\delta_{s_ps_h}\nk_1\cdot\nk_2
+i(\nk_1\times\nk_2)\cdot\nsigma_{ph}]
=
\sum_{s_h}
\nh_T \cdot (\nk_1-\nk_2)(\nk_1\cdot\nk_2)
\nonumber\\
=
2 \nh_T \cdot (\nq-2\nk_2)[(\nq-\nk_2)\cdot\nk_2]
=
-4 (\nh_T \cdot\nk_2)(\nq\cdot\nk_2-\nk_2^2).
\end{eqnarray}
Then we obtain
\begin{eqnarray}
w^T_{c\pi} & = & 
4t_h\delta_{t_pt_h} 
\frac{G_E^h}{m_N} 
\frac{f^2}{m_\pi^2}F_1^V
\int \frac{d^3k}{(2\pi)^3}
\frac{
-4 (\nh_T \cdot\nk_2)(\nq\cdot\nk_2-\nk_2^2)
}
{(\nk_1^2+m_{\pi}^2)(\nk_1^2+m_{\pi}^2)}.
\end{eqnarray}

\subsubsection{Magnetization-$\Delta$}

Inserting Eq. (\ref{deltaph}) into Eq. (\ref{mdsum}) we have
\begin{eqnarray}
w^T_{m\Delta}  &=&  
4t_h\delta_{t_pt_h}
\frac{G_M^h}{2m_N}
C_\Delta
\int \frac{d^3k}{(2\pi)^3} \nonumber \\
&&
\sum_{s_ps_h}
i(\nq\times\nsigma_{hp})
\cdot 
\left[
i\nq\times 
\left(
\frac{\nk_1^2\nsigma_{ph}+(\nsigma_{ph}\cdot\nk_1)\nk_1}
     {\nk_1^2+m_\pi^2}
+\frac{\nk_2^2\nsigma_{ph}+(\nsigma_{ph}\cdot\nk_2)\nk_2}
     {\nk_2^2+m_\pi^2}
\right)
\right] \nonumber \\
\end{eqnarray}
We need the following spin sums. The first one is similar to Eq. (\ref{qssk})
\begin{eqnarray}
\sum_{s_ps_h}
(\nq\times\nsigma_{hp})
\cdot 
(\nq\times\nsigma_{ph}) 
=4q^2
\end{eqnarray}
The second sum required is
\begin{eqnarray}
\sum_{s_ps_h}(\nq\times\nsigma_{hp})
\cdot (\nq\times \nk_1)
(\nsigma_{ph}\cdot\nk_1)
&=&
\sum_{s_ps_h}
[q^2 \nsigma_{hp}\cdot \nk_1-(\nq\cdot\nk_1)(\nq\cdot\nsigma_{hp})]
(\nsigma_{ph}\cdot\nk_1)
\nonumber\\
&&
\kern -2cm
=q^2 \sum_{s_ps_h}(\nsigma_{hp}\cdot \nk_1) (\nsigma_{ph}\cdot\nk_1)
-(\nq\cdot\nk_1)\sum_{s_ps_h}(\nq\cdot\nsigma_{hp})(\nsigma_{ph}\cdot\nk_1)
\nonumber\\ 
&&
\kern -2cm
=2q^2 k_1^2 -2(\nq\cdot\nk_1)^2
\end{eqnarray}
where we have used twice Eq. (\ref{asigbsig}).
Using these results the $m\Delta$ response function is
\begin{equation}
w^T_{m\Delta}  =  
-4t_h\delta_{t_pt_h}
\frac{G_M^h}{2m_N}
C_\Delta
\int \frac{d^3k}{(2\pi)^3}
\left(
\frac{6q^2k_1^2-2(\nq\cdot\nk_1)^2}
     {\nk_1^2+m_\pi^2}
+\frac{6q^2k_2^2-2(\nq\cdot\nk_2)^2}
     {\nk_2^2+m_\pi^2}
\right).
\end{equation}

\chapter{Spectral function and hadronic tensor}
\label{appF}

The spectral function is obtained by assuming plane waves for the
final nucleon. This assumption leads to a factorization approximation
for the response function within the impulse approximation, where the
current is considered to be one-body only. In this framework, the
response function can be factored into a product of the current matrix
element and the spectral function, which describes the distribution of
hole states in the nucleus. 

We assume that the initial nuclear state is a spin-zero nucleus at rest
with energy $E_i=M_A$, and wave function:
\begin{equation}
| i \rangle = | \Phi_0^{(A)} \rangle
\end{equation}
The final state correspond to a plane wave particle and a residual $A-1$ nucleus  
\begin{equation}
| f \rangle = | \Phi_{\alpha}^{(A-1)}, \np,s \rangle= a^\dagger_{\np,s} 
| \Phi_{\alpha}^{(A-1)} \rangle
\end{equation}
The label $\alpha$ denotes the quantum numbers of the daughter nucleus
in an excited state with excitation energy
$\epsilon_\alpha^{(A-1)}$. Then the final energy, neglecting the
recoil energy is,
\begin{equation}
E_f=m_N+ T_p+ M_{A-1}+ \epsilon_\alpha^{(A-1)}
\end{equation}
where $T_p=p^2/2m_N$. Then the diagonal component of the hadronic tensor is,
\begin{eqnarray}
W^{\mu\mu} &=& 
\sum_{\alpha\np s} |\langle \Phi_{\alpha}^{(A-1)}, \np,s|J^{\mu}(\nq) | \Phi_0^{(A)} \rangle|^2
\delta(E_i+\omega-E_f) \nonumber \\
\end{eqnarray}
Assuming that the current is a one-body operator and ignoring the
final nucleon spin for simplicity we have,
\begin{eqnarray}
W^{\mu\mu}
&=& \nonumber \\
&&
\kern -1cm 
\sum_{\alpha\np} |\langle \Phi_{\alpha}^{(A-1)}| a_{\np} \int d^3k
J^{\mu}(\nq+\nk,\nk) a^\dagger_{\nq+\nk}a_{\nk}| \Phi_0^{(A)}
\rangle|^2 \nonumber \\
 && 
\kern -1cm
\delta(M_A + \omega -m_N- T_p- M_{A-1}-
\epsilon_\alpha^{(A-1)})
\end{eqnarray}
Using the commutation properties of the creation and annihilation
operators,
\begin{equation}
a_{\np} a^\dagger_{\nq+\nk}= \delta(\np-\nq-\nk)-a^\dagger_{\nq+\nk}a_{\np},
\end{equation}
and assuming that the final particle momentum is large enough to neglect
high-momentum components in the initial wave function (as the dominant
contribution comes from momenta below the Fermi momentum),
$a_{\np}| \Phi_0^{(A)}\rangle \simeq 0$, then the hadronic tensor is
\begin{eqnarray}
W^{\mu\mu}
&=& \nonumber \\
&&
\kern -1cm 
\sum_{\alpha}\int d^3p |\langle \Phi_{\alpha}^{(A-1)}|
J^{\mu}(\np,\np-\nq) a_{\np-\nq}| \Phi_0^{(A)} \rangle|^2 \nonumber \\
 && 
\kern -1cm
\delta(M_A + \omega -m_N- T_p- M_{A-1}-
\epsilon_\alpha^{(A-1)})
\end{eqnarray}
Introducing the separation energy $S=M_{A-1}+m_N-M_A>0$ and the missing energy $E_m=\omega-T_p$ then
\begin{equation}
W^{\mu\mu}=
\int d^3p |J^{\mu}(\np,\np-\nq) |^2 S(\np-\nq,E_m)\nonumber \\
\end{equation}
where the one-hole spectral function is defined as,
\begin{equation}
S(\nh,E) = \sum_{\alpha,s} |\langle \Phi_{\alpha}^{(A-1)}|
a_{\nh,s}| \Phi_0^{(A)} \rangle|^2 
\delta(E-S-\epsilon_\alpha^{(A-1)}). \nonumber \\
\end{equation}



\chapter{Isospin Summations in CCQE matrix element }
\label{appG}
Here we provide the sums over the isospin index \(t_k\) of the
appearing in the 1p1h MEC matrix element. 
We first note that the MEC can be written as
a linear combination of  $\tau^{(1)}_\pm$ and $\tau^{(2)}_\pm$
and $i[ \ntau^{(1)} \times \ntau^{(2)}]_\pm$
\begin{equation}
j_{2b}^\mu= \tau^{(1)}_\pm j_1^\mu+ \tau^{(2)}_\pm j_2^\mu + 
i[ \ntau^{(1)} \times \ntau^{(2)}]_\pm j_3^\mu,
\end{equation}
where the 2b currents $j_1^\mu$,  $j_2^\mu$, and $j_3^\mu$
are isospin-independent. 
To be specific, we will consider the case of the \( (+) \) component
of the isospin current, corresponding to $N \rightarrow P$ transitions.
\begin{equation}
\tau_+= \tau_1 + i\tau_2 =
\begin{pmatrix}
0 & 2 \\
0 & 0
\end{pmatrix},
\end{equation}
\begin{equation}
i[\ntau^{(1)} \times \ntau^{(2)}]_+=
\tau_+^{(1)}\tau_3^{(2)}
-\tau_3^{(1)}\tau_+^{(2)},
\end{equation}
from where we obtain
\begin{eqnarray}
\tau_+|P\rangle = 0,
&&
\tau_+|N\rangle = 2|P\rangle 
\label{tau}
\end{eqnarray}
\begin{eqnarray}
  i[\ntau^{(1)} \times \ntau^{(2)}]_+  |NP\rangle &=& 2|PP\rangle,
  \nonumber\\
i[\ntau^{(1)} \times \ntau^{(2)}]_+  |NN\rangle &=& -2|PN\rangle+2|NP\rangle,
,\nonumber\\
i[\ntau^{(1)} \times \ntau^{(2)}]_+  |PP\rangle &=&0 
\nonumber\\
i[\ntau^{(1)} \times \ntau^{(2)}]_+  |PN\rangle &=&-2|PP\rangle
\label{tautau}
\end{eqnarray}
From these elementary results, it is straightforward
to compute the isospin sums. The results are the following.

\paragraph{Direct terms.}
\begin{equation} \label{g1}
\sum_{t_k=\pm 1/2} 
\langle P t_k|  \tau^{(1)}_+  | N t_k\rangle
=4,
\end{equation}
\begin{equation} \label{g2}
\sum_{t_k} 
\langle Pt_k| \tau^{(2)}_z  | Pt_k\rangle
=0,
\end{equation}
\begin{eqnarray}
\sum_{t_k} 
\langle Pt_k| i[\ntau^{(1)} \times \ntau^{(2)}]_+  | Nt_k\rangle 
&=&0. 
\label{g33}
\end{eqnarray}

\paragraph{Exchange terms.}
\begin{equation} \label{g4}
\sum_{t_k} 
\langle Pt_k|  \tau^{(1)}_+  | t_kN\rangle
=2,
\end{equation}
\begin{equation} \label{g5}
\sum_{t_k} 
\langle Pt_k| \tau^{(2)}_+  | t_kN\rangle
=2,
\end{equation}
\begin{eqnarray}
\sum_{t_k} 
\langle Pt_k| i[\ntau^{(1)} \times \ntau^{(2)}]_+  | t_kN\rangle 
=-4. 
 \label{g6}
\end{eqnarray}
Similar results are obtained for the $P\rightarrow N$ transition with the
$(-)$ isospin components.

After performing this sums the effective one-body matrix element
$j_{2b}(\mathbf{p},\mathbf{h})$ induced by the two-body current, Eq.
(\ref{effectiveOB}), can be written as sum of direct minus exchange currents
\begin{equation}
j_{2b}^{\mu}(\np,\nh) =
j_{2b}^{\mu}(\np,\nh)_{\rm dir}- 
j_{2b}^{\mu}(\np,\nh)_{ex}, 
\end{equation}
where
\begin{eqnarray}
j_{2b}^{\mu}(\np,\nh)_{\rm dir}
&=&
\frac{1}{V}\sum_\nk\sum_{s_k}
 4j_1^{\mu}(\np,\nk,\nh,\nk)
\\
j_{2b}^{\mu}(\np,\nh)_{\rm ex}
&=&
\frac{1}{V}\sum_\nk\sum_{s_k}
\left[
2j_1^{\mu}(\np,\nk,\nk,\nh)
\right.
\left.
+2j_2^{\mu}(\np,\nk,\nk,\nh)
   -4j_3^{\mu}(\np,\nk,\nk,\nh)
  \right].
\label{exchange}
\end{eqnarray}
\subsection*{Relation between CC and EM
interference responses in the vector sector}

We demonstrate that the interference responses between the one-body
and two-body currents in CC neutrino scattering in the vector sector
are exactly twice the corresponding electromagnetic responses when
both proton and neutron emission are summed. This relation holds in
symmetric nuclear matter and follows from the isovector nature of both
the CC and the electromagnetic two-body currents.

Indeed, the one-body electromagnetic and CC vector
currents written as isospin operators are
\begin{eqnarray}
  j^\mu_{1b,e}&=&s^\mu+f^\mu\tau_z \\
  j^\mu_{1b,V}&=&f^\mu\tau_+
\end{eqnarray}
where $s^\mu$ and $f^\mu$ are the isoscalar and isovector 1b currents,
respectively.  Therefore the electromagnetic current in a 1p1h
excitation between isospin states $|t_h\rangle$ and $|t_p\rangle$ is
\begin{equation}
    \langle t_p|j^\mu_{1b,e}(\np,\nh)|t_h\rangle=\delta_{t_pt_h}
    \left[s^\mu(\np,\nh)+2t_hf^\mu(\np.\nh)\right]
\end{equation}
and the CC vector current in a neutron-to-proton transition is
\begin{equation}
    \langle P|j^\mu_{1b,V}(\np,\nh)|N\rangle=2t_hf^\mu(\np.\nh).
\end{equation}
The two-body MEC currents are written in a similar
way
\begin{eqnarray}
  j^\mu_{2b,e}&=&\tau^{(1)}_zj^\mu_1+\tau^{(2)}_zj^\mu_2+
  i[\ntau^{(1)} \times \ntau^{(2)}]_zj^\mu_3 \\
  j^\mu_{2b,V}&=&\tau^{(1)}_\pm j^\mu_1+\tau^{(2)}_\pm j^\mu_2+
  i[\ntau^{(1)} \times \ntau^{(2)}]_\pm j^\mu_3
\end{eqnarray}
where the 2b vector currents $j^\mu_1$, $j^\mu_2$ and $j^\mu_3$ are
isospin- independent. Since the two-body current $j^\mu_{2b,V}$ has no
axial component, the corresponding effective one-body current contains
only the exchange term, as given in (\ref{exchange}). Thus in a
transition where h is neutron and p is proton, we have
\begin{equation}
    \langle P|j^\mu_{2b,V}(\np,\nh)|N\rangle=-j_{2b}^\mu(\np,\nh)_{ex}
\end{equation}
where $j_{2b}^\mu(\np,\nh)_{ex}$ is given in Eq. (\ref{exchange}) with only the
vector part. On the other hand the electromagnetic current for a transition from isospin $t_h$ to $t_p$ is given by
\begin{equation}
  \langle t_p|j^\mu_{2b,e}(\np,\nh)|t_h\rangle=
  -\delta_{t_pt_h}t_h j_{2b}^\mu(\np,\nh)_{ex}
\end{equation}
Let us compute a generic contribution to the interference tensor
$w_{1b2b}^{\mu\nu}$ of Eq. (\ref{w1b2b}), such as $j^\mu_{1b}j^\mu_{2b}$, with
the one-body and two-body currents. In the case of neutrino
scattering the vector part is
\begin{equation}\label{b9}
  \langle P|j^\mu_{1b,V}(\np,\nh)|N\rangle^*
  \langle P|j^\nu_{2b,V}(\np,\nh)|N\rangle= 2f^{\mu *} (j_{2b}^\nu)_{ex}
\end{equation}
In the em case it is
\begin{equation}
  \langle t_p|j^\mu_{1b,e}(\np,\nh)|t_h\rangle^*
  \langle t_p|j^\nu_{2b,e}(\np,\nh)|t_h\rangle=
  -\delta_{t_pt_h}(s^\mu+2t_hf^\mu)t_h(j_{2b}^\nu)_{ex}
\end{equation}
Performing the sum over isospin in the em case,
\begin{equation}\label{b11}
  \sum_{t_pt_h=\pm 1}
  \langle t_p|j^\mu_{1b,e}(\np,\nh)|t_h\rangle^*
  \langle t_p|j^\nu_{2b,e}(\np,\nh)|t_h\rangle=
  -f^{\mu *}(j_{2b}^\nu)_{ex}
\end{equation}
This factor of two between the CC (\ref{b9}) and EM (\ref{b11})
expressions thus applies to all interference response functions
considered in this work.

\chapter{Non relativistic reduction of the axial $\Delta$ current}
\label{appH}

Here we compute the non relativistic axial $\Delta$ current to leading
order in the $1/m_N$ expansion. The non-relativistic reduction follows
the same procedure as that used for the vector current in Appendix
\ref{appC}. We start by writing the weak forward and backward
relativistic currents (\ref{deltaF}, \ref{deltaB}) in the form
\begin{eqnarray}
  J^\mu_{\Delta F}= U_F A^\mu \\ 
  J^\mu_{\Delta B}= U_B B^\mu 
\end{eqnarray}
where,
\begin{eqnarray}
  A^\mu
  &=&
\bar{u}(p'_{2})\gamma^{5}\slashed{k}_{2}u(p_{2})
\bar{u}_{s'_1}(p'_{1})k_{2}^\alpha G_{\alpha \beta}(p_{1}+Q)
C_{5}^{A}(Q)g^{\beta\mu}
u_{s_1}(p_{1})+
(1 \leftrightarrow 2) 
\end{eqnarray}
\begin{eqnarray}
  B^\mu
  &=&
\bar{u}(p'_{2})\gamma^{5}\slashed{k}_{2}u(p_{2})
\bar{u}_{s'_1}(p'_{1})k_{2}^\beta C_{5}^{A}(Q)g^{\mu\alpha}
G_{\alpha\beta}(p'_{1}-Q)
u_{s_1}(p_{1})+
(1 \leftrightarrow 2) \label{h4}
\end{eqnarray}
In the following, we omit the explicit spinors and reduce the
expressions to leading order. The $\pi NN$ vertex is common in both
$A^\mu$ and $B^\mu$ tensors,
\begin{equation}
  \gamma^{5}\slashed{k}_{2} \rightarrow -\gamma^{5}\gamma^{i}k_{2}^{i}
  \rightarrow \nk_{2} \cdot \nsigma^2
\end{equation}
where we've used the non relativistic reduction of the gamma matrices:
$\gamma^{5}\gamma^{i} \rightarrow -\sigma_i$ and
$\slashed{k}_{2}=\gamma^{0}k_{2}^{0}-\gamma^{i}k_{2}^{i} \rightarrow
-\gamma^{i}k_{2}^{i}$.

\subsubsection{Foward $A^\mu$ tensor}

Using
$k_{2}^\alpha \rightarrow (0,\textbf{k}_{2})$ the $\Delta \pi N$
vertex becomes,
\begin{equation}
  k_{2}^\alpha G_{\alpha \beta}g^{\beta\mu} \rightarrow k_{2}^{k}G_{kl}g^{lj}
\end{equation}
Writing the $\Delta$ propagator in the static limit in the form
\begin{equation}
  G_{ij} \rightarrow \frac{1}{m-m_{\Delta}}(\delta_{ij}+\frac{1}{3}\gamma_i\gamma_j)
\end{equation}
 and inserting the spatial components of the
$\Delta$ propagator, we obtain the $A^i$ components
 \begin{eqnarray}
   \kern -0.3cm
  A^i \sim k_{2}^{k}G_{kl}g^{lj}  &\rightarrow&
  k_{2}^{k}\frac{1}{m-m_{\Delta}}(\delta_{kl}+\frac{1}{3}\gamma_k\gamma_l)
  g^{lj}
  =\frac{-1}{m-m_{\Delta}}
  k_{2}^{k}(\frac{2}{3}\delta_{kl}-\frac{i}{3}\epsilon_{klm}\sigma_m)\delta^{lj} \nonumber \\
  &=& \frac{-1}{m-m_{\Delta}}
  \big(\frac{2}{3}k_{2}^{j}+\frac{i}{3}\epsilon_{jkm}k_{2}^{k}\sigma_m \big)
\end{eqnarray}
where we have used $g_{ij}=-\delta_{ij}$, the reduction of gamma matrices 
\begin{equation}
\gamma_k\gamma_l  =  
\begin{bmatrix}
0 & \sigma_k \\
-\sigma_k & 0 
\end{bmatrix}
\begin{bmatrix}
0 & \sigma_l \\
-\sigma_l & 0 
\end{bmatrix}
\rightarrow -\sigma_k\sigma_l
\end{equation}
and
\begin{equation}
  \delta_{kl}+\frac{1}{3}\gamma_k\gamma_l=\delta_{kl}-\frac{1}{3}\sigma_k\sigma_l
  =\frac{2}{3}\delta_{kl}-\frac{i}{3}\epsilon_{klm}\sigma_m.
\end{equation}
The $A_i$ components can be written in vector form
\begin{eqnarray}
  \nA\Rightarrow \frac{-1}{m-m_{\Delta}}\Big(\frac{2}{3}\nk_{2}+\frac{i}{3}\nk_{2}\times \nsigma^1 \Big)
\end{eqnarray}
\subsubsection{Backward $B^\mu$ tensor}
Similarly, from Eq. (\ref{h4})
\begin{eqnarray}
   \kern -0.3cm
   B^i \sim k_{2}^k g^{il}G_{lk}  &\rightarrow&
   k_{2}^{k}g^{il}\frac{1}{m-m_{\Delta}}(\delta_{lk}+\frac{1}{3}\gamma_l\gamma_k)
   =\frac{-1}{m-m_{\Delta}}k_{2}^{k}\delta^{il}
   (\frac{2}{3}\delta_{lk}-\frac{i}{3}\epsilon_{lkm}\sigma_m) \nonumber \\
   &=&
  \frac{-1}{m-m_{\Delta}}( \frac{2}{3}k_{2}^{i}-\frac{i}{3}\epsilon_{ikm}k_{2}^{k}\sigma_m)
\end{eqnarray}
Moreover, in vector notation, $\nB$ can be expressed as:
\begin{equation}
  \nB\Rightarrow \frac{-1}{m-M_{\Delta}}
  \Big( \frac{2}{3}\nk_{2}-\frac{i}{3}\nk_{2}\times\nsigma^1 \Big)
\end{equation}

\subsubsection{Total $\Delta$ Current}
Using Eqs. \ref{UF} and \ref{UB} for the isospin operators $U_F$ and $U_B$
we can write the axial $\Delta$ current in the form
\begin{eqnarray}
  \nj_{\Delta A}=-\frac{2}{9}\sqrt{\frac{3}{2}}\frac{ff^{*}}{m_{\pi}^{2}}C_{5}^{A}\frac{1}{m_{\Delta}-m}
  \Biggl(4\tau_{+}^{(2)}\frac{(\nk_{2} \cdot \nsigma^2)\nk_{2}}{\nk_2^2+m^{2}_\pi}+4\tau_{+}^{(1)}\frac{(\nk_{1} \cdot \nsigma^1)\nk_{1}}{\nk_1^2+m^{2}_\pi}\nonumber \\
  +\left[\ntau^{(1)}\times\ntau^{(2)}\right]_{+}
  \biggl\{\frac{(\nk_{2} \cdot \nsigma^2)(\nk_{2}\times\nsigma^1)}{\nk_2^2+m^{2}_\pi}-\frac{(\nk_{1} \cdot \nsigma^1)(\nk_{1}\times\nsigma^2)}{\nk_1^2+m^{2}_\pi}\biggl\}\Biggl)
\end{eqnarray}

\chapter{Multipole expansion of the BG equation}
\label{appI}

In this appendix we perform the  multipole expansion of
the Bethe-Goldstone equation presented in chapter 6, section 6.2. We
start with the free wave function expansion with total angular
momentum $JM$ and a well-defined spin \( S \) is
\begin{eqnarray}
\frac{1}{(2\pi)^{3/2}}   e^{i\np\cdot\nr} \chi_{S\mu} =
   A\sum_{JM}{\sum_{lm_{l}}{i^{l}Y^{*}_{lm_{l}}(\hat{\np})}
     \langle lm_{l}S\mu|JM \rangle j_{l}(pr) \mathcal{Y}_{lSJM}(\hat{\nr})},
   \label{I1}
 \end{eqnarray}
with $A=\sqrt{2/\pi}$.
The spinor functions $\mathcal{Y}_{lSJM}(\hat{\nr})$ are 
obtained by coupling the angular momentum \( l \) and spin \( S \)
bases through the Clebsch-Gordan coefficients,
\begin{equation}
 \mathcal{Y}_{lSJM}(\hat{\nr})=\sum_{m\mu}\langle l m S \mu|JM \rangle Y_{lm}(\hat{\nr}) |S\mu \rangle.
\end{equation}
The inverse relation is,
\begin{equation}
Y_{lm}(\hat{\nr}) |S\mu \rangle=\sum_{JM}\langle l m S \mu|JM \rangle \mathcal{Y}_{lSJM}(\hat{\nr}).
\end{equation}
The integral Bethe-Goldstone equation with well-defined momentum 
$\np$, coupled to total spin $S$ with projection $\mu$ defined in
Eq. (\ref{bethegfin}) is
\begin{eqnarray}\label{I4}
  |\psi, \np S\mu \rangle = |\np S \mu \rangle + \int d^3\np' \frac{Q(\nP,\np')}{p^{2}-p'^{2}}
  \sum_{\mu'} |\np'S \mu'\rangle \langle \np' S \mu'| m_N V|\psi \rangle 
\end{eqnarray}
We assume that the potential commutes with \( S^2 \), \( J^2 \), and \(
J_z \), but not with the angular momentum \( L^2 \). The action of this
potential over the spinor functions multiplied by a radial function is
\begin{equation}
 Vj_{l}(pr) \mathcal{Y}_{lSJM}(\hat{\nr})=\sum_{l'}V_{l'l}^{SJ}(r)j_{l}(pr) \mathcal{Y}_{l'SJM}(\hat{\nr}).
\end{equation}
If we define $U_{l'l}^{SJ}(r)=m_N V_{l'l}^{SJ}(r)$, this leads us to,
\begin{eqnarray} \label{I6}
 m_N V |\np S \mu \rangle =  A\sum_{JM}{\sum_{ll'm}{i^{l'}Y^{*}_{l'm}(\hat{\np})}
    \langle l'm S\mu|JM \rangle U_{ll'}^{SJ}(r)j_{l'}(pr) \mathcal{Y}_{lSJM}(\hat{\nr})}
\end{eqnarray}
Looking at the multipole expansion of the free wave function,
Eq. (\ref{I1}), and considering how the potential acts on these
spinor-angular functions in Eq.(\ref{I6}), it becomes natural to anticipate that the
solution of the Bethe-Goldstone equation should have a similar
structure. Therefore, we propose the ansatz
\begin{equation}
  | \psi \rangle =
   A\sum_{JM}{\sum_{ll'm}{i^{l'}Y^{*}_{l'm}(\hat{\np})}
     \langle l'm S\mu|JM \rangle \phi_{ll'}^{SJ}(r) \mathcal{Y}_{lSJM}(\hat{\nr})}.
   \label{anz}
\end{equation}
We calculate the matrix element of \( V \) acting on the state \(
\langle \np'S \mu' | \), that appears in Eq.(\ref{I4})
\begin{eqnarray}
  \langle \np' S \mu' |V |\psi \rangle =
  A^2 \Big \langle \sum_{J_1M_1}{\sum_{l_1l'_1m_1}{i^{l'_1}Y^{*}_{l_1'm_1}(\hat{\np}')}
    \langle l_1'm_{1}S\mu'|J_1M_1 \rangle j_{l_1'}(p'r) U_{l_1l_1'}^{SJ}(r) \mathcal{Y}_{l_1SJ_1M_1}(\hat{\nr})} \Big | \nonumber \\
  &&
  \kern -12.5cm
  \times
 \Big | \sum_{JM}{\sum_{ll'm}{i^{l'}Y^{*}_{l'm}(\hat{\np})}
     \langle l'm S\mu|JM \rangle \phi_{ll'}^{SJ}(r) \mathcal{Y}_{lSJM}(\hat{\nr})} \Big \rangle
\end{eqnarray}
Using that $\langle
\mathcal{Y}_{l_1SJ_1M_1}(\hat{\nr})|\mathcal{Y}_{lSJM}(\hat{\nr})\rangle=\delta_{l_1l}\delta_{J_1J}
\delta_{M_1M}$ we reduce the matrix element to
\begin{eqnarray}
  \langle \np' S \mu' |V |\psi \rangle &=&
  A^2  \sum_{JM} \sum_{l_1'm_1}\sum_{ll'm} i^{l'-l'_1}Y_{l_1'm_1}(\hat{\np}')Y^{*}_{l'm}(\hat{\np})
  \nonumber \\
  &&
    \langle l_1'm_{1}S\mu'|JM \rangle \langle l'm S\mu|JM \rangle 
    \int dr \; r^2 j_{l_1'}(p'r) U_{ll_1'}^{SJ}(r) \phi_{ll'}^{SJ}(r).
\end{eqnarray}
We use this result in Eq. (\ref{I4}), written as
\begin{eqnarray}
  \Delta \psi \equiv
  \int d^3\np' \frac{Q(\nP,\np')}{p^{2}-p'^{2}}
  \sum_{\mu'} |\np'S \mu'\rangle \langle \np' S \mu'| m_N V|\Psi \rangle 
  &=& |\psi, \np S\mu \rangle - |\np S\mu \rangle.
\end{eqnarray}
Then
\begin{eqnarray}
  \Delta \psi&=&
  A^3 \int d^3\np' \frac{Q(\nP,\np')}{p^{2}-p'^{2}} \sum_{\mu'} \sum_{J_2M_2} \sum_{l_2m_2}
  i^{l_2}Y^{*}_{l_2m_2}(\hat{\np}')
  \langle l_2m_{2}S\mu'|J_2M_2 \rangle j_{l_2}(p'r)\mathcal{Y}_{l_2SJ_2M_2}(\hat{\nr})
  \nonumber \\
   &&
   \sum_{JM}\sum_{l_1'm_1}\sum_{ll'm} i^{l'-l'_1}Y_{l_1'm_1}(\hat{\np}')Y^{*}_{l'm}(\hat{\np})
   \langle l_1'm_{1}S\mu'|JM \rangle \langle l'm S\mu|JM \rangle \nonumber \\
   &\times&
  \int dr'r'^2 j_{l_1'}(p'r') U_{ll_1'}^{SJ}(r')\phi_{ll'}^{SJ}(r'). \label{I11}
\end{eqnarray}
Using the orthonormality of the  spherical harmonics
\begin{equation}
  \int Y^*_{l_2m_2}(\hat{\np}')Y_{l_1'm_1}(\hat{\np}') d \hat{\np}'=\delta_{l_2l_1'}\delta_{m_2m_1}.
\end{equation}
and the orthonormality relations of the Clebsch-Gordan coefficients,
\begin{equation}
  \sum_{\mu}\sum_{m_1} \langle l_1'm_{1}S\mu'|J_2M_2 \rangle  \langle l_1'm_{1}S\mu'|JM \rangle = \delta_{J_2J}\delta_{M_2M},
\end{equation}
equation (\ref{I11}) becomes,
\begin{eqnarray}
  \Delta \psi &=&  A^3\sum_{JM} \sum_{l_1'} \sum_{ll'm} i^{l'}  Y^{*}_{l'm}(\hat{\np}) \langle l'm S \mu|JM \rangle
  \mathcal{Y}_{l_1'SJM}(\hat{\nr}) \nonumber \\
  &\times&
  \int dr' r'^2 \int dp' p'^{2} \frac{Q(\nP,\np')}{p^{2}-p'^{2}}
  j_{l_1'}(p'r)j_{l_1'}(p'r') U_{ll_1'}^{SJ}(r')\phi_{ll'}^{SJ}(r'), \label{I13}
\end{eqnarray}
Finally  we define the radial Green's fuction:
\begin{equation}
  G_{l_1}(r,r')= A^2 \int dp' p'^2 \frac{Q(\nP,\np')}{p^{2}-p'^{2}} j_{l_1}(p'r)j_{l_1}(p'r'),
\end{equation}
and Eq.(\ref{I13}), replacing \( l \) with \( l_1 \), becomes
\begin{eqnarray} \label{I16}
  \Delta \psi = A \sum_{JM} \sum_{ll'm} i^{l'}  Y^{*}_{l'm}(\hat{\np}) \langle l'm S \mu|JM \rangle
  \mathcal{Y}_{lSJM}(\hat{\nr})
 \sum_{l_1}  \int dr' \; r'^2  G_{l}(r,r')  U_{l_1l}^{SJ}(r')\phi_{l_1l'}^{SJ}(r') .
\end{eqnarray}
Comparing with the ansatz proposed, Eq.(\ref{anz}), and the multipole
expansion of the free wave from which we started, we observe that
$\Delta \psi$ shares the same angular-spin
structure, 
\begin{eqnarray}
\Delta \psi = A \sum_{JM} \sum_{ll'm} i^{l'}  Y^{*}_{l'm}(\hat{\np}) \langle l'm S \mu|JM \rangle \Delta\phi_{ll'}^{SJ}(r) \mathcal{Y}_{lSJM}(\hat{\nr}).
\end{eqnarray}
Therefore, the radial functions must satisfy the integral equation
\begin{eqnarray}
 \Delta\phi_{ll'}^{SJ}(r)= \int dr' \; r'^2  G_{l}(r,r') \sum_{l_1'}  U_{l_1l}^{SJ}(r')\phi_{l_1l'}^{SJ}(r').
\end{eqnarray}
This ensures that the full wave function
\begin{equation}
   |\psi \rangle = |\np S \mu \rangle + |\Delta \psi\rangle,
\end{equation}
satisfies the Bethe-Goldstone equation.

\chapter{Solution of the Green's Function integral}
\label{appJ}

To solve the system of equations derived from the multipole expansion
of the Bethe--Goldstone equation, it is necessary to evaluate the
reduced Green's function, which we rewrite here for reference:

\begin{equation}
\hat{G}_{l}(r,r^\prime)=
\frac{2}{\pi} \int^{\infty}_0 dp^\prime\; \hat{j}_l(p^\prime r)\,
\frac{\overline{Q}(P,p^\prime)}{p^2-p^{\prime\,2}}\,
\hat{j}_l(p^\prime r^\prime).
\end{equation}
This Green’s function is symmetric and depends on spherical Bessel
functions—oscillatory in nature through sine and cosine terms—as well
as on the Pauli blocking function, which restricts the integration
domain. Moreover, the integrand exhibits a pole at \( p = p' \). To
properly handle this singularity, we proceed by dividing the integral
into regions according to the definition of the Pauli function:
\[ \int_0^\infty = \int_0^{q_0}+\int_{q_0}^{q_1}+\int_{q_1}^\infty, \]
\[ q_0=\sqrt{k_F^2-\frac{P^2}{4}} < q_1 =k_F+\frac{P}{2}, \]
Substituting these limits, we obtain
\begin{eqnarray} \label{j2}
  \hat{G}_{l}(r,r^\prime)
  = \frac{2}{\pi} \int_{q_0}^{q_1}  dp^\prime\; \hat{j}_l(p^\prime r)
\frac{\overline{Q}(P,p^\prime)}{p^2-p^{\prime\,2}}\,
\hat{j}_l(p^\prime r^\prime)
+
\frac{2}{\pi}\int_{q_1}^{\infty}  dp^\prime\; 
\frac{\hat{j}_l(p^\prime r)\;\hat{j}_l(p^\prime r^\prime) }{p^2-p^{\prime\,2}}
\end{eqnarray}
Since the Pauli blocking function is defined piece-wise, the integral
naturally separates into three parts. The first term vanishes, as can
be seen from the definition in Eq.~\ref{Qbar_function}. The second
integral corresponds to the region \( q_0 < p' < q_1 \), while the
third extends from \( q_1 \) to \( \infty \). Both of these integrals
contain poles in the denominator. To avoid the singularity at \( p' = p \),
the Cauchy principal value prescription is applied to the second
integral in Eq.~\ref{j2} (the first integral does not include the pole
within its domain).

For convenience, we can transform the Cauchy integral as follows:
\begin{eqnarray}
  \mathcal{I}&=&
  \frac{2}{\pi}\mathcal{P}\int_{q_1}^{\infty}  dp^\prime\; 
  \frac{\hat{j}_l(p^\prime r)\;\hat{j}_l(p^\prime r^\prime) }{p^2-p^{\prime\,2}}
  \nonumber \\
  &=&
  \frac{2}{\pi}\mathcal{P}\int_{0}^{\infty} dp^\prime\; 
  \frac{\hat{j}_l(p^\prime r)\;\hat{j}_l(p^\prime r^\prime) }{p^2-p^{\prime\,2}}
  -
  \frac{2}{\pi}\mathcal{P}\int_{0}^{q_1} dp^\prime\; 
  \frac{\hat{j}_l(p^\prime r)\;\hat{j}_l(p^\prime r^\prime) }{p^2-p^{\prime\,2}}
  \nonumber \\
  &=&
  \frac{1}{p}\hat{j}_{l}(pr_{\textless})\hat{y}_{l}(pr_{\textgreater})
  - \frac{2}{\pi}\mathcal{P}\int_{0}^{q_1} dp^\prime \; 
  \frac{\hat{j}_l(p^\prime r)\;\hat{j}_l(p^\prime r^\prime)}{p^2-p^{\prime\,2}},
\end{eqnarray}
where we define $r_{\textless}=\min(r',r)$,
$r_{\textgreater}=\max(r',r)$ and the function $\hat{y}_l$ is a
reduced Bessel function of the second kind, also known as the Neumann
function. From Eq.~\ref{j2}, we are thus left with an integral that
must be evaluated using the Cauchy principal value theorem, as
explained below.

\section*{Principal value theorem}
The principal value integral arises when evaluating integrals with
singularities, such as:
\begin{equation}
I \equiv {\cal P} \int_a^b \frac{f(x)}{x - x_0} dx,
\end{equation}
where \( a < x_0 < b \), and \( f(x) \) is a continuous function with a well-defined derivative at \( x = x_0 \). The singularity at \( x = x_0 \) prevents direct evaluation of the integral. Instead, it is defined in terms of a limiting process:
\begin{equation}
{\cal P} \int_a^b \frac{f(x)}{x - x_0} dx = \lim_{\epsilon \to 0} \left( \int_a^{x_0 - \epsilon} \frac{f(x)}{x - x_0} dx + \int_{x_0 + \epsilon}^b \frac{f(x)}{x - x_0} dx \right).
\end{equation}
This definition ensures that the integral remains well-defined despite the singularity.

A suitable approach for numerical computation involves rewriting the
integral in the following form by adding and subtracting \( f(x_0) \)
in the numerator:
\begin{equation}
I = {\cal P}\int_a^b \frac{f(x)-f(x_0)+f(x_0)}{x-x_0} dx.
\end{equation}
This can be separated into two integrals:
\begin{equation}
I = \int_a^b \frac{f(x)-f(x_0)}{x-x_0} dx 
+ f(x_0) {\cal P} \int_a^b \frac{1}{x-x_0} dx.
\end{equation}
The first integral no longer has a singularity at \( x_0 \) if \( f(x)
\) is differentiable, making it suitable for numerical evaluation
using Simpson’s rule or another standard quadrature method. The second
integral is the principal value of \( \frac{1}{x - x_0} \), which can
be computed analytically.

We assume that  $0<a<x_0<b$. We have
\begin{equation}
{\cal P} \int_a^b \frac{1}{x-x_0} dx=
\lim_{\epsilon\rightarrow0}\int_a^{x_0-\epsilon}\frac{1}{x-x_0} dx
+\int_{x_0+\epsilon}^{b}\frac{1}{x-x_0} dx
\end{equation}

In the first integral $x-x_0<0$. To compute the first integral we change variable $t=x_0-x >0$. Then $dt=-dx$ and

\begin{equation}
\int_a^{x_0-\epsilon}\frac{1}{x-x_0} dx=
\int_{x_0-a\textless}^{\epsilon}\frac{dt}{t}= \ln\epsilon-\ln(x_0-a)
\end{equation}
The second integral is directly integrable with
\begin{equation}
\int_{x_0+\epsilon}^{b}\frac{1}{x-x_0} dx= \left.\ln(x-x_0)\right|_{x_0+\epsilon}^b
=\ln(b-x_0)-\ln\epsilon
\end{equation}
Then
\begin{equation}
{\cal P} \int_a^b \frac{1}{x-x_0} dx=
\ln\epsilon-\ln(x_0-a)+\ln(b-x_0)-\ln\epsilon
= -\ln(x_0-a)+\ln(b-x_0)
= \ln\frac{b-x_0}{x_0-a}.
\end{equation}
Finally we obtain the general formula for the principal value
\begin{equation}
{\cal P} \int_a^b \frac{f(x)}{x - x_0} dx=
\int_a^b \frac{f(x)-f(x_0)}{x-x_0} dx 
+ f(x_0)\ln\frac{b-x_0}{x_0-a}.
\end{equation}
Once this result is obtained using the Cauchy method, the total result
for the Green’s function integral in Equation \ref{j2} is 
\begin{eqnarray}
  \hat{G}_{l}(r,r^\prime)
  = \frac{2}{\pi} \int_{q_0}^{q_1}  dp^\prime\; \hat{j}_l(p^\prime r)
\frac{\overline{Q}(P,p^\prime)}{p^2-p^{\prime\,2}}\,
\hat{j}_l(p^\prime r^\prime)
+ \frac{1}{p}\hat{j}_{l}(pr_{\textless})\hat{y}_{l}(pr_{\textgreater})\nonumber \\ 
&&
\kern -9cm
  -\frac{2}{\pi}\Big( \int_0^{q_{1}} \frac{f(p')-f(p)}{p-p'}dp^\prime\;- f(p)\ln\Big|\frac{q_1-p}{p}\Big| \Big).
\end{eqnarray}
with

\begin{eqnarray}
  f(p')&=&\frac{\hat{j}_l(p^\prime r)\hat{j}_l(p^\prime r^\prime)}{p+p'} \\
f(p)&=&\frac{\hat{j}_l(pr)\hat{j}_l(pr^\prime)}{2p}.
\end{eqnarray}

\end{appendix}

\end{document}